Eike-Manuel Edelmann

# Spiking Neural Networks for Communication Systems: Encoding Schemes, Learning Algorithms, and Equalization Techniques

**Forschungsberichte aus dem Institut für Nachrichtentechnik des Karlsruher Institut für Technologie**
Herausgeber: Prof. Dr. rer. nat. Friedrich Jondral,
Prof. Dr.-Ing. Laurent Schmalen

Band 1 Marcel Kohl
**Simulationsmodelle für die Bewertung von Satellitenübertragungsstrecken im 20/30 GHz Bereich**

Band 2 Christoph Delfs
**Zeit-Frequenz-Signalanalyse: Lineare und quadratische Verfahren sowie vergleichende Untersuchungen zur Klassifikation von Klaviertönen**

Band 3 Gunnar Wetzker
**Maximum-Likelihood Akquisition von Direct Sequence Spread-Spectrum Signalen**

Band 4 Anne Wiesler
**Parametergesteuertes Software Radio für Mobilfunksysteme**

Band 5 Karl Lütjen
**Systeme und Verfahren für strukturelle Musteranalysen mit Produktionsnetzen**

Band 6 Ralf Machauer
**Multicode-Detektion im UMTS**

Band 7 Gunther M. A. Sessler
**Schnell konvergierender Polynomial Expansion Multiuser Detektor mit niedriger Komplexität**

Band 8 Henrik Schober
**Breitbandige OFDM Funkübertragung bei hohen Teilnehmergeschwindigkeiten**

Band 9 Arnd-Ragnar Rhiemeier
**Modulares Software Defined Radio**

Band 10 Mustafa Mengüç Öner
**Air Interface Identification for Software Radio Systems**

**Forschungsberichte aus dem Institut für Nachrichtentechnik des Karlsruher Institut für Technologie**
Herausgeber: Prof. Dr. rer. nat. Friedrich Jondral,
Prof. Dr.-Ing. Laurent Schmalen

Band 11 Fatih Çapar
**Dynamische Spektrumverwaltung und elektronische Echtzeitvermarktung von Funkspektren in Hotspotnetzen**

Band 12 Ihan Martoyo
**Frequency Domain Equalization in CDMA Detection**

Band 13 Timo Weiß
**OFDM-basiertes Spectrum Pooling**

Band 14 Wojciech Kuropatwiński-Kaiser
**Messung von MIMO-Kapazitäten mit einem auf GSM-Komponenten basierenden Demonstrator**

Band 15 Piotr Rykaczewski
**Kompensierung hochfrequenzbedingter I/Q-Fehler in Software Defined Radio**

Band 16 Michael Eisenacher
**Kompensierung der stochastischen Eigenschaften von Ultra-Wideband-Signalen**

Band 17 Clemens Klöck
**Auction-based Medium Access Control**

Band 18 Martin Henkel
**Architektur eines DRM-Empfängers und Basisbandalgorithmen zur Frequenzakquisition und Kanalschätzung**

Band 19 Stefan Edinger
**Mehrträgerverfahren mit dynamisch-adaptiver Modulation zur unterbrechungsfreien Datenübertragung in Störfällen**

Band 20 Volker Blaschke
**Multiband Cognitive Radio-Systeme**

**Forschungsberichte aus dem Institut für Nachrichtentechnik des Karlsruher Institut für Technologie**

Herausgeber: Prof. Dr. rer. nat. Friedrich Jondral,
Prof. Dr.-Ing. Laurent Schmalen

Band 21 Ulrich Berthold
**Dynamic Spectrum Access using OFDM-based Overlay Systems**

Band 22 Sinja Brandes
**Suppression of Mutual Interference in OFDM-based Overlay Systems**

Band 23 Christian Körner
**Cognitive Radio – Kanalsegmentierung und Schätzung von Periodizitäten**

Band 24 Tobias Renk
**Cooperative Communications: Network Design and Incremental Relaying**

Band 25 Dennis Burgkhardt
**Dynamische Reallokation von spektralen Ressourcen in einem hierarchischen Auktionssystem**

Band 26 Stefan Nagel
**Portable Waveform Development for Software Defined Radios**

Band 27 Hans-Ulrich Dehner
**Interferenzuntersuchungen für inkohärente Multiband Ultra-Breitband (UWB) Übertragung**

Band 28 Maximilian Hauske
**Signalverarbeitung für optoelektronische Sensoren**

Band 29 Jens Elsner
**Interference Mitigation in Frequency Hopping Ad Hoc Networks**

Band 30 Georg Vallant
**Modellbasierte Entzerrung von Analog/Digital-Wandler-Systemen**

**Forschungsberichte aus dem Institut für Nachrichtentechnik**
**des Karlsruher Institut für Technologie**
Herausgeber: Prof. Dr. rer. nat. Friedrich Jondral,
Prof. Dr.-Ing. Laurent Schmalen

Band 31 Martin Braun
**OFDM Radar Algorithms in Mobile Communication Networks**

Band 32 Michael Sebastian Mühlhaus
**Automatische Modulationsartenerkennung in MIMO-Systemen**

Band 33 Michael Schwall
**Turbo-Entzerrung: Implementierungsaspekte für Software Defined Radios**

Band 34 Ralph Tanbourgi
**Diversity Combining under Interference Correlation in Wireless Networks**

Band 35 Florian Engels
**Multidimensional Frequency Estimation with Applications in Automotive Radar**

Band 36 Noha El Gemayel
**Smart Passive Localization Using Time Difference of Arrival**

Band 37 Ankit Kaushik
**Performance Analysis of Cognitive Radio Systems with Imperfect Channel Knowledge**

Band 38 Johannes Fink
**Software Defined Radar für den Nahbereich**

Band 39 Sebastian Koslowski
**Synchronisation und Entzerrung in Filterbank-Multicarrier-Empfängern**

Band 40 Felix Wunsch
**Drahtloses Low Power Wide Area Network bei 2,4 GHz**

Band 41 Eike-Manuel Edelmann
**Spiking Neural Networks for Communication Systems: Encoding Schemes, Learning Algorithms, and Equalization Techniques**

# Vorwort des Herausgebers

Künstliche Intelligenz, insbesondere das maschinelle Lernen mit künstlichen neuronalen Netzen (ANNs), ermöglicht es, moderne Kommunikationssysteme bis an ihre theoretischen Leistungsgrenzen zu führen. Diese hohe Komplexität geht jedoch mit einem steigenden Energieverbrauch einher. Gepulste neuronale Netze (engl. Spiking Neural Networks (SNNs)) stellen eine neue Generation neuronaler Netze dar, die vom energieeffizienten, ereignisgesteuerten Arbeitsprinzip des menschlichen Gehirns inspiriert sind. Durch ihre zeitliche Dynamik und die Verarbeitung von Informationen in Form kurzer binärer Signale („Spikes“) bieten sie großes Potenzial für energieeffiziente, echtzeitfähige Signalverarbeitung. Die bisherigen Anwendungen gepulster neuronaler Netze beschränken sich weitestgehend auf die Bild- und Videoverarbeitung. In der Kommunikationstechnik sind diese Netze bisher nahezu unbekannt.
Die vorliegende Arbeit untersucht SNN-basierte Empfänger und verschiedene Lernregeln und Kodierungsschemata, um den Übergang klassischer Nachrichtentechnik in die Welt gepulster Signale zu ermöglichen. Darüber hinaus werden SNN-basierte Empfänger für optische Übertragungssysteme mit Intensitätsmodulation und Direktdetektion entwickelt. Solche Übertragungssysteme sind insbesondere in modernen Datenzentren, wie sie für die künstliche Intelligenz benötigt werden, unabdingbar. Kombinationen aus verschiedenen Netzwerkarchitekturen und Kodierungen sollen sowohl eine hohe Entzerrungsleistung als auch eine geringe Spike-Rate ermöglichen. Insbesondere werden Verbesserungen sowohl in Bezug auf Komplexität als auch auf Leistungsfähigkeit erreicht. Abschließend wird ein neuartiger Optimierungsansatz vorgestellt, der ohne klassische Backpropagation auskommt. Dieser nutzt Reinforcement Learning und hat das Ziel, die Laufzeit, Komplexität und Spike-Anzahl pro Inferenz signifikant zu reduzieren, ohne die Leistungsfähigkeit zu beeinträchtigen. Insgesamt hat die Arbeit das Ziel, einen Entwurfs- und Optimierungsrahmen für SNN-basierte Empfänger zu entwerfen und damit den Weg für zukünftige energieeffiziente Kommunikationssysteme auf Basis neuartiger SNN-Architekturen zu ebnen. Als weitere Anwendung wird auch noch kurz die Ressourcenzuweisung in Mobilfunknetzen mit Reinforcement Learning verbessert.

Karlsruhe, im Februar 2026
Laurent Schmalen

# Spiking Neural Networks for Communication Systems: Encoding Schemes, Learning Algorithms, and Equalization Techniques

Zur Erlangung des akademischen Grades eines

**DOKTORS DER INGENIEURWISSENSCHAFTEN (Dr.-Ing.)**

von der KIT-Fakultät für Elektrotechnik und Informationstechnik des
Karlsruher Instituts für Technologie (KIT)

angenommene

**DISSERTATION**

von

**Eike-Manuel Edelmann, M.Sc.**
Geburtsname: Bansbach

# Kurzfassung

Künstliche Intelligenz, insbesondere maschinelles Lernen mit künstlichen neuronalen Netzen (ANNs), bietet Lösungsansätze für die wachsende Komplexität moderner Kommunikationssysteme und ermöglicht Datenübertragung nahe der theoretischen Grenze. Die steigende Komplexität geht jedoch mit einem hohen Energieverbrauch und somit mit energieintensiven Systemen einher.

Gepulste neuronale Netze (SNNs) sind vom menschlichen Gehirn inspirierte Modelle, die durch ereignisgesteuerte Mechanismen energieeffiziente Signalverarbeitung in Echtzeit ermöglichen. Sie unterscheiden sich von ANNs durch ihre inhärente zeitliche Dynamik, sowie durch ihre Informationsverarbeitung in Form kurzer binärer Pulse. Offene Herausforderungen sind insbesondere die Wahl geeigneter Lernregeln und neuronaler Kodierungen zur Konvertierung realer Signale in Pulse.

Diese Arbeit untersucht den Entwurf SNN-basierter Empfänger für verrauschte sowie frequenzselektive zeitinvariante Kanäle. Der erste Teil fokussiert sich auf die Entwicklung eines SNN-basierten Detektors für einen verrauschten Kanal. Untersucht werden drei Lernregeln: eine biologisch inspirierte und zwei gradientenbasierte, wobei die gradientenbasierte Lernregel "Backpropagation through time mit Ersatzgradienten" als vielversprechende Lernregel identifiziert wird. Weiterhin werden verschiedene neuronale Kodierungen untersucht, wobei drei vielversprechende Kandidaten identifiziert werden, z.B. Quantization encoding (QE).

Der zweite Teil widmet sich SNN-basierten Entzerrern und Demappern. Zwei Architekturen, mit und ohne Entscheidungsrückkopplung, werden jeweils mit den drei neuronalen Kodierungen kombiniert. Für das Modell einer nicht-kohärenten optischen Übertragung werden die Ansätze bezüglich ihrer Leistungsfähigkeit und der Anzahl generierter Pulse verglichen. Der Einsatz von Entscheidungsrückkopplung und QE ermöglicht leistungsfähige Entzerrer mit einer geringen Anzahl an generierten Pulsen. Bemerkenswerterweise übertreffen SNN-basierte Entzerrer ANN-basierte deutlich hinsichtlich ihrer Leistungsfähigkeit.

Der dritte Teil nutzt Methoden des verstärkenden Lernens (RL) um eine neue Lernregel für SNNs sowie der neuronalen Kodierung herzuleiten. Es wird ein RL-basierter Update-Algorithmus eingeführt, der keine Backpropagation benötigt. Für den SNN-basierten Entzerrer und Demapper werden mittels des neuen Algorithmus die Parameter der neuronalen Kodierung optimiert, wodurch ohne Leistungseinbußen die Laufzeit, Komplexität und Anzahl der pro Inferenz generierter Pulse erheblich reduziert wird.

Diese Arbeit leistet einen Beitrag zum erfolgreichen Entwurf von SNN-basierten Empfängern. Durch die Diskussion zentraler Herausforderungen erleichtert sie zukünftige Fortschritte im Entwurf und Einsatz energieeffizienter Empfänger auf Basis von SNNs.

# Abstract

Artificial intelligence, especially machine learning with artificial neural networks (ANNs), provides solutions for the growing complexity of modern communication systems, enabling communications close to the theoretical limits. This complexity, however, increases power consumption, making the systems energy-intensive.

Spiking neural networks (SNNs) represent a novel generation of neural networks inspired by the highly efficient human brain. By emulating its event-driven and energy-efficient mechanisms, SNNs enable low-power, real-time signal processing. They differ from ANNs in two key ways: they exhibit inherent temporal dynamics and process and transmit information as short binary signals called spikes. Despite their promise, major challenges remain, e.g., identifying optimal learning rules for task-specific training and effective neural encoding —the translation of real-world signals into spikes.

This thesis investigates the design of SNN-based receivers for the additive white Gaussian noise (AWGN) channel and nonlinear time-invariant frequency-selective channels. The first part focuses on designing an SNN-based detector for the AWGN channel and examines three update rules: one biologically plausible update rule and two gradient-based approaches. Backpropagation through time with surrogate gradients is identified as a promising update rule due to its ability to leverage established machine learning methods. We further study different neural encoding schemes that vary fundamentally and identify three promising candidates, e.g., quantization encoding (QE).

The second part addresses SNN-based equalizers and demappers. Two architectures, with and without decision feedback, are combined with the three neural encodings. Given the model of the intensity modulation with direct detection link, we compare the methods based on equalization performance and spike count. Using decision feedback and QE achieves both strong performance and low spike counts. Notably, SNN-based equalizers significantly outperform ANN-based counterparts.

The third part leverages reinforcement learning (RL) methods to derive a new learning rule for SNNs as well as for neural coding. It introduces policy gradient-based update (PGU), an RL-based update algorithm that requires no backpropagation. Using PGU, encoding parameters for the SNN-based equalizer and demapper are optimized, drastically reducing runtime, complexity, and spikes per inference while maintaining performance.

This thesis contributes a successful design and optimization framework for SNN-based receivers. By addressing key challenges in SNN optimization, it facilitates future advances in the design and deployment of energy-efficient SNN receivers.

# Acknowledgements

This thesis marks the culmination of five years as research assistant at the Communications Engineering Lab (CEL) at the Karlsruhe Institute of Technology (KIT). I would like to take the opportunity to thank the people who contributed to the success of this work.

First, I am deeply grateful to my supervisor Prof. Dr.-Ing. Laurent Schmalen for his continuous support of my work. I especially thank him for his numerous ideas, valuable suggestions, thorough reviews, and words of encouragement whenever an approach failed.

Furthermore, I would like to thank Dr.-Ing. Holger Jäkel for always having an open door and a willing ear. Thank you for the countless insightful discussions. An additional sincere thank you goes to my colleague Alexander von Bank, without whom this work would probably not have come to fruition. True to the German saying, “Zu zweit ist man weniger alleine”, it was a great pleasure to conduct research with you on a shared topic!

Scientific progress requires more than just scientists. My deepest thanks go to Kimberley Bender and Anna-Katharina Reiser, who supported me in all administrative tasks. I would also like to thank all my colleagues for creating an enjoyable and supportive working environment. My sincere thanks go to all colleagues I had the opportunity to exchange ideas with, in particular Luca Schmid, Benedikt Geiger, Daniel Gil, Jonathan Mandelbaum, and Marcus Müller. I also owe special thanks to Vincent Lauinger, Johannes Voigt, and Lars Gölz. Their proof-reading of parts of the manuscript resulted in many valuable improvements.

I would like to express my appreciation to the students who made significant contributions to my work. In particular, I would like to mention Jan Sternagel, Max Müller, Jonathan Ebert, Victor Eliachevitch, and Yigit Kiyak.

A person is the sum of all the impressions they have been fortunate to experience in a lifetime. At this point, I would like to extend my heartfelt thanks to my family, especially my parents, and friends. Your upbringing and all the time we have shared form the foundation of this achievement.

The person who probably suffered the most during this thesis is my daughter, who always lent me a listening ear during our long walks. Please forgive me for the long monologues about topics that are (for now) still boring to you. Dear Doro, thank you for your patience, support, encouraging words, and for always being there when I needed it most. Together, we made it through a tough year. Looking back, I could not be prouder of us.

It saddens me to write this, but, lastly this work is dedicated to the memory of a valued colleague, whose friendship and kindness I was fortunate to receive and will always remember with gratitude.

# Contents

# Notations, Symbols and Acronyms

## Notations

### Vectors and Matrices

| | |
|---|---|
| $x$ | lower case letters represent a scalar |
| $\boldsymbol{x}$ | bold lower case letters represent a vector |
| $\mathbf{X}$ | non-italic bold upper case letters represent a matrices |
| $X$ | italic upper case letters represent an RV |
| $\boldsymbol{X}$ | bold italic upper case letters represent a matrix containing RVs |

### Sets and Intervals

| | |
|---|---|
| $\mathcal{A}$ | set of actions of an MDP |
| $\mathcal{B}_{\mathrm{e}}$ | set of samples for MaL evaluation (evaluation batch) |
| $\mathcal{B}_{\mathrm{pol}}$ | set of samples processed per $\boldsymbol{\theta}$ for PGU |
| $\mathcal{B}_{\mathrm{t}}$ | set of samples used for MaL training (training batch) |
| $\mathbb{C}$ | set of complex numbers |
| $\mathcal{D}$ | replay memory |
| $\mathcal{H}$ | set of hidden (latent) neurons |
| $\mathcal{K}_{\mathrm{UE}}$ | set of UEs |
| $\mathbb{N}$ | set of natural numbers, excluding 0 |
| $\mathbb{N}_0$ | set of natural numbers, including 0 |
| $\mathcal{N}$ | set of neurons with $\mathcal{N} = \mathcal{Z} \cap \mathcal{H}$ |
| $\mathcal{N}_{k_{\mathrm{UE}}}$ | set of PRBs assigned to an UE |
| $\mathcal{N}_{\mathrm{PRB}}$ | set of PRBs in an allocation scenario |
| $\mathcal{P}_j$ | set of presynaptic neurons connected to the $j$th neuron |
| $\mathbb{R}$ | set of real numbers |
| $\mathbb{R}^+$ | set of non-negative real numbers |
| $\mathbb{R}^-$ | set of non-positive real numbers |
| $\mathcal{R}$ | set of rewards of an MDP |
| $\mathcal{S}$ | set of states of an MDP |
| $\mathfrak{S}$ | set of environment initializations used for DRL |
| $\mathfrak{S}_{\text{eval}}$ | set of environment initializations used for DRL evaluation |
| $\mathfrak{S}_{\text{test}}$ | set of environment initializations used for DRL testing |
| $\mathfrak{S}_{\text{train}}$ | set of environment initializations used for DRL training |
| $\mathcal{X}$ | set of transmit symbols |
| $\mathcal{Z}$ | set of observable neurons |
| $\emptyset$ | empty set |

## Constants

| | |
|---|---|
| e | Euler constant, $\mathrm{e} \approx 2.718\ldots$ |
| $\pi$ | Pi, $\pi \approx 3.142\ldots$ |

## Miscellaneous Mathematical Functions and Operators

| | |
|---|---|
| $\mathcal{B}(p)$ | Bernoulli distribution with probability $p$ |
| $\delta(\cdot)$ | Dirac delta function |
| $D(\cdot\vert\vert\cdot)$ | Kullback-Leibler divergence |
| $\mathbb{E}_X\{X\}$ | expectation w.r.t. $X$ |
| $\exp(\cdot)$ | exponential function, $\exp(x) = \mathrm{e}^x$ |
| $f(\cdot)$ | arbitrary function |
| $H(\cdot)$ | Heaviside step function |
| $\Im\{\cdot\}$ | imaginary part of a complex-valued number |
| $\boldsymbol{I}$ | unity matrix |
| ln | natural logarithm, i.e. logarithm to the base e |
| log | logarithm to the base 10, $\log(x) = \log_{10}(x) = \frac{\ln(x)}{\ln(10)}$ |
| $\mathcal{N}(\mu, \sigma^2)$ | normal distribution with mean $\mu$ and variance $\sigma^2$ |
| $p_X(x)$ | probability of observing $x$ |
| $p_{Y\vert X}(y\vert x)$ | conditional probability of observing $y$ given $x$ |
| $\Re\{\cdot\}$ | real part of a complex-valued number |
| $\sigma(x)$ | sigmoid function $\sigma(x) = \frac{1}{1+\mathrm{e}^x}$ |
| $\sigma_\eta(x)$ | sigmoid function parameterized by $\eta$ |
| $\mathcal{U}[a, b]$ | uniform distribution with lower bound $a \in \mathbb{R}$ and upper bound $b \in \mathbb{R}$ |
| $\nabla_{\boldsymbol{\theta}}$ | nabla w.r.t. $\boldsymbol{\theta}$ |
| $\|\cdot\|_p$ | $p$-norm |

# Symbols

## Latin Symbols and Variables

| | |
|---|---|
| $A_+$ | scaling factor of potentiation in STDP |
| $A_-$ | scaling factor of depression in STDP |
| $a_{k_{\mathrm{UE}},l}$ | age of packet inside $l$th buffer slot of $k_{\mathrm{UE}}$th UE |
| $A_t$ | action taken at time $t$ in an MDP |
| $B$ | RV modeling the transmit bit |
| $B_{\mathrm{PRB}}$ | bandwidth of a PRB in OFDMA systems |
| $\boldsymbol{b}$ | transmit bit-sequence |
| $\hat{\boldsymbol{b}}$ | estimated received bit-sequence |
| $c$ | scaling factor of 16-QAM |
| $c_{k_{\mathrm{UE}}}$ | CQI of $k_{\mathrm{UE}}$th UE |
| $e_{k_{\mathrm{UE}},l}$ | size (in bits) of packet inside $l$th buffer slot of $k_{\mathrm{UE}}$th UE |
| $G_t$ | (discounted) return |
| $H_{\mathrm{CD}}$ | transfer function of CD of an optical fiber |
| $\boldsymbol{H}_{\kappa}$ | spike behavior of latent neurons at time instance $\kappa$ |
| $\boldsymbol{H}_{\leq\kappa}$ | spike behavior of latent neurons till time instance $\kappa$ |
| $i$ | class label of a classification task |
| $\hat{i}$ | estimated class label |
| $i(t)$ | continuous-time synaptic current |
| $i[\kappa]$ | discrete-time synaptic current |
| $j$ | index of connected neuron |
| $J_{\boldsymbol{\theta}}()$ | objective function parameterized by $\boldsymbol{\theta}$ |
| $k$ | discrete-time of a communications system |
| $K$ | number of discrete-time steps an SNN is simulated |
| $k_{\mathrm{UE}}$ | index of UEs |
| $K_{\mathrm{UE}}$ | number of UEs |
| $\mathcal{L}$ | loss function |
| $L_{\mathrm{BS}}$ | number of slots inside the buffer of the BS |
| $\ell$ | discrete time index of the impulse response of FIR filters |
| $\ell_{\max}$ | number of samples of a raised-cosine FIR filter |
| $\ell_{\mathrm{off}}$ | offset of a raised-cosine FIR filter in samples |
| $\ell(\boldsymbol{\theta})$ | performance metric of $\boldsymbol{\theta}$ w.r.t. to a given task |
| $\ell_{\mathrm{diff}}(\boldsymbol{\theta}, \tilde{\boldsymbol{\theta}})$ | performance difference of $\boldsymbol{\theta}$ and $\tilde{\boldsymbol{\theta}}$ |
| $L_{\mathrm{CD}}$ | length of an optical fiber |
| $\mathcal{L}_{\leq K}(\boldsymbol{\theta})$ | log-likelihood observing $\boldsymbol{Z}_{\leq K-1}$ given $\boldsymbol{\theta}$ |
| $\ell_{\boldsymbol{\theta}_{\mathcal{Z}}}(\boldsymbol{Z}_{\leq K-1})$ | learning signal |
| #MAC | number of MAC operations per inference |
| $N_{\mathrm{n}}$ | RV modeling a noise sample |
| $N_{\mathrm{c}}$ | number of classes of a classification task |
| $n_{\mathrm{emb}}$ | output dimension of embedding |
| $N_{\mathrm{enc}}$ | number of spike signals output by a neural encoder |
| $N_{\mathrm{episodes}}$ | number of episodes a DRL agent is optimized |
| $n_{\mathrm{fb}}$ | number of feedback taps of DFE |
| $n_{\mathrm{ff}}$ | number of feedforward taps of DFE |
| $N_{\mathrm{hid}}$ | number of hidden neurons |
| $N_{\mathrm{in}}$ | number of input neurons/ spike signals input to SNN |
| $N_{\mathrm{NOMA}}$ | number of NOMA resources per subcarrier |
| $N_{\mathrm{out}}$ | number of output neurons |
| $N_{\mathrm{PRB}}$ | number of PRBs in an OFDMA scenario |
| $n_{\mathrm{PRB}}$ | index of PRBs in an OFDMA scenario |
| $N_{\mathrm{s},j}$ | number of spikes generated by the $j$th neuron |
| $N_{\theta}$ | number of learnable model parameters |
| $n_{\mathrm{tap}}$ | number of taps used for equalization |
| $o_j$ | output of $j$th neuron |
| $p_{\mathrm{drop}}$ | dropout probability of dropout layers |
| $p_{\mathrm{fire}}$ | firing probability of the SRM |
| $\mathbf{P}_{\mathrm{rand}}$ | random permutation matrix |
| $P_{\mathrm{RX}}$ | received power |
| $P_{\mathrm{TX}}$ | transmit power |
| $P_{\mathrm{Tx,max},k_{\mathrm{UE}}}$ | maximum transmit power of $k_{\mathrm{UE}}$th UE |
| $P_{\mathrm{Tx},k_{\mathrm{UE}},n_{\mathrm{PRB}}}$ | transmit power of $k_{\mathrm{UE}}$th UE on $n_{\mathrm{PRB}}$th PRB |
| $p_{\boldsymbol{\theta}}(X)$ | probability of observing $X$ given $\boldsymbol{\theta}$ |
| $q_{\pi}(s,a)$ | action-value function when following policy $\pi$ |
| $\mathfrak{q}_{k_{\mathrm{UE}}}$ | QoS identifier of $k_{\mathrm{UE}}$th UE |
| $Q_{\pi}(s,a)$ | $Q$ function when following policy $\pi$ |
| $Q_{\boldsymbol{\theta}}(s,a)$ | $Q$ function parameterized by $\boldsymbol{\theta}$ |
| $R_{\mathrm{bit}}$ | data rate (bit/s) |
| $R_{\mathrm{cap}}$ | reward at which dynamics DRL training is stopped |
| $\underline{r}_{k_{\mathrm{UE}}}$ | target data rate of $k_{\mathrm{UE}}$th UE |
| $r_{k_{\mathrm{UE}},n_{\mathrm{PRB}}}$ | data rate achieved by $k_{\mathrm{UE}}$th user equipments (UEs) using the $n_{\mathrm{PRB}}$th PRB |
| $r^{(\mathrm{MICKI})}$ | reward obtained by MICKI |
| $R_{\mathrm{sym}}$ | baudrate |
| $R_{t+1}(s_t,a_t)$ | reward obtained after taking $a_t$ in $s_t$ in an MDP |
| $r^{(\mathrm{TFRA})}$ | reward predefined by TFRA environment |
| $\boldsymbol{s}_{k_{\mathrm{UE}}}$ | state of $k_{\mathrm{UE}}$th UE |
| $\boldsymbol{s}'_{k_{\mathrm{UE}}}$ | state of $k_{\mathrm{UE}}$th UE after compression by ENNs |
| $S_t$ | state of an MDP at time $t$ |
| $t$ | continuous-time |
| $t_0$ | initial time |
| $t_{\mathrm{enc},z}$ | spike timing obtained by RFE |
| $T_{\mathrm{env}}$ | maximum simulation time of an MDP |
| $T$ | continuous-time simulation time of SNN |
| $t_{\mathrm{pre}}$ | spike timing of a presynaptic neuron |
| $t_{\mathrm{post}}$ | spike timing of a postsynaptic neuron |
| $T_{\mathrm{target}}$ | update interval of $\boldsymbol{\theta}$ in DRL setting |
| $v_j(t)$ | continuous-time membrane potential of $j$th neuron |
| $v_j[\kappa]$ | discrete-time membrane potential of $j$th neuron |
| $v_j[\kappa]^{(m)}$ | discrete-time membrane potential of $j$th neuron and $m$th sample |
| $v_{\text{out},j}(t)$ | continuous-time membrane potential of $j$th output neuron |
| $v_{\text{out},j}[\kappa]$ | discrete-time membrane potential of $j$th output neuron |
| $v_{\text{rest}}$ | resting potential of the membrane |
| $v_{\text{th}}$ | spiking threshold of the membrane potential |
| $v_{\pi}(s)$ | state-value function when following policy $\pi$ |
| $\boldsymbol{x}$ | vector of transmit symbols |
| $X$ | RV modelling the transmit symbol |
| $\hat{X}$ | RV modelling the estimate of the transmit symbol |
| $\boldsymbol{x}_{\text{a},m_{\text{NOMA}}}$ | vector of probabilities w.r.t. the action space $\mathcal{A}$ |
| $\boldsymbol{X}_{\text{a}}$ | matrix of stacked $\boldsymbol{x}_{\text{a},m_{\text{NOMA}}}$, $m_{\text{NOMA}} = 1, \ldots, N_{\text{NOMA}}$ |
| $x_{\text{enc}}$ | scalar value $x$ to be encoded using neural encoding |
| $\boldsymbol{x}_{\text{enc}}$ | bit-pattern output by the quantizer of TE/QE |
| $x_{\text{enc,min}}$ | minimal value of $x_{\text{enc}}$ |
| $x_{\text{enc,max}}$ | maximal value of $x_{\text{enc}}$ |
| $\boldsymbol{y}$ | vector of symbols received at channel output |
| $\tilde{\boldsymbol{y}}$ | vector output by an equalizer |
| $Y$ | RV modelling the received symbol |
| $Z_{\text{avg}}$ | average number of hidden layer spikes per inference |
| $Z_{\text{out}}[\kappa]$ | discrete-time RV modeling the spike at time-instant $\kappa$ |
| $Z_j^{(m)}[\kappa]$ | discrete-time RV modeling the spike at time-instant $\kappa$ of $j$th neuron and $m$th sample |
| $z_{\text{enc},j}(t)$ | $j$th continuous-time spike signal output by a neural encoder |
| $z_{\text{enc},j}[\kappa]$ | $j$th discrete-time spike signal output by a neural encoder |
| $\mathbf{Z}_{\text{enc}}$ | matrix of spikes output by a neural encoder |
| $z_{\text{hid},j}(t)$ | continuous-time spike signal output by the $j$th hidden layer neuron |
| $z_{\text{in}}(t)$ | continuous-time spike signal input to a neuron |
| $z_{\text{in},j}(t)$ | $j$th continuous-time spike signal input to a neuron or SNN |
| $z_{\text{in},j}[\kappa]$ | $j$th discrete-time spike signal input to a neuron or SNN |
| $\boldsymbol{Z}_{\kappa}$ | spike behavior of observable neurons at time instance $\kappa$ |
| $\boldsymbol{Z}_{\leq\kappa}$ | spike behavior of observable neurons till time instance $\kappa$ |
| $\boldsymbol{Z}_{\leq\kappa}^{(\mathcal{P}_j \cup \{j\})}$ | spike behavior till time instance $\kappa$ of presynaptic neurons of $j$th neuron and neuron $j$ itself |
| $z_{\text{out}}(t)$ | continuous-time spike signal output by a neuron |
| $z_{\text{out}}[\kappa]$ | discrete-time spike signal output by a neuron |
| $z_{\text{out},j}(t)$ | continuous-time spike signal output by the $j$th output layer neuron |
| $z_{\text{out},j}[\kappa]$ | discrete-time spike signal output by the $j$th output layer neuron |
| $Z_{\text{out}}[\kappa]$ | discrete-time RV modeling the output spike at time-instance $\kappa$ |

## Greek Symbols and Variables

| | |
|---|---|
| $\alpha_{\text{damp}}$ | dampening factor of PGU |
| $\alpha[\ell]$ | discrete-time impulse response of the feedforward filter of the SRM model |
| $\alpha_{\text{r}}$ | regularization loss |
| $\alpha_{\text{r},n}$ | parameters of regularization loss |
| $\beta_{\text{CD}}$ | dispersion coefficient of an optical fiber |
| $\beta_{\text{down}}$ | downsampling factor |
| $\beta[\ell]$ | discrete-time impulse response of the feedback filter of the SRM model |
| $\beta_{\text{RRC}}$ | roll-off factor of RRC filter |
| $\beta_{\text{up}}$ | upsampling factor |
| $\gamma$ | spike amplification for training of probabilistic spiking neural networks (SNNs) |
| $\gamma_{\text{G}_\text{t}}$ | discount rate |
| $\Delta_{\text{Q}}$ | quantization step size of TE/QE |
| $\Delta\theta$ | synapse weight change |
| $\Delta t$ | sampling time of a discrete-time system |
| $\Delta t_{\text{enc}}$ | interspike interval of BE |
| $\Delta t_{\text{f}}$ | interspike interval of STDP |
| $\Delta j$ | width of $j$th linear field of RFE |
| $\varepsilon_{\text{Q}}$ | greedy-factor of $Q$-learning |
| $\zeta$ | interference power (caused by the TFRA environment) |
| $\eta$ | steepness of sigmid function $\sigma_\eta(\cdot)$ |
| $\theta_j$ | synapse weight connecting the $j$th neuron |
| $\boldsymbol{\theta}$ | vector of learnable parameters of an adaptable model |
| $\boldsymbol{\theta}^*$ | vector of optimal parameters of an adaptable model |
| $\tilde{\boldsymbol{\theta}}$ | vector of parameter variations used for PGU |
| $\tilde{\boldsymbol{\Theta}}$ | RV modeling $\tilde{\boldsymbol{\theta}}$ |
| $^{(\mathrm{in})}$ | real-valued weights connection the input layer to the hidden layer |
| $^{(\mathrm{out})}$ | real-valued weights connection the hidden layer to the output layer |
| $\kappa$ | discrete-time index of a discrete-time variable |
| $\kappa_{\mathrm{enc},z}$ | discrete-time spike timing of RFE |
| $\lambda$ | wavelength |
| $\mu_j$ | center/mean of $j$th linear/Gaussian RFE |
| $\nu$ | learning rate |
| $\xi$ | inference power (caused by UEs on same PRB) |
| $\xi_{\mathrm{up}}$ | soft-update factor of DDPG |
| $\pi$ | policy in a DRL setting |
| $\pi_{\boldsymbol{\theta}}$ | policy in a DRL setting, parameterized by $\boldsymbol{\theta}$ |
| $\pi^*$ | optimal policy in a DRL setting |
| $\rho_{\boldsymbol{\theta}}$ | deterministic policy, parameterized by $\boldsymbol{\theta}$ |
| $\sigma_j$ | standard deviation of $j$th Gaussian RFE |
| $\sigma_{\mathrm{n}}^2$ | variance of AWGN |
| $\sigma_{\pi}^2$ | variance of Gaussian policy |
| $\sigma_{\pi,\mathrm{c}}^2$ | variance of Gaussian policy coupled to $\ell(\boldsymbol{\theta})$ |
| $\tau_{\mathrm{m}}$ | time constant of the membrane potential |
| $\tau_{\mathrm{rec}}$ | time constant of the recurrent behavior of the SRM |
| $\tau_{\mathrm{s}}$ | time constant of the synaptic current |
| $\tau_{+}$ | time constant of potentation |
| $\tau_{+}$ | time constant of depression |
| $\boldsymbol{\phi}$ | set of learnable parameters of an adaptable model |

## Acronyms

| | |
|---|---|
| **AI** | Artificial Intelligence |
| **ANN** | Artificial Neural Network |
| **AWGN** | Additive White Gaussian Noise |
| **BE** | Bernoulli Encoding |
| **BER** | Bit Error Rate |
| **BP** | Backpropagation |
| **BPSK** | Binary Phase-Shift-Keying |
| **BPTT** | Backpropagation Through Time |
| **BS** | Base Station |
| **BSI** | Buffer State Information |
| **CD** | Chromatic Dispersion |
| **CE** | Cross-Entropy |
| **CPU** | Central Processing Unit |
| **CQI** | Channel Quality Indicator |
| **DDPG** | Deep Deterministic Policy Gradient |
| **DFE** | Decision Feedback Equalization |
| **DL** | Deep Learning |
| **DNN** | Deep Neural Networks |
| **DQL** | Deep $Q$-Learning |
| **DQN** | Deep $Q$-Network |
| **DRL** | Deep Reinforcement Learning |
| **ELBO** | Evidence Lower Bound |
| **ELENA-SNN** | Enlarge-Likelihood-Each-Notable-Amplitude-SNN |
| **ELM** | Extreme Learning Machine |
| **EM** | Expectation Maximization |
| **ENN** | Encoder Neural Network |
| **EOTM** | End-Of-Time Membrane-potential |
| **FIR** | Finite Impulse Response |
| **FPGA** | Field Programmable Gate Array |
| **GBR** | Guaranteed Bit Rate |
| **GD** | Gradient Descent |
| **GPU** | Graphics Processing Unit |
| **IF** | Integate-and-Fire |
| **i.i.d.** | independent and identically distributed |
| **IM/DD** | Intensity Modulation with Direct Detection |
| **IQR** | Interquartile Range |
| **ISI** | Inter-Symbol Interference |
| **KL** | Kullback-Leibler |
| **LCD** | Low Chromatic Dispersion |
| **LDPC** | Low-Density Parity-Check |
| **LE** | Large Environment |
| **LI** | Leaky Integrate |
| **LIF** | Leaky Integrate-and-Fire |
| **MAC** | Multiply-Accumulate |
| **MaL** | Machine Learning |
| **MAP** | Maximum A Posteriori |
| **MC-NOMA** | Multi-Carrier NOMA |
| **MDP** | Markov Decision Process |
| **MF** | Matched Filter |
| **MICKI** | Mimicking Learning |
| **ML** | Maximum Likelihood |
| **ML-ELENA-SNN** | Multi-Level ELENA-SNN |
| **MNIST** | Modified National Institute of Standards and Technology Database |
| **MOTM** | Max-of-Time Membrane Potential |
| **MSE** | Mean Squared Error |
| **NFE** | No Feedback Equalization |
| **NN** | Neural Network |
| **NOMA** | Non-Orthogonal Multiple Access |
| **NPFCA** | NOMA-Uplink-PFCA |
| **NPS** | No Packet Shuffling |
| **OFDMA** | Orthogonal Frequency-Division Multiple Access |
| **PAM** | Pulse Amplitude Modulation |
| **PD** | Photo Diode |
| **PDB** | Packet Delay Budget |
| **PDF** | Probability Density Function |
| **PFCA** | Proportional Fair-Channel Aware |
| **PGU** | Policy Gradient-based Update |
| **PGT** | Policy Gradient Theorem |
| **PRB** | Physical Resource Block |
| **QAM** | Quadrature Amplitude Modulation |
| **QE** | Quantization Encoding |
| **QI** | QoS Identifier |
| **QoS** | Quality of Service |
| **RE** | Rate Encoder |
| **ReLU** | Rectified Linear Unit |
| **RFE** | Receptive Field Encoding |
| **RL** | Reinforcement Learning |
| **RNN** | Recurrent Neural Network |
| **ROC** | Rank-Order-Encoding |
| **RPS** | Random Packet Shuffling |
| **RRC** | Root-Raised-Cosine |
| **RRiT** | Round-Robin if Traffic |
| **RV** | Random Variable |
| **SDR** | Symbol Disagreement Rate |
| **SE** | Small Environment |
| **SER** | Symbol Error Rate |
| **SG** | Surrogate Gradient |
| **SIC** | Successive Interference Cancellation |
| **SINR** | Signal-to-Inference-plus-Noise-Ratio |
| **SNN** | Spiking Neural Network |
| **SPA** | Sum-Product Algorithm |
| **SPS** | Sorted Packet Shuffling |
| **SRD** | Spike Rate Decoding |
| **SRM** | Spike Response Model |
| **SSMF** | Standard Single-Mode Fiber |
| **STDP** | Spike-Timing-Dependent Plasticity |
| **TE** | Ternary Encoding |
| **TEAS** | Time Encoder Asynchronous Spikes |
| **TESS** | Time Encoder Synchronous Spikes |
| **TFRA** | Time-Frequency Resource Allocation |
| **TTFS** | Time-To-First-Spike Encoding |
| **UE** | User Equipment |
| **w.r.t.** | with respect to |

# 1 Introduction

Telecommunications are fundamental to today's connected world, enabling the exchange of information across long distances and transforming how people, businesses, and governments interact. Communication networks overcome geographic boundaries, driving economic progress, knowledge sharing, and social cohesion [PdFC+25]. Artificial intelligence (AI), specifically machine learning (MaL), deep learning (DL), and reinforcement learning (RL), have emerged as a groundbreaking technology, offering innovative solutions to address the increasing computational complexity of modern communication systems [ANKC25]. Typical examples include the allocation of limited radio resources for enhanced resource utilization enabled by RL [PdFC+25], and the mitigation of device and channel impairments through MaL [CVYA21]. In these examples, artificial neural networks (ANNs) are employed as the underlying computational model. However, the performance of ANN-based receivers typically depends on their complexity [FOS+21, Fig. 1], leading to power-hungry systems when implemented on digital electronics.

In contrast, the human brain only consumes tens of watts of power on average [Kas21, p. 664], and hence, it remains unchallenged in terms of its energy efficiency [CSSZ22]. By taking inspiration from the human brain, spiking neural networks (SNNs) opt to mimic its behavior and energy-efficiency [NMZ19]. In SNNs, neurons and synapses exhibit time-dependent dynamics and communicate information in an event-based way via discrete binary events, so-called spikes [Mas97]. Since spikes are only exchanged when information is processed, the energy consumption of SNNs can be a fraction of that of comparable ANNs [AHMK21]. When emulated on so-called neuromorphic hardware, SNNs enable fast and low-power signal processing [dLST+17, Fig. 1]. Thus, SNNs promise low-power AI.

When designing and optimizing SNN-based systems, two major issues persist: The first challenge is the still unresolved question of an efficient update rule for SNNs [ZBC+21], which determines how the weights of the synapses connecting the neurons are adjusted during training. For ANNs, the backpropagation (BP) algorithm combined with an optimization algorithm, e.g., gradient descent (GD) or Adam, has proven to be an efficient method for optimizing the parameters of the computational model [RHW86]. Yet, there is currently no de facto standard learning algorithm for SNNs [ZBC+21]. Various update rules exist, ranging from biologically inspired update rules to gradient-based update rules. For biologically inspired update rules, such as spike-timing-dependent plasticity (STDP) [BP98], the information for the adaptation of the synapses is only locally available [TGK+19]. Consequently, this locality limits the training of multi-layer SNNs [ZG18]. A gradient-based update rule inspired by the BP algorithm is backpropagation through time (BPTT) combined with surrogate gradients (SGs) [NMZ19]. The gradient of the error signal observed at the SNN output, which quantifies the contribution of each synaptic weight to the error, is backpropagated through both the spatial and temporal dimensions of the SNN using a time-discretized model. However, when modeling spiking neurons, their spiking nature introduces a non-differentiable activation function. By approximation of the true gradient with a differentiable function, SGs overcome this problem [ZG18]. While BPTT with SG enables the application of well-established MaL techniques, its memory requirements grow linearly with the number of time steps [ZBC+21], limiting the time duration of stimuli input to the SNN. Another gradient-based update rule is the optimization of probabilistic SNNs [JSGG19]. Using a time-discretized model of the SNN and stochastic neuron spiking, the model parameters are updated to maximize the probability of generating a target spike pattern at the output. Although the approach overcomes the locality constraint of STDP and does not require gradient approximations using SGs, apart from [JSGG19], there is little literature on probabilistic SNNs, particularly regarding applications on neuromorphic hardware. In contrast, BPTT with SGs has been successfully applied to analog/mixed signal neuromorphic hardware [CSSZ22].

The second issue when designing and optimizing SNNs is the conversion from real-world data into spikes, the so-called neural encoding. While for digital signal processing, discrete-time sampling and quantization are applied to convert real-world data into a representation that can be processed by digital electronics, there is no standardized rule for neural encoding. This is partly because the exact mechanisms of biological neural encoding have yet to be conclusively determined, with ongoing debates about whether the rate of multiple spikes, the absolute timing of a single spike, or the relative timing of multiple spikes encodes the information in the human brain [Bre15], resulting in various neural encoding schemes [AHMK21]. Depending on the given problem statement to be solved using SNNs and the update rule applied, an appropriate encoding must be selected and subsequently parameterized. Consequently, when compared to the design and optimization of ANN-based systems, the design and optimization of SNN-based systems come with major challenges.

This thesis proposes and explores multiple strategies for the design and implementation of SNN-based receivers in communication systems. First, three different update rules and various neural encoding schemes are investigated. As a toy example, the transmission of a 16-quadrature amplitude modulation (QAM) over an additive white Gaussian noise (AWGN) channel is considered, and an SNN-based detector is designed and optimized. Based on the findings, we then design an SNN-based equalizer and demapper for the mitigation of both linear and non-linear channel impairments. We introduce and compare various variations of the SNN-based equalizer, differing in multiple aspects, such as the use of decision feedback and various neural encodings, and demonstrate the superior performance of SNN-based equalizers and demappers when compared to their ANN-based counterparts and when compared to classical equalizers.

We furthermore discuss the impact of the neural encoding parameters on the performance of the system, which is limited by the ability of the neural encoding to transform real-world signals into meaningful spike patterns. This ability depends on the parameters of the neural encoding, which are typically determined heuristically. If chosen properly, the performance of the system significantly improves. Using lessons from RL, we develop a method for the joint optimization of the neural encoding parameters and the SNN. Given the SNN-based equalizer and demapper, we demonstrate that the joint optimization can identify appropriate encoding parameters, leading to high-performing systems while reducing dependence on carefully tuned neural encoding parameters. Moreover, joint optimization allows for a more compact system design, resulting in reduced inference durations, smaller SNN architectures, and a reduced number of spikes per inference, while maintaining performance.

**Structure of the Thesis**

Ch. 2 lays the theoretical foundations of SNNs. First, the temporal dynamics of biological neurons are discussed and different spiking neuron models are introduced. Afterwards, a brief overview of neural encoding schemes is given, with a more detailed description of the schemes used in this thesis. Neural decoding, which is the conversion from spike signals into real-world signals, is also introduced, and various schemes are discussed. Subsequently, a brief overview of update rules for SNNs is provided, followed by a detailed description of the three update rules examined in this thesis. Finally, a brief overview of simulation software for SNNs, as well as neuromorphic systems, is provided.

In Ch. 3, the three update rules introduced in Ch. 2 are investigated in the context of the transmission of a 16-QAM disturbed by AWGN. An SNN-based detector is designed and optimized. Furthermore, different neural encodings are investigated. As a result, an update rule and three different neural encodings are determined, which will be used for the rest of the thesis.

Next, Ch. 4 introduces the task of equalization and reviews traditional equalizers, such as linear equalizers and the decision feedback equalizer. Afterwards, two setups of SNN-based equalizers and demappers are introduced, differing in their feedback of already classified symbols. The resulting setups are combined with three different neural encodings. Given the model of a non-coherent fiber-based optical transmission, the various equalizers are compared with respect to (w.r.t.) their bit error rate (BER) and average number of generated spikes per inference.

Ch. 5 presents key theoretical concepts of RL. As an illustrative application of RL, the resource allocation problem in mobile communication networks is introduced. For two distinct allocation problems, we design and optimize an RL agent, and demonstrate its superiority compared to benchmark schemes.

In Ch. 6, we exploit RL to optimize the neural encoding. To update parameters that cannot be updated using gradient-based approaches, we derive the policy gradient-based update (PGU) based on RL theory. We investigate the optimization of SNNs using PGU and successfully demonstrate the application of the PGU to determine the neural encoding parameters, resulting in reduced inference durations, smaller SNN architectures, and a reduced number of spikes per inference.

Finally, we summarize the results obtained throughout this thesis and provide an outlook on future work.

Parts of the results of this thesis have been pre-published in the following references: [BES21], [BKS22], [BvBS23], [vBES23], [vBES24], [AEvB+25], [EvBS25], [SLN+25], [vBEMS25], [vBEM+25]. These references are marked by an underlined label, i.e., [______], throughout the thesis.

# 2 Spiking Neural Networks

This chapter lays the theoretical foundations of spiking neural networks (SNNs). The spiking neuron is introduced and various network architectures that interconnect spiking neurons in different ways are briefly discussed. Afterwards, neural encoding and neural decoding techniques are presented, followed by a short review of update rules for the optimization of SNNs. Finally, recent tools for the simulation and emulation of SNNs in software and hardware are reviewed.

In general, SNNs are continuous-time systems. However, when simulating SNNs in software, discrete-time implementations play a major role. Since this thesis uses discrete-time implementations, if applicable, both the continuous-time and the discrete-time models are introduced.

## 2.1 The Spiking Neuron

### 2.1.1 The Biological Neuron

The biological neuron consists of three functionally distinct parts: dendrites, soma, and axon, as illustrated in Fig. 2.1. The dendrites function as collectors for incoming signals from upstream connected neurons, the soma as the central processing unit, and the axon and its terminals as distributors of the output signal. The signals exchanged between neurons consist of short electrical pulses, the so-called *action potentials* or *spikes*, with a typical duration of 1-2 ms. Since all spikes are identical in shape, information is not encoded in their waveform. In a sequence of spikes, a so-called *spike signal*, the information is carried in both the number and the timing of spikes [GK02, Sec. 1.1].

The *synapse* is the neural connection between axon terminals of upstream connected neurons and the following dendrites. Consequently, neurons upstream of the synapse are called *presynaptic neurons*, and neurons downstream of the synapse are called *postsynaptic neurons*. Spikes arriving at the synapse from the axon terminals are called *presynaptic spikes*. If a presynaptic spike is passed to the synapse, the synapse is activated, triggering a chain of biochemical processing steps, which opens ion channels at the dendrites. A current of ions flows into the cell, the so-called *postsynaptic current* or *synaptic current* [GK02, Sec. 1.1]. Since the ions carry a charge from outside the soma into the soma, the *membrane potential*, indicating the difference in electrical potential between the exterior and interior of the cell, is charged or discharged. Hence, the soma integrates the incoming signals over time. In parallel, a small leakage current continuously discharges the membrane potential towards the *resting potential* [HH52].

If the membrane potential exceeds a *threshold*, an output spike is generated. The output spike is passed along the axon to the axon terminals, activating the synapses of downstream connected neurons. Furthermore, the membrane potential is reset to its *reset potential*. This spiking behavior of the neuron can be interpreted as non-linear signal processing [GK02, Sec. 1.1.1].

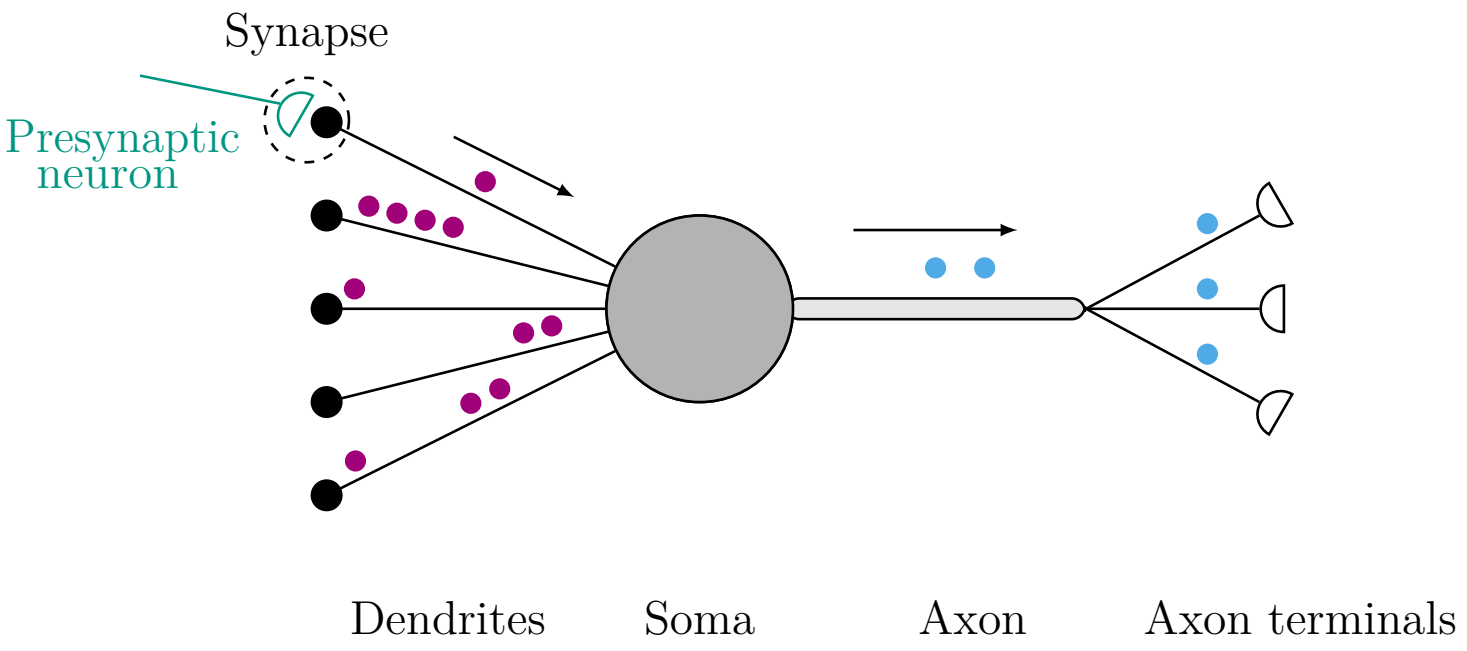

**Figure 2.1:** Schematic representation of a biological neuron, based on [BVM+19, Fig. 1(a)]. Purple dots denote incoming spikes, blue dots denote output spikes.

### 2.1.2 Models of Spiking Neurons

#### Overview

Hodgkin and Huxley developed a mathematical model that captures the behavior of biological neurons by accounting for the contributions of distinct ions [HH52]. Although this model is biologically plausible, it is computationally complex [TGK+19]. Many simplifications exist, e.g., the spike response model (SRM) [JTG04], the Izhikevich neuron model [Izh03], and the leaky integrate-and-fire (LIF) neuron model [BvR99]. The reduced complexity of the simplified models comes at the expense of a reduced biological plausibility. Fig. 2.2 shows a comparison of the different neuron models in terms of their biological plausibility and computational complexity.

The LIF neuron model captures the intuitive properties of the biological neuron without suffering from large complexity [TGK+19]. Due to its low complexity, it is widely used in computational neuroscience [NMZ19, Izh04]. While most neuron models exhibit deterministic spike behavior, the SRM exhibits non-deterministic spike behavior [JTG04]. In the following, the LIF neuron model, its variations, and the SRM will be briefly introduced.

#### The LIF Neuron Model

The three main properties of biological neurons captured by the LIF neuron model are the temporal integration of incoming spikes, the gradual leakage of the membrane potential, and the non-linear spiking behavior. Let $v(t) \in \mathbb{R}$ denote the membrane potential and $i(t) \in \mathbb{R}$ denotes the synaptic current at time $t \in \mathbb{R}^+$. The temporal dynamics of the LIF neuron model can be described by two ordinary differential equations [NMZ19]

$$\frac{\mathrm{d}v(t)}{\mathrm{d}t} = -\frac{1}{\tau_\mathrm{m}}\left(v(t) - v_\mathrm{rest}\right) + \frac{1}{\tau_\mathrm{m}} i(t)\,, \tag{2.1}$$

$$\frac{\mathrm{d}i(t)}{\mathrm{d}t} = -\frac{i(t)}{\tau_\mathrm{s}} + \sum_j \theta_j z_{\mathrm{in},j}(t)(t)\,. \tag{2.2}$$

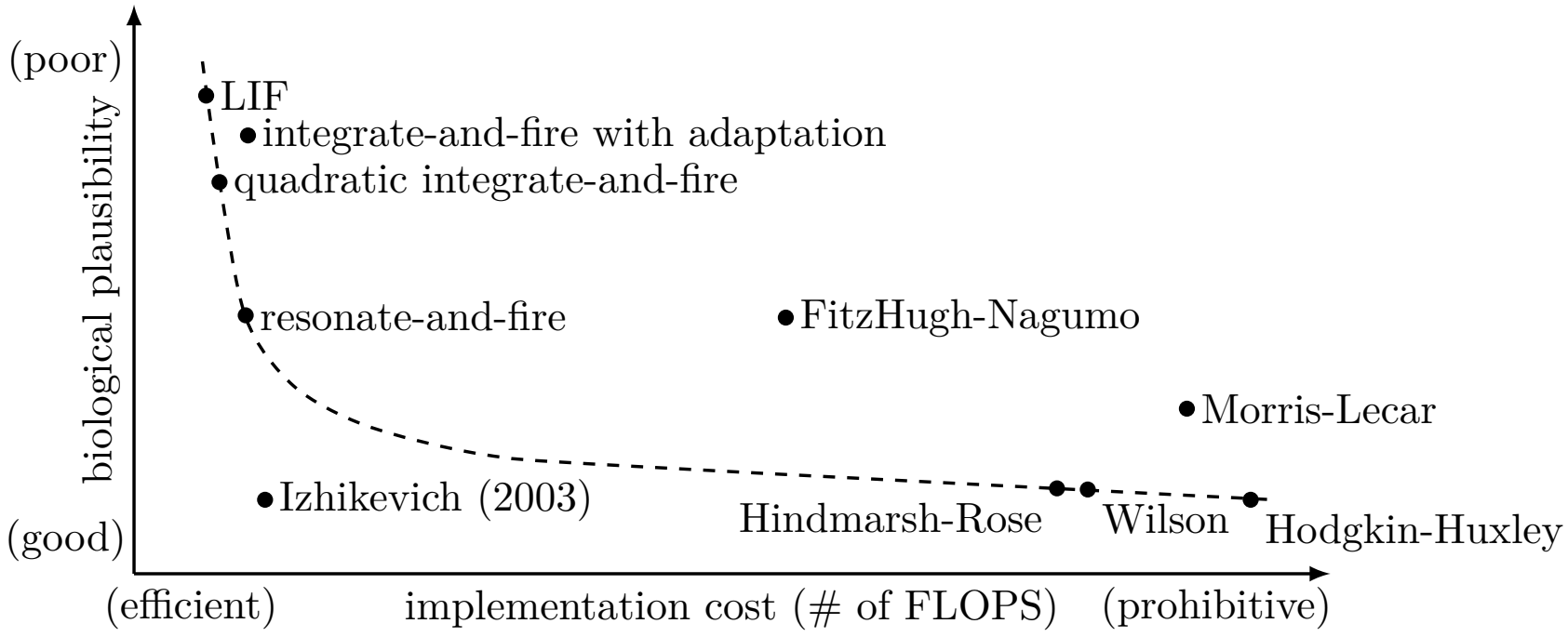

**Figure 2.2:** Comparison of different neuron models in terms of their biological plausibility and computational complexity, as in [Izh04, Fig. 2].

While the time constant $\tau_\mathrm{m} \in \mathbb{R}^+$ controls the leakage of the membrane potential $v(t)$, the time constant $\tau_\mathrm{s} \in \mathbb{R}^+$ models the closing of the ion channels, directly affecting the synaptic current $i(t)$. Furthermore, $v_\mathrm{rest} \in \mathbb{R}$ denotes the resting potential, and $z_{\mathrm{in},j}(t) \in \{0, 1\}$ denotes the input spike signal received by the $j$th connected neuron upstream. It is scaled by the adaptable weight $\theta_j \in \mathbb{R}$, which models the strength of the respective synapse. We model $z_{\mathrm{in},j}(t)$ as

$$z_{\mathrm{in},j}(t) = \sum_n \delta(t - t_{\mathrm{z},n})\,, \tag{2.3}$$

where $t_{\mathrm{z},n} \in \mathbb{R}$ denotes the exact time of the $n$th spike and $\delta(\cdot)$ the Dirac delta function.

The solution of these ordinary differential equations, and therefore the dynamics of the LIF neuron, can be approximated using the forward Euler method [CSSZ22]. Assume that $v(t_0)$ and $i(t_0)$ are some arbitrary initial values at $t_0 = 0$. Using numerical integration with a fixed integration step size $\Delta t$, the differential equations can be solved, leading to a discrete-time system defined at time instants $t = \kappa \Delta t$, $\kappa \in \mathbb{N}$. The discrete-time dynamics of the LIF neuron can be expressed as [CSSZ22, NMZ19][1]

$$v[\kappa+1] = v[\kappa] \cdot \mathrm{e}^{-\frac{\Delta t}{\tau_\mathrm{m}}} + i[\kappa] \cdot \mathrm{e}^{-\frac{\Delta t}{\tau_\mathrm{m}}} , \tag{2.4}$$

$$i[\kappa+1] = i[\kappa] \cdot \mathrm{e}^{-\frac{\Delta t}{\tau_\mathrm{s}}} + \sum_j \theta_j z_{\mathrm{in},j}[\kappa] . \tag{2.5}$$

We can interpret $\Delta t$ as the sampling time of the system; hence, $v[\kappa]$ and $i[\kappa]$ are the membrane potential and synaptic current at $t := \kappa\Delta t$. Both the integration of incoming spike signals $z_{\mathrm{in},j}[\kappa]$ over time and the leakage of the membrane potential $v[\kappa]$ are realized by (2.4) and (2.5).

The spiking behavior can be modeled using the Heaviside step function $H(\cdot)$ and the threshold $v_\mathrm{th}$ of a spiking neuron. If $v[\kappa]$ exceeds $v_\mathrm{th}$, the neuron fires an output spike

$$z_\mathrm{out}[\kappa] = H(v[\kappa] - v_\mathrm{th}) = \begin{cases} 1, & \text{if} \quad v[\kappa] > v_\mathrm{th} , \\ 0, & \text{otherwise} . \end{cases} \tag{2.6}$$

After firing a spike, the membrane voltage $v[\kappa]$ is reset to the reset potential $v_\mathrm{reset}$ by $v[\kappa] \leftarrow v_\mathrm{reset}$. Within this thesis, the resting potential and the reset potential are used synonymously; hence $v_\mathrm{reset} = v_\mathrm{rest}$.

Fig. 2.3 shows two examples of the temporal dynamics of the LIF neuron model when stimulated by a spike signal. Incoming spikes are denoted by purple and output spikes by blue dots. An incoming spike induces a synaptic current $i(t)$, which increases sharply and then decreases exponentially. The membrane potential $v(t)$ is charged by the synaptic current and also decreases exponentially over time. If $v(t)$ exceeds the threshold $v_\mathrm{th}$, an output spike is fired and $v(t)$ is reset. When comparing Fig. 2.3(a) with Fig. 2.3(b), the impact of $\tau_\mathrm{m}$ is demonstrated. With increasing $\tau_\mathrm{m}$, the exponential decay of $v(t)$ slows, resulting in more output spikes.

### The LI Neuron Model

A variation of the spiking LIF neuron model is the non-spiking leaky integrate (LI) neuron model. The LI neuron model follows the same dynamics as the LIF neuron model (see (2.1) and (2.2)); however, its ability to generate an output spike is deactivated. Therefore, the LI neuron acts as a leaky integrator of the input to the neuron, without resetting the membrane potential. Fig. 2.4 exemplarily shows the behavior of a LI neuron. Due to the very large chosen time constant $\tau_\mathrm{m}$, the exponential decay of $v[\kappa]$ in the diagram shown appears almost linear. After the arrival of the last input spike, the exponential decay can be observed initially.

[1] Note that square brackets $[\cdot]$ indicate discrete-time signals.

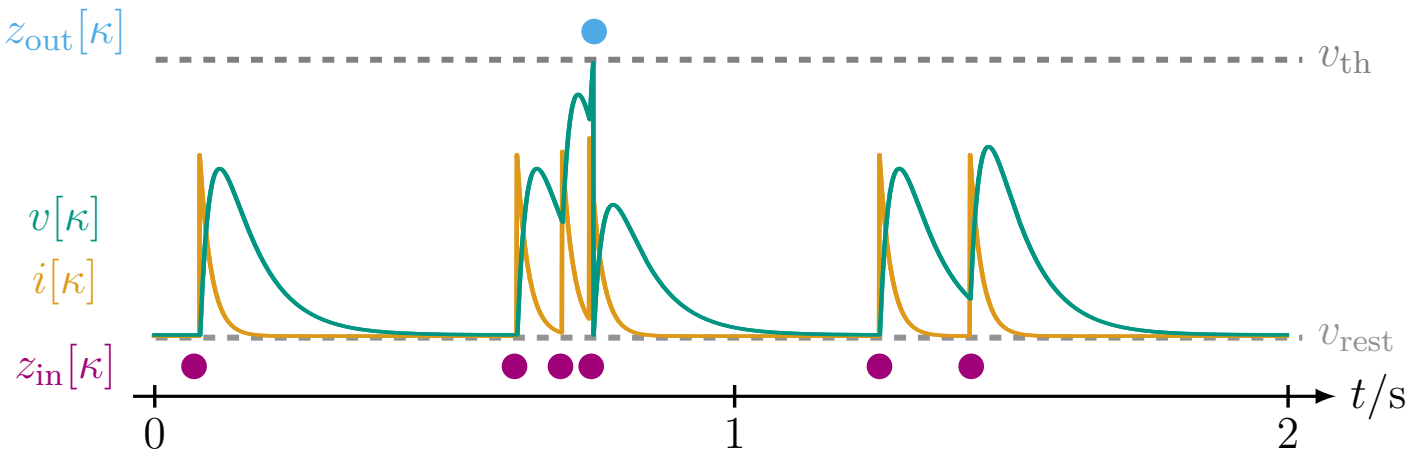

**(a)** Dynamics for $\tau_{\text{m},1} = 100\,\text{ms}$.

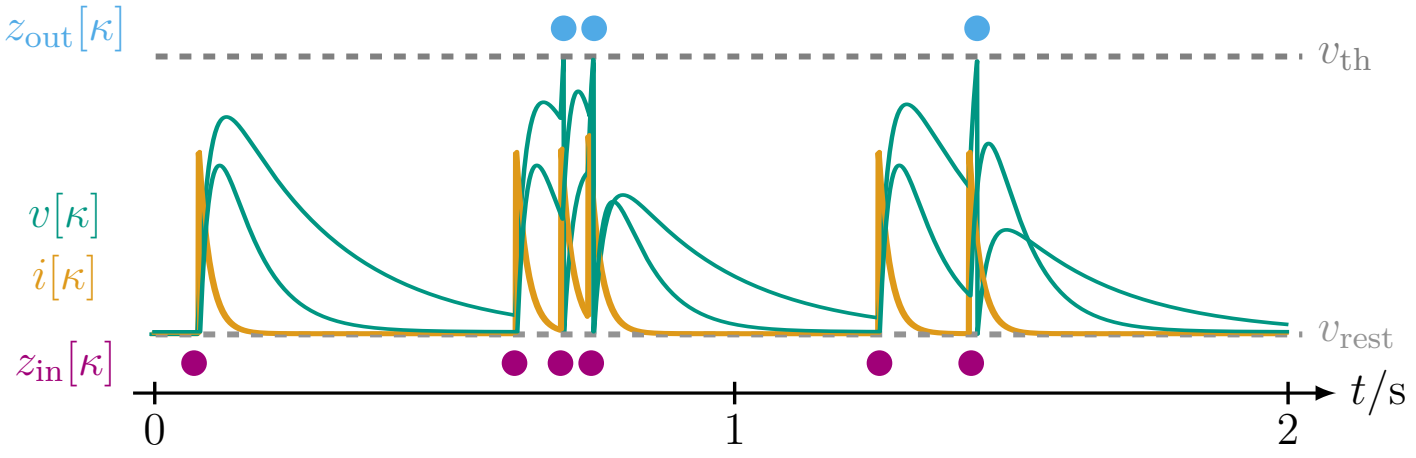

**(b)** Dynamics for $\tau_{\text{m},2} = 3 \cdot \tau_{\text{m},1} = 300\,\text{ms}$.

**Figure 2.3:** Exemplary temporal dynamics of the LIF neuron model, following (2.4) and (2.5), with $\Delta t = 1\,\text{ms}$, $\tau_{\text{s}} = 30\,\text{ms}$, and different $\tau_{\text{m}}$.

### The Spike Response Model

One criticism of the LIF and LI neuron models is their oversimplification, particularly due to the instantaneous reset of the membrane potential [SFM+22]. Moreover, neuroscience has shown that the spike behavior of a neuron is not deterministic, but is subject to stochastic influences [GK02, Sec. 4.2]. A generalization of the spiking LIF model that incorporates stochasticity is the SRM. Instead of a deterministic activation function, e.g., the Heaviside step function, at each time instant $\kappa$ the output of the neuron can be modelled by a random variable (RV) $Z_{\text{out}}[\kappa] \in \{0, 1\}$, thereby generating a random process. The RV follows a Bernoulli distribution with firing probability $p_{\text{fire}}[\kappa]$, i.e., $Z_{\text{out}}[\kappa] \sim \mathcal{B}(p_{\text{fire}}[\kappa])$, where all RVs are assumed to be independent. A discrete-time implementation of the SRM is depicted in Fig. 2.5. First, each input spike signal $z_{\text{in},j}[\kappa] \in \{0, 1\}$, $j \in \{1, 2, \dots, J\}$, is passed through a causal finite impulse response (FIR) filter with discrete-time impulse response $\alpha_j[\ell]$ and multiplied by the synaptic weight $\theta_j^{(\text{f})} \in \mathbb{R}$. Finally, the $J$ signals are summed up. The so-called *feed-forward filters* $\alpha_j[\ell]$ capture the memory characteristics of the membrane potential. One approach is the so-called *exponential filter*, which is defined as [JSS21, JSGG19]

$$\alpha[\ell] = \mathrm{e}^{-\frac{\ell \Delta t}{\tau_{\text{m}}}}\,, \tag{2.7}$$

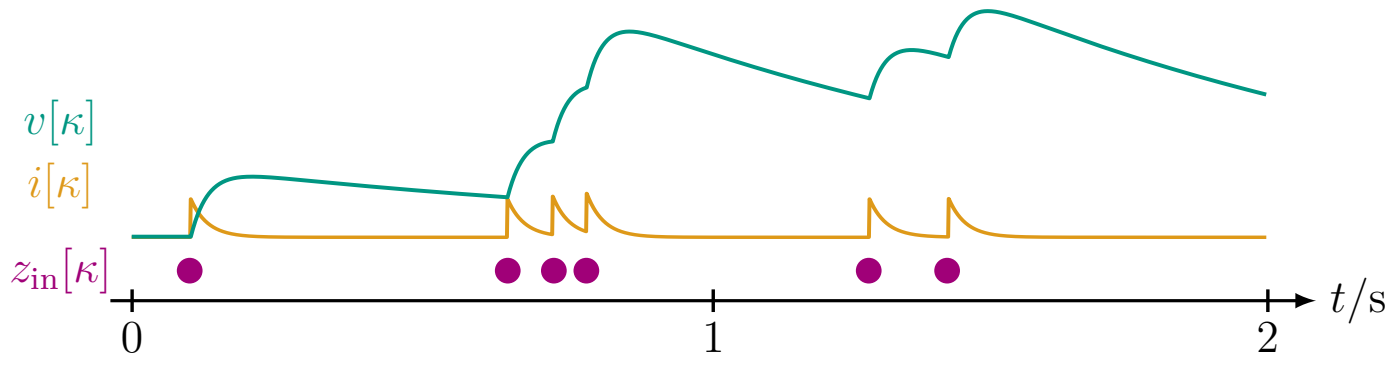

**Figure 2.4:** Exemplary dynamics of a non-spiking LI neuron, following (2.4) and (2.5), with $\Delta t = 1\,\mathrm{ms}$, $\tau_\mathrm{s} = 30\,\mathrm{ms}$, and $\tau_\mathrm{m} = 3\,\mathrm{s}$.

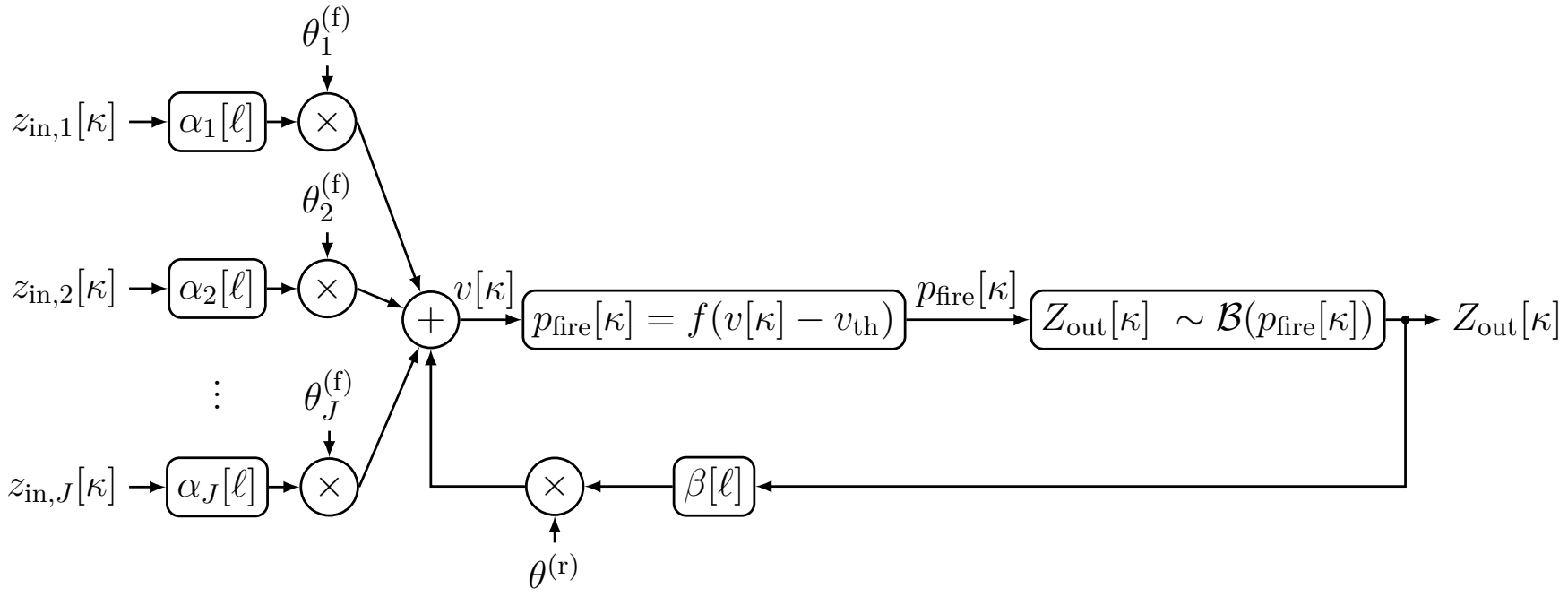

**Figure 2.5:** Block diagram of the SRM, based on [JSS21, JSGG19]. $\alpha[\ell]$ and $\beta[\ell]$ denote FIR filters.

with $\tau_\mathrm{m} \in \mathbb{R}^+$ being time-constants assigned to the $j$th filter. In general, $\alpha[\ell]$ can be of arbitrary shape, providing neuron models with complicated dynamics and rich temporal information [SFM$^+$22]. A second type of feed-forward filter used in this thesis is the delayed raised-cosine filter, motivated by findings that such filters can model the behavior of retinal ganglion cells in response to visual stimuli [PPU$^+$05]. Delayed raised-cosine filters are given by

$$\alpha[\ell]^{(\ell_\mathrm{off})} = \begin{cases} \frac{\cos\left(\pi\left(2\frac{\ell-\ell_\mathrm{off}}{\ell_\mathrm{max}}-1\right)\right)+1}{2}, & \text{if } \ell \in \{\ell_\mathrm{off}, 1, \dots, \ell_\mathrm{max} + \ell_\mathrm{off}\}\,, \\ 0, & \text{otherwise}\,, \end{cases} \tag{2.8}$$

where $\ell_\mathrm{off} \in \mathbb{N}_0$ denotes the delay they introduce, and $\ell_\mathrm{max} + 1$, $\ell_\mathrm{max} \in \mathbb{N}_0$ the width of the raised-cosine pulse in samples.

The feedback filter $\beta[\ell]$ models the reset mechanism of the neuron. Its impulse response can be expressed as [JSS21, GKNP14]

$$\beta[\ell] = -\mathrm{e}^{-\frac{\ell \Delta t}{\tau_\mathrm{rec}}}\,, \tag{2.9}$$

where $\tau_{\text{rec}} \in \mathbb{R}^+$ denotes the time constant characterizing the refractory behavior of the neuron. By summing up all filtered spike signals, the membrane potential $v[\kappa]$ is obtained by [JSS21, SFM+22]

$$v[\kappa] = \sum_{j=1}^{J} \theta_j^{(\text{f})} (z_{\text{in},j} * \alpha)[\kappa] + \theta^{(\text{r})} (z_{\text{out}} * \beta)[\kappa - 1] , \tag{2.10}$$

where $\theta^{(\text{r})} \in \mathbb{R}$ is the recurrent weight, and $z_{\text{out}}[\kappa] \in \{0, 1\}$ is a realization of the RV modeling the output spike $Z_{\text{out}}[\kappa] \in \{0, 1\}$. Based on $v[\kappa]$, the firing probability

$$p_{\text{fire}}(Z_{\text{out}}[\kappa] = 1 | v[\kappa]) = f(v[\kappa]) \tag{2.11}$$

is obtained using a monotonically increasing function $f : \mathbb{R} \to [0, 1]$, e.g., the sigmoid function [JSS21]

$$p_{\text{fire}}(Z_{\text{out}}[\kappa] = 1 | v[\kappa]) := \sigma(v[\kappa]) = \frac{1}{1 + \mathrm{e}^{-(v[\kappa] - v_{\text{th}})}} , \tag{2.12}$$

shifted by the threshold $v_{\text{th}}$. Hence, the probability that an output spike is fired increases with $v[\kappa]$.

To enable more complex neuron dynamics, the neuron model in Fig. 2.5 can be extended to multiple feed-forward filters per input spike signal $z_{\text{in},j}[\kappa]$ [PPU+05], e.g., multiple raised-cosine filters with different time-shifts $\ell_{\text{off}}$ can be applied in parallel. Consequently, the membrane potential can be written as

$$v[\kappa] = \sum_{j=1}^{J} \sum_{m=0}^{\ell_{\text{off,max}}} \left( z_{\text{in},j} * \alpha_j^{(m)} \right) [\kappa] \cdot \theta_j^{(\text{f},m)} + \theta^{(\text{r})} (z_{\text{out}} * \beta)[\kappa - 1] , \tag{2.13}$$

where $\alpha_j^{(m)}[\ell]$ denotes the $m$th, $m \in \mathbb{N}$, feed-forward filter linked to the $j$th input signal, $\theta_j^{(\text{f},m)} \in \mathbb{R}$ the $m$th, synapse weight linked to the $j$th input signal, and $\ell_{\text{off,max}} \in \mathbb{N}_0$ the maximum time shift.

## 2.2 Spiking Neural Networks

An SNN is a directed, possibly cyclic, network of spiking neurons [JSS21]. Different network topologies exist [SPP+17], see Fig. 2.6(a) for an illustration. Although neurons can be connected arbitrarily, we typically group neurons into layers [SFM+22]. We distinguish between three different types of layers: the input layer, the hidden layers, and the output layer. The input and output layers can be interpreted as the interface of the SNN. At the output layer, the processed data is read out. In addition, one or more hidden layers can be inserted between the

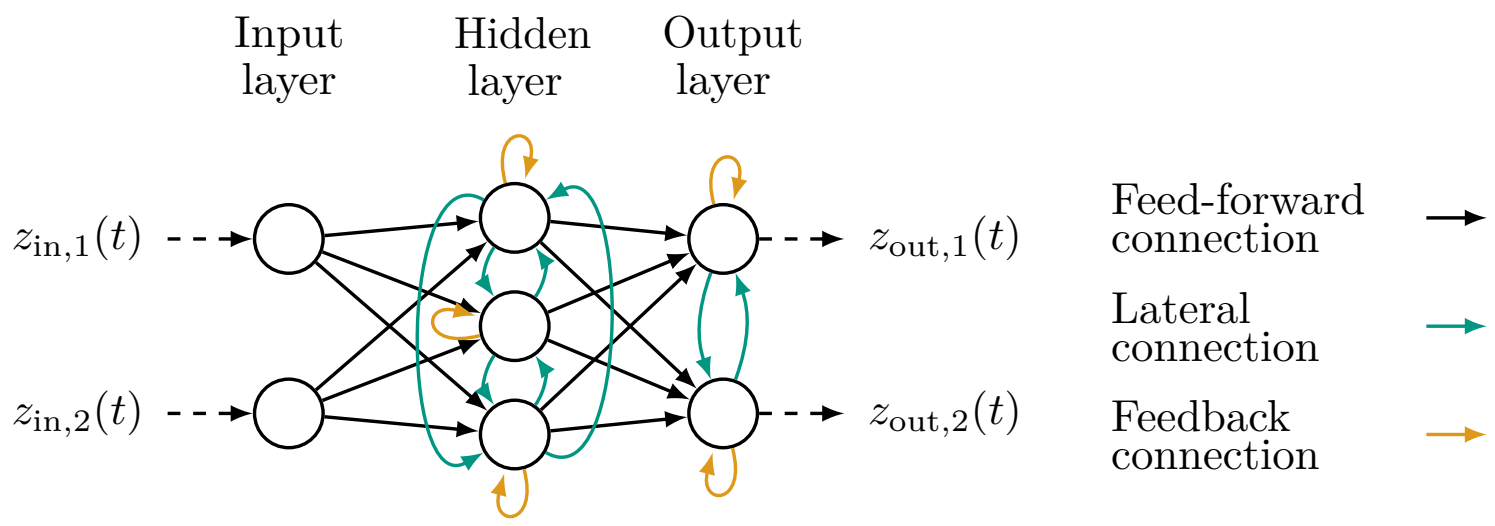

**(a)** Architecture of an SNN.

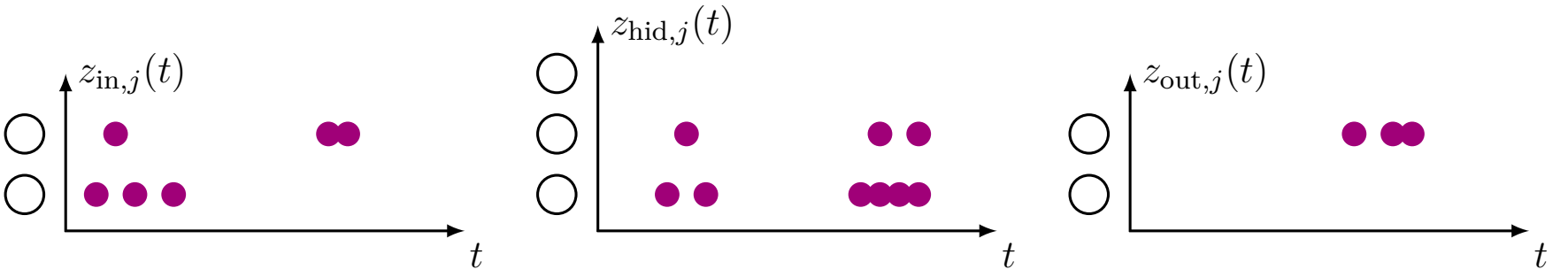

**(b)** Exemplary spike signals.

**Figure 2.6:** (a) Architecture of an SNN with a single hidden layer and $N_{\text{in}} = 2$ input, $N_{\text{hid}} = 3$ hidden, and $N_{\text{out}} = 2$ output layer neurons. Synapses connecting neurons can be distinguished into feed-forward, lateral and feedback connections. Solid lines denote adaptable connections, dashed lines solely visualize the signal flow.
(b) Exemplary spike signals emitted by the different neurons of the input, hidden and output layer of (a). White hollow dots denote the neurons of (a).

input and output layers. The number of neurons of the different layers is given by $N_{\text{in}} \in \mathbb{N}$ for the input, $N_{\text{hid}} \in \mathbb{N}$ for the hidden, and $N_{\text{out}} \in \mathbb{N}$ for the output layer. Fig. 2.6(a) shows an SNN with a single hidden layer and $N_{\text{in}} = 2$ input neurons, $N_{\text{hid}} = 3$ hidden neurons, and $N_{\text{out}} = 2$ output neurons. If all layers are connected in a feed-forward manner, a feed-forward SNN is obtained [SPP+17]. Due to the internal dynamics of the LIF neuron, an SNN inherently includes implicit recurrent connections. In addition, explicit recurrent connections can be incorporated, such as lateral connections and feedback connections [SPP+17]. While lateral connections either excite or inhibit neighboring neurons to regulate their firing [DC15], feedback connections enable a neuron to self-excite or self-inhibit. Thus, the output of a neuron affects not only neurons in downstream layers but also all connected neurons within its own layer, including itself. For classification tasks utilizing various datasets (Randman, MNIST, SHD, RawHD, RawSC), it has been demonstrated that the inclusion of explicit recurrent connections can lead to improved classification accuracy [ZN21, Fig. 3].

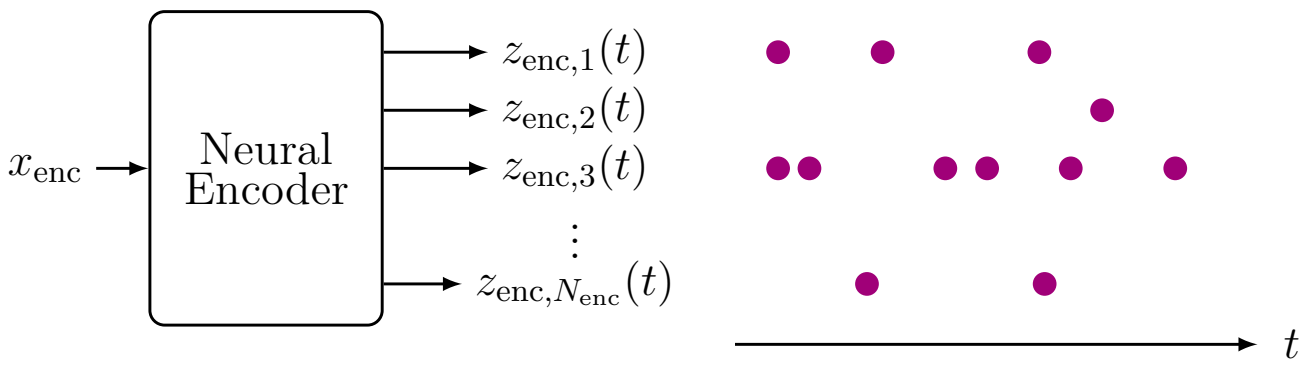

**Figure 2.7:** Exemplary neural encoding, which converts an arbitrary real-world input $x_{\text{enc}}$ into $N_{\text{enc}}$ parallel spike signals $z_{\text{enc},j}(t) \in \{0, 1\}$ with $j = 1, 2, \ldots, N_{\text{enc}}$. On the right, an exemplary spike pattern output by the neural encoder is displayed, where each purple dot denotes a spike.

As discussed in Sec. 2.1.2, different neuron models exist. If the SNN is implemented using the probabilistic SRM, the model is referred to as probabilistic SNN. If a deterministic neuron model is used, e.g., the spiking LIF neuron, we simply refer to the model as SNN.

## 2.3 Neural Encoding

### 2.3.1 Overview

Before an SNN can process real-world data $x_{\text{enc}}$, which can be, e.g., an image, a time signal, a video, or a complex or real-valued number, *neural encoding* needs to map it to spike signals [AHMK21]. Fig. 2.7 shows an exemplary neural encoder, which converts an arbitrary input $x_{\text{enc}}$ into $N_{\text{enc}} \in \mathbb{N}$ parallel spike signals $z_{\text{enc},j}(t) \in \{0, 1\}$, where $j = 1, 2, \ldots, N_{\text{enc}}$. Depending on the characteristics of the input, such as the dimensionality of images or the time dependence for both audio signals and video, a suitable neural encoding needs to be chosen [AHMK21].

Most encoding techniques are inspired by biological findings. The literature identifies three main encoding techniques: rate encoding [DA05], temporal encoding [AHMK21], and population encoding [PKK20]. However, within the neuroscience community, the nature of neural encoding remains a topic of intense debate [RT11, Bre15, AHMK21], and no clear consensus has been reached yet. It remains uncertain whether individual spikes and neurons encode information independently or whether information is instead conveyed through temporal and spatial correlations among spikes and across neuronal populations [DA05, Sec. 1.5]. Tab. 2.1 gives an overview of a few representative encoding techniques. So far, the spike signals have been unipolar, i.e., $z_{\text{enc},j}(t) \in \{0, 1\}$. However, some encoding techniques, such as temporal contrast methods, generate bipolar spike signals, i.e., $z_{\text{enc},j}(t) \in \{-1, 0, +1\}$.

Most of the techniques in Tab. 2.1 are designed to encode static data, e.g., complex-valued numbers or images. Hence, they convert static data into spike signals, which exhibit a time dimension. Within this thesis, we denote by $T \in \mathbb{R}^+$ the continuous-time duration of the spike signal output when encoding static data. When implemented in discrete time, $K \in \mathbb{N}$ denotes the maximum number of discrete time steps. For encoding time-dependent data, e.g., video or audio signals, the data can be directly converted into spike signals by using, e.g., temporal contrast methods. Hence, the time dimension of the data is preserved and is directly correlated with the time dimension of the spike signals. Since some encodings output multiple spike signals in parallel, we denote $N_{\text{enc}} \in \mathbb{N}$ as the number of spike signals output by the encoder. For $N_{\text{enc}} = 1$, we simply write $z_{\text{enc}}(t)$, neglecting the $j$ in the subscript.

**Table 2.1:** Literature review of popular neural encoding schemes.

| | | |
|---|---|---|
| Rate encoding | | Bernoulli encoding [DA05, Sec. 1.4]<br>Exponential encoding [DA05, Hee00, Sec. 1.4] |
| Temporal encoding | Temporal delay | Time-to-first-spike encoding [GK02, Sec. 1.6.1]<br>Phase encoding [GK02, Sec. 1.6.2] |
| | Temporal contrasts | Threshold-based encoding [DL07, PKK20]<br>Step-forward encoding [PKK20]<br>Moving-window encoding [PKK20] |
| | Temporal stimulation | Hough-spiker algorithm [HDK$^+$99]<br>Modified Hough-spiker algorithm [SV03]<br>Ben's spiker algorithm [SV03, PKK20] |
| Population encoding | | Population average rate [GK02, Sec. 1.7]<br>Receptive field encoding [BLK02]<br>Rank order encoding [TG98]<br>Spatial population encoding [DC15] |

In order to give the reader a clearer understanding of the various encoding techniques introduced in Tab. 2.1, their mechanisms are briefly outlined, with an emphasis on techniques that are applied in the thesis. We assume that the input to the neural encoding is real-valued and static, i.e., $x_{\text{enc}} \in \mathbb{R}$. We furthermore assume that $x_{\text{enc}}$ is bounded, hence $x_{\text{enc}} \in [x_{\text{enc,min}}, x_{\text{enc,max}}]$, with $x_{\text{enc,min}}$, $x_{\text{enc,max}} \in \mathbb{R}, x_{\text{enc,min}} < x_{\text{enc,max}}$.

### 2.3.2 Rate Encoding

#### Overview

In rate encoding, the input $x_{\text{enc}}$ is encoded in the spike rate of $z_{\text{enc}}(t)$, i.e., the number of spikes in a fixed time interval. Typically, the spike rate is proportional to the amplitude $|x_{\mathrm{enc}}|$. For example, the Poisson spike generator uses $x_{\mathrm{enc}}$ as a deterministic estimate of the firing rate and uses random processes for spike signal generation [DA05, Sec. 1.4].

Two approaches exist for realizing the Poisson spike generator [DA05, Sec. 1.4]. The first approach divides the interval of duration $T$ into $K \in \mathbb{N}$ shorter time intervals. Based on $x_{\mathrm{enc}}$, a firing probability $p_{\mathrm{fire}} \in [0, 1]$ is calculated, indicating whether a spike is fired or not within each interval. Hence, a Bernoulli experiment is conducted $K$ times. Consequently, this approach is referred to as Bernoulli encoding (BE). The second approach is exponential encoding, which is based on the time difference of two spikes. By sampling from an exponential distribution parameterized by $x_{\mathrm{enc}}$, multiple interspike intervals $\Delta t_{\mathrm{enc},i} \in \mathbb{R}^+$, $i \in \mathbb{N}$, are obtained. Based on $\Delta t_{\mathrm{enc},i}$, the spike sequence is constructed until the overall duration $T$ is reached.

Since the information is represented by multiple spikes, rate encoding is very robust. For high-speed processing and fast responses, rate encoding is not suitable, since a short observation window limits the resolution of the encoding [AHMK21].

#### Bernoulli Encoding

Due to its intrinsic discrete-time nature when converting static data into spike signals, BE is well suited for software implementation [DA05, Sec. 1.4]. BE divides the signal duration $T$ into $K \in \mathbb{N}$ short intervals $\Delta t$, i.e., $\Delta t = T/K$. Thus, the spike signal output by the encoder can be written as a discrete-time spike signal $z_{\mathrm{enc}}[\kappa]$, $\kappa = 0, 1, \ldots, K-1$. Using a monotonically increasing function $f : \mathbb{R} \to [0, 1]$, $x_{\mathrm{enc}}$ is mapped to a firing probability $p_{\mathrm{fire}} \in [0, 1]$. For each instant $[\kappa]$, the output of the BE follows a Bernoulli distribution with parameter $p_{\mathrm{fire}}$, i.e., $Z_{\mathrm{enc}}[\kappa] \sim \mathcal{B}(p_{\mathrm{fire}})$.

Within this thesis, we assume $x_{\mathrm{enc}} \geq 0$, and obtain $p_{\mathrm{fire}}$ by

$$p_{\mathrm{fire}} = \frac{x_{\mathrm{enc}}}{x_{\mathrm{enc,max}}} \,. \tag{2.14}$$

Fig. 2.8(a) illustrates the output of the BE, initialized with $x_{\mathrm{enc,min}} = 0$, $x_{\mathrm{enc,max}} = 1$, and $K = 10$, for input values $x_{\mathrm{enc}} = 0.1$ and $x_{\mathrm{enc}} = 0.9$.

### 2.3.3 Temporal Encoding

#### Overview

Temporal encodings techniques can be divided into three categories: temporal delay, temporal contrast, and temporal stimulation. While temporal delay methods encode static data, temporal contrast and temporal stimulation focus on the encoding of time-dependent data.

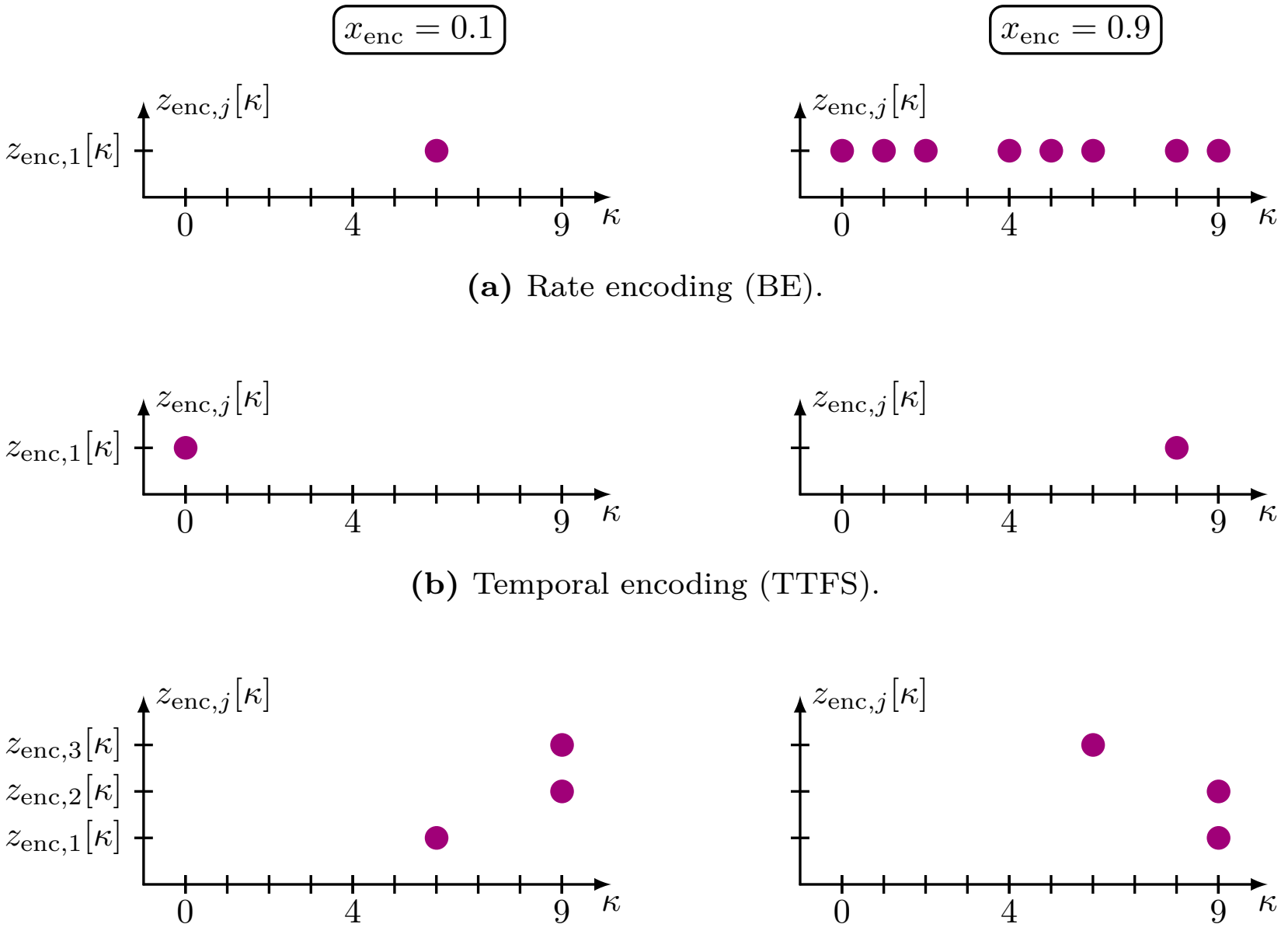

**(a)** Rate encoding (BE).

**(b)** Temporal encoding (TTFS).

**(c)** Population encoding (RFE).

**Figure 2.8:** Exemplary neural encoding for rate (BE), temporal (TTFS) and population rank encoding (Gaussian RFE) for input values $x_{\text{enc}} = 0.1$ and $x_{\text{enc}} = 0.9$, when initialized with $x_{\text{enc,min}} = 0$, $x_{\text{enc,max}} = 1$, and $K = 10$. All plots are obtained using a discrete-time implementation of the neural encodings. Purple dots denote spikes.

In temporal delay methods, the precise timing of a single spike relative to a reference, e.g, a reference spike or the start of the signal, encodes the information. The delay of the spike is directly proportional to $|x_{\text{enc}}|$ [DL07]. Temporal contrast techniques compare the value of a continuous-time signal with a baseline. If the signal deviates from the baseline by more than a defined threshold, a positive or negative spike is generated, and the baseline is updated accordingly [PKK20]. Temporal stimulation methods assume that a positive-only continuous-time signal can be obtained by convolving the spike signal with an FIR filter. Hence, temporal stimulation techniques generate spike signals that approximate the input signal after filtering [HDK+99, SV03].

### Time-to-first-spike Encoding

When employing time-to-first-spike encoding (TTFS), information is represented by the timing of the (first) spike relative to a reference [GK02, Sec. 1.6.1]. The TTFS encoder outputs a single spike signal, i.e., $N_{\text{enc}} = 1$, that consists of exactly one spike, fired at $t_{\mathrm{enc,z}} \in \mathbb{R}^+$. Given a continuous-time TTFS encoder, the timing $t_{\mathrm{enc},z}$ is obtained by

$$t_{\mathrm{enc,z}} = \frac{x_{\mathrm{enc}} - x_{\mathrm{enc,min}}}{x_{\mathrm{enc,max}} - x_{\mathrm{enc,min}}} T \,. \tag{2.15}$$

If the encoding is used within a discrete-time system, its timing $\kappa_{\mathrm{enc,z}} \in \mathbb{N}$ is given by

$$\kappa_{\mathrm{enc,z}} = \left\lfloor \frac{x_{\mathrm{enc}} - x_{\mathrm{enc,min}}}{x_{\mathrm{enc,max}} - x_{\mathrm{enc,min}}} (K - 1) \right\rfloor , \tag{2.16}$$

with

$$z_{\mathrm{enc}}[\kappa] = \begin{cases} 1\,, & \text{if} \quad \kappa = \kappa_{\mathrm{enc,z}}\,, \\ 0\,, & \text{otherwise}\,. \end{cases} \tag{2.17}$$

Fig. 2.8(b) shows the discrete-time TTFS encoding for $x_{\mathrm{enc}} = 0.1$ and $x_{\mathrm{enc}} = 0.9$, when initialized with $x_{\mathrm{enc,min}} = 0$, $x_{\mathrm{enc,max}} = 1$, and $K = 10$.

### 2.3.4 Population Encoding

#### Overview

In contrast to rate and temporal encoding techniques with $N_{\mathrm{enc}} = 1$, population encoding distributes the information across several spike signals $z_{\mathrm{enc},j}(t)$, $j \in \{1, 2, \ldots, N_{\mathrm{enc}}\}$, a so-called population. Several approaches exist to realize population encoding. Population average rate encodes $|x_{\mathrm{enc}}|$ in the firing rate of a neuron population. Because multiple neurons may fire simultaneously, a high resolution of the rate can be achieved even over short observation windows [GK02, Sec. 1.7]. Two alternative techniques based on the timing of spikes are receptive field encoding (RFE) and rank-order-encoding (ROC). For both, each generated spike signal $z_{\mathrm{enc},j}(t)$ consists of exactly one single spike. For RFE, the information is encoded in the relative timing between the spikes of different signals [BLK02] and for ROC the information is encoded in the order of firing within the population [TG98, AHMK21]. While RFE and ROC activate each neuron in the population, spatial population encoding only activates different subgroups of the population. The latter is implicitly performed in [DC15] in a rate encoding manner and is biologically motivated by the human retina [MM91]. In [DC15], the Modified National Institute of Standards and Technology database (MNIST) handwritten dataset, which is a set of $28 \times 28$-pixel images, is classified. For each pixel in an image, a spike signal is created using the intensity of the pixel and rate encoding techniques. Hence, the number of spike signals corresponds to the number of pixels. These spike signals are fed into the SNNs in parallel, such that each pixel is connected to a single neuron in the input layer. Different digits activate different subsets of input neurons, resulting in a spatial input pattern. Since [DC15] does not explicitly name this encoding scheme, we refer to it as *spatial population encoding.* Based on the ideas of spatial population encoding, we developed ternary encoding (TE) [BvBS23]. It converts a real-valued number into a bit pattern of $N_{\text{enc}}$ bits using a quantizer, where each bit determines the spiking behavior of one of the $N_{\text{enc}}$ spike signals output by the encoder.

### Receptive Field Encoding

RFE is a population encoding method, motivated by visual receptive fields [Mal25, Ch. 2]. In the retina, a ganglion cell is connected to multiple photoreceptor cells that span a two-dimensional plane, the receptive field. When individual photoreceptor cells are stimulated, the response, in terms of the number of spikes generated by the ganglion cell, can be measured relative to the position of the photoreceptor with respect to the center of the receptive field. As a result, the base function $f : \mathbb{R}^2 \to \mathbb{R}$ of the receptive field of the ganglion cell indicates the response of the ganglion cell based on the spatial position (Euclidean distance with respect to (w.r.t.) the field center) of the stimulus [Mal25, Fig. 2.1]. In the following, we only consider one-dimensional receptive fields case, $f : \mathbb{R} \to \mathbb{R}$. Instead of coupling the rate of the response to the Euclidean distance of the stimulus and field center, we can couple the timing $t_{\text{enc},z} \in \mathbb{R}^+$ of a spike to the Euclidean distance [BLK02]. Similar to TTFS, the spike signal $z_{\text{enc}}(t)$, emitted by the RFE consists of a single spike. The receptive field can be modeled as Gaussian-shaped [BLK02], logarithmically shaped [ABM+22a], or a piecewise linear function [ABS+23].

### Gaussian Receptive Fields

For a population of $N_{\text{enc}}$ neurons, $N_{\text{enc}}$ fields are applied in parallel, resulting in $N_{\text{enc}}$ spike signals. For the $j$th spike signal, $j = 1, \ldots, N_{\text{enc}}$, the timing $t_{\text{enc,z},j}$ is obtained by [DA05, (3.28)]

$$t_{\text{enc,z},j} = T \cdot \left(1 - \exp\left(-\frac{(x_{\text{enc}} - \mu_j)^2}{2\sigma_j^2}\right)\right), \tag{2.18}$$

where $\mu_j \in \mathbb{R}$ and $\sigma_j \in \mathbb{R}^+$ are the parameterization of the Gaussian-shaped base function characterizing the $j$th receptive field. The discrete-time RFE is obtained by the application of the ceil operator

$$\kappa_{\text{enc,z},j} = \left\lceil (K-1) \cdot \left(1 - \exp\left(-\frac{(x_{\text{enc}} - \mu_j)^2}{2\sigma_j^2}\right)\right) \right\rceil. \tag{2.19}$$

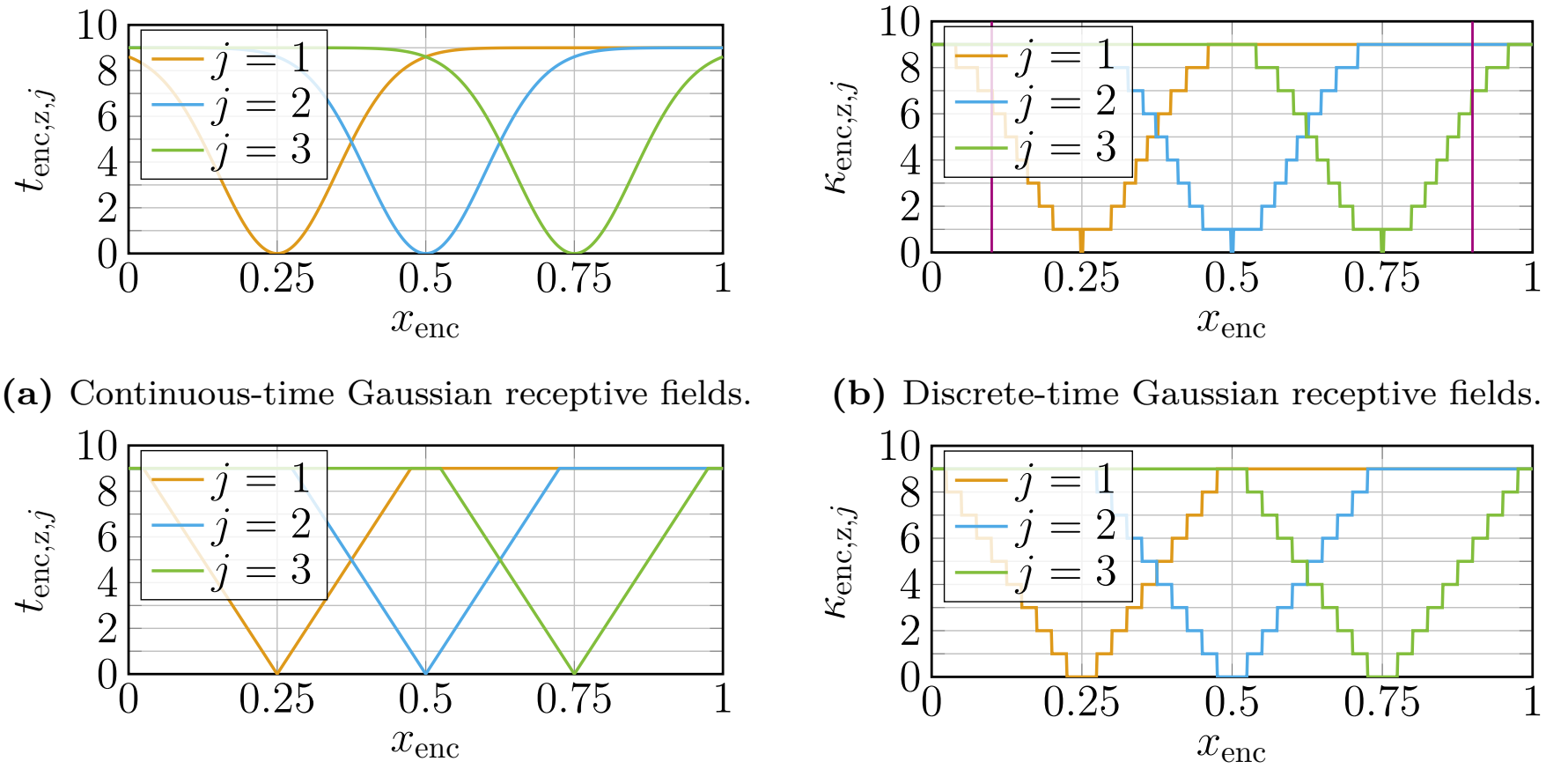

**(a)** Continuous-time Gaussian receptive fields.

**(b)** Discrete-time Gaussian receptive fields.

**(c)** Continuous-time linear receptive fields.

**(d)** Discrete-time linear receptive fields.

**Figure 2.9:** Gaussian and linear receptive fields for both time-continuous and discrete-time implementations. The field centers are given by $\mu_j = \frac{j}{4}$, $j = 1, 2, 3$. For Gaussian fields it is $\sigma_j = \sigma = 0.1$, for linear fields $\Delta_j = \Delta = 0.5$. Continuous-time implementations use $T = 9$, and discrete-time implementations use $K = 10$.

After $\kappa_{\mathrm{enc,z},j} \in \mathbb{N}$ is obtained, the $j$th spike signal $z_{\mathrm{enc},j}$ is generated through

$$z_{\mathrm{enc},j}[\kappa] = \begin{cases} 1\,, & \text{if } \kappa = \kappa_{\mathrm{enc,z},j}\,, \\ 0\,, & \text{otherwise}\,. \end{cases} \tag{2.20}$$

Fig. 2.9(a) shows three different Gaussian receptive fields with the parameters $\mu_j = \frac{j}{4}$, $j \in \{1, 2, 3\}$, $\sigma_j = \sigma = 0.1$, and $T = 9$. The closer $x_{\mathrm{enc}}$ is to the fiel center $\mu_j$, the smaller $t_{\mathrm{enc,z},j}$. Hence, an early spike indicates that the stimulus $x_{\mathrm{enc}}$ is close to the center $\mu_j$. In Fig. 2.9(b), the discrete-time receptive fields of Fig. 2.9(a) using $K = 10$ time steps are shown. In Fig. 2.8 we exemplarily encoded the values $x_{\mathrm{enc}} = 0.1$ and $x_{\mathrm{enc}} = 0.9$, with Fig. 2.8(c) illustrating the output of the discrete-time receptive fields of Fig. 2.9(a). To illustrate the connection between the two figures, we added purple lines to Fig. 2.9(a), indicating both values. For $x_{\mathrm{enc}} = 0.1$, we obtain $\kappa_{\mathrm{enc,z},1} = 7$, $\kappa_{\mathrm{enc,z},2} = 10$, and $\kappa_{\mathrm{enc,z},3} = 10$. For $x_{\mathrm{enc}} = 0.9$, we obtain $\kappa_{\mathrm{enc,z},1} = 10$, $\kappa_{\mathrm{enc,z},2} = 10$, and $\kappa_{\mathrm{enc,z},3} = 7$.

### Linear Receptive Fields

Instead of Gaussian-shaped functions, linear RFE applies (piecewise) linear functions. The spike timing $t_{\text{enc,z},j}$ is obtained by

$$t_{\text{enc,z},j} = \begin{cases} T \cdot \frac{2|x_{\text{enc}} - \mu_j|}{\Delta_j}, & \text{for} \quad \mu_j - \frac{\Delta_j}{2} \leq x_{\text{enc}} \leq \mu_j + \frac{\Delta_j}{2}\,, \\ T, & \text{otherwise}\,, \end{cases} \tag{2.21}$$

where $\Delta_j \in \mathbb{R}^+$ is the width and $\mu_j \in \mathbb{R}$ is the center of the $j$th field. Fig. 2.9(c) shows three linear receptive fields with $\mu_j = \frac{j}{4}$, $j \in \{1, 2, 3\}$, $\Delta_j = \Delta = 0.5$, and $T = 9$. By applying the floor operator, we obtain the discrete-time encoding

$$\kappa_{\text{enc,z},j} = \begin{cases} \left\lfloor \frac{2}{\Delta_j} |x_{\text{enc}} - \mu_j| \, (K-1) \right\rfloor, & \text{for} \quad \mu_j - \frac{\Delta_j}{2} \leq x_{\text{enc}} \leq \mu_j + \frac{\Delta_j}{2}\,, \\ K-1, & \text{otherwise}\,. \end{cases} \tag{2.22}$$

The obtained timing of the spikes is converted into spike signals following (2.20). The discrete-time linear receptive fields with the same parameters as in Fig. 2.9(c) and $K = 10$ are displayed in Fig. 2.9(d). All plots in Fig. 2.9 show that the closer the stimulus $x_{\text{enc}}$ is to the center $\mu_j$ of a field, the smaller $t_{\text{enc,j}}$.

While this thesis uses the term "linear receptive fields" to emphasize the linear shape of $t_{\text{enc,j}}$, in neuroscience, the term is used to describe the properties of a receptive field, which follows the rule of superposition [RVS97]: The response of the field to a sum of stimuli is equal to the sum of the responses of each stimuli.

### Ternary Encoding

TE is a novel population encoding method we introduced in [BvBS23]. The encoding is motivated by spatial population encoding. Depending on the input $x_{\text{enc}}$, only a subgroup of the population is activated. We assume that the range of $x_{\text{enc}}$ is centered around zero, thus $x_{\text{enc,min}} = -x_{\text{enc,max}}$.

Fig. 2.10(a) shows the TE encoder. The upper branch computes $|x_{\text{enc}}|$. Afterwards, a quantizer $Q : \mathbb{R}^+ \to \{0, 1\}^{N_{\text{enc}}}$ is applied. The quantizer executes both quantization and bit conversion and, thus, outputs a bit pattern $\boldsymbol{x}_{\text{enc}} \in \{0, 1\}^{N_{\text{enc}}}$. It is characterized by $x_{\text{enc,max}}$ and the number of quantization bits $N_{\text{enc}}$, which equals the number of output spike signals, and is designed for inputs $0 \leq |x_{\text{enc}}| \leq x_{\text{enc,max}}$. Let $Q' : \mathbb{R} \to \mathcal{Q}$ denote the characteristics of the quantizer without conversion to bits, where $\mathcal{Q} = \{0, \Delta_Q, 2\Delta_Q \ldots, (2^{N_{\text{enc}}} - 1)\Delta_Q\}$ denotes the set of quantization levels. With the quantization step size

$$\Delta_{\text{Q}} = \frac{x_{\text{enc,max}}}{2^{N_{\text{enc}}}}\,, \tag{2.23}$$

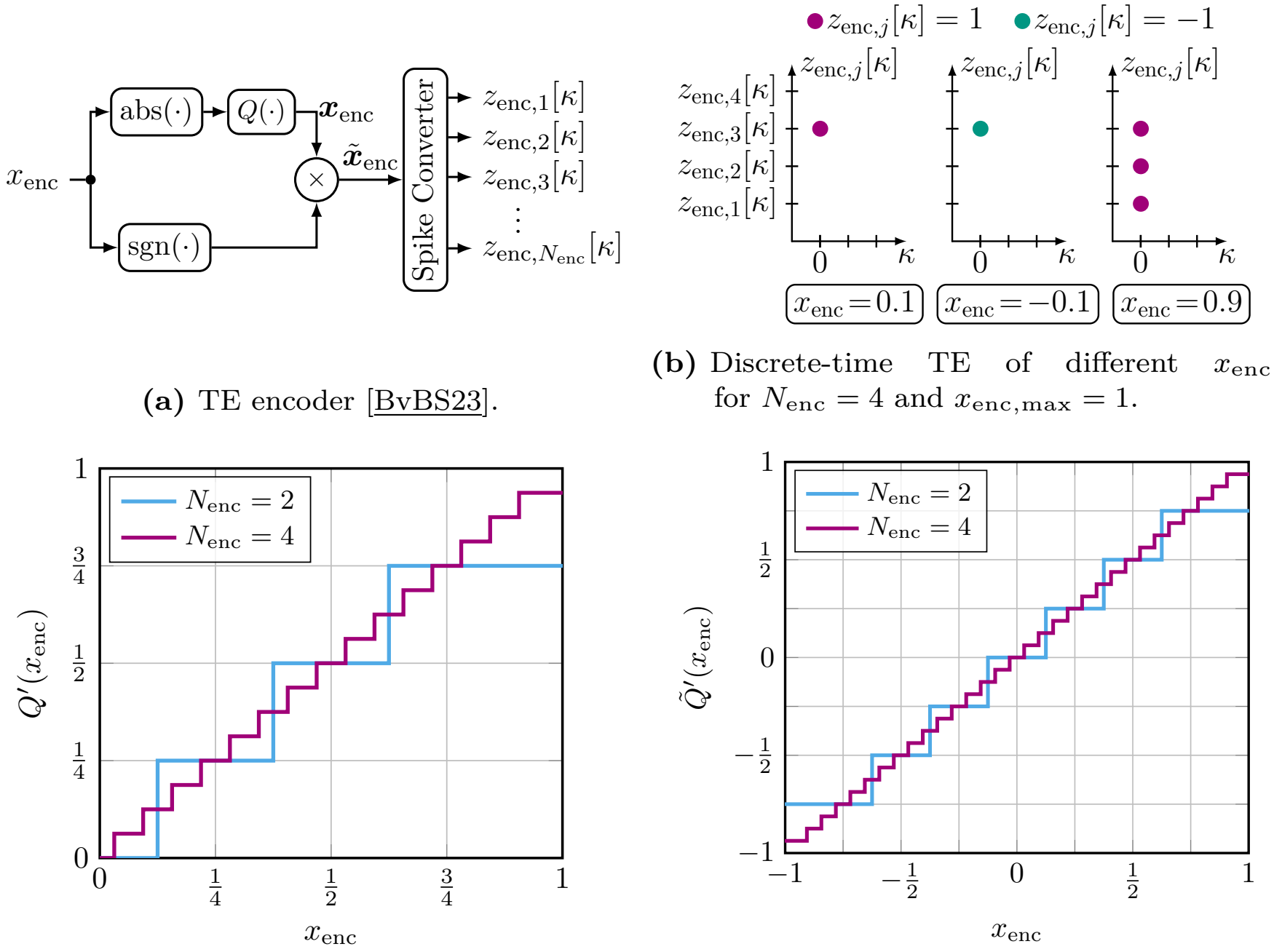

**(a)** TE encoder [BvBS23].

**(b)** Discrete-time TE of different $x_{\mathrm{enc}}$ for $N_{\mathrm{enc}} = 4$ and $x_{\mathrm{enc,max}} = 1$.

**(c)** Quantization characteristics.

**(d)** Equivalent quantization characteristics.

**Figure 2.10:** Block diagram, exemplary encoding output and characteristics of the TE.

the quantization characteristic without bit conversion is given by

$$Q' = \min\left(\left\lfloor \frac{|x_{\mathrm{enc}}| + \frac{\Delta_{\mathrm{Q}}}{2}}{\Delta_{\mathrm{Q}}} \right\rfloor,\ 2^{N_{\mathrm{enc}}} - 1\right) \cdot \Delta_{\mathrm{Q}}\,. \tag{2.24}$$

For the conversion of quantization levels into bit patterns $\boldsymbol{x}_{\mathrm{enc}} \in \{0,1\}^{N_{\mathrm{enc}}}$, a Gray mapping is used. The bit pattern $\boldsymbol{x}_{\mathrm{enc}}$ is then multiplied by $\operatorname{sign}(x_{\mathrm{enc}})$, which results in $\tilde{\boldsymbol{x}}_{\mathrm{enc}} \in \{-1,0,1\}^{N_{\mathrm{enc}}}$. Afterwards, $\tilde{\boldsymbol{x}}_{\mathrm{enc}}$ is converted into $N_{\mathrm{enc}}$ spike signals, where the $j$th element of $\tilde{\boldsymbol{x}}_{\mathrm{enc}}$ indicates whether a spike is fired at the beginning of the $j$th spike signal $z_{\mathrm{enc},j}(t)$. For a continuous-time system, we obtain

$$z_{\mathrm{enc},j}(0) = \tilde{x}_{\mathrm{enc},j}\,, \tag{2.25}$$

and for a discrete-time system, we obtain

$$z_{\mathrm{enc},j}[0] = \tilde{x}_{\mathrm{enc},j} , \tag{2.26}$$

where $\tilde{x}_{\mathrm{enc},j}$ denotes the $j$th element of $\tilde{\boldsymbol{x}}_{\mathrm{enc}}$.

Fig. 2.10(c) shows $Q'(|x_{\mathrm{enc}}|)$ for different $N_{\mathrm{enc}}$ and $x_{\mathrm{enc,max}} = 1$. Due to the multiplication by $\mathrm{sign}(x_{\mathrm{enc}})$, the TE can be fully characterized by its equivalent quantization characteristic $\tilde{Q}'(x_{\mathrm{enc}})$, which is illustrated in Fig. 2.10(d), and corresponds to the characteristic of a midtread quantizer.

Fig. 2.10(b) shows the exemplary encoding of $x_{\mathrm{enc}} = 0.1$, $x_{\mathrm{enc}} = -0.1$, and $x_{\mathrm{enc}} = 0.9$, for $x_{\mathrm{enc,max}} = 1$ and $N_{\mathrm{enc}} = 4$. While purple dots denote $z_{\mathrm{enc},j}[\kappa] = 1$, green dots denote $z_{\mathrm{enc},j}[\kappa] = -1$, and $z_{\mathrm{enc},j}[\kappa] = 0$ if no dot is plotted. For $x_{\mathrm{enc}} = 0.1$ and $x_{\mathrm{enc}} = -0.1$ the same subpopulation of neurons is activated. However, the sign of the spikes is inverted. When comparing $x_{\mathrm{enc}} = 0.1$ and $x_{\mathrm{enc}} = 0.9$, a different subpopulation is activated.

### Quantization Encoding

Another novel population encoding, which overcomes the ternary character of TE, is quantization encoding (QE). Its setup is displayed in Fig. 2.11(a). A quantizer $Q : \mathbb{R} \to \{0, 1\}^{N_{\mathrm{enc}}}$ converts $x_{\mathrm{enc}} \in \mathbb{R}$ into $N_{\mathrm{enc}}$ binary values. Similar to TE, we assume that the range of $x_{\mathrm{enc}}$ is centered around zero, thus $x_{\mathrm{enc,min}} = -x_{\mathrm{enc,max}}$. In contrast to TE, $x_{\mathrm{enc}}$ is directly fed to the quantizer. The quantizer is a midrise quantizer, defined by

$$Q(x_{\mathrm{enc}}) = \Delta_{\mathrm{Q}} \left( \left\lfloor \frac{x_{\mathrm{enc}}}{\Delta_{\mathrm{Q}}} \right\rfloor + \frac{1}{2} \right) , \tag{2.27}$$

where

$$\Delta_{\mathrm{Q}} = \frac{x_{\mathrm{enc,max}}}{2^{N_{\mathrm{enc}}-1}} . \tag{2.28}$$

Fig. 2.11(b) displays the characteristics of $Q(x_{\mathrm{enc}})$ for $x_{\mathrm{enc,max}} = 1$ and different $N_{\mathrm{enc}}$. At the output of the quantizer, the bit pattern $\boldsymbol{x}_{\mathrm{enc}} \in \{0, 1\}^{N_{\mathrm{enc}}}$ is transformed into spike signals by a spike converter, analogous to (2.25) and (2.26).

While TE is bipolar and emits both positive and negative spikes, QE solely emits positive spikes and thus has a unipolar character. Comparing the characteristics of QE (Fig. 2.11(b)) and the characteristics of TE (Fig. 2.10(d)), TE has more quantization levels than QE for identical $N_{\mathrm{enc}}$. For QE, the number of quantization levels is $2^{N_{\mathrm{enc}}}$, while for TE it is $2^{N_{\mathrm{enc}}+1} - 1$. Thus, for a fixed $N_{\mathrm{enc}}$, TE has a higher resolution than. However, to apply TE on neuromorphic hardware, the hardware needs to support spikes of negative polarity.

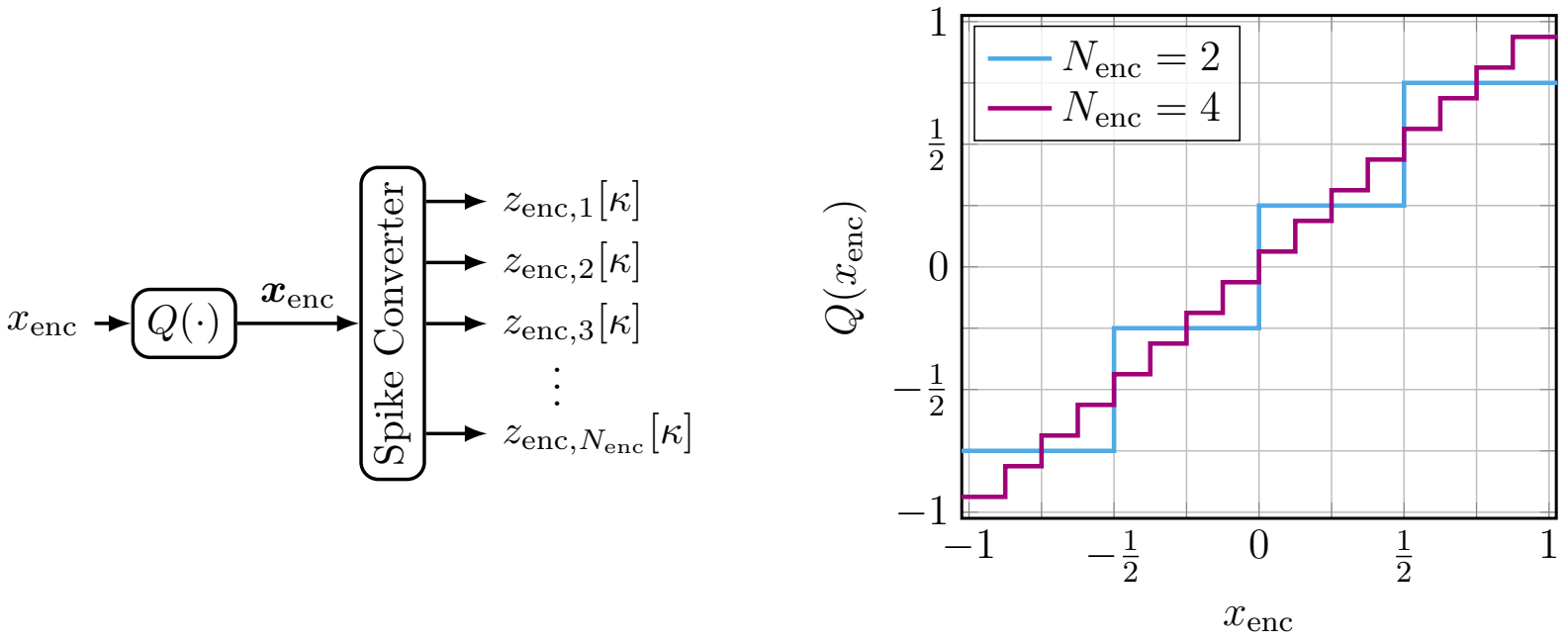

**(a)** Block diagram of QE. **(b)** Quantization characteristics $Q(x_{\text{enc}})$.

**Figure 2.11:** Block diagram and quantization characteristics $Q(\cdot)$ of QE.

## 2.4 Neural Decoding

While neural encoding converts real-world data into the spike domain, *neural decoding* transforms spike signals into a useful representation [MPC⁺24]. Motivated by neural encoding techniques, several neural decoding techniques can be identified in the literature [LBCS24]. The techniques can be categorized based on the type of neurons used in the output layer: either spiking neurons (e.g., the LIF neuron) or non-spiking neurons (e.g., the LI neuron).

To provide a more intuitive understanding of the different neural encoding schemes, we assume a classification task with $N_\text{c}$ distinct classes, where $i = 1, \ldots, N_\text{c}$ denotes the class labels. To solve the classification task using an SNN, its output layer typically consists of $N_{\text{out}} = N_\text{c}$ output neurons, where each neuron represents a class. We furthermore let $z_{\text{out},j}(t)$, respectively $z_{\text{out},j}[\kappa]$, denote the spike signal emitted by the $j$th output neuron with $j = 1, \ldots, N_{\text{out}}$. Furthermore, $v_{\text{out},j}(t)$, respectively $v_{\text{out},j}[\kappa]$, denote the membrane potential of the output neurons.

### 2.4.1 Neural Decoding Using Spiking Neurons

If the output layer contains spiking neurons, e.g., LIF neurons or the SRM model, spikes are assumed to contain the information needed for classification. In the following, a rate-based and a temporal-based approach are introduced.

#### Spike Rate Decoding

To perform a classification task, the spikes emitted by each neuron in the output layer are counted. The estimated class label $\hat{\imath}$ is then determined by selecting the class associated with the output neuron that generated the highest number of

spikes:

$$\hat{\imath} = \underset{j=1,2,\ldots,N_\mathrm{c}}{\arg\max} \int_{t=0}^{T} z_{\mathrm{out},j}(t)\,\mathrm{d}t\,, \tag{2.29}$$

respectively

$$\hat{\imath} = \underset{j=1,2,\ldots,N_\mathrm{c}}{\arg\max} \sum_{\kappa=0}^{K-1} z_{\mathrm{out},j}[\kappa]\,. \tag{2.30}$$

Spike rate decoding has been successfully applied to classify the MNIST dataset [DC15].

**Time-to-first-spike Decoding**
For TTFS decoding, the first spike fired contains all the information and indicates $\hat{\imath}$. Let $t_{\mathrm{out},\mathrm{z},j}$ (respectively $\kappa_{\mathrm{out},\mathrm{z},j}$) denote the time of the first spike fired by the $j$th output neuron. We can write the classification as

$$\hat{\imath} = \underset{j=1,2,\ldots,N_\mathrm{c}}{\arg\min}\ t_{\mathrm{out},\mathrm{z},j}\,, \tag{2.31}$$

or, in the discrete-time case, as

$$\hat{\imath} = \underset{j=1,2,\ldots,N_\mathrm{c}}{\arg\min}\ \kappa_{\mathrm{out},\mathrm{z},j}\,. \tag{2.32}$$

### 2.4.2 Neural Decoding Using Non-spiking Neurons

Instead of output spikes, the membrane potential of non-spiking neurons, e.g., the LI neuron, can be exploited. For spiking LIF neurons, the possible values of the membrane potential $v(t)$ are limited to $v(t) \in [-\infty, v_\mathrm{th}]$. The membrane potential of a LI neuron can take any real number, $v(t) \in \mathbb{R}$.

**End-of-time Membrane-potential Decoding**
For end-of-time membrane-potential (EOTM) decoding, the SNN is simulated for duration $T$ (or $K$ discrete time steps). The output neuron that exhibits the highest membrane potential $v_{\mathrm{out},j}(T)$ indicates the respective class label [BvBS23]

$$\hat{\imath} = \underset{j=1,2,\ldots,N_\mathrm{c}}{\arg\max}\ v_{\mathrm{out},j}(T)\,, \tag{2.33}$$

respectively

$$\hat{\imath} = \underset{j=1,2,\ldots,N_\mathrm{c}}{\arg\max}\ v_{\mathrm{out},j}[K]\,. \tag{2.34}$$

#### Max-of-time Membrane-potential Decoding

Functionally, max-of-time membrane potential (MOTM) is similar to EOTM. However, the maximum of $v_{\text{out},j}(t)$, respectively $v_{\text{out},j}[\kappa]$, is taken over time. The neuron that exhibits the highest maximum membrane potential during a predefined interval indicates the estimated class label $\hat{\imath}$ [CSSZ22]:

$$\hat{\imath} = \underset{j=1,2,\ldots,N_\text{c}}{\arg\max} \max_{t\in[0,T]} v_{\text{out},j}(t)\,, \tag{2.35}$$

respectively

$$\hat{\imath} = \underset{j=1,2,\ldots,N_\text{c}}{\arg\max} \max_{\kappa=0,1,\ldots,K-1} v_{\text{out},j}[\kappa]\,. \tag{2.36}$$

#### Comparison of Decoding Strategies

In [LBCS24], spike rate, TTFS, EOTM and MOTM decoding are compared when solving a classification task. EOTM and MOTM decoding perform best. For a small $T$ (respectively $K$), EOTM and MOTM decoding achieve good classification performance, outperforming spike rate and TTFS decoding. Furthermore, when EOTM and MOTM are used, a low number of spikes is emitted by the neurons of the hidden layer. Thus, EOTM and MOTM decoding seem to be a proper choice for fast and low-energy classification.

## 2.5 Update Algorithms

When solving a specific task, a neural network (NN) needs the appropriate connectivity [ZBC+21]. For artificial neural networks (ANNs), the backpropagation (BP) algorithm [RHW86] in combination with an update algorithm, such as gradient descent, has enabled efficient optimization of the parameters of an ANN, launching the era of deep learning (DL). The BP algorithm computes the gradient of the loss function with respect to the network parameters by propagating an error signal backward through the network, thereby quantifying the contribution of each parameter to the overall error. In a supervised learning task, the loss function typically quantifies the difference between the actual network output and the desired output. For ANNs, the backpropagation of the gradient is guaranteed by well-behaved differentiable activation functions inside the neurons, such as the rectified linear unit (ReLU) or hyperbolic tangent function. However, the derivative of the Heaviside step function $H(\cdot)$ is zero almost everywhere, except at the discontinuity, where it is not well-defined. Consequently, gradient-based optimization fails in SNNs [ZBC+21].

Several alternative update algorithms exist for SNNs, differing in both biological plausibility and computational efficiency. However, the choice of update algorithm remains the subject of ongoing debate [ZBC+21]. The following section offers a brief overview of various selected update algorithms. Afterwards, the three update algorithms investigated in this thesis are discussed in more detail.

### 2.5.1 Biologically Plausible Update Algorithms

Donald Hebb introduced *Hebb's rule* [Heb49], a theory on how neurons strengthen their connection depending on their relative firing activity. Let two neurons A and B be connected via a synapse, where postsynaptic neuron B receives spike signals from presynaptic neuron A. Hebb claimed that if the firing of the presynaptic neuron A regularly contributes to the firing of the postsynaptic neuron B, the connecting synapse between both should be strengthened. Hebb's rule is often summarized as "neurons that fire together, wire together" [Heb49].

Hebb's rule ignores the relative timing of spikes. A refined version of Hebb's rule is *spike-timing-dependent plasticity (STDP)*, which was observed during in vitro studies with human brain cells [BP98]. Again, two neurons are connected via a synapse. Both neurons fire a single spike with firing times $t_{\mathrm{pre}}$ and $t_{\mathrm{post}}$, respectively. The relative firing time between both neurons is calculated by $\Delta t_{\mathrm{f}} = t_{\mathrm{post}} - t_{\mathrm{pre}}$. In Fig. 2.12, the synaptic modification $\Delta\theta$ as a function of $\Delta t_{\mathrm{f}}$ is sketched, which has been experimentally verified in [FTR+06]. If the postsynaptic neuron spikes later than the presynaptic neuron, i.e., $t_{\mathrm{post}} > t_{\mathrm{pre}}$ and $\Delta t_{\mathrm{f}} > 0$, it is likely that the spike of the presynaptic neuron contributed to the firing of the postsynaptic neuron, and the synapse is strengthened by $\Delta\theta \sim 1/\Delta t_{\mathrm{f}}$ (potentiation). If both spikes occur very closely in time, the synapse is significantly strengthened. As the time difference $\Delta t_{\mathrm{f}}$ increases, the synapse undergoes only a weak update. In contrast, if $t_{\mathrm{post}} < t_{\mathrm{pre}}$ and thus $\Delta t_{\mathrm{f}} < 0$, the spike of the presynaptic neuron does not contribute to the spike of the postsynaptic neuron. The synapse is weakened, where the change in the strength of the synapse is proportional to $\Delta\theta \sim -1/\Delta t_{\mathrm{f}}$ (depression). The STDP-based update $\Delta\theta$ of a synapse can be described by [FTR+06]

$$\Delta\theta(\Delta t_{\mathrm{f}}) = \begin{cases} A_{+}\mathrm{e}^{-\frac{\Delta t_{\mathrm{f}}}{\tau_{+}}}, & \text{if } \Delta t_{\mathrm{f}} > 0\,, \\ A_{-}\mathrm{e}^{-\frac{\Delta t_{\mathrm{f}}}{\tau_{-}}}, & \text{if } \Delta t_{\mathrm{f}} < 0\,, \\ 0, & \text{if } \Delta t_{\mathrm{f}} < 0\,, \end{cases} \tag{2.37}$$

where $A_{+} \in \mathbb{R}^{+}$ and $A_{-} \in \mathbb{R}^{-}$ denote scaling factors, and $\tau_{+} \in \mathbb{R}^{+}$ and $\tau_{-} \in \mathbb{R}^{+}$ are the time constants of potentiation and depression, respectively.

Both Hebb's rule and STDP are motivated by biology. However, for both, the information for the adaptation of the synapse is only locally available [TGK+19]. Hence, training is limited to supervised learning without hidden units [ZG18], or unsupervised learning, which leads to a degradation of accuracy [LDP16, FG16].

In [RW02], different in vitro and in vivo studies are summarized, suggesting that synapse updates depend not only on the timing of the spikes but also on a global signal, which is given by the level of different neurotransmitters. For example, at a high dopamine level, synapses are rather strengthened. At a weak dopamine level, the synapses are weakened. Inspired by the findings above, *three-factor rules* were introduced. Three-factor rules can overcome the locality of STDP by introducing a global signal, e.g., a "reward" or "punishment" signal, indicating the performance of the SNN on a specific task [FG16]. Hence, synapse updates depend on three factors: the timing $t_\mathrm{pre}$ of the spike fired by the presynaptic neuron, the timing $t_\mathrm{post}$ of the spike fired by the postsynaptic neuron, and the global signal.

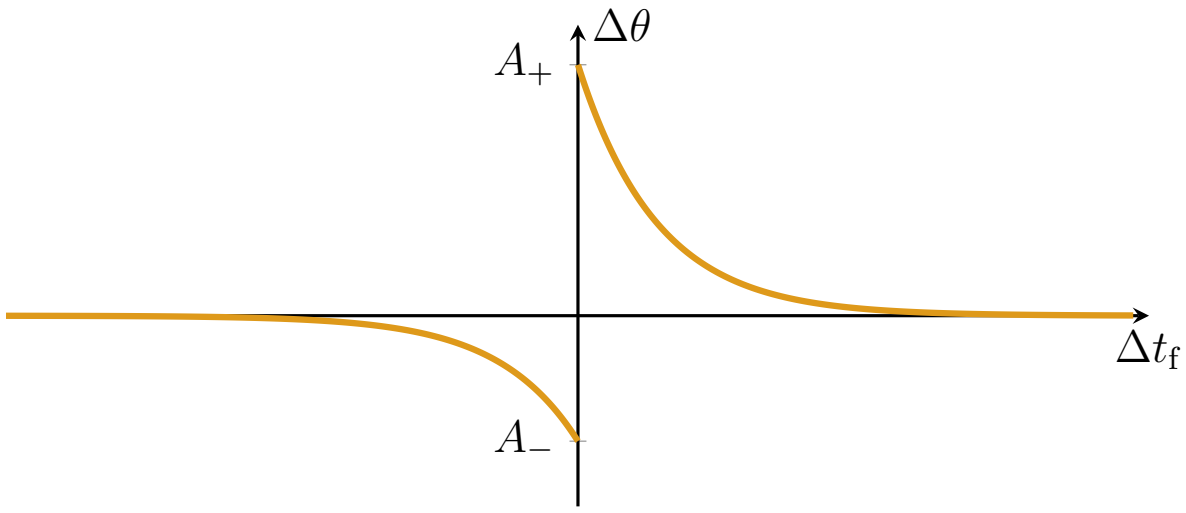

**Figure 2.12:** Sketch of the synaptic modification $\Delta\theta$ as a function of the relative firing time $\Delta t_\mathrm{f} = t_\mathrm{post} - t_\mathrm{pre}$ [FTR$^+$06, Fig. 1C]. For $\Delta t_\mathrm{f} > 0$ the synapse is strengthened (potentiation), for $\Delta t_\mathrm{f} < 0$ the synapse is weakened (depression).

We can summarize that biologically plausible update algorithms are limited by either their locality (Hebb's rule, STDP) or their heuristic approach and slow convergence (three-factor rules). In [ZBC$^+$21], the authors discuss the future of SNNs. Instead of biologically plausible updating algorithms, gradient-based updating algorithms seem to be a promising alternative.

### 2.5.2 Gradient-based Update Algorithms

In ANNs, the de facto standard for updating neural connections and addressing the spatial credit assignment problem is gradient-based optimization, typically implemented via the BP algorithm combined with methods such as gradient descent. To enable the BP of gradient information through the network, neural activation functions with well-defined derivatives are required [ZBC$^+$21]. For SNNs, the neural activation function is often modeled by the Heaviside step function $H(\cdot)$, see (2.6). Its derivative poses a problem because it is zero almost everywhere, except at $v = v_\mathrm{th}$, where it is not well-defined [ZG18, NMZ19]. In recent years, different approaches for gradient-based optimization have been introduced, which circumvent the gradient of the Heaviside step function in various ways.

One approach is the ANN-to-SNN conversion [DNB$^+$15]. The ReLU function used in ANNs can be seen as a firing rate approximation of a spiking LIF neuron without leakage, a so-called *integate-and-fire (IF)* neuron. Given an input $x \in \mathbb{R}$, the output of the ReLU function is given by $f(x) = \max(0, x)$. When charging an IF neuron with a constant current of amplitude $x$, the neuron outputs a signal with a rate that is directly proportional to $x$, provided $x > 0$. For negative inputs $x < 0$, no output spike is generated at all. Due to the similar behavior, instead of directly optimizing an SNN, an ANN with an alike architecture, the ReLU function as the activation function, and zero bias is optimized as a proxy. Afterwards, the optimized weights are transferred to an SNN consisting of spiking IF neurons. For the classification of the handwritten MNIST dataset, the SNN updated using ANN-to-SNN conversion nearly achieves similar performance as the ANN [DNB$^+$15].

It is important to emphasize that the ANN-to-SNN conversion relies on spike rates instead of precise spike timings. Hence, the ANN-to-SNN conversion is not suitable for processing temporally encoded data. In the following, two gradient-based methods are discussed, both of which account for spike timing.

### Surrogate Gradients

Using the BP algorithm, the global nature of the reward in three-factor learning rules can be eliminated [ZG18]. Due to BP, the impact of each synapse weight on the reward (error) can be tracked, making the reward weight-specific, not global. To enable BP, $H(v(t) - v_{\text{th}})$ is approximated by a smooth differentiable auxiliary function $\sigma_\eta(v(t) - v_{\text{th}})$ [ZBC$^+$21], where $\sigma_\eta : \mathbb{R} \to \mathbb{R}$ is a monotonic function that increases steeply close to $v = v_{\text{th}}$, and its derivative peaks at $v = v_{\text{th}}$. An example is the fast sigmoid function [ZG18]

$$\sigma_\eta(v - v_{\text{th}}) = \frac{1}{2} \cdot \frac{\eta \cdot (v - v_{\text{th}})}{1 + |\eta \cdot (v - v_{\text{th}})|} + \frac{1}{2}, \tag{2.38}$$

where $\eta \in \mathbb{R}^+$ defines the steepness of the sigmoid function. The derivative is given by [ZG18, CSSZ22]

$$\sigma'_\eta(v - v_{\text{th}}) := \frac{\partial \sigma_\eta(v - v_{\text{th}})}{\partial v} = \frac{1}{2} \cdot \frac{\eta}{(1 + |\eta(v - v_{\text{th}})|)^2}. \tag{2.39}$$

Fig. 2.13 shows both the Heaviside function $H(\cdot)$ and the fast sigmoid function $\sigma_\eta(\cdot)$, as well as their derivatives for $v_{\text{th}} = 1$. For the Heaviside function, the derivative is zero everywhere except at $v = v_{\text{th}}$, where it is not well-defined. The derivative of $\sigma_\eta(\cdot)$ is well-defined everywhere. Furthermore, its steepness can be controlled by $\eta$. Since the gradient of $H(\cdot)$ is approximated by a surrogate, the method is referred to as surrogate gradient (SG).

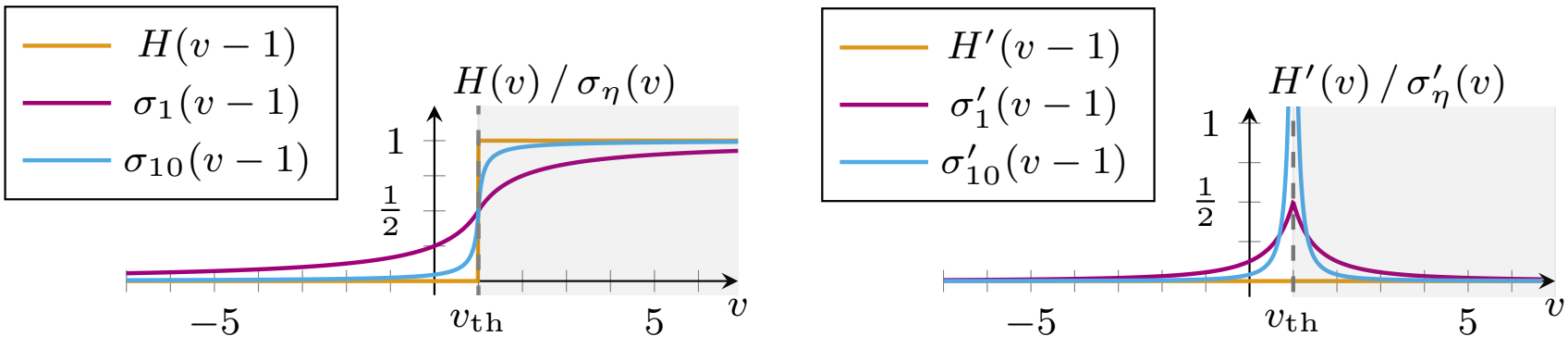

**(a)** **(b)**

**Figure 2.13:** (a) The Heaviside step function $H(\cdot)$ and the fast sigmoid function $\sigma_\eta(\cdot)$ with different steepness controlled by $\eta$, and $v_{\text{th}} = 1$.
(b) Derivative of $H(\cdot)$ and $\sigma_\eta(\cdot)$ for $v_{\text{th}} = 1$. The gray dashed line marks the spiking threshold $v_{\text{th}}$, whereas the gray shaded region indicates a domain that is inaccessible as a result of the spike-generation mechanism. The figure is based on [NMZ19, Fig. 3].

### Backpropagation Through Time with Surrogate Gradients

Methods based on SG enable the optimization of SNNs using standard (and well-studied) machine learning (MaL) software and techniques [ZBC+21]. For ANNs with a recurrent structure, so-called recurrent neural networks (RNNs), the backpropagation through time (BPTT) algorithm enables the gradient flow through the recurrent structure [GBC16, Ch. 10.2]. To achieve this, the computational graph of an RNN is unrolled over time, and the gradient is backpropagated through the unrolled structure.

SNNs feature two forms of recurrence: explicit recurrence, stemming from recurrent synaptic connections, and implicit recurrence, arising from the intrinsic temporal dynamics of spiking neurons. Similar to RNNs, the computational graph of an SNN can be unrolled over time. Afterwards, the BP algorithm can be applied. To enable gradient flow, SG can be applied, resulting in BPTT with SG [NMZ19, BSS+18].

When updating SNNs using BPTT with SG, the forward path of the SNN is simulated on a discrete-time grid. For the simulation of an LIF neuron, (2.4), (2.5) and (2.6) are implemented. The equations describe the temporal dynamics and can be converted into a computational graph, as illustrated in Fig. 2.14(a). The input spike signal $\boldsymbol{z}_{\text{in}}[\kappa] \in \{0,1\}^J$ from $J$ presynaptic neurons denotes the spikes at time instant $\kappa$, which are scaled by the respective weights of the synapse $\boldsymbol{\theta} \in \mathbb{R}^J$, and then summed up. To better visualize the temporal dependencies, the computational graph of the LIF neuron can be unrolled in time, as illustrated in Fig. 2.14(b).

Akin to the LIF neuron, an SNN can be described by its computational graph, where the unrolled graph displays spatial and temporal dependencies inside a neuron and between different neurons [NMZ19]. Similar to the optimization of RNNs and ANNs, a loss function $\mathcal{L}(\cdot)$ quantifies the error between the desired and actual output w.r.t. the learnable model parameters. For each time instant of the network, the error can be backpropagated not only from the output layer to the input layer, but also through the unrolled network and, thus, through time [GBC16, Ch. 10.2]. BPTT with SG enables the use of autodifferentiation tools and, hence, the use of standard MaL toolkits [NMZ19]. Despite its lack of biological plausibility, BPTT with SG is discussed as a promising candidate for future optimization of SNNs [ZBC+21].

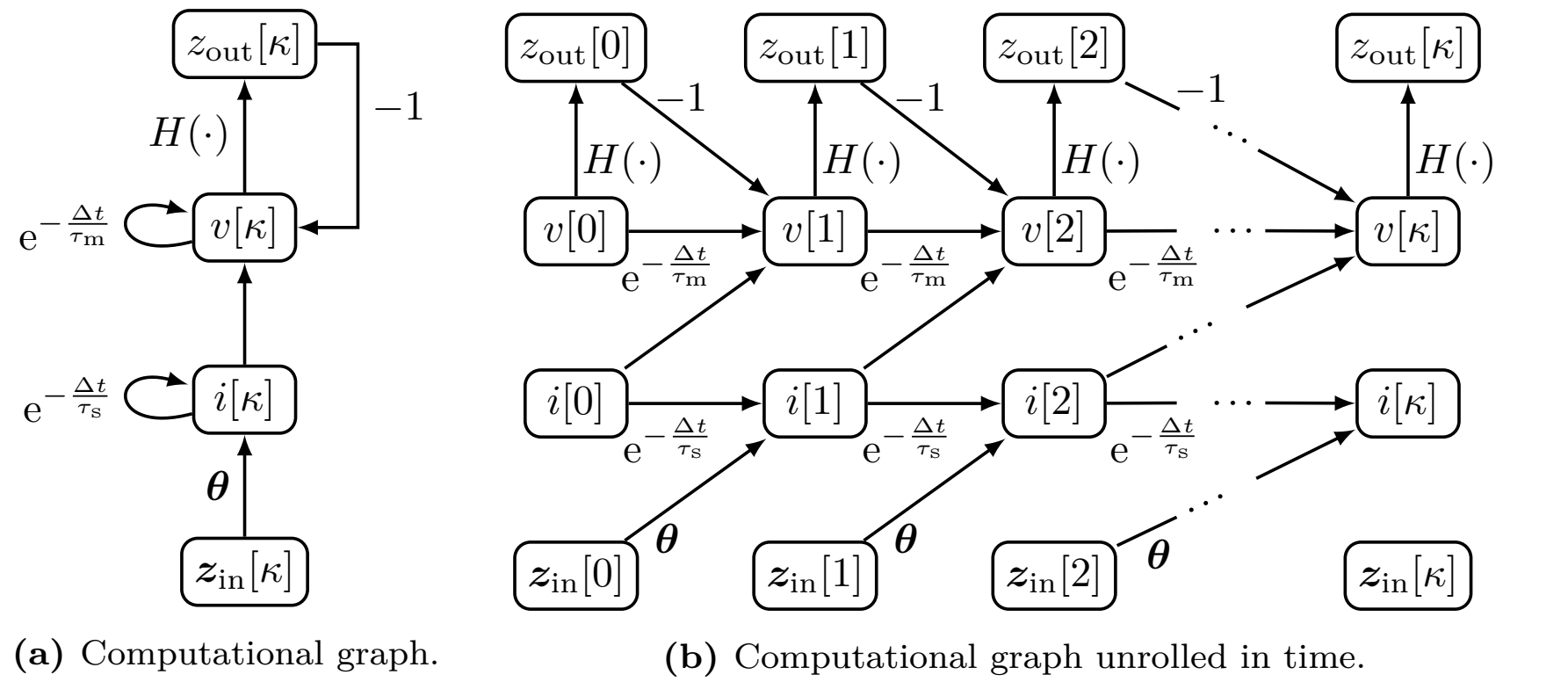

**Figure 2.14:** (a) Computational graph of a LIF neuron, realizing (2.4), (2.5) and (2.6). (b) Computational graph of a LIF neuron, unrolled in time [NMZ19, Fig. 2].

### 2.5.3 Update Algorithm for Probabilistic Spiking Neural Networks

The update rules discussed so far assumed a deterministic neuron model. By replacing the deterministic model with the SRM, a probabilistic SNN is obtained, which can updated using gradient-based methods [JSGG19]. For probabilistic SNNs, the spiking behavior of each neuron follows a probability distribution $p_{\mathrm{fire}}(Z_{\mathrm{out}}[\kappa]|v[\kappa])$, see (2.11). Thus, the spike signals emitted by all neurons can be modeled as a jointly distributed binary random process, and learning consists of finding the model parameters that maximize the likelihood of observing a desired realization of this process.

We distinguish between *fully observed* and *partially observed* probabilistic SNNs. If all neurons follow a predefined and thus known spike pattern, we call the SNN fully observed. If, e.g., the SNN consists of an input and output layer, where the spike signals of the input layer are defined by the encoded input data and the spike signals of the output layer are predefined by some desired spike signals (target), the SNN is fully observed. In contrast, if the SNN is extended by at least one hidden layer, the spike signals of the hidden neurons are not predefined and hence unknown. We call these neurons *latent* neurons, and the resulting SNN is partially observed. Let $\mathcal{Z}$ be the set of observable neurons, and $\mathcal{H}$ be the set of latent (hidden) neurons. Then, $\mathcal{N} = \mathcal{Z} \cup \mathcal{H}$ denotes the set of neurons of the SNN. Consequently, if $\mathcal{H} = \emptyset$, the SNN is fully observed. We now briefly motivate and discuss the update rule for probabilistic SNNs [JSGG19]. Therefore, we assume a supervised learning task with the training data set $\mathcal{B}_\mathrm{t}$ containing $|\mathcal{B}_\mathrm{t}|$ training samples.

**Fully observed SNN**

First, we assume a probabilistic fully observed SNN, hence $\mathcal{H} = \emptyset$. The input layer is directly connected to the output layer, with the connecting weights defined by the parameter vector $\boldsymbol{\theta} \in \mathbb{R}^d$, where the dimensionality $d \in \mathbb{N}$ depends on the architecture of the SNN, e.g., if recurrent connections are applied. The spiking of the $j$th neuron at time instant $\kappa$ is modeled by the RV $Z_j[\kappa] \in \{0,1\}$, with $j = 1, 2, \ldots, N_\mathrm{in} + N_\mathrm{out}$, and $\kappa = 0, 1, \ldots, K-1$. Let[2]

$$\boldsymbol{Z}_\kappa = (Z_1[\kappa], Z_2[\kappa], \ldots, Z_j[\kappa], \ldots, Z_{N_\mathrm{in}+N_\mathrm{out}}[\kappa]) \,, \quad \boldsymbol{Z}_\kappa \in \{0,1\}^{(N_\mathrm{in}+N_\mathrm{out})} \tag{2.40}$$

denote a vector of RVs modeling the spiking at time instant $\kappa$. We then define the spike history $\boldsymbol{Z}_{\leq K-1}$ of all neurons at all time steps as

$$\boldsymbol{Z}_{\leq K-1} \coloneqq (\boldsymbol{Z}_0, \boldsymbol{Z}_1, \ldots, \boldsymbol{Z}_{K-1}) \,, \quad \boldsymbol{Z}_{\leq K-1} \in \{0,1\}^{(N_\mathrm{in}+N_\mathrm{out})\times(K)} \,, \tag{2.41}$$

and

$$\boldsymbol{Z}_{\leq \kappa} \coloneqq (\boldsymbol{Z}_0, \boldsymbol{Z}_1, \ldots, \boldsymbol{Z}_{\kappa}) \,, \quad \boldsymbol{Z}_{\leq \kappa} \in \{0,1\}^{(N_\mathrm{in}+N_\mathrm{out})\times(\kappa+1)} \tag{2.42}$$

as the spike history up to time instant $\kappa$. We furthermore define

$$p_{\boldsymbol{\theta}}(\boldsymbol{Z}_{\leq K-1}) \coloneqq p(\boldsymbol{Z}_{\leq K-1}|\boldsymbol{\theta}) \tag{2.43}$$

as the probability $p_{\boldsymbol{\theta}}(\boldsymbol{Z}_{\leq K-1}) \in [0,1]$ of observing $\boldsymbol{Z}_{\leq K-1}$ given the model parameters $\boldsymbol{\theta}$. The objective is to find the parameter vector $\boldsymbol{\theta}$ that transforms the input spikes into the desired output spikes, thereby enabling the observation of $\boldsymbol{Z}_{\leq K-1}$. Using the maximum likelihood (ML) criterion, we can obtain the optimal parameters $\boldsymbol{\theta}^*$ as

$$\boldsymbol{\theta}^* = \underset{\boldsymbol{\theta}\in\mathbb{R}^d}{\arg\max}\ p_{\boldsymbol{\theta}}(\boldsymbol{Z}_{\leq K-1}) \,. \tag{2.44}$$

Instead of maximizing $p_{\boldsymbol{\theta}}(\boldsymbol{Z}_{\leq K-1})$, we can maximize the log-likelihood

$$\mathcal{L}_{\leq K-1}(\boldsymbol{\theta}) \coloneqq \ln(p_{\boldsymbol{\theta}}(\boldsymbol{Z}_{\leq K-1})) \,, \tag{2.45}$$

[2]For compactness, the time index in probabilistic SNNs is written as a subscript.

and obtain $\boldsymbol{\theta}^*$ as

$$\boldsymbol{\theta}^* = \arg\max_{\boldsymbol{\theta}\in\mathbb{R}^d} \mathcal{L}_{\leq K-1}(\boldsymbol{\theta})\,. \tag{2.46}$$

Since $p_{\boldsymbol{\theta}}(\boldsymbol{Z}_{\leq K-1}) \in [0,1]$ and since the logarithm is a strictly monotonically increasing function, (2.44) and (2.46) are equivalent. To maximize $\mathcal{L}_{\leq K-1}(\boldsymbol{\theta})$, the parameters $\boldsymbol{\theta}$ can be updated using gradient ascent

$$\boldsymbol{\theta} \leftarrow \boldsymbol{\theta} + \nu \cdot \nabla_{\boldsymbol{\theta}} \mathcal{L}_{\leq K-1}(\boldsymbol{\theta})\,, \tag{2.47}$$

where $\nu \in [0,1]$ is the learning rate. We can obtain the gradient $\nabla_{\boldsymbol{\theta}} \mathcal{L}_{\leq K-1}(\boldsymbol{\theta})$ by

$$\nabla_{\boldsymbol{\theta}} \mathcal{L}_{\leq K-1}(\boldsymbol{\theta}) = \nabla_{\boldsymbol{\theta}} \ln(p_{\boldsymbol{\theta}}(\boldsymbol{Z}_{\leq K-1})) \tag{2.48}$$

$$= \sum_{\kappa=0}^{K-1} \sum_{j\in\mathcal{Z}} \nabla_{\boldsymbol{\theta}_j} \ln\Big(p_{\boldsymbol{\theta}_j}\Big(Z_j[\kappa]\Big|v_j[\kappa]\Big)\Big)\,, \tag{2.49}$$

where we exploit that the probability of observing $\boldsymbol{Z}_{\leq K-1}$ can be expressed as the product of the individual probabilities $p_{\boldsymbol{\theta}_j}(Z_j[\kappa]|v_j[\kappa])$ of all neurons $j \in \mathcal{Z}$ at all time steps $\kappa$. For each neuron $j$, we can now calculate the gradient w.r.t. its feed-forward weights $\boldsymbol{\theta}_j^{(\mathrm{f})}$ and its recurrent weight $\theta_j^{(\mathrm{r})}$. It is given by

$$\nabla_{\boldsymbol{\theta}_j^{(\mathrm{f})}} \ln\Big(p_{\boldsymbol{\theta}_j}\Big(Z_j[\kappa]\Big|v_j[\kappa]\Big)\Big) = (\boldsymbol{z}_{\mathrm{in},j} * \alpha)[\kappa] \cdot (Z_j[\kappa] - \sigma(v_j[\kappa]))\,, \tag{2.50}$$

$$\nabla_{\theta_j^{(\mathrm{r})}} \ln\Big(p_{\boldsymbol{\theta}_j}\Big(Z_j[\kappa]\Big|v_j[\kappa]\Big)\Big) = (z_{\mathrm{out},j} * \beta)[\kappa-1] \cdot (Z_j[\kappa] - \sigma(v_j[\kappa]))\,, \tag{2.51}$$

where $\boldsymbol{z}_{\mathrm{in},j}[\kappa] \in \{0,1\}^J$, $J \in \mathbb{N}^+$, is a vector containing the input signals received by the $J$ presynaptic neurons of neuron $j$, and $z_{\mathrm{out},j}[\kappa] \in \{0,1\}$ the spike signal fed back from neuron $j$ itself. A detailed derivation is given in App. A.1.

If the presynaptic neurons do not input any spikes, i.e., $(\boldsymbol{z}_{\mathrm{in},j} * \alpha)[\kappa] = \mathbf{0}$ respectively $(z_{\mathrm{out},j} * \beta)[\kappa] = 0$, the gradient obtained by (2.50) and (2.51) equals zero. Thus, the respective synapse is not updated. Next we investigate the update when a presynaptic spike is fired. Remember that $Z_j[\kappa] \in \{0,1\}$ models the target spike behavior and $\sigma(v_j[\kappa]) \in [0,1]$ denotes the probability of firing. For both $Z_j[\kappa] = 1$ and high values of $\sigma(v_j[\kappa])$ as well as $Z_j[\kappa] = 0$ and low values of $\sigma(v_j[\kappa])$, the neuron is close to the desired behavior. Hence, the resulting gradient update is small, and presynaptic spikes correctly contribute to the firing of an output spike. However, if $Z_j[\kappa] = 1$ and $\sigma(v_j[\kappa])$ is small, the gradient is positive, increasing the respective weights and therefore the impact of incoming presynaptic spikes on the membrane potential $v[\kappa]$. In contrast, if $Z_j[\kappa] = 0$ and $\sigma(v_j[\kappa])$ is large, the gradient becomes negative, decreasing the respective weights and thus the impact of incoming presynaptic spikes. The update resembles Hebb's principle: if presynaptic spikes are unlikely to induce a desired output spike, the respective weights are increased. Moreover, if presynaptic spikes are likely to induce an unwanted output spike, the respective weights are decreased.

For batch optimization using the $|\mathcal{B}_\mathrm{t}|$ samples from the training set $\mathcal{B}_\mathrm{t}$, we can modify (2.47) to

$$\boldsymbol{\theta} \leftarrow \boldsymbol{\theta} + \nu \frac{1}{|\mathcal{B}_\mathrm{t}|} \sum_{m\in\mathcal{B}_\mathrm{t}} \sum_{\kappa=0}^{K-1} \sum_{j\in\mathcal{Z}} \nabla_{\boldsymbol{\theta}_j} \ln\Big(p_{\boldsymbol{\theta}_j}\left(Z_j^{(m)}[\kappa]\Big|v_j^{(m)}[\kappa]\right)\Big) , \tag{2.52}$$

where the superscript $m$ assigns the respective variable to the $m$th sample of the training set.

**Partially observed SNN**

Next, the model contains hidden neurons, i.e., $\mathcal{H} \neq \emptyset$. Using the same notation as for $\boldsymbol{Z}_{\leq K-1}$, we define $\boldsymbol{H}_{\leq K-1} \in \{0,1\}^{|\mathcal{H}|\times K}$ as a matrix of RVs modeling the spike history of latent neurons. Consequently, $\boldsymbol{H}_\kappa$ models the spike behavior of all hidden neurons at time instant $\kappa$. Note that $\boldsymbol{H}_{\leq K-1}$ has $2^{|\mathcal{H}|\cdot K}$ possible realizations.

Again, we can formulate the learning problem as maximizing the probability of observing the desired output spike pattern $\boldsymbol{Z}_{\leq K-1}$, see (2.44). The gradient of (2.46) w.r.t. $\boldsymbol{\theta}$ is obtained by

$$\nabla_{\boldsymbol{\theta}} \mathcal{L}_{\leq K-1}(\boldsymbol{\theta}) = \mathbb{E}_{p_{\boldsymbol{\theta}}(\boldsymbol{H}_{\leq K-1}|\boldsymbol{Z}_{\leq K-1})}\Big\{\nabla_{\boldsymbol{\theta}} \ln(p_{\boldsymbol{\theta}}(\boldsymbol{Z}_{\leq K-1}, \boldsymbol{H}_{\leq K-1}))\Big\} . \tag{2.53}$$

We can compute the posterior by

$$p_{\boldsymbol{\theta}}(\boldsymbol{H}_{\leq K-1}|\boldsymbol{Z}_{\leq K-1}) = \frac{p_{\boldsymbol{\theta}}(\boldsymbol{H}_{\leq K-1}, \boldsymbol{Z}_{\leq K-1})}{p_{\boldsymbol{\theta}}(\boldsymbol{Z}_{\leq K-1})} \tag{2.54}$$

$$= \frac{p_{\boldsymbol{\theta}}(\boldsymbol{H}_{\leq K-1}, \boldsymbol{Z}_{\leq K-1})}{\sum_{\boldsymbol{H}'_{\leq K-1}} p_{\boldsymbol{\theta}}(\boldsymbol{H}'_{\leq K-1}, \boldsymbol{Z}_{\leq K-1})} , \tag{2.55}$$

where the marginal distribution $p_{\boldsymbol{\theta}}(\boldsymbol{Z}_{\leq K-1})$ is obtained via marginalization over all $\boldsymbol{H}'_{\leq K-1} \in \{0,1\}^{|\mathcal{H}|\times K}$. With an increasing number of hidden neurons $N_\mathrm{hid} = |\mathcal{H}|$ and time instants $K$, the combinatorial space becomes too large, and the marginal distribution is computationally infeasible.

To overcome this problem, *variational inference* can be used [JSGG19]. Following the derivation of [JSGG19], the update rules for batch optimization using the training set $\mathcal{B}_\mathrm{t}$ are obtained by (see App. A.2)

$$\boldsymbol{\theta}_\mathcal{Z} \leftarrow \boldsymbol{\theta}_\mathcal{Z} + \nu_{\boldsymbol{\theta}_\mathcal{Z}} \frac{1}{|\mathcal{B}_\mathrm{t}|} \sum_{m\in\mathcal{B}_\mathrm{t}} \sum_{\kappa=0}^{K-1} \sum_{j\in\mathcal{Z}} \nabla_{\boldsymbol{\theta}_j} \ln\Big(p_{\boldsymbol{\theta}_j}\left(Z_j^{(m)}[\kappa]\Big|v_j^{(m)}[\kappa]\right)\Big) \tag{2.56}$$

$$\boldsymbol{\theta}_{\mathcal{H}} \leftarrow \boldsymbol{\theta}_{\mathcal{H}} + \nu_{\boldsymbol{\theta}_{\mathcal{H}}} \frac{1}{|\mathcal{B}_{\mathrm{t}}|} \sum_{m \in \mathcal{B}_{\mathrm{t}}} \ell^{(m)}_{\boldsymbol{\theta}_{\mathcal{Z}}}(\boldsymbol{Z}_{\leq K-1}) \sum_{\kappa=0}^{K-1} \sum_{j \in \mathcal{H}} \nabla_{\boldsymbol{\theta}_j} \ln\Big(p_{\boldsymbol{\theta}_j}\left(Z_j^{(m)}[\kappa] \Big| v_j^{(m)}[\kappa]\right)\Big) , \tag{2.57}$$

where the parameter vector $\boldsymbol{\theta}$ is divided into parameters $\boldsymbol{\theta}_{\mathcal{Z}}$ linked to observable neurons and parameters $\boldsymbol{\theta}_{\mathcal{H}}$ linked to hidden neurons. Again, the superscript $m$ assigns the respective variable to the $m$th sample of the training set, and $\nu_{\boldsymbol{\theta}_{\mathcal{Z}}} \in [0, 1]$ and $\nu_{\boldsymbol{\theta}_{\mathcal{H}}} \in [0, 1]$ denote learning rates. Furthermore, the so-called learning signal $\ell_{\boldsymbol{\theta}_{\mathcal{Z}}}(\boldsymbol{Z}_{\leq K-1}) \in \mathbb{R}$ is given by

$$\ell_{\boldsymbol{\theta}_{\mathcal{Z}}}(\boldsymbol{Z}_{\leq K-1})^{(m)} = \sum_{\kappa=0}^{K-1} \sum_{j \in \mathcal{Z}} \ln p_{\boldsymbol{\theta}_j} \left(\left(Z_j^{(m)}[\kappa] \Big| v_j^{(m)}[\kappa]\right)\right) . \tag{2.58}$$

The comparison of (2.56) with (2.52) reveals that the update of the observable neurons does not change. However, the update of hidden neurons is scaled by the learning signal $\ell_{\boldsymbol{\theta}_{\mathcal{Z}}}(\boldsymbol{Z}_{\leq K-1})$, which acts as a global reward, indicating the plausibility of a hidden layer spike pattern $\boldsymbol{H}_{\leq K-1}$. When comparing the update rules with the biological update rules, it is worth noting that observable neurons are updated by a Hebbian-like update rule, whereas hidden neurons are updated by a three-factor rule, where the Hebbian term is scaled by the global reward signal $\ell_{\boldsymbol{\theta}_{\mathcal{Z}}}(\boldsymbol{Z}_{\leq K-1})$.

## 2.6 Implementation and Emulation of SNNs

Due to the wide variety of neuron models and update algorithms, an increasing number of software frameworks support the implementation of SNNs. In parallel, the availability of neuromorphic hardware for emulating SNNs is also growing. The following section provides a brief overview of state-ofthe-art simulation software, including the neuron models and update algorithms they support, as well as currently available neuromorphic hardware. While digital systems simulate the analog dynamics of SNNs, analog or mixed-signal solutions emulate the dynamics by representing them as physical units [CSSZ22].

### 2.6.1 Software Frameworks

The software packages for simulation of SNNs can be divided into two categories: The first category aims to construct realistic models of neurobiological systems. These software packages are useful for rebuilding large networks of spiking neurons and therefore for better understanding the dynamics of the mammalian brain, e.g. `BRIAN2` [SBG19], `Nengo` [BBH+14], and `nest` [GD07]. All three packages support a variety of different neuron models and synaptic models, although they are not intended for conducting optimization. The second category is intended for the optimization of SNNs. These software packages are commonly built on top of *PyTorch* [PGM+19] or *tensorflow* [AAB+15]. Common frameworks are `BindsNET` [HSK+18], `NengoDL` [Ras18], `SpikingJelly` [FCD+23], `snntorch` [EWN+23], `Lava` [Int24], and `Norse` [PP21].

**Table 2.2:** Comparison of software frameworks for simulation and optimization of SNNs.

| Framework | `BindsNET` | `Norse` | `SpikingJelly` | `snntorch` | `NengoDL` | `Lava` |
|---|---|---|---|---|---|---|
| Neurons | LIF, SRM | LIF | LIF | LIF | LIF | LIF |
| Update Rule | STDP | BPTT/SG | BPTT/SG | BPTT/SG | BP | STDP |
| Built upon | PyTorch | PyTorch | PyTorch | PyTorch | tensorflow | PyTorch |
| Backend/ Platform | CPU/GPU | CPU/GPU | CPU/GPU | CPU/GPU | Loihi, SpiNNaker CPU/GPU | Loihi CPU/GPU |

Tab. 2.2 compares the different software frameworks with respect to the supported neurons, the update rule, the autodifferentiation software package they are built on, and the supported backend, e.g., central processing units (CPUs), graphics processing units (GPUs), or neuromorphic hardware [MVSK+23]. The comparison is based on [MVSK+23, Tab. 1]. Tab. 2.2 is limited to neuron types and update rules that are used in this thesis.

In the following chapter of this thesis, we investigate the optimization of SNNs using STDP, BPTT with SG, and probabilistic SNNs. To update SNNs using STDP, we chose `BindsNET`, as it includes an STDP learning rule and computationally efficient LIF neurons. We did not select `Lava`, as it had not yet been published at the time of our investigation into STDP update mechanisms.

To investigate BPTT with SG, we chose `Norse`. It is built directly on top of PyTorch, making it very natural for PyTorch users to integrate spiking neurons seamlessly into existing PyTorch models. Furthermore, it comes with detailed documentation. When comparing the runtime of the different software frameworks that support BPTT with SG, `Norse` stands out due to its fast computation [LHS+23].

To implement probabilistic SNNs, the SRM is required. Since `BindsNET` supplies the SRM, we build our probabilistic SNNs using `BindsNET`, and implement the update rules as derived in the respective section.

### 2.6.2 Neuromorphic Hardware

SNNs can be emulated on neuromorphic hardware. In recent years, several neuromorphic chips were developed [PSHP23, SFM+22]: TrueNorth [ASC+15], Loihi [DSL+18], Loihi2 [OFR+21], SpiNNaker [FLP+13, MHF19], BrainScaleS-2 [PBC+22], Tianjic [DWL+20], NeuronFlow [MYC+20], and Akida [SFM+22, ICK+22]. The systems can be divided into digital systems (TrueNorth, Loihi, Loihi2, Tijanic, SpiNNaker, Neuronflow, Akida) and analog/mixed signals systems (BrainScaleS) [SFM+22, ICK+22]. With the exception of SpiNNaker, digital systems are implemented in an event-based manner [ICK+22]. While most support on-device training using STDP (Loihi, Loihi2, SpiNNaker, Akida, BrainScaleS) or SG with BPTT (Loihi2, BrainScaleS), some require off-device training (TrueNorth, Tijanic, NeuronFlow) [ICK+22]. For further reading, the interested reader is referred to [ICK+22, SFM+22].

# 3 Spiking Neural Network-based Detector for the AWGN Channel

In this chapter, we apply SNNs to the detection of complex-valued symbols corrupted by additive white Gaussian noise (AWGN), which is a fundamental problem in communications engineering. Since both the update mechanisms of SNNs and the employed encoding techniques are active topics of research, different update rules and neural encoding methods are explored. Specifically, we investigate STDP, BPTT with SGs, and the update rule for probabilistic SNNs. To simulate the SNNs, we use the libraries `BindsNET` [HSK+18] and `Norse` [PP21]. All simulations are based on a discrete-time implementation of the SNN, executed over $K \in \mathbb{N}$ discrete time steps.

## 3.1 The AWGN Channel and its Optimal Detector

A fundamental problem in communication engineering is the reliable detection of transmit symbols disturbed by AWGN, as illustrated in Fig. 3.1. Let $X \in \mathbb{C}$ be a RV, representing a complex-valued transmit symbol. The symbol $x$ is drawn from a predefined finite constellation $\mathcal{X} \subset \mathbb{C}$, i.e., $x \in \mathcal{X}$. After transmission over an AWGN channel, the received signal $y \in \mathbb{C}$ can be modeled by

$$y = x + n_{\mathrm{n}}\,, \quad N_{\mathrm{n}} \sim \mathcal{CN}\big(0, \sigma_{\mathrm{n}}^2\big)\,, \tag{3.1}$$

where the RV $N_{\mathrm{n}} \in \mathbb{C}$ models circularly-symmetric complex AWGN with zero mean and noise power $\sigma_{\mathrm{n}}^2$.

To obtain an estimate $\hat{x}$ of the transmit symbol, the observed channel output $y$ is forwarded to the detector. For the AWGN channel, the optimal detector w.r.t. the detection probability is the maximum a posteriori (MAP) detector [PS08, Sec. 4.1], which computes a MAP estimation of $\hat{x}$ based on the observation $y$

$$\hat{x}(y) = \arg\max_{x \in \mathcal{X}} P_{X|Y}(x|y)\,, \tag{3.2}$$

where $P_{X|Y}(x|y)$ denotes the conditional probability of $X = x$ given the observation $Y = y$. Assuming the transmit symbols are distributed uniformly, i.e., $P_X(x) = \frac{1}{|\mathcal{X}|}$, the MAP criterion simplifies to the ML criterion [PS08, Sec. 4.1]

$$\hat{x}(y) = \arg\max_{x \in \mathcal{X}} f_{Y|X}(y|x)\,, \tag{3.3}$$

where $f_{Y|X}(y|x)$ denotes the conditional probability density function (PDF) of receiving $y$ given that $x$ was transmitted. Thus, for uniformly distributed

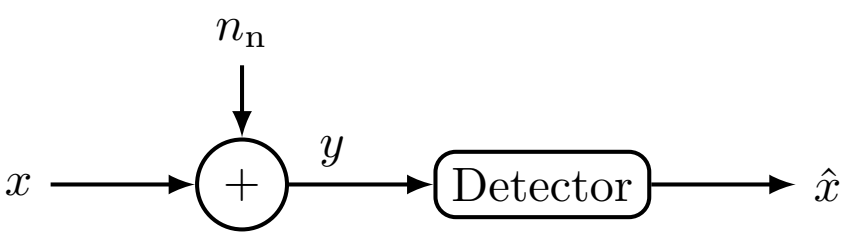

**Figure 3.1:** The AWGN channel.

transmit symbols, the ML detector is the optimal detector w.r.t. the symbol error rate (SER).

In this chapter, we consider two different modulation schemes, binary phase-shift-keying (BPSK) and 16-quadrature amplitude modulation (QAM), which define the constellation $\mathcal{X}$. Transmit bits are generated independent and identically distributed (i.i.d.), following a uniform distribution. Using a bit mapper, the transmit bits $b \in \{0, 1\}$ are converted into the transmit symbols. Hence, transmit symbols follow a uniform distribution, and the ML detector serves as a benchmark for optimal detection w.r.t. the bit error rate (BER). Furthermore, the transmit symbols are normalized to the average symbol energy $E_{\mathrm{s,avg}} = 1$. For BPSK, the set of transmit symbols $\mathcal{X}$ is given by

$$\mathcal{X} = \{-1, +1\} . \tag{3.4}$$

The BPSK-symbols $x$ are mapped by $x = (-1)^b$, where $b$ denotes the transmit bit. For 16-QAM, $\mathcal{X}$ is given by

$$\mathcal{X} = \{m \cdot c + \mathrm{j}n \cdot c\}, \qquad (m, n) \in \{-3, -1, +1, +3\}^2 , \tag{3.5}$$

where the scaling factor $c = \frac{\sqrt{10}}{10}$ ensures $E_{\mathrm{s,avg}} = 1$. When using a 16-QAM, the bit mapping fulfills the requirements of Gray coding.

The ML detector given by (3.3) returns $\hat{x}$. To obtain an estimation of the transmit bits $\hat{\boldsymbol{b}} \in \{0, 1\}^{\log_2(|\mathcal{X}|)}$, the bit mapping is reversed. Because the bit mapping is bijective, each transmitted symbol can be uniquely associated with a corresponding bit sequence.

## 3.2 Design of an SNN-based Detector

An SNN-based detector for symbol detection in the AWGN channel is displayed in Fig. 3.2. First, the channel observation $y \in \mathbb{C}$ is fed to the neural encoding, which converts $y$ into $N_{\mathrm{in}}$ discrete-time spike signals $\boldsymbol{z}_{\mathrm{in},j} \in \{0, 1\}^K$, $j = 1, \dots, N_{\mathrm{in}}$, where each spike signal consists of $K$ samples. Afterwards, the obtained spike signals are forwarded to the input layer of the SNN, where each spike signal is passed to exactly one input neuron. Thus, the number of input layer neurons equals the number of spike signals output by the neural encoding, $N_{\mathrm{in}} = N_{\mathrm{enc}}$.

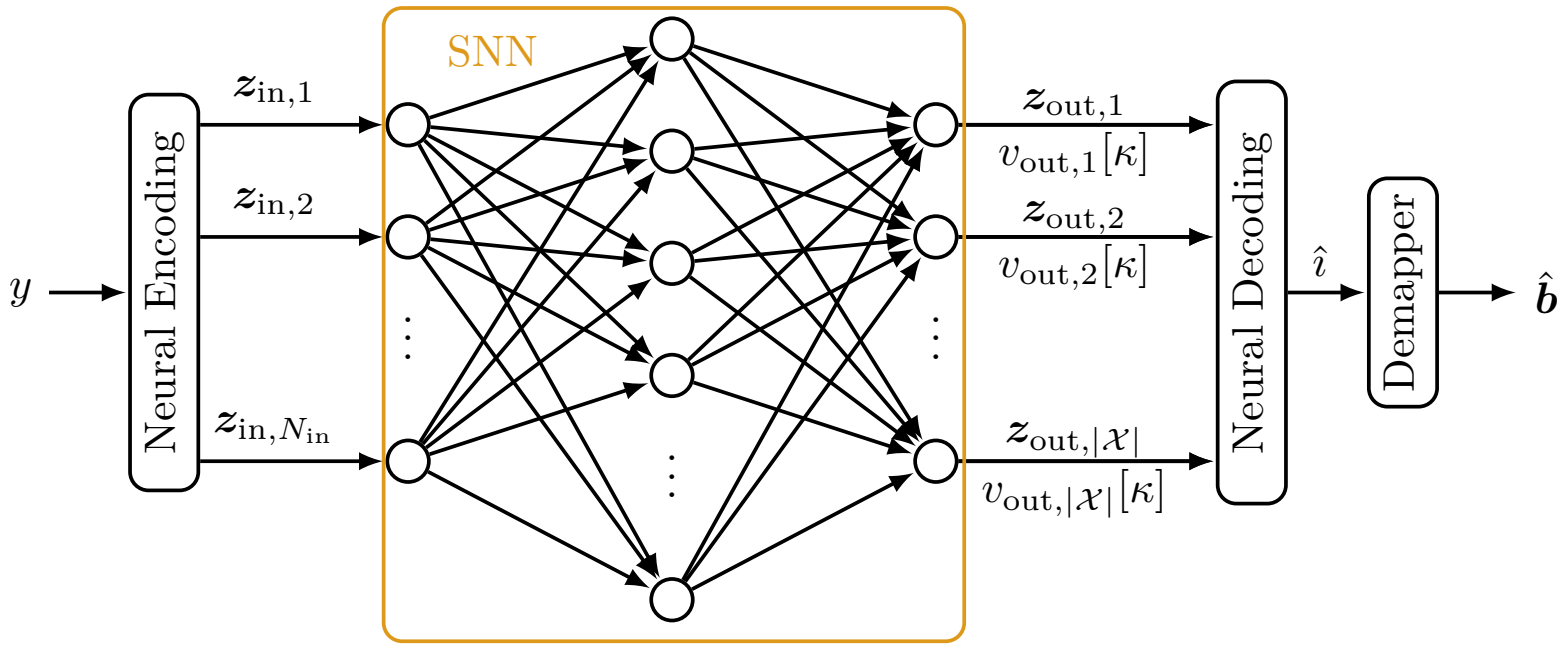

**Figure 3.2:** Design of the SNN-based detector with $N_{\mathrm{in}}$ input neurons, a single hidden layer and $N_{\mathrm{out}} = |\mathcal{X}|$ output neurons.

Unless specified otherwise, the SNN follows a feed-forward architecture with a single hidden layer, and does not incorporate explicit recurrent connections.

The output layer consists of $N_{\mathrm{out}} = |\mathcal{X}|$ neurons, each representing a transmit symbol $x \in \mathcal{X}$. Both the spike signals $\boldsymbol{z}_{\mathrm{out},j} \in \{0,1\}^K$, j $= 1, 2, \ldots, |\mathcal{X}|$, and the membrane potential $v_{\mathrm{out},j}[\kappa]$ of the output neurons are passed to the neural decoding. The neural decoding determines the estimated class label $\hat{\imath} \in \{1, \ldots, |\mathcal{X}|\}$, of the transmit symbol based on the spike signals, the membrane potentials, or both. A demapper then converts the estimated class labels $\hat{\imath}$ into the estimated transmit bits $\hat{\boldsymbol{b}}$.

Based on the AWGN channel model and the detector architecture shown in Fig. 3.2, we specify, implement, and evaluate various neural encoding and decoding schemes, as well as different update rules in the following sections.

## 3.3 Optimization Using STDP

We first investigate the update of SNN-based detectors for BPSK and 16-QAM using STDP. To update, we apply the STDP update rule with the LIF neuron model, provided by `BindsNET` [HSK+18]. Since STDP is a local learning rule, we iteratively develop a method to successfully update SNNs with a single hidden layer using STDP. By optimizing an SNN-based detector for BPSK and 16-QAM symbols disturbed by AWGN, we demonstrate the success of our proposed approach. We furthermore design and investigate several neural encoding schemes that are based on BE and TTFS.

### 3.3.1 Supervised Learning without Hidden Layers

As discussed in Sec. 2.5, STDP is a biologically plausible, local learning rule and is intended to be unsupervised [PP18]. To investigate STDP, we optimize an SNN to correctly classify noiseless BPSK symbols. Fig. 3.3 shows the setup. We apply BE to convert $x$ into spike signals. Since BE is limited to the encoding of positive values, two BEs are applied: The first encodes $\max(0, x)$, and the second encodes $\max(0, -x)$. Thus, depending on the sign of $x$, either the BE in the upper branch or the BE in the lower branch of Fig. 3.3 is activated. We implement the SNN without a hidden layer. The SNN consists of two input neurons, one per BE, and two output neurons, each assigned to a BPSK symbol $x \in \{-1, +1\}$. The neurons are connected by four trainable weights $\theta_{i,j} \in \mathbb{R}$, where $\theta_{i,j}$ connects the $j$th input neuron to the $i$th output neuron. The weights are randomly initialized by $\theta_{i,j} \overset{iid}{\sim} \mathcal{N}\left(\frac{1}{2}, 2\right)$, which is a good trade-off between weights that are too small and weights that are too large. While the former overly attenuate spikes, the latter excessively amplify them. Using spike rate decoding (SRD) and the number of counted spikes per output neuron, the class label $\hat{\imath}$ is estimated, and demapped into the corresponding estimated bit $\hat{b} \in \{0, 1\}$.

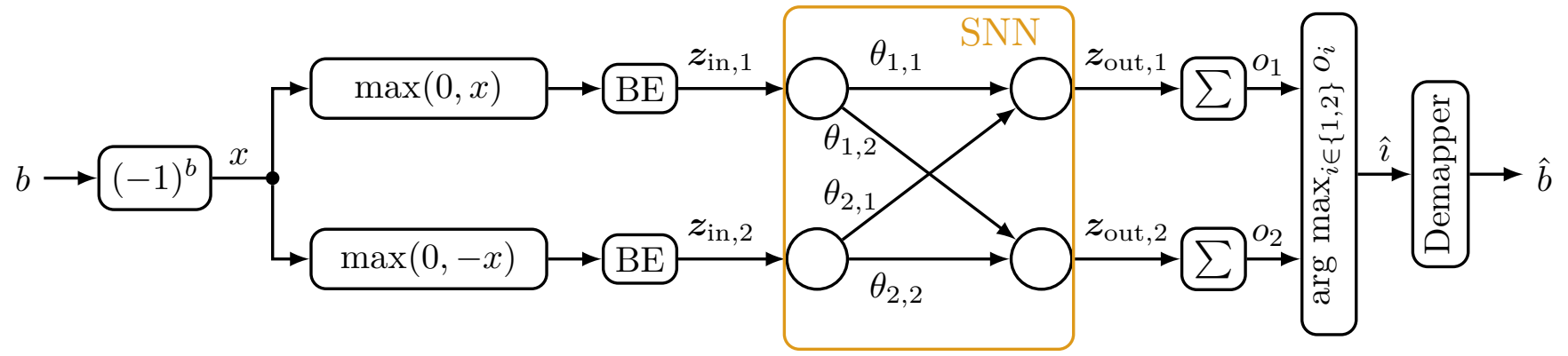

**Figure 3.3:** SNN structure used for the classification of noise-free BPSK symbols.

For the given setup, we independently optimized 100 SNNs using 100 BPSK symbols. After processing a single symbol, STDP is applied and the synapses are updated. Since the BPSK symbols are noiseless, 100 BPSK symbols have demonstrated to be sufficient for successful optimization. After optimization, each SNN is evaluated using $10^5$ BPSK symbols. An average classification accuracy of 65.1% is achieved. When output spikes are generated, synaptic connections are randomly strengthened or weakened, leading to arbitrary accuracy. Moreover, if no output spike is observed, the synaptic weights remain unchanged. When using STDP as update rule, the precise timing of spikes emitted by the presynaptic and postsynaptic neurons is required, see (2.37). Consequently, if neither a presynaptic nor a postsynaptic spike is observed, STDP cannot successfully update the synaptic weight.

To overcome this issue, `BindsNET` provides the `clamp`-function, which forces a neuron to spike at each discrete time step. While the internal dynamics of the neuron are calculated as in (2.4) and (2.5), the `clamp`-function does not apply the Heaviside function $H(\cdot)$, compare (2.6). Instead, it always assumes the firing of an output spike, without the reset of the membrane potential. When applying, a rate-based neural decoding, e.g., SRD, the `clamp`-function can be used to force the $i$th output neuron to generate output spikes, where $i \in \{1, 2, \ldots, |\mathcal{X}|\}$ is the index of the true transmit symbol. More formally, we can write the `clamp`-function

as

$$z_{\mathrm{out},j}[\kappa] = \begin{cases} 1, & \text{if } j = i,\ \forall \kappa \in \{0, \ldots, K-1\}\,, \\ 0, & \text{otherwise}\,, \end{cases} \tag{3.6}$$

where $z_{\mathrm{out},j}[\kappa] \in \{0,1\}$ denotes the spike signal output by the $j$th output neuron at time instant $\kappa$. Using the `clamp`-function, the STDP update rule can be transformed into a supervised learning problem, since output neurons are forced to emit spikes Based on the timing of presynaptic spikes (input spikes) and postsynaptic spikes (output spikes), STDP can update the weights accordingly. Using STDP and the `clamp`-function, we also optimized 100 SNNs, and achieved a classification accuracy of 100%.

We furthermore initialized 100 SNNs with $\theta_{i,j} = 0$, $\forall i, j$. Thus, in the initialized SNN no information is passed from the input layer to the output layer at all. After optimization, all SNNs achieved an average classification accuracy of 100%. We observed that using the `clamp`-function to fix the output spike signals during optimization establishes new connections. Based on our observations, we conclude that the `clamp`-function enables the successful optimization of SNNs without a hidden layer.

### 3.3.2 Supervised Learning with a Hidden Layer

Adding a hidden layer increases the capacity of a NN, and thus makes it more powerful [ALL19]. To investigate the impact of a hidden layer on the optimization using STDP and the `clamp`-function, we expand the SNN of Fig. 3.3 by a hidden layer with $N_{\mathrm{hid}} \in \mathbb{N}$ neurons. Using the `clamp`-function, we fix the spike signals of the output neurons. We update both $\boldsymbol{\theta}^{(\mathrm{in})}$ and $\boldsymbol{\theta}^{(\mathrm{out})}$ using STDP, where $\boldsymbol{\theta}^{(\mathrm{in})} \in \mathbb{R}^{N_{\mathrm{hid}} \times N_{\mathrm{in}}}$ denotes the matrix containing all synapse weights connecting the input and hidden layer, and $\boldsymbol{\theta}^{(\mathrm{out})} \in \mathbb{R}^{N_{\mathrm{out}} \times N_{\mathrm{hid}}}$ the matrix containing all synapse weights connecting the hidden and output layer. As before, all weights are randomly initialized by $\boldsymbol{\theta}^{(\mathrm{in})} \sim \mathcal{N}\left(\frac{1}{2}, 2\right)^{N_{\mathrm{hid}} \times N_{\mathrm{in}}}$ and $\boldsymbol{\theta}^{(\mathrm{out})} \sim \mathcal{N}\left(\frac{1}{2}, 2\right)^{N_{\mathrm{out}} \times N_{\mathrm{hid}}}$.

For various $N_{\mathrm{hid}}$ and $K = 50$, we optimized 20 SNNs, each using $= 100$ (noiseless) BPSK symbols. As before, after the processing of a single symbol, the weights are updated using STDP. Afterwards, we evaluated each optimized SNN using $10^5$ symbols and calculated their average classification accuracy. Fig. 3.4(a) shows the classification accuracy as a function of $N_{\mathrm{hid}}$. Although accuracy generally increases with $N_{\mathrm{hid}}$, the trend is not strictly monotonic and shows some fluctuations. Even for large $N_{\mathrm{hid}}$, a significant percentage of misclassifications is observed. Since STDP is a local learning rule, the input weights $\boldsymbol{\theta}^{(\mathrm{in})}$ are updated independently of the output weights $\boldsymbol{\theta}^{(\mathrm{out})}$ and vice versa. We conjecture that this uncoordinated update harms optimization and, hence, the classification performance of the SNN. Moreover, it provides an explanation for the fluctuations observed in Fig. 3.4(a). As a consequence, we either need to coordinate the updates, or only update one layer.

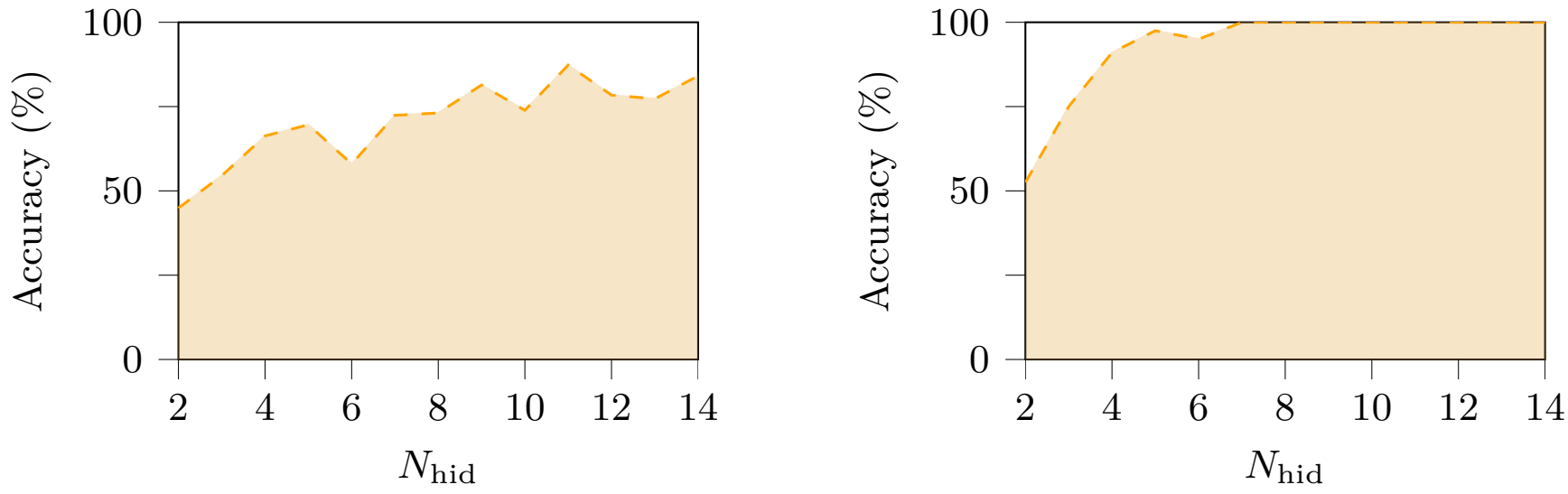

**(a)** Input and output weights are updated using STDP.

**(b)** Input weights are fixed, output weights are updated using STDP.

**Figure 3.4:** Average classification accuracy versus number of hidden layer neurons $N_{\text{hid}}$ for SNNs with a single hidden layer, updated using STDP. In (a) both input and output weights are updated using STDP, while in (b) the input weights are fixed and merely the output weights are updated using STDP.

In classical MaL, the extreme learning machine (ELM) [HZS06] successfully applies the latter approach to feed-forward ANN with a single hidden layer: both the input $\boldsymbol{\theta}^{(\text{in})}$ and output weights $\boldsymbol{\theta}^{(\text{out})}$ are randomly initialized, but the input weights are fixed and not learned. After processing a batch of samples, only the output weights are updated. Inspired by the ELM, we propose to fix $\boldsymbol{\theta}^{(\text{in})}$ and only update $\boldsymbol{\theta}^{(\text{out})}$ using STDP. While the fixed $\boldsymbol{\theta}^{(\text{in})}$ is responsible for distributing the input information among the neurons of the hidden layer in a deterministic manner, the adaptable $\boldsymbol{\theta}^{(\text{out})}$ is responsible for realizing the desired output spike signals. Following this, we also optimized 20 SNNs for various $N_{\text{hid}}$, evaluated each using $|\mathcal{B}_{\text{e}}| = 10^5$ symbols, and then calculated the average classification accuracy. Fig. 3.4(b) shows the classification accuracy over $N_{\text{hid}}$ if the ELM approach is used. For $N_{\text{hid}} \geq 7$, a classification accuracy of 100% is reached.

We can summarize that when updating both $\boldsymbol{\theta}^{(\text{in})}$ and $\boldsymbol{\theta}^{(\text{out})}$ using STDP, the update of both layers is uncoordinated, and at least for $N_{\text{hid}} < 14$ the optimized SNN is not able to achieve a classification accuracy of 100%. However, when we fix $\boldsymbol{\theta}^{(\text{in})}$ and solely update $\boldsymbol{\theta}^{(\text{out})}$ using STDP, we can increase the accuracy to 100%. We conclude that when using STDP and SNNs with a single hidden layer, fixing the input weights $\boldsymbol{\theta}^{(\text{in})}$ and merely updating the output weights $\boldsymbol{\theta}^{(\text{out})}$ improves the optimization.

### 3.3.3 Detection of BPSK Symbols Disturbed by AWGN

We now test the proposed training setup by optimizing an SNN-based detector for BPSK symbols disturbed by AWGN. The neural encoder in Fig. 3.3 is based on rate coding. To investigate STDP when using temporal encoding, we furthermore design two different temporal encodings, which are both based on TTFS. For better differentiation, we refer to them as rate encoder (RE), time encoder synchronous spikes (TESS) and time encoder asynchronous spikes (TEAS). Their block diagrams and characteristics are displayed in Fig. 3.5.

The setup of RE is displayed in Fig. 3.5(a). Similar to the encoding of Fig. 3.3, both $\max(0, y)$ and $\max(0, -y)$ are fed to a BE. For both BEs, we set $y_{\mathrm{enc,max}} = 1$, which corresponds to the point of saturation of the input of the BE encoder, see Sec. 2.3.2. We can characterize the RE by defining $r_1 \in \{0, 1\}$ and $r_2 \in \{0, 1\}$ as the average spike rate of the first and second BE, which is the expected number of generated spikes when run for $K$ discrete time steps. For $r_1 = 0$, we expect the first BE to not output a spike at all. For $r_1 = K$, we expect the first BE to spike at each time instant, thus $K$ spikes in total. The resulting characteristic of the RE is also displayed in Fig. 3.5(a). Depending on the sign of the input $y$, one of the BEs is active and generates spikes, while the other remains idle. With increasing amplitude $|y|$, the rate of the active BE also increases.

The two temporal coding approaches TESS and TEAS are based on TTFS encoding, where each neural encoding contains two TTFS encoding blocks, see Fig. 3.5(b) and Fig. 3.5(c). Thus, each spike signal consists of exactly one spike fired at time $t_{\mathrm{enc,z},j}$, $j = 1, 2$. Due to different preprocessing applied to $y$, TESS and TEAS differ in their characteristics, which describe the spike timings $t_{\mathrm{enc,z},j}$ as a function of $y$. The setup and continuous-time characteristics of TESS are displayed in Fig. 3.5(b). The preprocessing of TESS is designed such that for $y = 0$ both input spikes are fired at the same time, i.e., $t_{\mathrm{enc,z},2} = t_{\mathrm{enc,z},1} = \frac{K}{2}$. With increasing amplitude $|y|$, the time difference between both spikes increases, where the sign of $y$ defines the order of the spikes. The setup and continuous-time characteristics of TEAS are displayed in Fig. 3.5(c). Its preprocessing is designed such that for $y = -1$ both spikes are fired at the same time, i.e., $t_{\mathrm{enc,z},2} = t_{\mathrm{enc,z},1} = \frac{K}{2}$. As $y$ increases or decreases, the time difference between the two spikes changes accordingly, while the firing order indicates whether $y$ is increasing or decreasing.

RE and TESS are inspired by the ML decision for BPSK and the AWGN channel. The ML decision is given by

$$y \underset{\hat{b}=0}{\overset{\hat{b}=1}{\lessgtr}} 0\,. \tag{3.7}$$

For RE, $y = 0$ marks the value at which the spike activity of both input signals is switched. For TESS, at $y = 0$ the temporal firing order of the two input spikes is switched. Hence, RE and TESS seem to already implement the ML decision, and the task of the SNN is to identify which input neuron spikes at all, and which neuron spikes first. In contrast, for TEAS both curves in Fig. 3.5(c) intersect at $y = -1$. The design of TEAS is motivated by the idea that a single input spike cannot trigger an output spike. However, two input spikes sufficiently correlated in time can trigger an output spike. Thus, for TEAS we reduce the number of output neurons to $N_{\mathrm{out}} = 1$. Instead of using SRD, we apply the detection rule

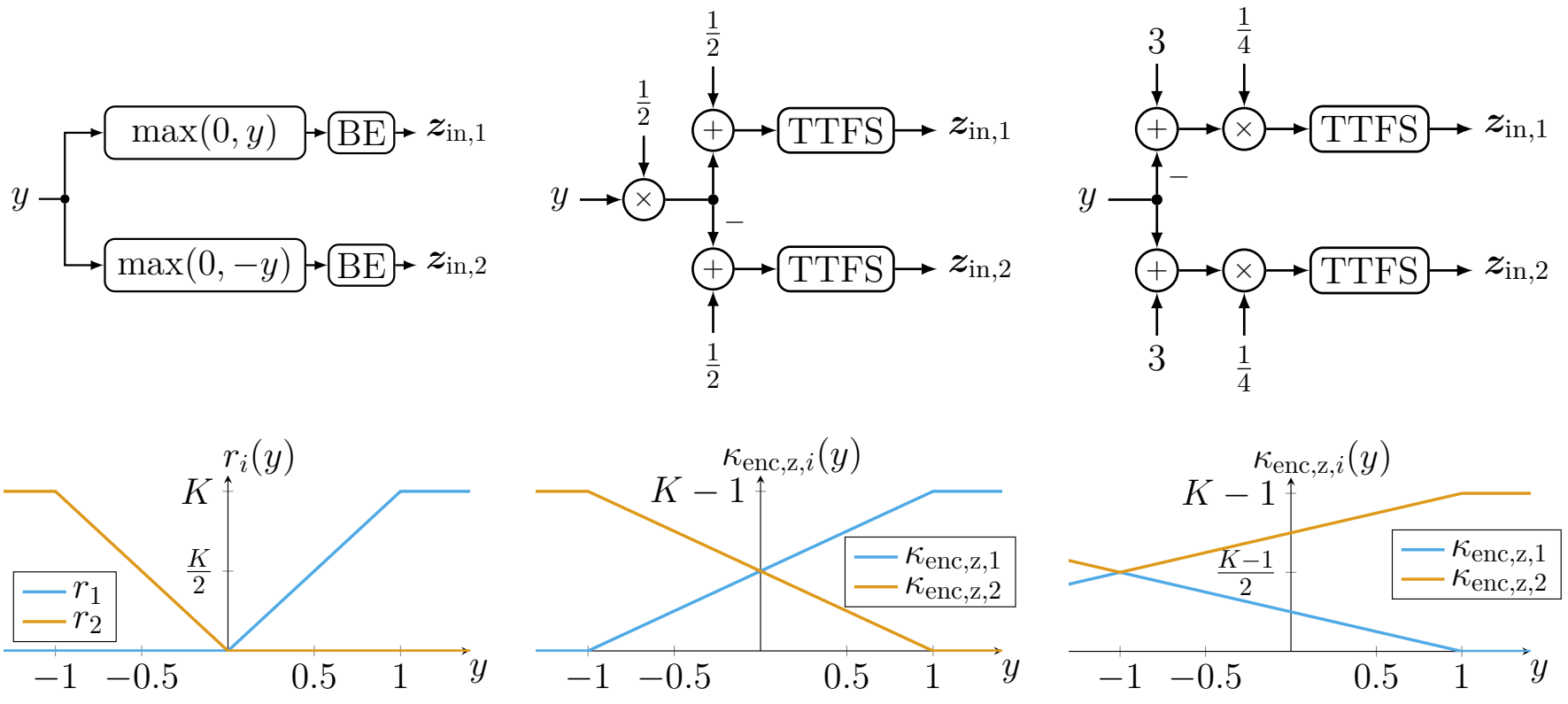

**(a)** RE. **(b)** TESS. **(c)** TEAS.

**Figure 3.5:** Block diagram (top row) and resulting characteristic (bottom row) of the RE, TESS, and TEAS.

by

$$\hat{b} = \begin{cases} 0, & \text{if } \sum_{\kappa} z_{\text{out}}[\kappa] = 0\,, \\ 1, & \text{if } \sum_{\kappa} z_{\text{out}}[\kappa] \geq 1\,, \end{cases} \tag{3.8}$$

where $z_{\text{out}}[\kappa]$ is the spike signal emitted by the output neuron. If at least one output spike is observed, the input spikes are close enough in time to trigger output spikes, indicating that $x = -1$ respectively $b = 1$ was transmitted. Fig. 3.7 shows the modified setup for TEAS, and Fig. 3.6 the setup for RE or TESS. For all encodings, the SNN consists of $N_{\text{in}} = 2$ input neurons and a single hidden layer with $N_{\text{hid}} = 8$ neurons. When using RE and TESS, the received bit $\hat{b}$ is obtained through SRD followed by demapping. In contrast, when using TEAS, the occurrence of at least one output spike indicates the detection of $\hat{b} = 1$. As before, all weights are initialized by $\boldsymbol{\theta}^{(\text{in})} \sim \mathcal{N}\left(\frac{1}{2}, 2\right)^{N_{\text{hid}} \times N_{\text{in}}}$ and $\boldsymbol{\theta}^{(\text{out})} \sim \mathcal{N}\left(\frac{1}{2}, 2\right)^{N_{\text{out}} \times N_{\text{hid}}}$. The input weights $\boldsymbol{\theta}^{(\text{in})}$ are fixed; only the output weights $\boldsymbol{\theta}^{(\text{out})}$ are updated using STDP and the `clamp`-function. It is important to note that all encodings are implemented in discrete time using $K = 50$. All networks are optimized using $10^4$ samples at a fixed $E_{\text{b}}/N_0 = 10\,\text{dB}$. After each sample, the SNN is updated. For evaluation, $10^6$ samples are used.

Fig. 3.8(a) shows the performance of the SNN-based detectors using the three different encodings. While TESS and RE exhibit a gap compared to the ML performance, TEAS achieves performance comparable to the ML detector. This is unexpected, considering that RE and TESS inherently implement the ML decision.

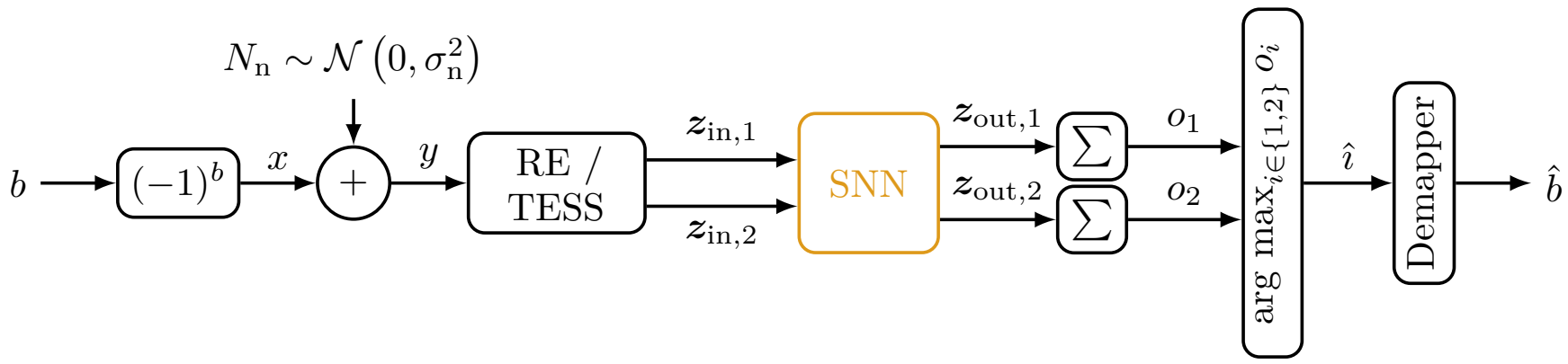

**Figure 3.6:** Setup of the SNN-based detector used for the classification of BPSK symbols with AWGN, when using RE and TESS as neural encoding.

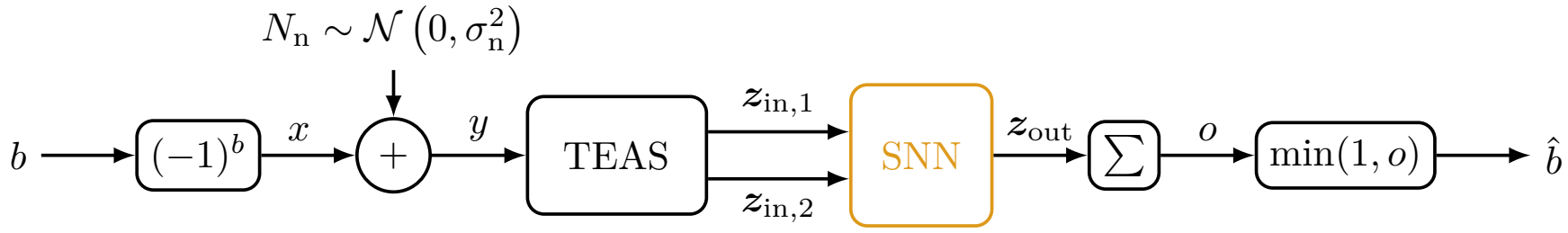

**Figure 3.7:** Setup of the SNN-based detector using TEAS for classification of BPSK symbols with AWGN, when using TEAS as neural encoding.

We can further investigate the errors by comparing the symbol decision obtained by the ML detector with the symbol decision obtained by the SNN-based detector. We define the symbol disagreement rate (SDR) as

$$\mathrm{SDR} = \frac{\sum_{m=1}^{|\mathcal{B}_e|} \mathbb{1}\left(\hat{x}_{\mathrm{ML}}^{(m)} \neq \hat{x}_{\mathrm{SNN}}^{(m)}\right)}{|\mathcal{B}_e|}\,, \tag{3.9}$$

where $\hat{x}_{\mathrm{ML}}$ is the symbol decision obtained by the ML detector, $\hat{x}_{\mathrm{SNN}}$ the symbol decision obtained by the SNN-based detector, and

$$\mathbb{1}\left(\hat{x}_{\mathrm{ML}}^{(m)} \neq \hat{x}_{\mathrm{SNN}}^{(m)}\right) = \begin{cases} 0\,, & \text{if } \hat{x}_{\mathrm{ML}}^{(m)} = \hat{x}_{\mathrm{SNN}}^{(m)}\,, \\ 1\,, & \text{if } \hat{x}_{\mathrm{ML}}^{(m)} \neq \hat{x}_{\mathrm{SNN}}^{(m)}\,. \end{cases} \tag{3.10}$$

For each encoding, Fig. 3.8(b) shows the SDR for specific values of $y$. For RE, the errors are concentrated around $y = 0$. This is intuitive, as around $y = 0$ the probability of generating spikes is rather low, resulting in insufficient excitation of the SNN. With increasing amplitude of $y$, the BE is more likely to generate a sufficient amount of spikes to excite the hidden layer, and thus forward spikes to the output layer. This behavior also explains the Gaussian-like shape via the central limit theorem. In general, the BER increases for values $y$ close to zero, which can be attributed to the probabilistic spiking behavior of the BE.

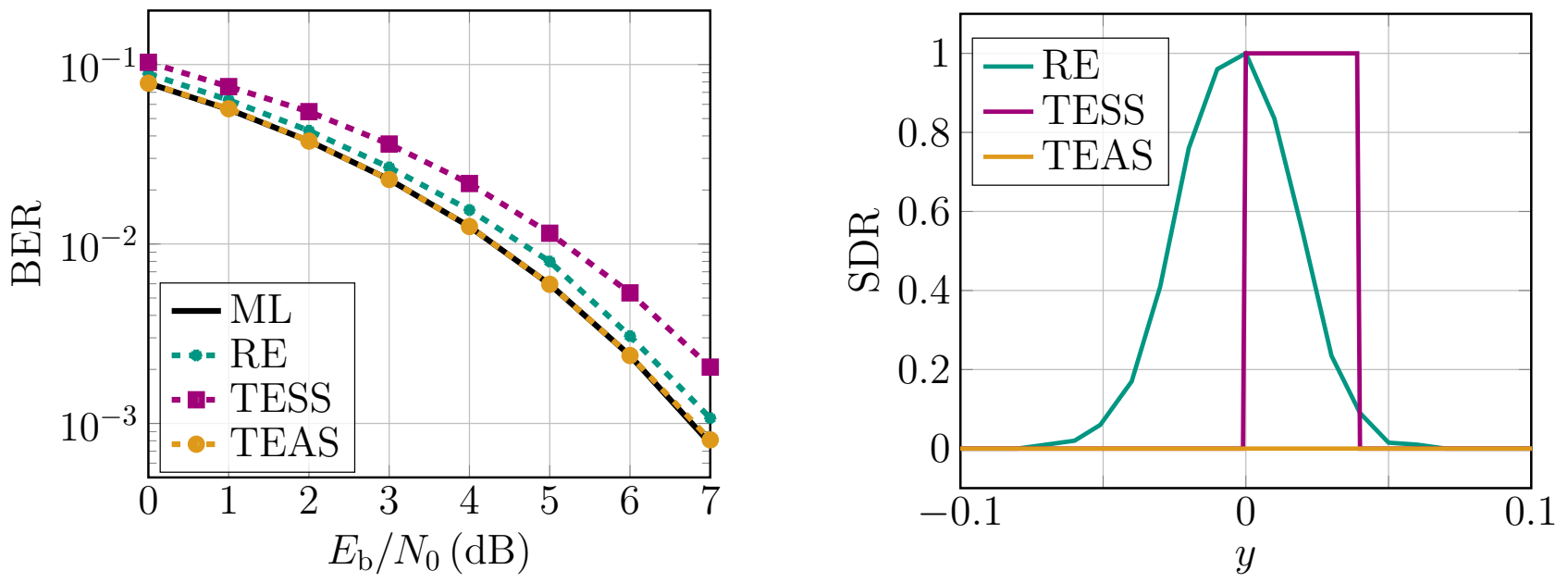

**(a)** BER curves. **(b)** SDR curves.

**Figure 3.8:** Performance of the SNN-based detector for BPSK with AWGN and in dependence of the applied encoding.

In contrast, TESS and TEAS are purely deterministic, resulting in sharp edges and an SDR of either zero or one. For TESS, the errors are observed for $y \in [0, 0.04]$. Within the interval $y \in [-0.04, 0.04]$, both input spikes of TESS are fired at the same discrete time step, i.e., $\kappa_{\mathrm{enc,z,1}} = \kappa_{\mathrm{enc,z,2}}$. Therefore, when identical spike signals are input to the SNN, the resulting bit decisions should differ, compare (3.7). Consequently, the SNN is optimized to decide for either $\hat{b} = 0$ or $\hat{b} = 1$, resulting in an error in the positive or negative half of the interval $[-0.04, 0.04]$. Thus, the discrete-time nature of the implemented TESS yields the ambiguity interval $[-\frac{2}{K}, \frac{2}{K}]$. Using TEAS, the SDR equals zero for all investigated values of $y$. Hence, the SNN only generates output spikes for $y < 0$, see (3.8). This suggests that the SNN is able to optimize its weights in order to only generate output spikes if the time difference $\kappa_{\mathrm{enc,z,2}} - \kappa_{\mathrm{enc,z,}j1}$ of both spikes is below $\frac{K-1}{2}$, which is the time difference obtained for $y = 0$.

We can summarize that for discrete-time simulations, the performance of both RE and TESS is limited by the number of discrete time steps $K$. The time discretization results in the ambiguous interval for TESS. With TEAS, the weights are updated such that the performance limitation caused by time discretization is eliminated. Hence, temporal coding based on TEAS seems a suitable choice for further investigations.

### 3.3.4 Detection of 16-QAM Symbols Transmitted over an AWGN Channel

Next, we extend TEAS and design an SNN-based detector for 16-QAM. Two novel issues arise: First, for 16-QAM, the transmit symbols $x \in \mathbb{C}$, and hence the received values $y \in \mathbb{C}$ are complex-valued, with $y = y_{\mathrm{I}} + \mathrm{j}y_{\mathrm{Q}}$. We propose to encode $y_{\mathrm{I}}$ and $y_{\mathrm{Q}}$ independently. Second, the detector must distinguish between different signal levels since $\Re\{x\} \in \{-3c, -c, c, 3c\}$.

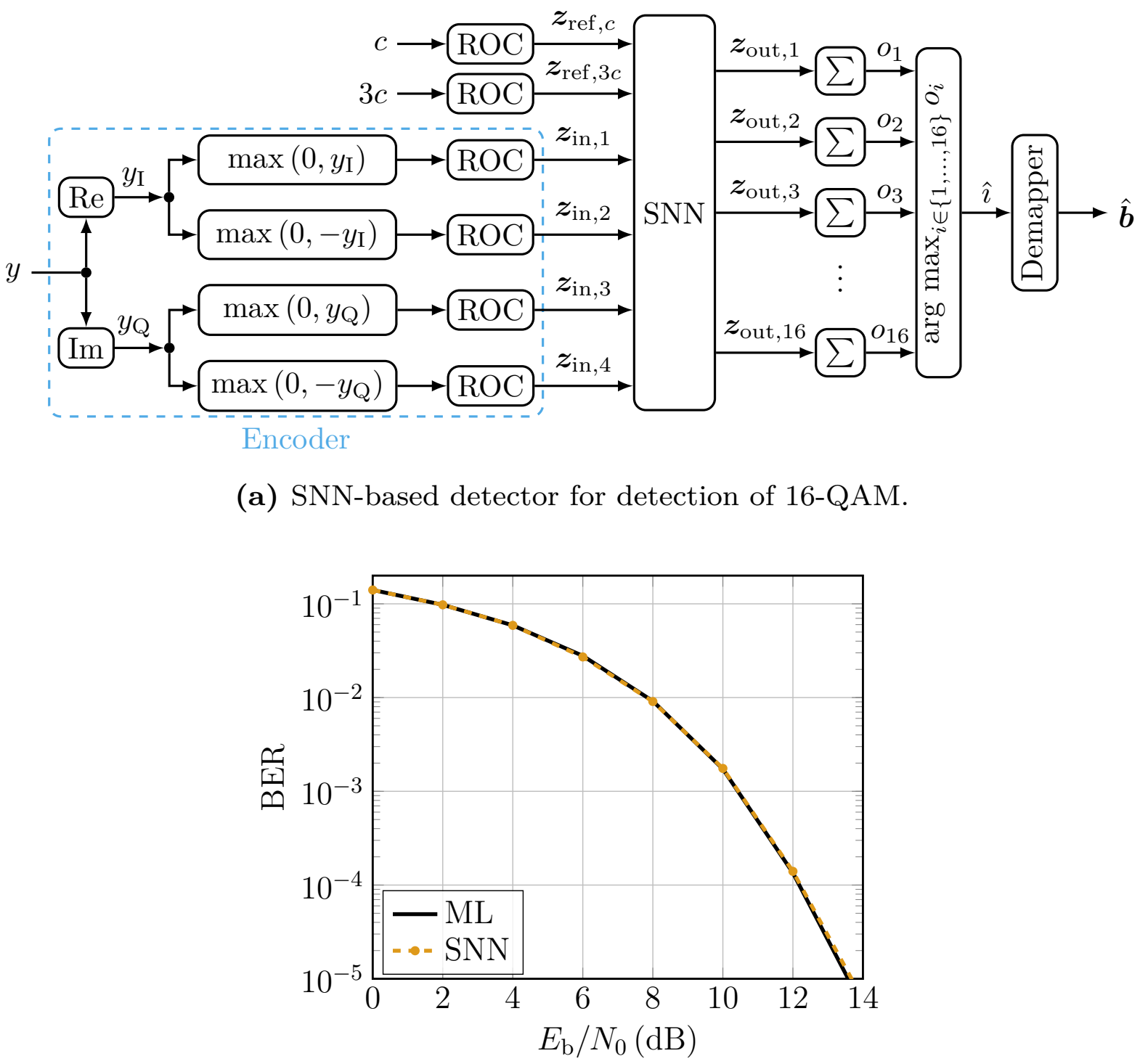

**(a)** SNN-based detector for detection of 16-QAM.

**(b)** BER curves.

**Figure 3.9:** Block diagram and BER of the SNN-based detector for 16-QAM.

We solve these issues with the setup displayed in Fig. 3.9(a): As before, we represent the sign of $y_\mathrm{I}$ ($y_\mathrm{Q}$) through the relative firing order of two independent TTFSs, each driven by $\max(0, y_\mathrm{I})$ and $\max(0, -y_\mathrm{I})$ ($\max(0, y_\mathrm{Q})$ and $\max(0, -y_\mathrm{Q})$). To distinguish between the amplitudes $c$ and $3c$, we generate reference spike signals $\boldsymbol{z}_{\mathrm{ref},c} \in \{0,1\}^K$ and $\boldsymbol{z}_{\mathrm{ref},3c} \in \{0,1\}^K$ that encode $c$ and $3c$ respectively. Inspired by TEAS, the SNN needs to correctly recognize temporal correlations of spikes obtained by the encoding of $y$ and spikes generated by the reference spike signals.

The SNN contains a single hidden layer with $N_\mathrm{hid} = 8$ neurons. At the output layer, each of the $N_\mathrm{out} = 16$ output neurons corresponds to a symbol class $i \in \{1, \dots, 16\}$ and SRD is used to estimate the index $\hat{i}$ of the transmit symbol. A demapper recovers the transmit bit sequence $\hat{\boldsymbol{b}} \in \{0,1\}^4$. Again, the input weights are fixed and only the output weights are optimized using STDP and the `clamp`-function. As before, the weights of the SNN are initialized by $\boldsymbol{\theta}^{(\mathrm{in})} \sim \mathcal{N}\left(\frac{1}{2}, 2\right)^{N_\mathrm{hid} \times N_\mathrm{in}}$ and $\boldsymbol{\theta}^{(\text{out})} \sim \mathcal{N}\left(\frac{1}{2}, 2\right)^{N_\text{out} \times N_\text{hid}}$. The network is simulated for $K = 50$ discrete time steps, optimized using $|\mathcal{B}_\text{t}| = 10^4$ samples at a fixed $E_\text{b}/N_0 = 16\,\text{dB}$, and evaluated using $|\mathcal{B}_\text{e}| = 10^6$ samples. After each sample, the weights are updated. The results are displayed in Fig. 3.9(b). The proposed SNN-based detector achieves equal performance as the ML detector.

#### Main Findings

We successfully demonstrated the optimization of SNNs using STDP and the `clamp`-function provided by `BindsNET` for the detection of BPSK and 16-QAM symbols disturbed by AWGN. Based on BE and TTFS, we designed and investigated three different neural encodings. However, all encodings were manually designed, limiting scalability and generalizability. Furthermore, some incorporate *a priori* knowledge, such as the ML decision boundary or amplitude levels $c$ and $3c$ of the noise-free data, which makes it hard to adapt the encoding to unknown data. We remark that the optimization is limited to the output weights $\boldsymbol{\theta}^{(\text{out})}$.

## 3.4 Optimization Using SG and BPTT

As a second update rule, we explore BPTT combined with SG to implement an SNN-based detector for the classification of noisy 16-QAM symbols.

The neural encoders presented in Sec. 3.3 were developed with significant use of prior knowledge about the data. Since they are built upon encoding schemes that output a single spike signal, e.g., BE and TTFS, we observed that the main task of the design lies in the proper arrangement of the independent spike signals. As a result, the encodings are tailored to a specific constellation set, e.g., the detection of 16-QAM, and cannot be easily adapted to different tasks. In contrast, population encoding schemes, such as RFE, TE, and QE, output multiple spike signals, which jointly represent the encoded information. This section investigates different population encoding schemes, namely Linear RFE, Gaussian RFE, the novel TE [BvBS23] and the novel QE.

For simulation of the SNNs, we use `Norse` [PP21], which implements BPTT with SG and the LIF neuron model. All SNNs consist of a single hidden layer with $N_\text{hid} = 16$ LIF neurons. Unless stated otherwise, we use the default LIF neuron parameters of `Norse`, i.e., $\tau_\text{m} = 10\,\text{ms}$, $\tau_\text{s} = 5\,\text{ms}$, and $v_\text{th} = 1$. The output layer of the SNN consists of $N_\text{out} = 16$ LI neurons with $\tau_\text{m} = 100\,\text{ms}$ and $\tau_\text{s} = 5\,\text{ms}$. EOTM neural decoding is applied. We use a learning rate of $\nu = 10^{-3}$, and a training batch size of $|\mathcal{B}_\text{t}| = 10^4$ samples. All networks are optimized for 20 000 epochs, with one update step per epoch, at a fixed $E_\text{b}/N_0 = 10\,\text{dB}$. For each epoch, new training data is created. The evaluation is carried out for various values of $E_\text{b}/N_0$, with a batch size of $|\mathcal{B}_\text{e}| = 10^6$ samples. The inphase component $y_\text{I} = \Re\{y\}$ and the quadrature component $y_\text{Q} = \Re\{y\}$ of the received signal $y$ are encoded independently for all encoders. Consequently, two RFEs, TEs, or QEs are applied in parallel, the first encoding $y_\text{I}$, and the second encoding $y_\text{Q}$.

### 3.4.1 Receptive Field Encoding

Each of the two RFEs employs two receptive fields, resulting in $N_{\text{enc}} = 2$ spike signals. Thus, the number of SNN input signals is given by $N_{\text{in}} = 2 \cdot N_{\text{enc}} = 4$. Each spike signal is composed of $K = 9$ discrete time steps. Due to its internal dynamics, the response of an output layer LIF neuron to an input stimulus is delayed when simulating with the neuron parameters described above. Hence, we add two empty time steps to the spike signal we obtain from the RFE, resulting in spike signals $\boldsymbol{z}_{\text{enc}} = \begin{pmatrix} \boldsymbol{z}_{\text{enc,RFE}} & 0 & 0 \end{pmatrix} \in \{0, 1\}^{11}$ of length $K = 11$, where $\boldsymbol{z}_{\text{enc,RFE}} \in \{0, 1\}^{9}$ is the spike signal output by the RFE.

#### Linear RFE

To demonstrate the impact of the parameterization of RFE, we initialize three different linear RFEs. First, we choose both field centers to $\mu_1 = -c$ and $\mu_2 = -c$, and a constant field width $\Delta_1 = \Delta_2 = 18c$. The obtained combination of RFE and SNN is denoted as $\text{SNN}_{[-c,\, c]\,,18c}$. We furthermore vary the parameters of the RFE twice: First, with equal field width ($\Delta_1 = \Delta_2 = 18c$), but shifted field centers $\mu_1 = -\frac{c}{2}$ and $\mu_2 = \frac{c}{2}$, denoted as $\text{SNN}_{[-c/2,\, c/2]\,,18c}$. The second with equal field centers ($\mu_1 = -c$, $\mu_2 = c$), but an increased width $\Delta_1 = \Delta_2 = 24c$, denoted as $\text{SNN}_{[-c,\, c]\,,24c}$.

The resulting continuous- and discrete-time characteristics of the different RFEs are displayed in Figs. 3.10(a)-3.10(c). Red vertical lines denote the decision boundaries of the ML detector for 16-QAM, when the real or imaginary part of the received value $y$ is viewed separately. They are given by $y_{\text{ML},1} = -2c$, $y_{\text{ML},2} = 0$, and $y_{\text{ML},3} = 2c$. The performance of the obtained SNN-based decoders is shown in Fig. 3.10(d).

The $\text{SNN}_{[-c,\, c]\,,18c}$ achieves performance comparable to the ML detector. However, its variations perform significantly worse. This is caused by the discrete-time implementation of the RFE and poor parameterization. For $\text{SNN}_{[-c/2,\, c/2]\,,18c}$, we examine the decision boundary located at $y_{\text{ML},2} = 0$ in Fig. 3.10(b). If $y$ is small and $y > 0$, the resulting spike timings are given by $\kappa_{\text{enc,z},1}(y) = 0$ and $\kappa_{\text{enc,z},2}(y) = 0$. If $y$ is small and $y < 0$, we also obtain $\kappa_{\text{enc,z},1}(y) = 0$ and $\kappa_{\text{enc,z},2}(y) = 0$. For $-\frac{c}{2} \leq y \leq \frac{c}{2}$, the encoding returns identical spike timings, which hinders the SNN from making correct classifications. For the decision boundaries at $y_{\text{ML},1} = -2c$ and $y_{\text{ML},3} = 2c$, the same consideration applies. If the RFE is parameterized as $\text{SNN}_{[-c,\, c]\,,24c}$, the ambiguous region around $y_{\text{ML},2}$ is reduced to $-\frac{c}{3} \leq y \leq \frac{c}{3}$, see Fig. 3.10(c). Compared to $\text{SNN}_{[-c/2,\, c/2]\,,18c}$, this results in improved performance. When using the parameters as given by $\text{SNN}_{[-c,\, c]\,,18c}$, the ambiguous region around the decision boundaries is eliminated completely, enabling near-ML performance.

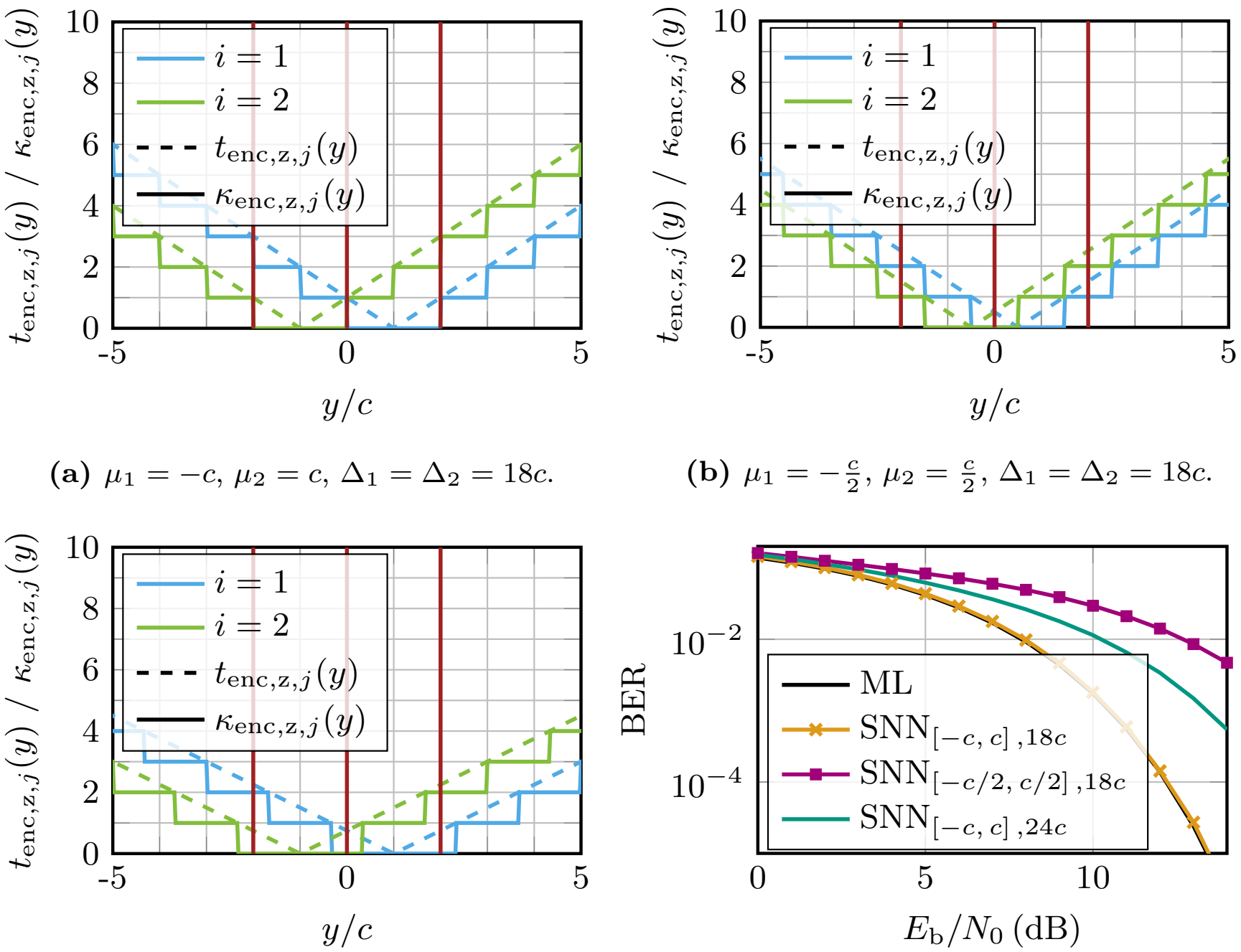

**(a)** $\mu_1 = -c$, $\mu_2 = c$, $\Delta_1 = \Delta_2 = 18c$.

**(b)** $\mu_1 = -\frac{c}{2}$, $\mu_2 = \frac{c}{2}$, $\Delta_1 = \Delta_2 = 18c$.

**(c)** $\mu_1 = -c$, $\mu_2 = c$, $\Delta_1 = \Delta_2 = 24c$.

**(d)** BER curves of the three linear RFEs.

**Figure 3.10:** (a)-(c): Continuous-time and discrete-time characteristics of the linear RFE for different parameterizations. Decision boundaries of the ML decoder are indicated by red vertical lines.
(d): BER curves obtained using the three different RFEs.

### Gaussian RFE

Next we initialize and investigate three different Gaussian RFEs with $N_\mathrm{enc} = 2$. For the first RFE, the field centers are again placed at $\mu_1 = -c$ and $\mu_2 = c$. From the findings above, the variances $\sigma_1$ and $\sigma_2$ of the discrete-time fields must be chosen such that values below and above the decision boundaries result in different spike timings. Thus, we need to ensure that at each decision boundary, at least one function $\kappa_{\mathrm{enc,z},j}(y)$, $j = 1, 2$, changes its values, i.e, ensure a different output for $y > y_{\mathrm{ML},n}$ and $y \leq y_{\mathrm{ML},n}$, $n \in \{1, 2, 3\}$. That is, for $y = y_{\mathrm{ML},n}$, the continuous-time function $t_{\mathrm{enc,z},j}(y)$ is required to equal some arbitrary integer $A \in \{1, \ldots, K-1\}$, and $T = K$. If this is fulfilled, the discrete-time function $\kappa_{\mathrm{enc,z},j}(y)$ will then exhibit a step at $y = y_{\mathrm{ML},n}$. For any decision boundary $y_{\mathrm{ML},n}$ and the $j$th receptive field, we can write

$$t_{\mathrm{enc,z},j}(y = y_{\mathrm{ML},n}, \mu_j, \sigma_j, T) = T \cdot \exp\left(-\frac{(y_{\mathrm{ML},n} - \mu_j)^2}{2\sigma_j^2}\right) \stackrel{!}{=} A\,. \qquad (3.11)$$

We can solve for $\sigma_j$ leading to

$$\sigma_j = \sqrt{-\frac{(y_{\mathrm{ML},n} - \mu_j)^2}{2\ln\left(\frac{A}{T}\right)}}\,. \qquad (3.12)$$

For $A = 8$, $T = K = 9$, $\mu_1 = -c$, and $y_{\mathrm{ML},2} = 0$, we obtain $\sigma_1 \approx 0.6515$. Due to the symmetry of the decision regions of the 16-QAM and the symmetry of both field centers at $\mu_1 = -c$ and $\mu_2 = c$, we choose $\sigma_2 = \sigma_1$. The SNN and encoding configured with these parameters is referred to as $\mathrm{SNN}_{[-c,\,c]\,,\sigma_1}$. Fig. 3.11(a) shows the resulting continuous-time and discrete-time characteristics for the mentioned parameters, where red lines again visualize the decision boundaries obtained by the ML detector.

To demonstrate the impact of the parameterization, we shift the field centers to $\mu_1 = -\frac{c}{2}$ and $\mu_2 = \frac{c}{2}$, which we denote as $\mathrm{SNN}_{[-1/2c,\,1/2c]\,,\sigma_1}$. Fig. 3.11(b) shows the corresponding continuous-time and discrete-time characteristics. For the decision boundary $y_{\mathrm{ML},2} = 0$, an ambiguous region can be spotted, returning alike spike timings $\kappa_{\mathrm{enc,z},j}$ for values larger and smaller than $y_{\mathrm{ML},2}$.

To explore the impact of additional receptive fields, we increase $N_{\mathrm{enc}}$ to 4. We choose $\mu_1 = -3c$, $\mu_2 = -c$, $\mu_3 = 1c$, $\mu_4 = 3c$, and $\sigma_j = \sigma \approx 0.6515$, $j \in \{1, 2, 3, 4\}$. The setup is denoted as $\mathrm{SNN}_{[-3c,\,-c,\,c,\,3c]\,,\sigma_1}$, and its resulting continuous-time and discrete-time characteristics are displayed in Fig. 3.11(c).

As Fig. 3.11(d) shows, $\mathrm{SNN}_{[-c,\,c]\,,\sigma_1}$ and $\mathrm{SNN}_{[-3c,\,-c,\,c,\,3c]\,,\sigma_1}$ achieve near-ML performance. Due to the ambiguous intervals, $\mathrm{SNN}_{[-1/2c,\,1/2c]\,,\sigma_1}$ yields poor performance. When comparing $\mathrm{SNN}_{[-c,\,c]\,,\sigma_1}$ and $\mathrm{SNN}_{[-3c,\,-c,\,c,\,3c]\,,\sigma_1}$, we observe that $\mathrm{SNN}_{[-3c,\,-c,\,c,\,3c]\,,\sigma_1}$ exhibits faster convergence during training, which we attribute to its higher resolution resulting from a larger number of fields, compare Fig. 3.11(a) and Fig. 3.11(c). However, it comes at the cost of doubling the input signals to $N_{\mathrm{in}} = 2 \cdot N_{\mathrm{enc}} = 8$.

We conclude that for the given task, the combination of SNNs updated using BPTT with SG, along with linear as well as Gaussian RFEs is able to achieve near-ML performance. Due to the time discretization, the obtained characteristics of the RFEs consist of multiple intervals, which result in the output of equal spike signals for all values within an interval. Thus, when designing an RFE, its parameterization has a significant impact on the overall performance. Using prior knowledge about the data, e.g., the decision boundaries obtained by the ML criterion, the RFE can be initialized properly. In the absence of prior knowledge,

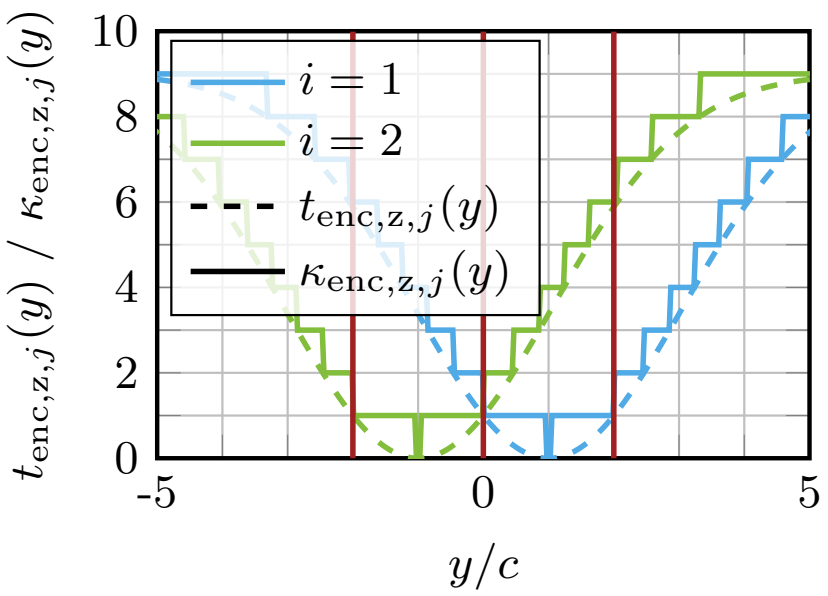

**(a)** $\alpha_{\mathrm{c}} = [-c,\, c]$.

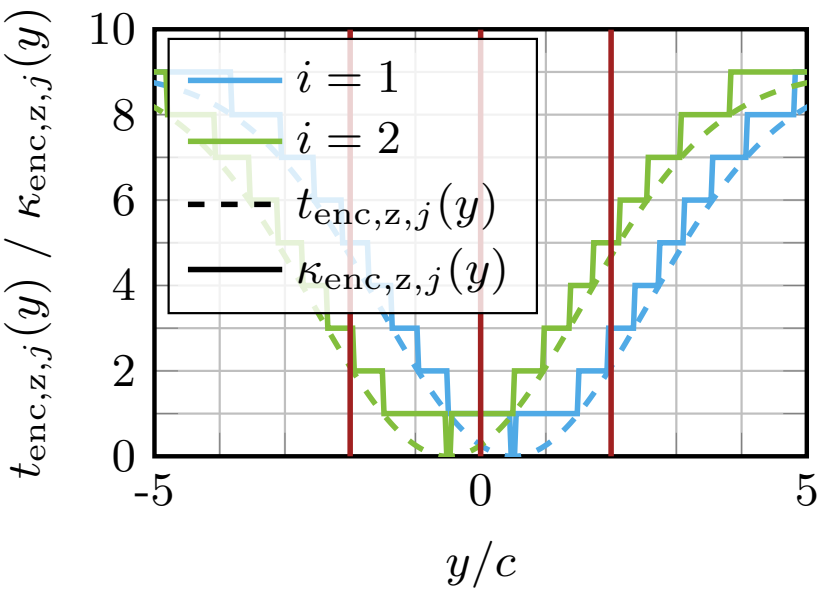

**(b)** $\alpha_{\mathrm{c}} = \left[-\frac{1}{2}c,\, \frac{1}{2}c\right]$.

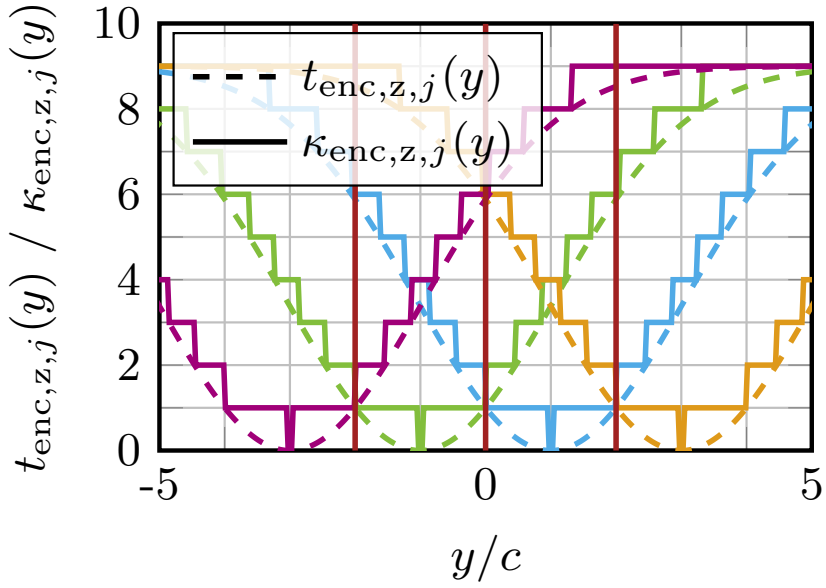

**(c)** $\alpha_{\mathrm{c}} = [-3c,\, -c,\, c,\, 3c]$.

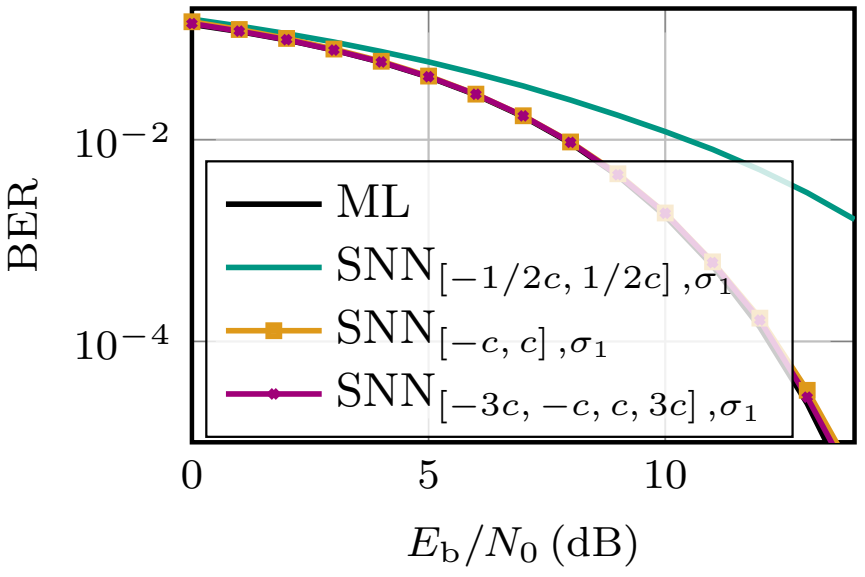

**(d)** BER of the three Gaussian RFEs.

**Figure 3.11:** (a)-(c): Continuous-time and discrete-time characteristics of the Gaussian RFE for different parameterizations. Decision boundaries of the ML decoder are indicated by red vertical lines.
(d): BER curves obtained using the three different RFEs.

the resolution of the RFE can be improved by increasing the number $N_{\mathrm{enc}}$ of receptive fields.

### 3.4.2 Ternary Encoding

Compared to RFE, TE emits all spikes at the first time step $\kappa = 0$. Afterwards, we append $K-1$ time steps without any spikes, which allows the SNN to react to the input. For all simulations using TE, we choose $K = 5$, which allows fast access to the output while ensuring a sufficient simulation time of the SNN. Further parameters of TE are given by the number $N_{\mathrm{enc}}$ of generated spike signals, which also indicates the number of bits of the applied quantizer, and the boundaries of the applied quantizer. For the SNN-based detector, we choose $y_{\mathrm{enc,max}} = 4c$ as the upper boundary of the quantizer. Fig. 3.12(a) shows the performance of the SNN-based detector combined with TE for varying $N_{\mathrm{enc}}$. As $N_{\mathrm{enc}}$ increases, the performance of the detector improves. Fig. 3.12(b) shows the characteristics of

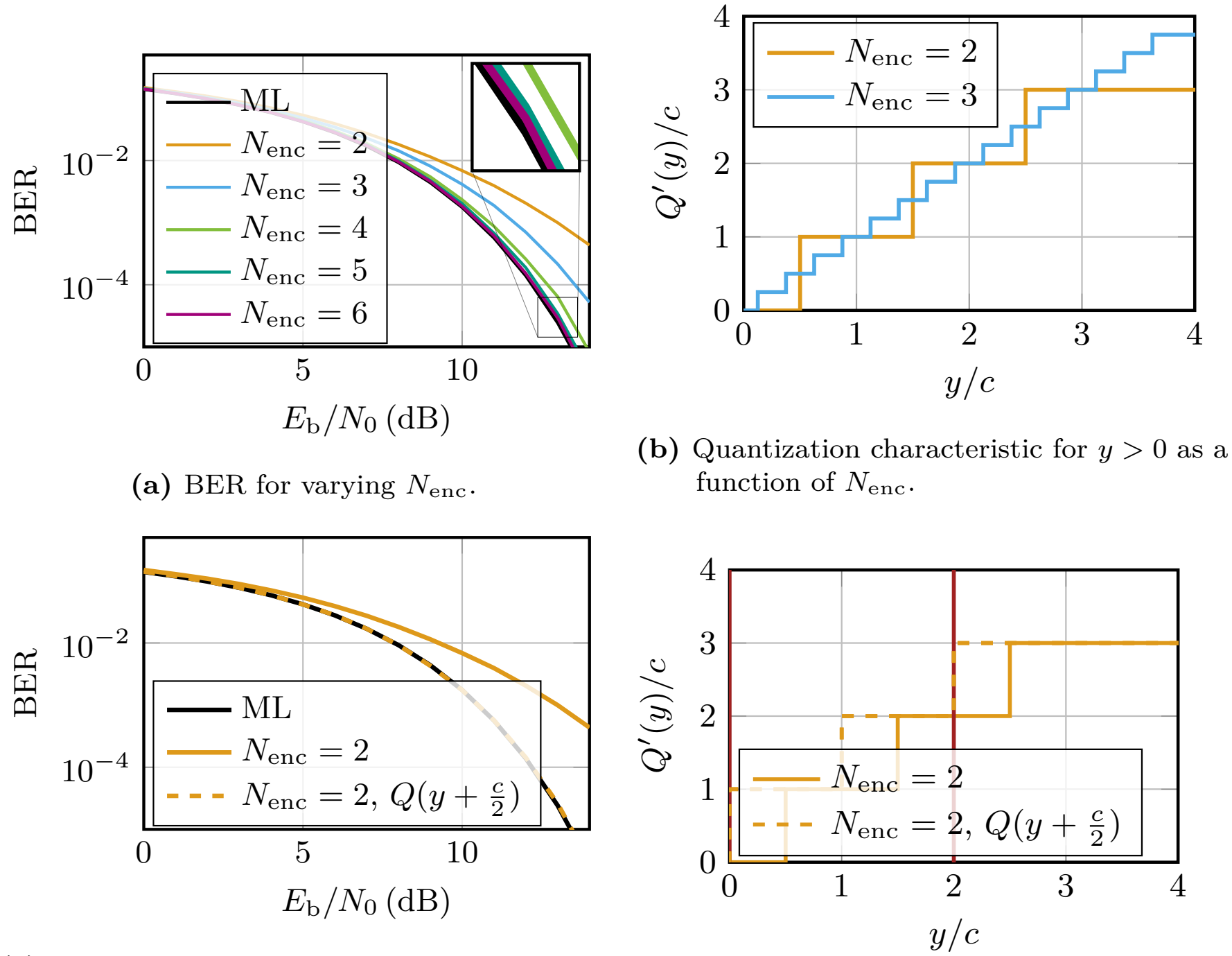

**(a)** BER for varying $N_{\mathrm{enc}}$.

**(b)** Quantization characteristic for $y > 0$ as a function of $N_{\mathrm{enc}}$.

**(c)** BER for $N_{\mathrm{enc}} = 2$ and optimized quantization characteristic.

**(d)** Quantization characteristic for $N_{\mathrm{enc}} = 2$ with and without optimized quantization.

**Figure 3.12:** Impact of $N_{\mathrm{enc}}$ on the BER and quantization characteristic of the TEs. Red lines in (d) indicate the decision boundaries of the ML detector.

the applied quantizer as a function of $N_{\mathrm{enc}}$. With increasing $N_{\mathrm{enc}}$, the resolution of the quantizer increases, resulting in a lower quantization error and, hence, enhanced performance. Thus, with increasing $N_{\mathrm{enc}}$, the resolution of TE and, as a result, the overall performance is increased.

Again, when designing the encoding, prior knowledge about the decision boundaries obtained by the ML detector can be included. Fig. 3.12(d) shows the characteristics $Q'(y)$ of the applied quantizer as a function of $N_{\mathrm{enc}} = 2$. We furthermore plot the characteristics if a bias of $\frac{c}{2}$ is applied prior to quantization, which can be modeled by $Q'(y + \frac{c}{2})$. As before, red lines denote the ML decision boundaries $y_{\mathrm{ML},2} = 0$ and $y_{\mathrm{ML},3} = 2c$. By adding the bias, the decision boundaries match the steps of the quantizer. Hence, for $y < y_{\mathrm{ML},n}$ and $y \geq y_{\mathrm{ML},n}$, $n = 1, 2, 3$, different quantization levels and thus spike signals are obtained. By adjusting the TE in this way, the system achieves near-ML performance even for $N_{\mathrm{enc}} = 2$, see Fig. 3.12(c).

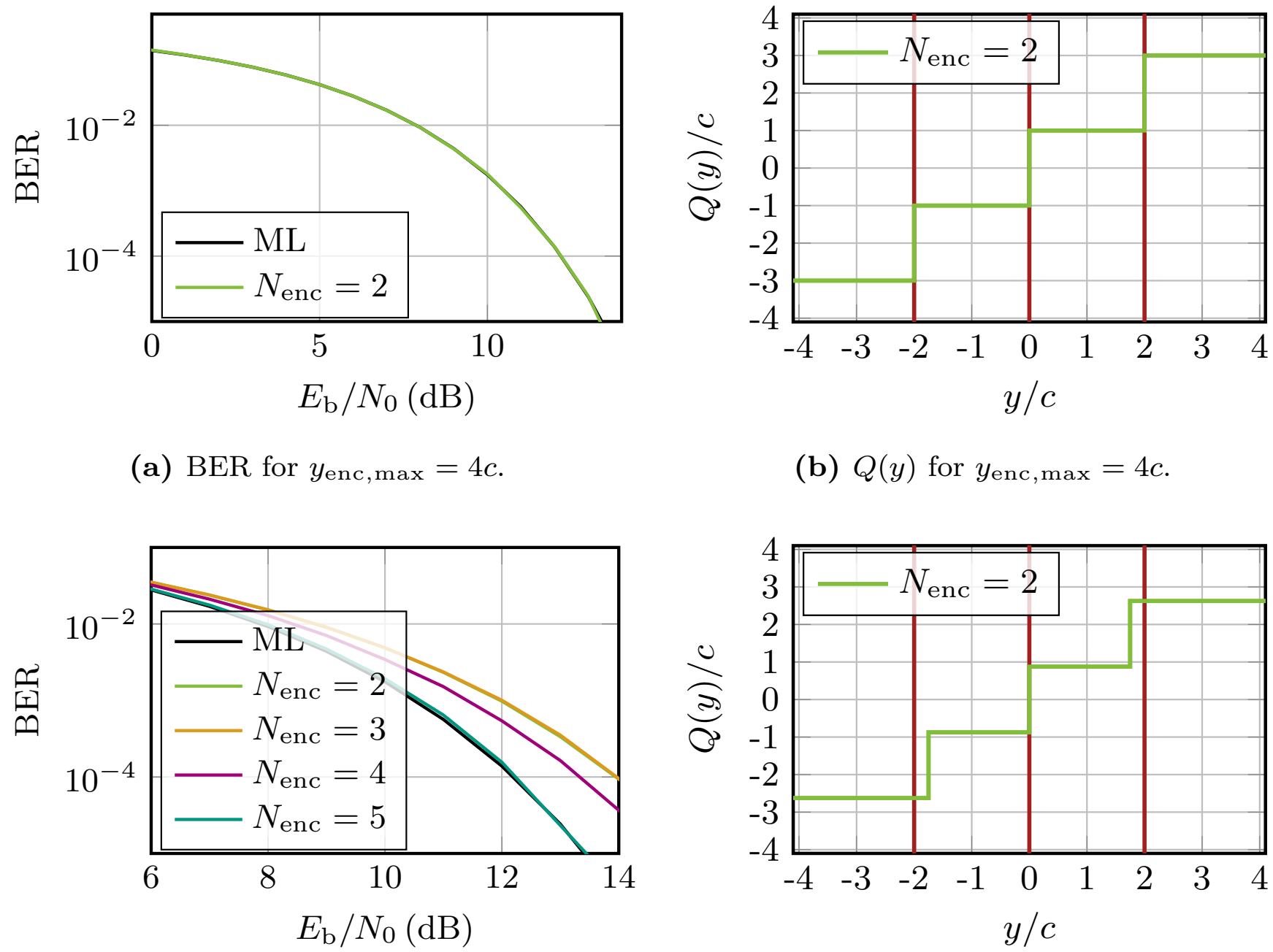

**(a)** BER for $y_{\mathrm{enc,max}} = 4c$.

**(b)** $Q(y)$ for $y_{\mathrm{enc,max}} = 4c$.

**(c)** BER for $y_{\mathrm{enc,max}} = \frac{7}{2}c$.

**(d)** $Q(y)$ for $y_{\mathrm{enc,max}} = \frac{7}{2}c$.

**Figure 3.13:** Impact of $N_{\mathrm{enc}}$ on the BER of QEs for $y_{\mathrm{enc,max}} = 4c$ and $y_{\mathrm{enc,max}} = \frac{7}{2}c$. For $N_{\mathrm{enc}} = 2$, the quantization characteristic $Q(y)$ is also given. Red lines indicate the decision boundaries obtained by the ML detector.

### 3.4.3 Quantization Encoding

We further introduce QE, another population-based scheme that relies on quantizing $y$. As with TE, we also initialize different QEs, varying in the number $N_{\mathrm{enc}}$ of spike signals output by the encoder. The upper boundary of the applied midrise quantizer $Q(y)$ is set to $y_{\mathrm{enc,max}} = 4c$.

Fig. 3.13(a) shows the obtained BER curves for $N_{\mathrm{enc}} = 2$, which achieves near-ML performance. For $N_{\mathrm{enc}} > 2$ we obtain similar curves, which we do not plot for the sake of clarity. Fig. 3.13(b) gives the respective quantization characteristic $Q(y)$. Since QE uses a midrise quantizer, $Q(y)$ has a step at $y_{\mathrm{ML},2} = 0$. By choosing $y_{\mathrm{enc,max}} = 4c$, for all $N_{\mathrm{enc}} \geq 2$, two more steps are at $y_{\mathrm{ML},1} = -2c$ and $y_{\mathrm{ML},1} = 2c$. Thus, the steps of $Q(y)$ coincide with the decision boundaries of the ML detector, and all approaches achieve near-ML performance.

To demonstrate the performance of the QE for a poorly chosen $y_{\mathrm{enc,max}}$, we also select $y_{\mathrm{enc,max}} = \frac{7}{2}c$. Fig. 3.13(d) shows the resulting quantization characteris-tic $Q(y)$ for $N_{\mathrm{enc}} = 2$, where the steps of $Q(y)$ do not further match with the decision boundaries obtained by the ML detector. Hence, we expect a penalty in terms of the BER. Fig. 3.13(c) shows the obtained BER curves. As expected, it performs poorly, however, with increasing $N_{\mathrm{enc}}$, the performance also improves. For $N_{\mathrm{enc}} = 5$, the ML performance is approximately achieved. Similar to TE, we conclude that increasing $N_{\mathrm{enc}}$ results in increased resolution of the quantizer, resulting in enhanced system performance.

#### Main Findings

In this section, we evaluated BPTT with SG combined with four different population-based neural encodings. All approaches were able to achieve performance comparable to that of the ML detector. During parameterization of the encodings, prior information about the decision boundaries of the ML detector can be incorporated, resulting in a minimization of the number $N_{\mathrm{enc}}$ of spike signals. If no prior information is available, by increasing $N_{\mathrm{enc}}$ the resolution of the encodings can be increased, which improves the performance of the SNN-based detector.

All SNNs optimized in this section were updated using BPTT and SG. Compared to STDP, BPTT with SG lacks biological plausibility. However, by exploiting the backpropagation algorithm, it enables the optimization of SNNs with hidden layers.

## 3.5 Optimization Using Probabilistic SNNs

Another update rule that overcomes the locality of STDP is the update rule for probabilistic SNNs, see Sec. 2.5.3. Thus, as a third update algorithm, we implement and investigate probabilistic SNNs and the corresponding update rules [JSGG19]. Since the spiking behavior of the probabilistic SNN is inherently non-deterministic, we first propose a method to convert the pre-trained probabilistic SNN into a deterministic counterpart. Afterwards, we optimize the SNN-based detector for noisy 16-QAM symbols using probabilistic SNNs, and compare the performance of both the probabilistic SNN with its deterministic counterpart. Since the SRM supports a variety of feed-forward filters, we furthermore construct the SRM using three different types of filters. As before, we optimize an SNN-based detector with a single hidden layer.

### 3.5.1 Conversion to Deterministic SNNs

Probabilistic SNNs apply the SRM as neuron model. However, for safety-critical applications, e.g., autonomous driving and medical applications, it may be desirable to obtain a deterministic and, thus, reproducible SNN behavior. In the following, we briefly derive a conversion rule from a probabilistic SNN into a corresponding deterministic SNN.

At the SRM output, the firing probability is given by $p_{\text{fire}}(Z[\kappa] = 1) = \sigma(v[\kappa])$, where $\sigma(\cdot)$ is the sigmoid function. To make the spiking behavior of neurons deterministic, we replace the sigmoid function $\sigma(\cdot)$ with the Heaviside function $H(\cdot)$, resulting in the same deterministic behavior as the LIF neuron model. Hence, $p_{\text{fire}}(Z[k] = 1|v[\kappa]) = H(v[\kappa] - v_{\text{th}})$.

However, we need to choose an appropriate value for $v_{\text{th}}$. We propose a strategy based on first-order statistics at the neuron level. After the optimization of the SNN, a batch $\mathcal{B}_{\text{t}}$ of $|\mathcal{B}_{\text{t}}|$ samples is processed. For the given batch and all spiking neurons of the hidden and output layer, we count the number of spikes $N_{\text{s},j} \in \mathbb{N}_0$ generated by neuron $j$, $j = 1, 2, \ldots, N_{\text{hid}} \cdot N_{\text{out}}$ by

$$N_{\text{s},j} = \sum_{\kappa=0}^{K-1} \sum_{m \in \mathcal{B}_{\text{t}}} z_j^{(m)}[\kappa] \, , \tag{3.13}$$

where $z_j^{(m)}[\kappa] \in \{0, 1\}$ indicates if the $j$th neuron has spiked at time instant $\kappa$ when processing the $m$th sample of $\mathcal{B}_{\text{t}}$. We further define $p_j^{(m)}[\kappa]$ as the probability that the $j$th neuron emits an output spike at the time step $\kappa$ when processing the $m$th sample. For each spiking neuron $j$, we introduce the first-order statistics $\mu$ as auxiliary variable. We differentiate between two categories: The first calculates the average membrane potential conditioned on the probability of firing a spike. The second calculates the average membrane potential conditioned on the probability of not firing a spike. Both are indicated by $\mu_{\text{s}}$ and $\mu_{\bar{\text{s}}}$, respectively, and are obtained by

$$\mu_{\text{s},j} = \frac{1}{N_{\text{s},j}} \sum_{\kappa=0}^{K-1} \sum_{m \in \mathcal{B}_{\text{t}}} p_j^{(m)}[\kappa] z_j^{(m)}[\kappa] \, , \tag{3.14}$$

$$\mu_{\bar{\text{s}},j} = \frac{1}{|\mathcal{B}_{\text{t}}|K - N_{\text{s},j}} \sum_{\kappa=0}^{K-1} \sum_{m \in \mathcal{B}_{\text{t}}} p_j^{(m)}[\kappa](1 - z_j^{(m)}[\kappa]) \, . \tag{3.15}$$

While $\mu_{\text{s},j}$ is the average neuron-specific firing probability if a spike is fired, $\mu_{\bar{\text{s}},n}$ is the average neuron-specific firing probability if no spike is fired. With this, we define the neuron-specific threshold $v_{\text{th},j}$ as

$$v_{\text{th},j} = \frac{\mu_{\text{s},j} + \mu_{\bar{\text{s}},j}}{2} \, . \tag{3.16}$$

We have also tested further heuristics incorporating second-order statistics. However, the given approach has proven to be the most robust and has performed the best.

### 3.5.2 Output Spike Amplification

Prior to the update of probabilistic SNNs, a rule for obtaining the target spike signal $\boldsymbol{Z}_{\leq K-1} \in \{0,1\}^{N_\text{out} \times K}$ needs to be defined. To enable a correct classification with a minimal number of output spikes, we construct the part of $\boldsymbol{Z}_{\leq K-1}$ representing the output neurons to consist of a single spike fired at the last time step, which indicates the correct class $i$

$$z_{\text{out},j}[\kappa] = \begin{cases} 1, & \text{if } \kappa = K \text{ and } j = i\,, \\ 0, & \text{else}\,, \end{cases} \tag{3.17}$$

where $z_{\text{out},j}[\kappa] \in \{0,1\}$ is the spike activity of the $j$th output neuron at time step $\kappa$, which is a realization of $Z_{\text{out},j}[\kappa]$. During optimization, we observed that the SNN tends to entirely suppress spiking activity in the output layer. Since the learning task is to have $N_\text{out} \cdot K - 1$ output samples $z_{\text{out},j}[\kappa]$ without a spike, and only a single output sample with a spike, a single spike does not have a great impact on the learning signal $\ell_{\boldsymbol{\theta}_\mathcal{Z}}(\boldsymbol{Z}_{\leq K-1})$. To enforce single spikes more strongly, we introduce an additional hyperparameter $\gamma \in \mathbb{R}$, $\gamma > 1$, and modify the learning signal of (2.58) to

$$\ell_{\boldsymbol{\theta}_\mathcal{Z}}(\boldsymbol{Z}_{\leq K-1}) = \sum_{\kappa=1}^{K} \sum_{j \in \mathcal{Z}} (\gamma z_{\text{out},j}[\kappa] + (1 - z_{\text{out},j}[\kappa]))\ \ln p_{\boldsymbol{\theta}_j}\Big(z_j^{(m)}[\kappa]\Big| v_j^{(m)}[\kappa]\Big)\,. \tag{3.18}$$

Hence, if $z_{\text{out},j}[\kappa] = 1$, its impact is amplified by $\gamma$.

### 3.5.3 Results

We now evaluate the SNN-based detector for the AWGN channel and 16-QAM when optimized using probabilistic SNNs. As neural encoding, we choose the linear RFE with the parameters discussed in Sec. 3.4.1: The inphase component $y_\text{I}$ and the quadrature component $y_\text{Q}$ of the received sample $y$ are encoded independently using $N_\text{enc} = 2$ receptive fields each. Based on Sec. 3.4.1, we choose the field centers to $\mu_1 = -c$ and $\mu_2 = c$, and a uniform width of $\Delta x_1 = \Delta x_2 = 18c$. The spike signals generated by the RFE consist of $K = 9$ time steps; however, by adding two empty time steps, the SNN is simulated for $K = 11$ time steps. The SNN consists of $N_\text{in} = 4$ input neurons, $N_\text{hid}$ hidden layer neurons, and $N_\text{out} = |\mathcal{X}| = 16$ output layer neurons. As neuron model, we apply the SRM with the exponential feedback filter given in (2.9) with $\tau_\text{rec} = \frac{1}{2}$ and $\Delta t = 1$. We implement the SRM using three types of feed-forward filters to investigate their influence:

- exponential filter,
- delayed raised-cosine filters,

- delay filters.

The exponential feed-forward filter is given by (2.7). We define $\tau_\mathrm{m} = 3$. With the exponential filter, we directly approximate the dynamics of the LIF neuron model with $\tau_\mathrm{s} = 0$. The delayed raised-cosine filters are given by (2.8), with $\ell_\mathrm{max} = K$, $\ell_\mathrm{off} = 0, 1, \ldots, \ell_\mathrm{off,max}$, and $\ell_\mathrm{off,max} = 10$. Thus, for each input spike signal, ten parallel feed-forward filters are used, each delayed by one sample. We furthermore define delay filters, which are delayed raised-cosine filters of width $\ell_\mathrm{max} = 1$. The impulse response of the delay filter can be expressed as $\alpha[\ell]^{(\ell_\mathrm{off})} = \delta[\ell - \ell_\mathrm{off}]$, and delays the incoming signal by $\ell_\mathrm{off}$. Since the $\ell_\mathrm{off}$th delay filter is connected to the $\ell_\mathrm{off}$th synapse weight $\theta^{(\ell_\mathrm{off})}$, see (2.13), we can interpret $\theta^{(\ell_\mathrm{off})}$ as the $\ell_\mathrm{off}$th tap of an FIR-filter, applied to the input spike signal $z_\mathrm{in}[\kappa]$. Thus, using delay filters, we can learn an optimal feed-forward filter with coefficients $\theta^{(\ell_\mathrm{off})}$. For the delay filters we use $\ell_\mathrm{off} = 0, 1, \ldots, \ell_\mathrm{off,max}$ with $\ell_\mathrm{off,max} = 10$.

For each type of feed-forward filter, we conduct training using the desired output spike pattern $\boldsymbol{Z}_{\leq K-1}$ with a single spike, see (3.17), and the modified learning signal (3.18). After training is finished, we convert the probabilistic SNN into its deterministic counterpart, obtained as described in Sec. 3.5.1. As neural decoding, we implement SRD and EOTM. The training aims to fire only a single spike indicating the correct class label $i$, and thus should be well suited for SRD. However, in the presence of multiple output spikes, SRD may fail; e.g., if two spikes are fired, each indicating another class label. To avoid such situations, we additionally investigated EOTM as neural decoding. In order to prevent output neurons to fire when using EOTM, we set their threshold to $v_\mathrm{th} = 100$. As a result, for each optimized SNN, we evaluate four different approaches:

- Probabilistic SNN with SRD (Prob.-SRD)
- Probabilistic SNN with EOTM (Prob.-EOTM)
- Deterministic SNN with SRD (Det.-SRD)
- Deterministic SNN with EOTM (Det.-EOTM)

For training, for each epoch a batch of $|\mathcal{B}_\mathrm{t}| = 100$ samples is processed, then the parameters of the SNN are updated. Overall, the SNN is optimized for 10 000 epochs. For evaluation, an evaluation batch size of $|\mathcal{B}_\mathrm{e}| = 10^7$ is chosen. Based on a hyperparameter search, we choose the number of neurons in the hidden layer to be $N_\mathrm{hid} = 100$. To obtain the output spike amplification $\gamma$ and learning rates $\nu = \nu_{\boldsymbol{\theta}_\mathcal{Z}} = \nu_{\boldsymbol{\theta}_\mathcal{H}}$, we also executed a hyperparameter search, conducted at a fixed $E_\mathrm{b}/N_0 = 5\,\mathrm{dB}$. To determine $\gamma$ and $\nu$, we independently optimized probabilistic SNNs with the respective hyperparameters, and then evaluated for all four final structures (Prob.-SRD, Prob.-EOTM, Det.-SRD, Det.-EOTM). Fig. 3.14(a) shows the BER performance as a function of $\gamma$ when using delay filters. While both SRD-based approaches are not able to achieve ML performance, both EOTM-based approaches achieve near-ML performance. For $\gamma > 300$, Prob.-EOTM and Det.-EOTM both achieve near-ML performance. Given that $\gamma$ modulates the learning signal, we select the smallest value that still yields near-ML performance for both Prob.-EOTM and Det.-EOTM, namely $\gamma = 300$.

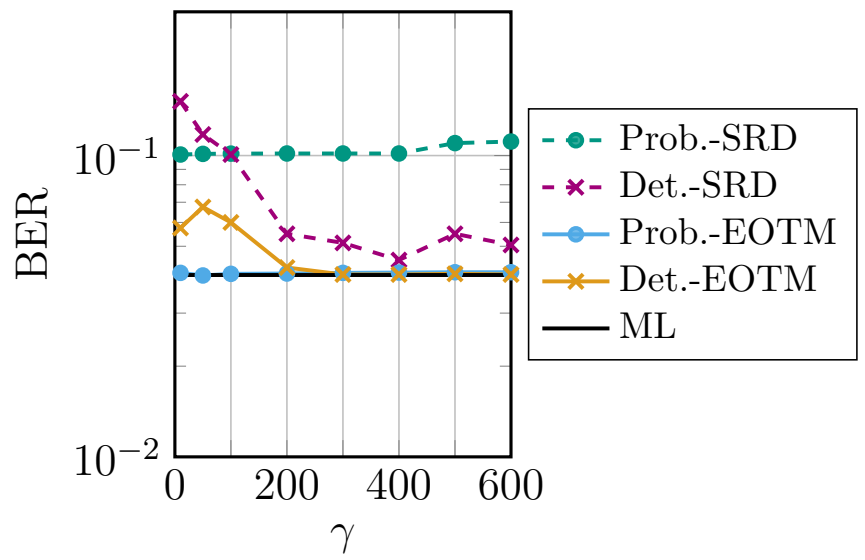

**(a)** Impact of $\gamma$ when using delay feed-forward filters.

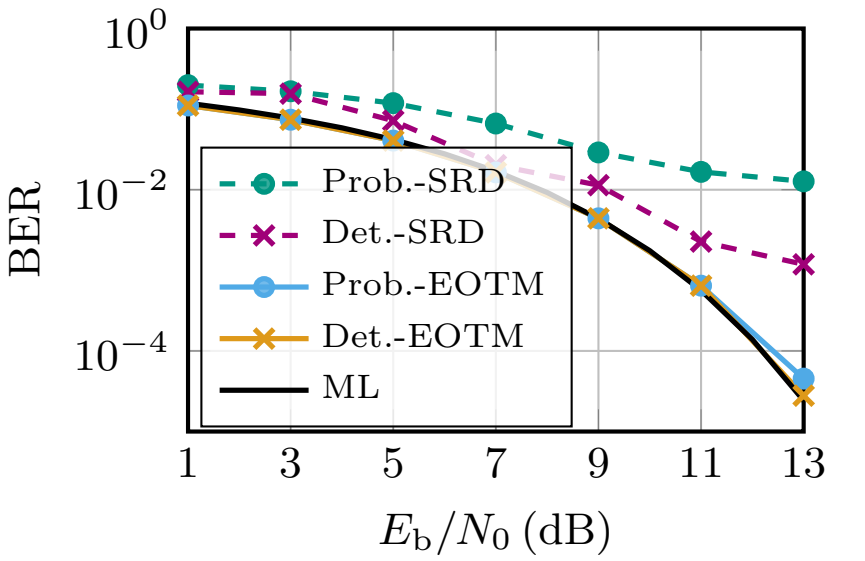

**(b)** BER-curve when using delay feed-forward filters.

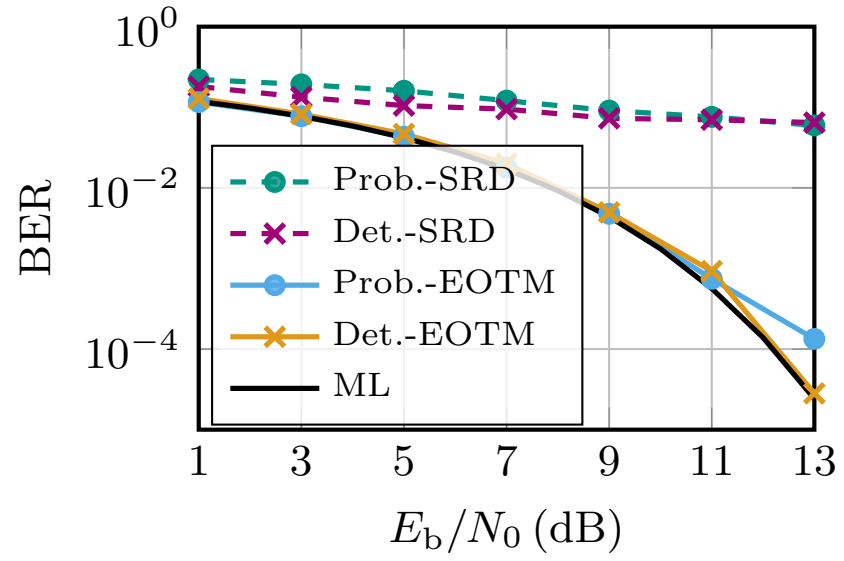

**(c)** BER-curve when using raised-cosine feed-forward filters.

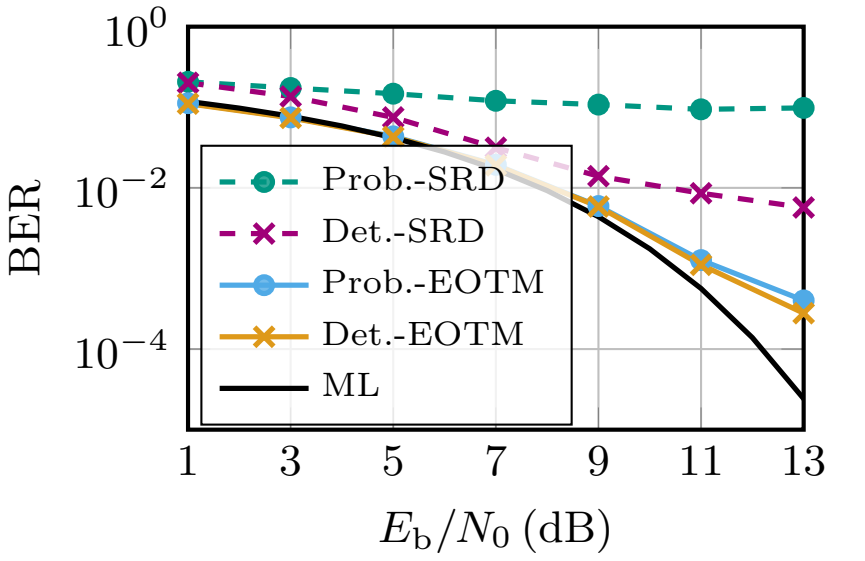

**(d)** BER-curve when using exponential feed-forward filters.

**Figure 3.14:** Performance of the SNN-based detector for 16-QAM updated using probabilistic SNNs.

Using the same approach, we obtain the output spike amplification $\gamma$ and the learning rates $\nu$ for the various feed-forward filters. For the exponential filter, we use $\gamma = 300$ and $\nu = \frac{1}{2}$, for the raised-cosine filters $\gamma = 200$ and $\nu = 1$, and for the delay filters $\gamma = 300$ and $\nu = 3$.

The resulting BER curves for all feed-forward filters are shown in Figs. 3.14(b)-3.14(d). For each $E_b/N_0$-value, an independent SNN is trained, which is then converted into the four final structures. For all feed-forward filters, the SNNs using EOTM perform significantly better than those using SRD. This is intuitive since SRD relies on single output spikes. If the output neuron indicating the true label is not sufficiently exhibited to emit an output spike, its membrane potential may still be large enough to indicate the correct class using EOTM. The same applies if two or more output neurons are sufficiently exhibited and both emit an identical number of spikes: Under SRD, the class labels represented by these spiking neurons are equally likely. However, their membrane potentials can differ, which allows correct classification using EOTM. Thus, EOTM proves to be more effective for a classification task than SRD. These findings are supported by [LBCS24], which investigated different neural decodings for a classification task, and demonstrated, that for $K < 15$, EOTM clearly outperforms SRD.

Using the raised-cosine and delay feed-forward filters, Det.-EOTM achieves near-ML performance. However, when using the exponential feed-forward filter, which approximates the LIF-neuron, none of the optimized SNNs achieves near-ML performance. This is not in line with our expectations, since we demonstrated in Sec. 3.4.1 that the combination of linear RFE and an SNN consisting of LIF neurons can achieve near-ML performance. The SRM model with an exponential feed-forward filter corresponds to an LIF neuron with $\tau_\mathrm{s} = 0$. When converting the probabilistic SNN into its deterministic counterpart, $v_\mathrm{th}$ of the SRM is fixed. Thus, both the SRM and LIF exhibit deterministic spike behavior. When comparing the different feed-forward filters, both the delayed raised-cosine and delay filters are able to delay the response to a single incoming spike up to $\ell_{\mathrm{off,max}}$ time steps: if a single spike is fed to a neuron at time step $\kappa = 0$, it can trigger an output spike at all time steps $\kappa = 0, \dots, \ell_{\mathrm{off,max}}$. However, the exponential filter lacks this property, since its response to a single incoming spike can only be immediate.

Due to the character of RFE, all spikes of the input signals are contained in the first 9 time steps, i.e., for $\kappa = 0, \dots, 8$, with the spikes carrying the most information being fired very early. In contrast, the only spike of $\boldsymbol{Z}_{\leq K-1}$ is expected at the last time step $\kappa = 10$. Thus, input spikes need to be delayed. We hypothesize that the missing ability of the exponential filter to delay spikes, combined with the properties of the used neural encoding and our definition of $\boldsymbol{Z}_{\leq K-1}$ results in suboptimal learning.

To validate this hypothesis, we modify the definition of $\boldsymbol{Z}_{\leq K-1}$ to

$$z_{\mathrm{out},j}[\kappa] = \begin{cases} 1, & \text{if } j = i,\ \kappa = 0, \dots, K-1\,, \\ 0, & \text{else}\,. \end{cases} \tag{3.19}$$

Instead of a single spike at the last time step $\kappa = K$, compare (3.17), the modified definition (3.19) enforces the $i$th output neuron to spike at all time steps $\kappa = 0, \dots, K-1$, with $i \in \{1, \dots, |\mathcal{X}|\}$ denoting the true class label. With the modified definition of $\boldsymbol{Z}_{\leq K-1}$, both Prob.-EOTM and Det.-EOTM, combined with the exponential filter as a feed-forward filter, achieve near-ML performance.

### Main Findings

In this section, we investigated the update of probabilistic SNNs. To convert probabilistic SNNs into deterministic SNNs, we introduced a method to determine $v_\mathrm{th}$ based on the statistical properties of all neurons. We demonstrated the successful optimization of probabilistic SNNs using exponential, delayed raised-cosine and delay filters as feed-forward filters, and discussed the necessity of a properly chosen desired output spike pattern $\boldsymbol{Z}_{\leq K-1}$.

## 3.6 Key Findings

In this chapter, we investigated three different update rules and several neural encodings. For all approaches, an SNN-based detector for the AWGN-channel using 16-QAM was successfully designed. However, we experienced that the update rules and neural encodings exhibit considerable differences in practical applicability and adaptability.

When optimizing SNNs using STDP, the optimization is limited to a single linear layer. Inspired by the ELM, we extended the SNN to a multilayer network. However, due to the locality of STDP, only the last layer is optimized. In contrast, probabilistic SNNs and SG with BPTT enable optimization across several layers. SNNs using LIF neurons and optimized using SG with BPTT are able to achieve near-ML performance. For probabilistic SNNs, exponential, shifted raised-cosine and delay feed-forward filters are also able to achieve ML performance. When both optimization approaches are compared, a major difference lies in the definition of the learning objective: whereas the BPTT-based update uses the cross-entropy (CE) loss, requiring only the correct class label $i$ as a target, probabilistic SNNs require a predefined spike pattern $\boldsymbol{Z}_{\leq K}$. Consequently, updating probabilistic SNNs requires a carefully chosen $\boldsymbol{Z}_{\leq K}$, which can pose a significant challenge. When using the CE loss, the spike timings of the output neurons do not matter, as long as a sufficient amount of output spikes is generated; respectively, the membrane potential is sufficiently charged. Furthermore, BPTT with SG enables the application of standard MaL techniques. The above findings support [ZBC+21], highlighting SG with BPTT as a promising candidate for an SNN update rule.

Additionally, we investigated several neural encoding strategies. In Sec. 3.3, we proposed several neural encodings, which are based on BE and TTFS. The obtained encodings were highly specified to the problem setting, incorporating prior information of the ML detector. Furthermore, the resolution of the encoding and hence the performance of the SNN-based detector highly depends on $K$. In contrast, the population encoding methods RFE, TE and QE have demonstrated a fast and efficient encoding, resulting in a reasonable amount of spike signals $N_{\text{enc}}$, and a reasonable number of discrete time steps $K$. Incorporating prior knowledge of the ML detector enabled minimization of both $N_{\text{enc}}$ and $K$. In contrast, increasing $N_{\text{enc}}$ enhances system performance, mitigating the impact of poorly initialized encodings.

# 4 Spiking Neural Network-based Equalization and Demapping

In the previous chapter, several neural encodings and update rules were investigated, identifying BPTT with SGs as a promising update rule, and RFE, TE, and QE as promising neural encodings. In this chapter, we apply these techniques to design an SNN-based equalizer and demapper.

## 4.1 Overview of Existing Work

Recent developments in SNN-based equalizers and demappers have led to two general structures for SNN-based equalizers and demappers: without decision feedback [ABM+22a, ABS+23], and with decision feedback [BvBS23, vBES23]. In the following, we briefly outline both approaches.

### 4.1.1 SNN-based Equalizer and Demapper without Decision Feedback

The SNN-based equalizer and demapper without decision feedback (NFE-SNN) receives a sequence of symbols, which is observed at the output of a frequency-selective channel, and estimates the transmitted bits [ABM+22a, ABS+23]. Thus, it combines the tasks of an equalizer and a demapper [PS08, Sec. 9.4]. For an intensity modulation with direct detection (IM/DD)-link, a non-coherent optical communication channel [Cha19], the NFE-SNN clearly outperforms linear equalizers [ABM+22a, ABS+23]. It can either be implemented in software using `Norse`, or it can be run on neuromorphic hardware, e.g., BrainScaleS-2 [ABM+22b, ABS+23]. To optimize the NFE-SNN on BrainScaleS-2, the hardware observables (membrane potential, synaptic current, spikes) are discretely sampled on a fixed temporal grid. Using `Norse` and BPTT with SG, the gradient is estimated and the parameters of the model on BrainScaleS-2 are updated. Both implementations yield similar results, with a minor performance penalty on the neuromorphic hardware. This suggests that `Norse` seems to be a proper tool for simulating SNNs, matching the dynamics of BrainScaleS-2. Furthermore, the NFE-SNN is able to mitigate nonlinear impairments of the transmission.

RFE and MOTM are used as neural encoding and decoding, respectively [ABS+23]. Due to the conversion of spike times into spike signals, the RFE is non-differentiable, and hence not optimizable using BPTT. To enable optimization, we proposed two methods to improve the neural encoding [vBES24, EvBS25]. The first approach replaces the binary RFE with a real-valued adaptable neural encoding that aims to reduce the number of spikes generated by the hidden layer and hence increase energy efficiency [vBES24]. Using a quantizer with $N_\mathrm{c}$ quanti-zation levels, the values to be encoded are mapped to one out of $N_\text{c}$ real-valued matrices of size $N_\text{in} \times K$. Due to the real-valued nature of the encoding, both the SNN and neural encoding can be jointly optimized in terms of classification accuracy and energy efficiency using BPTT with SG. To achieve the latter, the $\ell_1$-over-$\ell_2$ quasinorm ratio, measuring the sparsity of the encoding matrices, is added to the CE loss. We demonstrate that the number of generated spikes can be reduced while preserving the equalization and demapping performance of the NFE-SNN. However, due to the real-valued encoding matrices, the proposed encoding loses its binary spiking nature.

The second approach exploits reinforcement learning (RL) to update the parameters of the RFE [EvBS25]. As we demonstrated in Sec. 3.4.1, the initialization of the field centers and field width of the RFE can significantly affect performance. Thus, the parameters need to be determined heuristically, which is time-consuming since it requires the optimization of several networks. Using update methods inspired by RL, we can jointly optimize the field centers and field widths of the RFE together with the SNN. For poor initializations of the neural encoding, we demonstrate that the RL-based update is able to find proper encoding parameters. We furthermore demonstrate that the joint optimization enables a significant reduction of the dimension of input features $N_\text{enc}$ and runtime $K$ of the SNN, while preserving performance. A more detailed description and obtained results are given later in Ch. 6.

Additionally to the neural encoding, the neural decoding can be varied. When comparing the NFE-SNN combined with different neural decodings, membrane potential-based decodings, such as EOTM and MOTM, outperform spike-based decodings, such as TTFS and rate decoding [LBCS24]. The membrane potential-based decodings achieve lower latency and require less spiking activity from the hidden layer than the spike-based decodings.

For experimental IM/DD data obtained by the specific setup of [BSA$^+$23], the NFE-SNN performs similarly to a linear equalizer of equal size [BSA$^+$23]. Since the given channel is mainly impaired by linear distortion and noise, the NFE-SNN and the linear equalizer achieve the same performance, which implies that the nonlinear equalization capability of the NFE-SNN remains unused

### 4.1.2 SNN-based Equalizer and Demapper with Decision Feedback

The SNN-based equalizer and demapper with decision feedback (DFE-SNN) extends the NFE-SNN by the feedback of already decided symbols [BvBS23]. Thus, the estimate of the transmit bits is obtained based on the most recent symbols observed at the channel output, combined with the most recent symbol decisions. For three linear frequency-selective channels [PS08, p. 654], we demonstrated that the DFE-SNN outperforms linear equalizers and achieves similar performance as the classical decision feedback equalizer [BvBS23]. Replacing the linear frequency-selective channels with the IM/DD-link, the DFE-SNN is able to significantly outperform linear equalizers and the classical decision feedback equalizer [vBES23].

When comparing the NFE-SNN and the DFE-SNN on IM/DD links, where the non-linearity is the primary contributor to the distortion, both equalizers perform similarly. If the parameterization of the IM/DD link is chosen to match the parameters of the third and fourth generations of optical communication systems [Cha19, Tab. I], inter-symbol interference (ISI) caused by chromatic dispersion (CD) becomes more severe. For more severe ISI, the DFE-SNN clearly outperforms the NFE-SNN [vBES23].

For the IM/DD-link with more severe ISI, the DFE-SNN was implemented on a field programmable gate array (FPGA) in [MNHW24]. When comparing the DFE-SNN with an ANN-based equalizer and demapper of alike architecture, the DFE-SNN achieves noticeably better performance. However, the DFE-SNN requires 25 times less energy per multiply-accumulate (MAC) operation than the ANN-based equalizer and demapper. Yet, it requires 25 times as many MAC operations as the ANN-based equalizer and demapper. Thus, the overall energy consumption per processed bit is similar. It should be noted that the FPGA is digital and synchronous. The use of asynchronous analog hardware may further decrease energy consumption.

### 4.1.3 Comparison of the NFE-SNN and DFE-SNN

To motivate developers of neuromorphic hardware to design energy-efficient receivers, we provide a dataset consisting of the NFE-SNN and DFE-SNN together with the model of the IM/DD link and two parameterizations [AEvB+25]. When comparing the NFE-SNN [ABS+23] and the DFE-SNN [BvBS23], several key differences can be identified (based on the dataset [AEvB+25]). While the NFE-SNN uses a linear RFE as neural encoding and MOTM as neural decoding, the DFE-SNN uses TE and EOTM. Furthermore, the DFE-SNN applies recurrent connections, which are neglected for the NFE-SNN. Lastly, in order to minimize the number of generated spikes, the NFE-SNN applies a regularization term. As a result, the literature lacks a systematic and, hence, fair comparison of the NFE-SNN and the DFE-SNN.

In the following, for two parameterizations of the IM/DD-link [AEvB+25], we attempt to provide a fair and comprehensive comparison of the NFE-SNN and DFE-SNN.

## 4.2 Equalization Structures

### 4.2.1 Feed-forward Structures and Decision Feedback Structures

In band-limited channels, ISI may degrade the transmission quality [PS08, p. 597]. Channel equalization aims to compensate for ISI [PS08, p. 640]. Let $\boldsymbol{y} \in \mathbb{C}^{n_{\text{tap}}}$ denote a sequence of complex-valued symbols, received after transmission of a sequence of symbols $\boldsymbol{x} \in \mathcal{X}$. Hereby, $n_{\text{tap}} \in \mathbb{N}$ denotes the number of transmission symbols affected by ISI. Given the observation $\boldsymbol{y}$, the equalization returns a

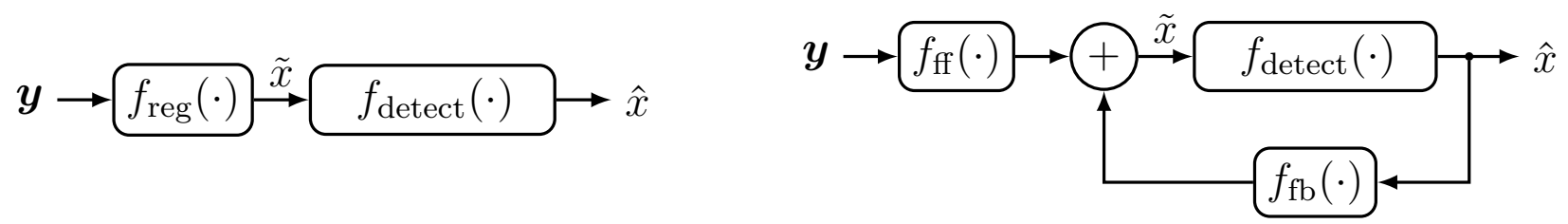

**(a)** Feed-forward structure.

**(b)** Feedback structure [PS08, p. 661].

**Figure 4.1:** Equalization structures when using regression.

correct estimate $\hat{x}$ of the transmit symbols $x$. Afterwards, a demapper can be applied, which converts $\hat{x}$ into an estimate $\hat{\boldsymbol{b}}$ of the transmitted bit sequence. We now introduce equalization using either regression or classification. For both approaches, feed-forward and decision-feedback methods exist, which we will discuss next.

### Equalization via Regression

Equalization via regression aims at recovering the transmit symbols $x$ as closely as possible and returns an estimate $\tilde{x} \in \mathbb{C}$. We can express the estimate as $\tilde{x} = f_{\text{reg}}(\boldsymbol{y})$, where $f_{\text{reg}} : \mathbb{C}^{n_{\text{tap}}} \to \mathbb{C}$ denotes an arbitrary function [PS08, Sec. 9.4]. Using the mean squared error (MSE) criterion, we can determine $f_{\text{reg}}(\cdot)$ by [PS08, Sec. 9.4-2]

$$f_{\text{reg}}(\boldsymbol{y}) = \underset{f:\mathbb{C}^{n_{\text{tap}}}\to\mathbb{C}}{\arg\min}\ \mathbb{E}_X\left\{|f_{\text{reg}}(\boldsymbol{y}) - x|^2\right\} . \tag{4.1}$$

To convert the estimate $\tilde{x}$ into a valid (received) value $\hat{x} \in \mathcal{X}$, a detector $f_{\text{detect}} : \mathbb{C} \to \mathcal{X}$ can be applied. Using the Lloyd-Max quantizer as detector, optimal quantization levels and decision boundaries are obtained, also minimizing the MSE [Llo82]. The resulting block diagram is displayed in Fig. 4.1(a). Since the setup does not incorporate any loops, we refer to it as a feed-forward structure. A well-known representative of the feed-forward structure is the linear equalizer [PS08, Sec. 9.4].

By adding a feedback loop, the decision feedback structure is obtained, whose performance is generally better than that of feed-forward structures [PS08, Sec. 9.5]. The general setup is displayed in Fig. 4.1(b). Compared to the feed-forward structure, a feedback loop is applied, which processes the already decided symbols by another arbitrary function $f_{\text{fb}}(\cdot)$. Consequently, we distinguish between the feed-forward filter $f_{\text{ff}} : \mathbb{C}^{n_{\text{ff}}} \to \mathbb{C}$ and the feedback filter $f_{\text{fb}} : \mathbb{C}^{n_{\text{fb}}} \to \mathbb{C}$, where $n_{\text{ff}} \in \mathbb{N}$ and $n_{\text{fb}} \in \mathbb{N}^+$ denote the dimension of the filter input. The estimate $\tilde{x} \in \mathbb{C}$ is obtained by

$$\tilde{x} = f_{\text{ff}}(\boldsymbol{y}) + f_{\text{fb}}(\hat{\boldsymbol{x}}) , \tag{4.2}$$

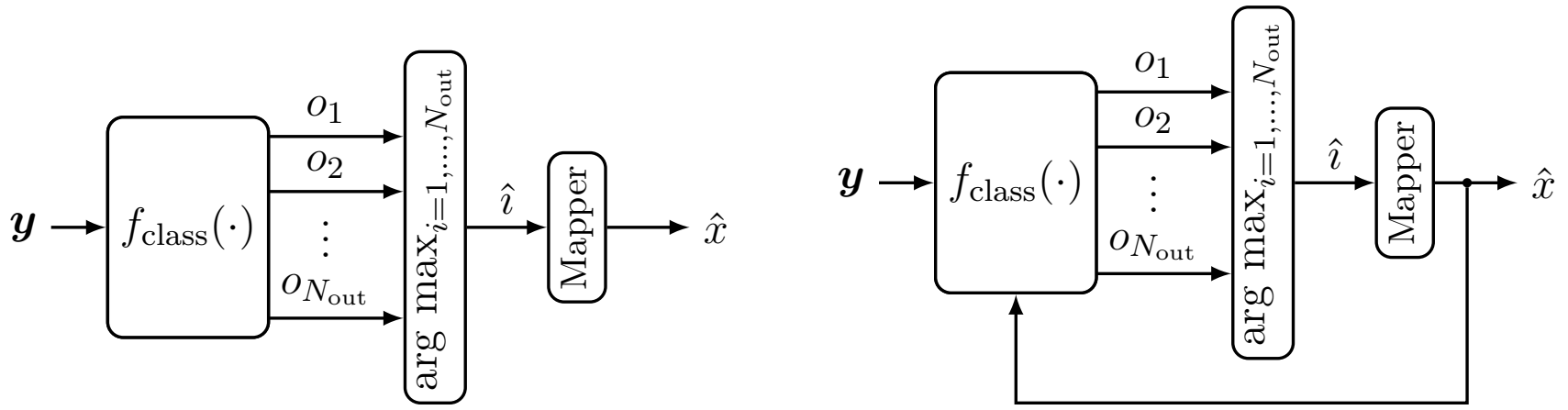

**(a)** Feed-forward structure. **(b)** Feedback structure.

**Figure 4.2:** Equalization structures when using classification.

where $\hat{\boldsymbol{x}} \in \mathcal{X}^{n_{\text{fb}}}$ denotes the sequence of the most recent estimates, and $\boldsymbol{y} \in \mathbb{C}^{n_{\text{ff}}}$ the sequence of the most recent received symbols. Both $f_{\text{ff}}(\cdot)$ and $f_{\text{fb}}(\cdot)$ can be obtained by using the MSE criterion.

### Equalization via Classification

We can also write the equalization task as a classification problem. Via a bijective mapping, we can represent each transmit symbol $x \in \mathcal{X}$ by an index $i \in \{1, \ldots, |\mathcal{X}|\}$. Let $f_{\text{class}} : \mathbb{C}^{n_{\text{tap}}} \to \mathbb{R}^{N_{\text{out}}}$ be an arbitrary function, where the number $N_{\text{out}}$ of output features is given by $N_{\text{out}} = |\mathcal{X}|$. Furthermore, let $o_i$, $i = 1, \ldots, N_{\text{out}}$ denote the $i$th output value of $f_{\text{class}}(\cdot)$. We can obtain the estimated class index $\hat{\imath}$ by

$$\hat{\imath} = \underset{i=1,\ldots,N_{\text{out}}}{\arg\max}\; o_i\,, \tag{4.3}$$

which we can demap to the respective estimated transmit symbol $\hat{x}$. Fig. 4.2(a) shows the respective block diagram. Again, a feedback loop can be added, incorporating the most recent decisions, see Fig. 4.2(b). To obtain $f_{\text{class}}(\cdot)$, we can apply the softmax function to the outputs $o_i$, which returns a probability distribution over the possible classes, and minimize the CE between the modified outputs and the desired distribution, obtained using the true label $i$ (ground truth).

## 4.2.2 Realization of Equalizers

To distinguish between the feed-forward and decision feedback structures, we refer to them as *no feedback equalization (NFE)* and *decision feedback equalization (DFE)*. To realize $f_{\text{reg}}(\cdot)$, $f_{\text{ff}}(\cdot)$, $f_{\text{fb}}(\cdot)$, and $f_{\text{class}}(\cdot)$, several approaches exist, which we shortly introduce:

### Regression Using FIR Filters

We can implement $f_{\text{reg}}(\cdot)$, $f_{\text{ff}}(\cdot)$, and $f_{\text{fb}}(\cdot)$ using FIR filters, where the filter

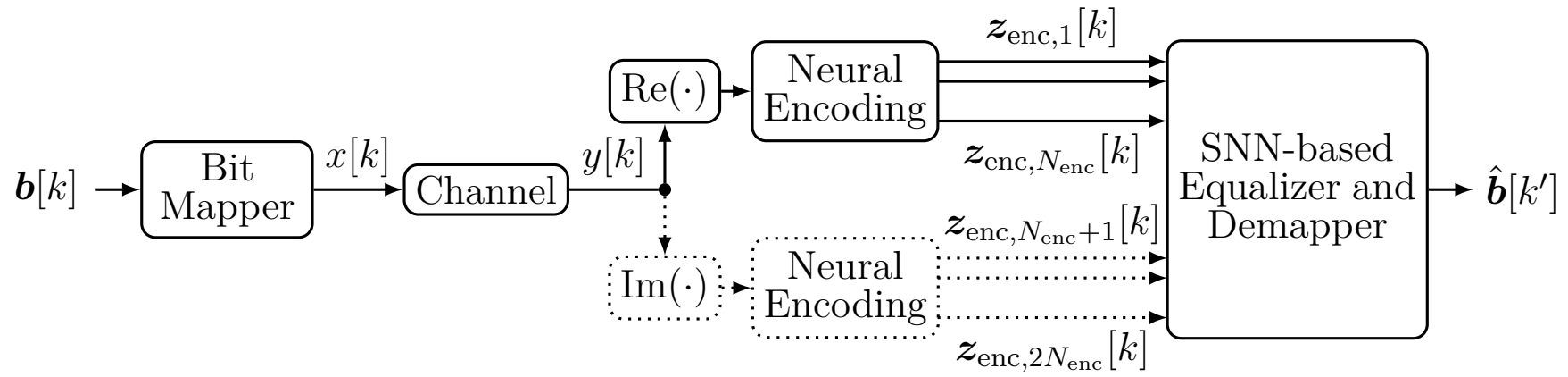

**Figure 4.3:** Block diagram of the communication link used with the NFE-SNN and DFE-SNN. By adding the dotted components, the setup can be extended to coherent communications [BvBS23].

taps are updated using the MSE criterion. We refer to these approaches as NFE-FIR and DFE-FIR.

**Regression and Classification Using ANNs**

Since ANNs are function approximators, we can use ANNs to implement the different equalization structures. We denote the regression-based equalization methods using ANNs as NFE-ANN$_{\mathrm{reg}}$ and DFE-ANN$_{\mathrm{reg}}$. When using the feedback structure, we can jointly approximate $f_{\mathrm{ff}}(\cdot)$ and $f_{\mathrm{fb}}(\cdot)$ by the same ANN. The parameters of the ANNs can be obtained using the MSE loss between $x$ and $\tilde{x}$.

When executing classification-based equalization using ANNs, we refer to both approaches as NFE-ANN$_{\mathrm{class}}$ and DFE-ANN$_{\mathrm{class}}$. For both approaches, the ANN approximates $f_{\mathrm{class}}(\cdot)$ and can be trained using the CE loss.

**Classification Using SNNs**

Classification-based equalization can also be carried out using SNNs [ABS$^+$23, BvBS23]. In the next section, we introduce the NFE-SNN [ABS$^+$23] and DFE-SNN [BvBS23].

## 4.3 SNN-based Equalization and Demapping

While the NFE-SNN is an SNN-based equalizer without a feedback structure [ABS$^+$23], the DFE-SNN is an SNN-based equalizer with decision feedback [BvBS23]. For both equalizers, the communication system is as depicted in Fig. 4.3. At each time instant $k \in \mathbb{N}_0$ of the system, uniformly distributed bits $\boldsymbol{b}[k] = \{0,1\}^{\log_2(|\mathcal{X}|)}$ are generated i.i.d. and mapped to a transmit symbol $x[k] \in \mathcal{X}$, which is then transmitted over a channel. At the channel output, $y[k]$ is observed. In the case of coherent communications, i.e., when $\mathcal{X} \subset \mathbb{C}$, both the inphase and the quadrature components are encoded independently using identical neural encoders. In the case of non-coherent communications, i.e., when $\mathcal{X} \subset \mathbb{R}$, the dashed components in Fig. 4.3 are not required and can be omitted. For each symbol, each neural encoding outputs $N_{\mathrm{enc}}$ spike signals $\boldsymbol{z}_{\mathrm{enc},j} \in \{0,1\}^K$, $j = 1, \ldots, N_{\mathrm{enc}}$. The spike signals are fed to the SNN-based equalizer and demapper, which conducts both equalization and bit demapping. Thus, it returns an estimate $\hat{\boldsymbol{b}}[k']$ of the transmit bit sequence $\boldsymbol{b}[k]$, where $k' = k - n_{\text{off}}$ indicates a latency of $n_{\text{off}} \in \mathbb{N}_0$ samples, which depends on the channel.

For a clearer representation when introducing the NFE-SNN and DFE-SNN, we define a compressed notation for the spike signals $\boldsymbol{z}_{\text{enc},j}$, and assume coherent communications. Furthermore, it is important to distinguish between the communication system time $k \in \mathbb{N}_0$ and the neural encoding time $\kappa = 0, 1, \ldots, K-1$. At each time instant $k$ of the communication system, the neural encoding generates $2N_{\text{enc}}$ spike signals with $K$ time steps. We can write the output of both neural encoders and therefore the input to the SNN-based equalizer and demapper as a binary matrix $\boldsymbol{Z}_{\text{enc}}[k] \in \{0,1\}^{2N_{\text{enc}} \times K}$ with

$$\boldsymbol{Z}_{\text{enc}}[k] := \begin{pmatrix} \boldsymbol{z}_{\text{enc},1}[k] \\ \boldsymbol{z}_{\text{enc},2}[k] \\ \vdots \\ \boldsymbol{z}_{\text{enc},2N_{\text{enc}}}[k] \end{pmatrix} \tag{4.4}$$

$$= \begin{pmatrix} z_{\text{enc},1}[k,1] & z_{\text{enc},1}[k,2] & \ldots & z_{\text{enc},1}[k,K] \\ z_{\text{enc},2}[k,1] & z_{\text{enc},2}[k,2] & \ldots & z_{\text{enc},2}[k,K] \\ \vdots & \vdots & \ldots & \vdots \\ z_{\text{enc},2N_{\text{enc}}}[k,1] & z_{\text{enc},2N_{\text{enc}}}[k,2] & \ldots & z_{\text{enc},2N_{\text{enc}}}[k,K] \end{pmatrix},$$

where $z_{\text{enc},j}[k,\kappa] \in \{0,1\}$ denotes the value of the $j$th spike signal at the $\kappa$th time step of the neural encoding, in dependence on the overall communication system time instant $k$, where $j = 1, \ldots, 2N_{\text{enc}}$. Consequently, $\boldsymbol{z}_{\text{enc},j}[k]$ denotes the $j$th discrete-time spike signal output by the neural encoding at overall communication system time instant $k$. For non-coherent communication, the same considerations apply; however, $\boldsymbol{Z}_{\text{enc}}[k] \in \{0,1\}^{N_{\text{enc}} \times K}$, since only one neural encoder is required.

Thus, at each time instant $k$, the neural encoding converts the received value $y[k]$ into a spike matrix $\boldsymbol{Z}_{\text{enc}}[k]$. Afterwards, $\boldsymbol{Z}_{\text{enc}}[k]$ is fed to the SNN-based equalizer and demapper, which is then run for $K$ time steps. After the $K$ time steps are finished, the output $\hat{\boldsymbol{b}}[k']$ is obtained, and the internal parameters of the SNN, e.g., the membrane potential and the synaptic currents of all neurons, are reset to their default values. Thus, after each estimate, the SNN is reset, and the communication system time $k$ is increased, i.e., $k \leftarrow k + 1$.

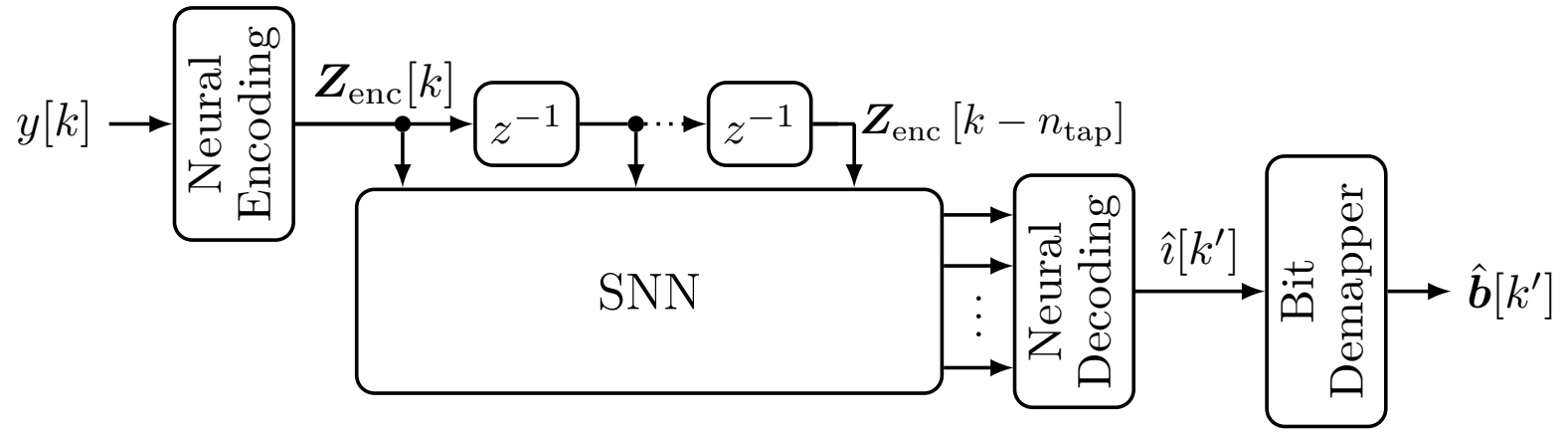

**Figure 4.4:** Block diagram of the NFE-SNN, based on [ABS+23].

### 4.3.1 SNN-based Equalization and Demapping without Decision Feedback

We now provide a more detailed description of the NFE-SNN [ABS+23]. The NFE-SNN is an SNN-based equalizer and demapper without a feedback structure, which is inspired by classification-based equalization. Its setup is displayed in Fig. 4.4. The neural encoding translates the received symbols $y[k]$ into the input spike matrix $\boldsymbol{Z}_{\text{enc}}[k]$. Both $\boldsymbol{Z}_{\text{enc}}[k]$ and the previous $n_{\text{tap}}$ spike matrices are provided to the SNN. Thus, the SNN receives $\boldsymbol{Z}_{\text{enc}}[k], \boldsymbol{Z}_{\text{enc}}[k-1], \ldots, \boldsymbol{Z}_{\text{enc}}[k-n_{\text{tap}}]$, representing the $n_{\text{tap}}$ most recent received symbols. After processing the input, the output of the SNN is fed to the neural decoding, which estimates $\hat{\imath}[k'] \in \{1, 2, \ldots, |\mathcal{X}|\}$. The bit demapper then obtains the received bits $\hat{\boldsymbol{b}}[k']$.

The number $N_{\text{in}}$ of input neurons of the SNN is determined by $N_{\text{in}} = 2 \cdot n_{\text{tap}} \cdot N_{\text{enc}}$ for coherent communications, and $N_{\text{in}} = n_{\text{tap}} \cdot N_{\text{enc}}$ for non-coherent communications.

### 4.3.2 SNN-based Equalization and Demapping with Decision Feedback

When adding a feedback loop to the NFE-SNN, we obtain the DFE-SNN [BvBS23], shown in Fig. 4.5. The DFE-SNN has two paths: a feed-forward path and a feedback path, consisting of $n_{\text{ff}}$ and $n_{\text{fb}}$ taps, respectively. While the feed-forward path contains the $n_{\text{ff}}$ most recent spike matrices $\boldsymbol{Z}_{\text{enc}}[k]$ representing the $n_{\text{ff}}$ most recent received values $y[k]$, the feedback path contains the $n_{\text{fb}}$ most recent, already decided symbols. In the feedback path, a one-hot encoder converts the estimated class label $\hat{\imath}[k]$ into a spike matrix $\boldsymbol{Z}_{\hat{\imath}}[k] \in \{0,1\}^{|\mathcal{X}| \times K}$. It is generated by

$$z_{\hat{\imath},j}[k, \kappa] = \begin{cases} 1, & \text{if } j = \hat{\imath} \text{ and } \kappa = 0\,, \\ 0, & \text{else}\,. \end{cases} \tag{4.5}$$

Thus, $\boldsymbol{Z}_{\hat{\imath}}[k]$ is all-zero, except for a single spike in the first column and $\hat{\imath}$th row, which indicates the estimated class label. As with $\boldsymbol{Z}_{\text{enc}}[k]$, the individual rows of $\boldsymbol{Z}_{\hat{\imath}}[k]$ can be interpreted as spike signals over time, which are fed to the SNN.

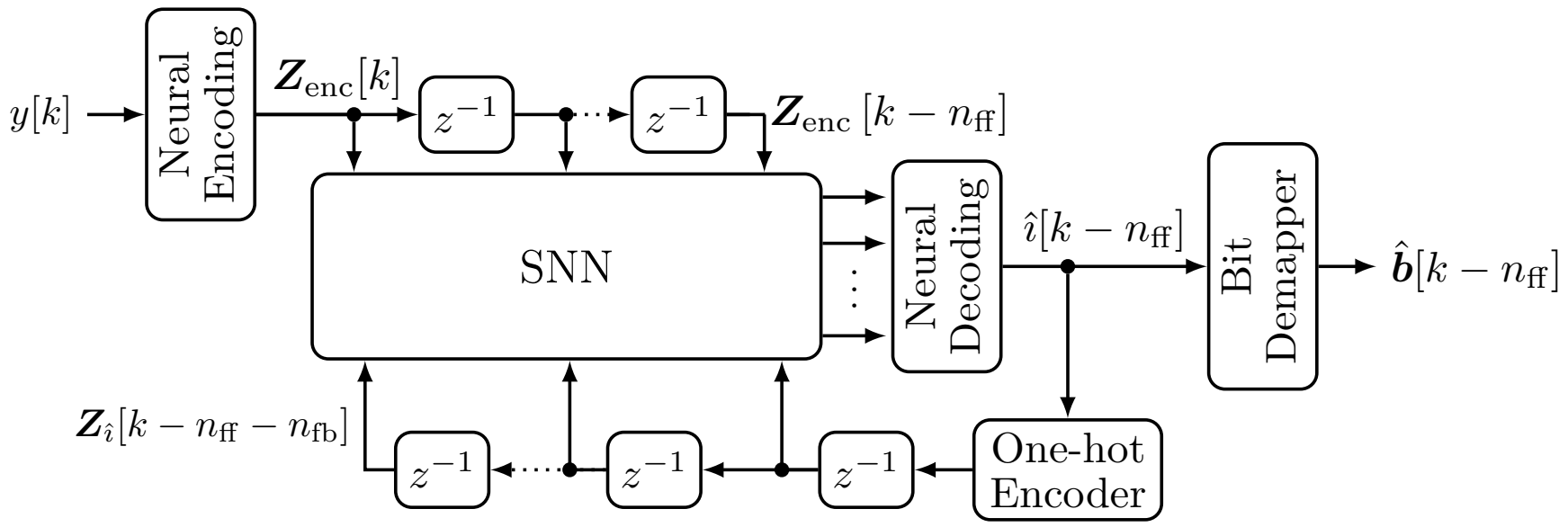

**Figure 4.5:** Block diagram of the DFE-SNN, based on [BvBS23].

With this design of the DFE-SNN, for coherent communications, the number of input neurons $N_{\mathrm{in}}$ of the SNN is given by

$$N_{\mathrm{in}} = 2N_{\mathrm{enc}}n_{\mathrm{ff}} + |\mathcal{X}|n_{\mathrm{fb}}\,, \tag{4.6}$$

and for non-coherent communication by

$$N_{\mathrm{in}} = N_{\mathrm{enc}}n_{\mathrm{ff}} + |\mathcal{X}|n_{\mathrm{fb}}\,. \tag{4.7}$$

## 4.4 Systematic Overview of SNN-based Equalizers and Demappers

In the following, we will briefly discuss modifications to both the optimization and the architecture of the NFE-SNN and DFE-SNN. Afterwards, we give a systematic overview of the resulting SNN-based equalizers and demappers, which we then compare with each other w.r.t. their performance given the IM/DD link.

### 4.4.1 Design Aspects

#### Regularization

The energy efficiency of SNNs depends on achieving a small number of spike events [YBW24]. To minimize the number of generated spikes within the SNN, during optimization we can apply a regularization term [ABS$^+$23, AEvB$^+$25]. It depends on the weights of the SNN and the number of generated spikes per hidden layer neuron. When minimizing the regularization term, the average number of spikes $Z_{\mathrm{avg}}$ emitted by the hidden layer can also be reduced. Given a batch $\mathcal{B}_{\mathrm{t}}$, the regularization loss $\alpha_{\mathrm{r}} \in \mathbb{R}^+$ is calculated by [AEvB$^+$25]

$$\alpha_{\mathrm{r}} = \alpha_{\mathrm{r},1}\frac{1}{\|\boldsymbol{\theta}^{(\mathrm{in})}\|_0}\sum_{\theta\in\boldsymbol{\theta}^{(\mathrm{in})}}\theta^2 + \alpha_{\mathrm{r},2}\frac{1}{\|\boldsymbol{\theta}^{(\mathrm{out})}\|_0}\sum_{\theta\in\boldsymbol{\theta}^{(\mathrm{out})}}\theta^2$$

$$+ \alpha_{\mathrm{r},3} \frac{1}{|\mathcal{H}|} \sum_{j \in \mathcal{H}} \left( \frac{1}{|\mathcal{B}_\mathrm{t}|} \sum_{m \in \mathcal{B}_\mathrm{t}} \left( \alpha_{\mathrm{r},4} - \sum_{\kappa=1}^{K} z_{\mathrm{hid},j}^{(m)}[\kappa] \right) \right)^2 . \tag{4.8}$$

The term can be divided into three summands, which are scaled by their respective parameter $\alpha_{\mathrm{r},n} \in [0,1]$, $n = 1,2,3$. The first two summands calculate the mean of the squared weights, where $\boldsymbol{\theta}^{(\mathrm{in})}$ denotes the vector of weights connecting the input and the hidden layer, and $\boldsymbol{\theta}^{(\mathrm{out})}$ denotes the vector of weights connecting the hidden and output layer. The third term returns the deviation of the actual number of generated spikes per hidden neuron from a predefined target $\alpha_{\mathrm{r},4} \in \mathbb{R}^+$, where $\mathcal{H}$ is the set of hidden neurons, $\mathcal{B}_\mathrm{t}$ the set of training samples, and $z_{\mathrm{hid},j}^{(m)}[\kappa] \in \{0,1\}$ indicates if the $j$th neuron of the hidden layer emitted a spike at time instant $\kappa$ when processing the $m$th sample. During optimization, we add the regularization loss $\alpha_\mathrm{r}$ to the CE-loss.

### Recurrent Connections

As outlined in Sec. 2.2, SNNs may exhibit lateral connections, where neurons within the same layer are interconnected, and feedback connections, where the output of a neuron is routed back to its input. In the following, we refer to these connections as recurrent connections. When solving classification tasks, adding recurrent connections can improve performance [ZN21, Fig. 3]. Consequently, adding recurrent connections to the NFE-SNN and DFE-SNN may be beneficial. On the other hand, adding recurrent connections comes at the cost of an increased number of adaptable parameters. When no recurrent connections are used, the number $N_\theta \in \mathbb{N}^+$ of learnable parameters is obtained by

$$N_\theta = N_\mathrm{hid} \left( N_\mathrm{in} + N_\mathrm{out} \right) . \tag{4.9}$$

If recurrent connections are applied in the hidden layer, $N_\theta$ is determined by

$$N_\theta = N_\mathrm{hid} \left( N_\mathrm{in} + N_\mathrm{hid} + N_\mathrm{out} \right) . \tag{4.10}$$

Since $N_\theta$ counts the number of parameters of the linear layer, it indicates the complexity of a model. Tab. 4.1 provides $N_\theta$ as a function of the architecture, where the subscript rec indicates that the hidden layer employs recurrent connections.

### Neural Encoding and Decoding

For both the NFE-SNN and DFE-SNN, we can apply different neural encodings. Since the linear RFE, TE, and QE demonstrated good performance and scalability in Ch. 3, we combine both the NFE-SNN and DFE-SNN with the three neural encoders. Based on the results of Sec. 3.5, where membrane potential-based neural decoding has proven to be more robust than spike-based neural decoding, we apply EOTM and MOTM as neural decoding. Both decodings apply the arg max-function to the output of the SNN. For training, we replace the arg max-function with the softmax-function.

**Table 4.1:** Number $N_\theta$ of parameters of different architectures.

| | $N_\theta$ |
|---|---|
| NFE-SNN | $N_{\text{hid}} \cdot (n_{\text{tap}} N_{\text{enc}} + N_{\text{out}})$ |
| NFE-SNN$_{\text{rec}}$ | $N_{\text{hid}} \cdot (n_{\text{tap}} N_{\text{enc}} + N_{\text{hid}} + N_{\text{out}})$ |
| DFE-SNN | $N_{\text{hid}} \cdot (n_{\text{ff}} N_{\text{enc}} + n_{\text{fb}} \lvert\mathcal{X}\rvert + N_{\text{out}})$ |
| DFE-SNN$_{\text{rec}}$ | $N_{\text{hid}} \cdot (n_{\text{ff}} N_{\text{enc}} + n_{\text{fb}} \lvert\mathcal{X}\rvert + N_{\text{hid}} + N_{\text{out}})$ |

### 4.4.2 Systematic Overview

To demonstrate the impact of the neural encoding, the neural decoding, regularization, and recurrent connections, we define various SNN-based equalizers and demappers. For both the NFE-SNN and DFE-SNN, we obtain 24 variations, each. In order to clearly differentiate the approaches, we define the following notation:

$$\text{structure}_{\text{enc, dec, reg, rec}} \tag{4.11}$$

While the parameter "structure" $\in$ {NFE − SNN, DFE − SNN} provides the general structure, the parameters "enc" $\in$ {RFE, TE, QE} and "dec" $\in$ {E, M} indicate the applied neural encoding and neural decoding, where "E" represents EOTM, and "M" represents MOTM. Furthermore, the parameter "reg" $\in$ { , R} indicates if regularization is applied, and the parameter "rec" $\in$ { , rec} indicates if recurrent connections are used. For a better understanding of the notation, Tab. 4.2 shows three example variations of the SNN-based equalizer and demapper.

## 4.5 Performance on IM/DD Links

### 4.5.1 The IM/DD Link

In high-speed non-coherent optical communications, IM/DD models with CD are crucial for modeling optical interconnects in passive optical networks, data centers, and system-on-chip interconnects [MTT+20, Sec. 4.5.4]. In IM/DD systems, increasing data rates intensify the impairments caused by CD, increasing the need for additional signal processing as a countermeasure.

The simplified setup of an IM/DD link is displayed in Fig. 4.6(a) [AEvB+25]. Transmit bits $\boldsymbol{b} \in \{0,1\}^{\log_2 |\mathcal{X}|}$ are mapped to transmit symbols $\boldsymbol{x} \in \mathcal{X}$, $\mathcal{X} \subset \mathbb{R}$, using Gray mapping. In the following, a 4-level pulse amplitude modulation (PAM4) with $|\mathcal{X}| = 4$ transmit symbols is used; thus, $\log_2(|\mathcal{X}|) = 2$ bits are mapped to one transmit symbol. Using a root-raised-cosine (RRC) filter with a roll-off factor $\beta_{\text{RRC}} \in [0, 1]$, the transmit symbols are pulse-shaped into a time-continuous signal and fed to the modulator. At the modulator, the intensity of a constant light source, e.g., a laser diode, is modulated according to the pulse-shaped transmit symbols. The modulated light is coupled into the optical fiber, which serves as the transmission channel. Inside the fiber, the frequency-dependent propagation speed results in CD, leading to a broadening of the transmit pulses and thus ISI. The transfer function of CD in the continuous frequency domain is given by [Cha19]

$$H_{\text{CD}}(f) = \mathrm{e}^{\mathrm{j}\frac{1}{2}\beta_{\text{CD}}(2\pi f)^2 L_{\text{CD}}}, \tag{4.12}$$

where $f \in \mathbb{R}$ is the deviation of the frequency from that of the laser, $L_{\text{CD}} \in \mathbb{R}^+$ the length of the fiber, and $\beta_{\text{CD}} \in \mathbb{R}$ the dispersion coefficient of the fiber when operated at the wavelength $\lambda \in \mathbb{R}^+$ of the laser. Thus, the amount of ISI depends on the length of the fiber, its dispersion coefficient, the laser wavelength, and the symbol rate.

**Table 4.2:** Examples of notation used for SNN-based equalizers and demappers.

| | $\text{NFE-SNN}_{\text{RFE,M}}$ | $\text{DFE-SNN}_{\text{TE,E,R}}$ | $\text{DFE-SNN}_{\text{QE,E,R,rec}}$ |
|---|---|---|---|
| Structure | NFE-SNN | DFE-SNN | DFE-SNN |
| Neural encoding | RFE | TE | QE |
| Neural decoding | MOTM | EOTM | EOTM |
| Regularization | | ✓ | ✓ |
| Recurrent connections | | | ✓ |

At the receiver, a photo-detector, e.g., a photo diode (PD), measures the optical power of the received signal. The measurement effectively squares the electric field, thereby introducing a nonlinear distortion. A second RRC filter with parameters equal to the parameters of the first one serves as matched filter (MF). The output of the MF is sampled and fed to the SNN-based equalizer and demapper, which outputs an estimate of the transmitted bit sequence. In addition to CD and the nonlinear distortion, thermal noise in the trans-impedance amplifier following the PD can be modeled as AWGN with variance $\sigma_{\text{n}}^2$.

The continuous-time IM/DD link, as shown in Fig. 4.6(a), can be modeled realistically using discrete-time simulations and oversampling $\beta_{\text{up}}$ [ABM+22a].

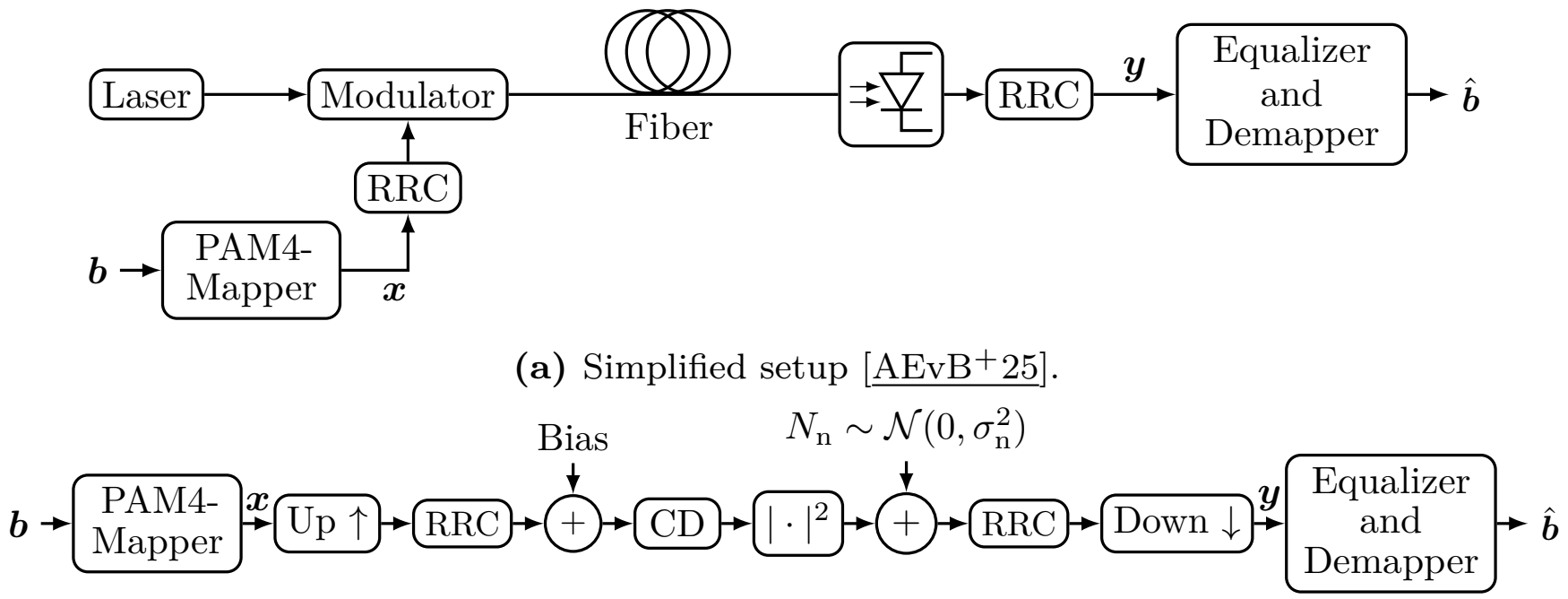

**(a)** Simplified setup [AEvB+25].

**(b)** System model [ABM+22a].

**Figure 4.6:** Simplified setup and system model of the IM/DD link.

**Table 4.3:** Parameter set of LCD and SSMF, as given in [AEvB+25].

| Parameter | LCD link | SSMF link |
|---|---|---|
| Set of transmit symbols $\mathcal{X}$ | $\{-3, -1, 1, 3\}$ | $\{0, 1, \sqrt{2}, \sqrt{3}\}$ |
| Upsampling factor $\beta_\mathrm{up}$ | 3 | 3 |
| Baudrate $R_\mathrm{sym}$ | 112 GBd | 50 GBd |
| Wavelength $\lambda$ | 1270 nm | 1550 nm |
| Dispersion coefficient $\beta_\mathrm{CD}$ | $-5\,\mathrm{ps\,nm^{-1}\,km^{-1}}$ | $-17\,\mathrm{ps\,nm^{-1}\,km^{-1}}$ |
| Fiber length $L_\mathrm{CD}$ | 4 km | 5 km |
| Roll-off factor $\beta_\mathrm{RRC}$ | 0.2 | 0.2 |
| Bias | 2.25 | 0.25 |

The resulting model is displayed in Fig. 4.6(b). A bias ensures that the signal coupled into the fiber is positive. While the squaring operation models the PD, AWGN models the thermal noise of the receiver. After the MF, the signal is downsampled with $\beta_\mathrm{down} = \beta_\mathrm{up}$.

For the simulation of the IM/DD link, we use two different sets of parameters, which are given in Tab. 4.3 [AEvB+25]. We refer to the two sets as low chromatic dispersion (LCD) and standard single-mode fiber (SSMF). The parameters of the LCD link are taken from [ABS+23]. Due to a low attenuation coefficient when operating at a wavelength of around $\lambda \approx 1550\,\mathrm{nm}$, the third and fourth generations of optical communication systems are typically operated around $\lambda \approx 1550\,\mathrm{nm}$ [Cha19, Tab. I]. Thus, for SSMF, we alter the wavelength $\lambda$, resulting in a larger dispersion coefficient $\beta_\mathrm{CD}$. Furthermore, for SSMF, the square root of the transmit symbols is applied prior to modulation to compensate for the photodiode nonlinearity, thereby approximating a linear input–output relationship. Note that the model of

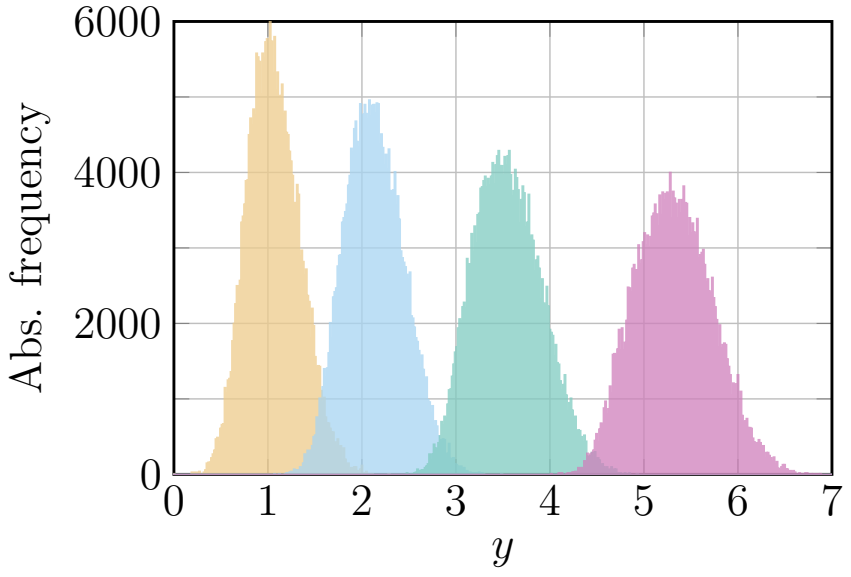

**(a)** LCD link.

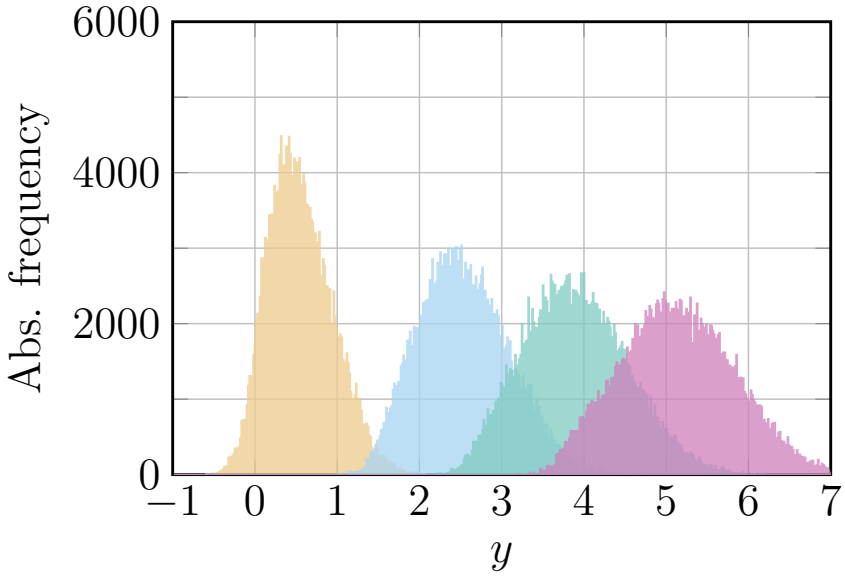

**(b)** SSMF link.

**Figure 4.7:** Histogram of received symbols at the channel output for $10^6$ transmit symbols and the parameters as given in Tab. 4.3. Both links were simulated for $\sigma_\mathrm{n}^2 = -20\,\mathrm{dB}$. Different colors indicate the affiliation to one of the four transmit symbols.

the IM/DD link, as introduced above, neglects the attenuation, which is implicitly taken into account by the AWGN.

Fig. 4.7 shows the histogram of received symbols $y$ for both the LCD and the SSMF models, with $\sigma_\mathrm{n}^2 = -20\,\mathrm{dB}$. Different colors indicate the affiliation to one of the four transmit symbols.

### 4.5.2 Hyperparameters

For both the NFE-SNN and DFE-SNN, we use SNNs with a single hidden layer with $N_\mathrm{hid}$ LIF neurons. The number of input neurons $N_\mathrm{in}$ depends on the equalizer type, the applied neural encoding, and the channel. Since EOTM and MOTM are used as neural decoding, the output layer uses $N_\mathrm{out} = |\mathcal{X}|$ (non-spiking) LI neurons. Furthermore, all linear layers connecting the neuron layers do not add any bias.

Tab. 4.4 shows the parameters of the neural encoding. For TE and QE, the parameters are determined heuristically, with $y_\mathrm{enc,max} = 3.5$. Since both TE and QE assume that the input data is centered around $y = 0$, we apply a direct current block before neural encoding to remove the mean of the input data [vBES23]. For RFE, the parameters are chosen to fit the parameters of [ABS+23, AEvB+25]. The field centers $\mu_j$ are initialized by $\mu_j = j \cdot \frac{y_\mathrm{enc,max}}{N_\mathrm{enc}}$, $j = 1, \dots, N_\mathrm{enc}$ with $y_\mathrm{enc,max} = 7$, and the field width by $\Delta_j = \Delta = 2$. Tab. 4.5 provides the parameters of the neurons, which are the default parameters suggested by `Norse` [PP21]. When using EOTM, we increase $\tau_\mathrm{m}$ of the LI neurons to nearly disable the leakage. For regularization, we use $\alpha_{\mathrm{r},1} = \alpha_{\mathrm{r},2} = 10^{-4}$, $\alpha_{\mathrm{r},3} = 5 \cdot 10^{-4}$, and $\alpha_{\mathrm{r},4} = 0.5$ [AEvB+25].

The parameters of the network architecture depend on the parameterizations of the IM/DD link. Tab. 4.6 displays the parameters, where we chose $n_\mathrm{ff} = \lceil n_\mathrm{tap}/2 \rceil$ and $n_{\text{fb}} = n_{\text{tap}} - n_{\text{ff}}$ [vBES23]. Optimization is executed using a training batch size of $\lvert\mathcal{B}_{\text{t}}\rvert = 2 \cdot 10^4$ samples at a noise power of $\sigma_{\text{n}}^2 = -20\,\text{dB}$. For evaluation, $\lvert\mathcal{B}_{\text{e}}\rvert = 10^7$ samples are used. If not stated otherwise, all parameters are fixed and not varied.

**Table 4.4:** Neural encoding parameters.

| | RFE | TE | QE |
|---|---|---|---|
| $N_{\text{enc}}$ | 10 | 8 | 8 |
| $K$ | 60 | 5 | 5 |

**Table 4.5:** Neuron parameters.

| | LIF | LI (MOTM) | LI (EOTM) |
|---|---|---|---|
| $\tau_{\text{m}}$ (ms) | 10 | 10 | 1000 |
| $\tau_{\text{s}}$ (ms) | 5 | 5 | 5 |
| $\Delta t$ (ms) | 1 | 1 | 1 |
| $v_{\text{th}}$ | 1.0 | – | – |
| $v_{\text{rest}}$ | 0 | – | – |

**Table 4.6:** SNN parameters.

| | LCD | SSMF |
|---|---|---|
| Number of equalizer taps $n_{\text{tap}}$ | 7 | 21 |
| Number of feed-forward taps $n_{\text{ff}}$ | 4 | 11 |
| Number of feedback taps $n_{\text{fb}}$ | 3 | 10 |
| Number of hidden neurons $N_{\text{hid}}$ | 40 | 60 |
| Training batch size $\lvert\mathcal{B}_{\text{t}}\rvert$ | $2 \cdot 10^4$ | $2 \cdot 10^4$ |
| Evaluation batch size $\lvert\mathcal{B}_{\text{e}}\rvert$ | $10^7$ | $10^7$ |
| Training noise power $\sigma_{\text{n}}^2$ | $-20\,\text{dB}$ | $-20\,\text{dB}$ |

### 4.5.3 Hyperparameter Impact Studies

In Sec. 4.4.2, we introduced several architectures for SNN-based equalizers and demappers. To reduce the number of different approaches, we investigate the performance of MOTM and EOTM and Afterwards fix the neural decoding. We also demonstrate the impact of the batchsize $\lvert\mathcal{B}_{\text{t}}\rvert$ used for training and the effect of the noise power $\sigma_{\text{n}}^2$ used during training.

To obtain a baseline and retain clarity given the large number of variations, we fix some hyperparameters while systematically varying others. We furthermore fix LCD as the parameterization of the IM/DD link and conduct the investigations using NFE-SNN$_{\text{RFE,M}}$, NFE-SNN$_{\text{RFE,M,rec}}$, NFE-SNN$_{\text{RFE,E}}$, and NFE-SNN$_{\text{RFE,E,rec}}$. First, we investigate the impact of neural decoding. Afterwards,

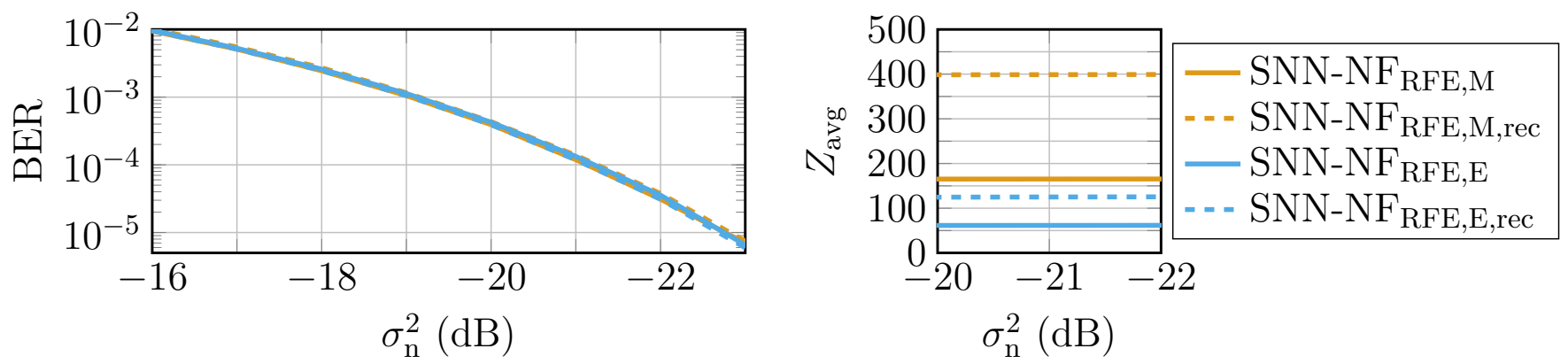

**(a)** BER and $Z_{avg}$.

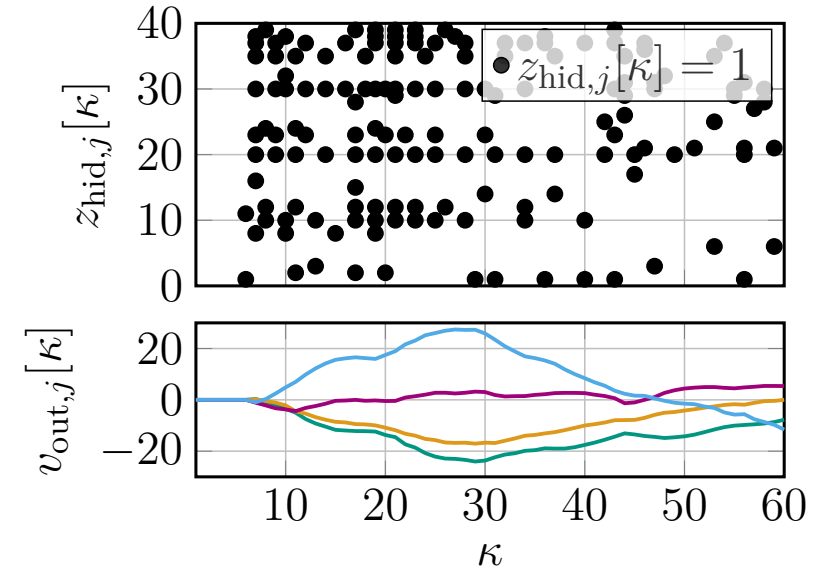

**(b)** NFE-SNN$_{RFE,M}$.

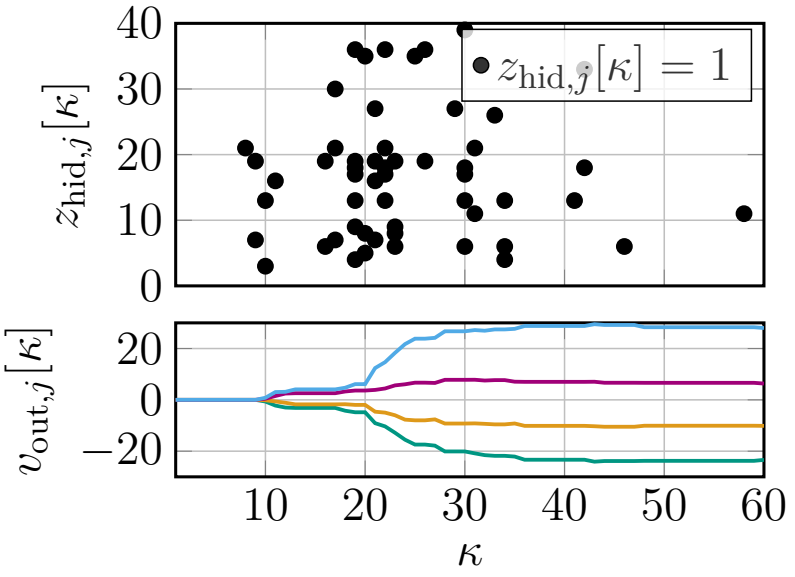

**(c)** NFE-SNN$_{RFE,E}$.

**Figure 4.8:** Comparison of variations of the NFE-SNN$_{RFE}$ without regularization.
(a): Achieved BER and $Z_{avg}$.
(b)-(c): Hidden layer spike pattern $z_{hid,j}[\kappa]$, $j = 1, \dots, N_{hid}$, and output layer membrane potential $v_{out,j}[\kappa]$ , $j = 1, \dots, N_{out}$, when processing the same sample. The colors of the curves denote the affiliation to one of the $N_{out} = 4$ output labels, where the blue line corresponds to the correct label.

we investigate the impact of the batch size $|\mathcal{B}_t|$ and the noise power $\sigma_n^2$ at which the training is carried out.

In addition to the BER, we define $Z_{avg} \in \mathbb{R}^+$ as a second performance measure. It indicates the average number of spikes generated by the hidden layer per processed sample and is obtained by

$$Z_{avg} = \frac{1}{|\mathcal{B}_t|} \sum_{m \in \mathcal{B}_t} \sum_{j \in \mathcal{H}} \sum_{\kappa=1}^{K} z_{hid,j}^{(m)}[\kappa] \,. \tag{4.13}$$

### Investigation of Neural Decoding

Fig. 4.8(a) compares the variations of the NFE-SNN$_{RFE}$. To directly spot the impact of neural decoding on the BER and average spike rate $Z_{avg}$, we neglect

regularization. For all approaches, the obtained $Z_{\text{avg}}$ is constant over the range of evaluated $\sigma_{\text{n}}^2$. Thus, for $Z_{\text{avg}}$, the plot shows only a selected range. In terms of BER, all tested approaches perform similarly. However, the average number of spikes $Z_{\text{avg}}$ greatly differs. Despite differing only in their neural decoding, NFE-SNN$_{\text{RFE,E}}$ yields a 64% reduction in $Z_{\text{avg}}$ compared to NFE-SNN$_{\text{RFE,M}}$. If recurrent connections are applied, $Z_{\text{avg}}$ is increased for both MOTM and EOTM. As before, if EOTM is used, $Z_{\text{avg}}$ is significantly smaller.

We can conclude that in terms of BER, all variations perform similarly, and both neural decodings are equally powerful. However, the use of EOTM results in the hidden layer emitting substantially fewer spikes per processed sample than with MOTM. Due to the leakage of the output layer, we suspect that when using MOTM, more spikes are necessary to produce unambiguous classification. To draw further conclusions about the difference of $Z_{\text{avg}}$, Fig. 4.8(b) and Fig. 4.8(c) display en exemplary spike pattern $z_{\text{hid},j}[\kappa]$, $j = 1, \ldots, N_{\text{hid}}$, of the hidden layer and the membrane potential $v_{\text{out},j}[\kappa]$, $j = 1, \ldots, 4$, of the output layer neurons for an arbitrary input sample, and the NFE-SNN$_{\text{RFE,M}}$ and NFE-SNN$_{\text{RFE,E}}$. Both process the same sample, which is highlighted by the fact that for both, the classification results in the choice of the class label indexed by the blue color.

For both, the hidden layer spikes that mostly contribute to the decision are emitted between $20 \leq \kappa \leq 28$. All spikes emitted Afterwards do not contribute to the decision anymore. For $\kappa > 28$, the NFE-SNN$_{\text{RFE,M}}$ has a high spiking activity, with membrane potentials that develop erratically. In contrast, for $\kappa > 28$, the NFE-SNN$_{\text{RFE,E}}$ exhibits a rather sparse spiking activity, where all membrane potentials maintain a nearly constant level. This is necessary, since the decision is obtained at $\kappa = K - 1$, and erratically developing membrane potentials, such as observed for the MOTM, would harm the decision. Thus, EOTM encourages the hidden layer to suppress spiking once the essential information has been processed. Since NFE-SNN$_{\text{RFE,E}}$ is able to reduce $Z_{\text{avg}}$ by approximately 64% compared to NFE-SNN$_{\text{RFE,M}}$, we suspect that the absence of leakage in the output layer already decreases the spike activity of the hidden layer.

From the results above, we can summarize: When comparing MOTM and EOTM with and without recurrent connections, they perform nearly alike in terms of BER. In terms of $Z_{\text{avg}}$, the different approaches greatly differ: The $Z_{\text{avg}}$ achieved by EOTM is significantly smaller than the $Z_{\text{avg}}$ achieved by MOTM. Adding recurrent connections, both MOTM and EOTM increase $Z_{\text{avg}}$. As the average spike rate $Z_{\text{avg}}$ is much lower for EOTM, we will henceforth concentrate on EOTM.

### Impact of Training Parameters on the Performance

Next, we investigate the impact of selected training parameters on performance. In the literature, the batch size that is applied during the optimization of SNN-based equalizers ranges from $|\mathcal{B}_{\text{t}}| = 100$ [ABS+23, AEvB+25] to $|\mathcal{B}_{\text{t}}| = 2 \cdot 10^5$ [vBES23, AEvB+25] samples per batch. Fig. 4.9(a) displays the impact of the training

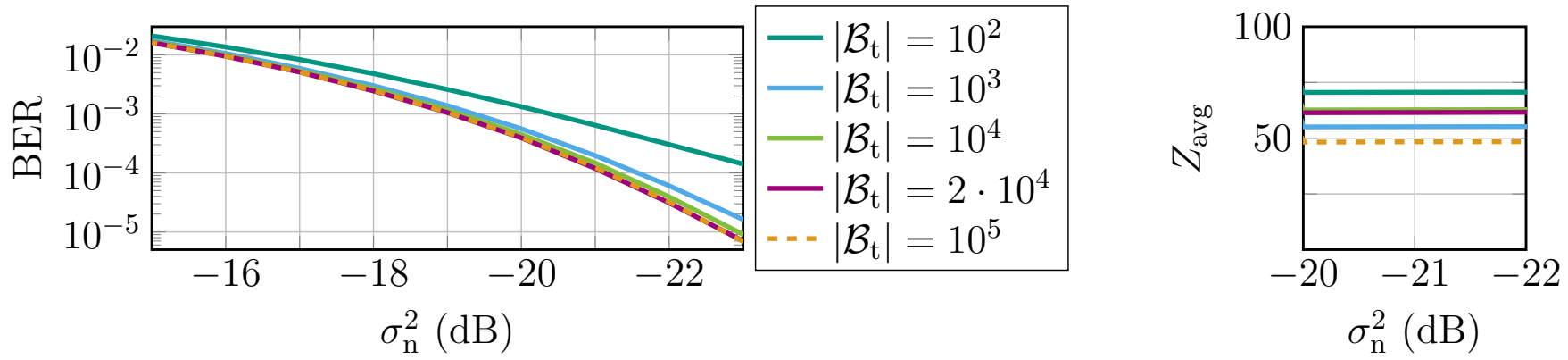

**(a)** Impact of the training batch size $|\mathcal{B}_t|$.

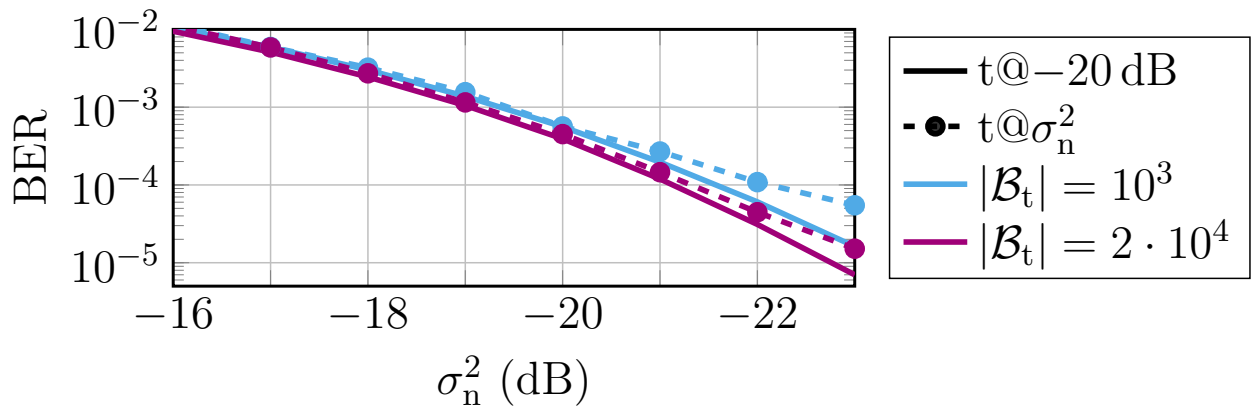

**(b)** Impact of the noise power $\sigma_n^2$ during training.

**Figure 4.9:** Impact of training parameters on the BER and $Z_{avg}$ for the NFE-SNN$_{RFE,E}$.

batch size $|\mathcal{B}_t|$ on the BER and $Z_{avg}$ for the NFE-SNN$_{RFE,E}$. With increasing batch size, performance also increases. As the curves obtained for $|\mathcal{B}_t| = 2 \cdot 10^4$ and $|\mathcal{B}_t| = 10^5$ coincide, we hypothesize that a batch size of $|\mathcal{B}_t| = 2 \cdot 10^4$ samples per update step is sufficient, and for $|\mathcal{B}_t| > 2 \cdot 10^4$, no more significant gains are achieved. In terms of $Z_{avg}$, there is no recognizable trend, and the optimized SNNs achieve an average spike rate in the range of $48 \leq Z_{avg} \leq 70$.

The literature also offers various approaches regarding the choice of the noise power $\sigma_n^2$ used during training. While $\sigma_n^2$ can be fixed during training [vBES23], it can also be varied during training [ABS+23], which results in noise power-specific SNNs. In Fig. 4.9(b), we compare the performance of the NFE-SNN$_{RFE,E}$ when trained at $\sigma_n^2 = -20$ dB (t@−20 dB), and when trained for each $\sigma_n^2$ (t@$\sigma_n^2$). We furthermore tested two different batch sizes, $|\mathcal{B}_t| = 10^3$ and $|\mathcal{B}_t| = 2 \cdot 10^5$. For both batch sizes, as the noise power increases, both training approaches perform similarly. However, as the noise power decreases, the training approach with a fixed noise power of $\sigma_n^2 = -20$ dB outperforms the training approach with noise power-specific SNN parameters. We furthermore observe that the deviation of both approaches increases with decreasing batch size. This indicates that, in low-BER regions, e.g., for $\sigma_n^2 = -22$ dB, the model encounters too few errors for sufficient training. With increasing batch size, the number of observed errors increases; however, we still see a gap for $|\mathcal{B}_t| = 2 \cdot 10^4$.

We can summarize that training at a fixed noise power of $\sigma_\mathrm{n}^2 = -20\,\mathrm{dB}$ results in superior performance compared to varying the noise power during training. We further observe that the SNN is able to generalize over a wide range of $\sigma_\mathrm{n}^2$, and a proper choice of $\sigma_\mathrm{n}^2$ used during training results in better performance than training multiple SNNs [BvBS23, vBES23]. Here, $\sigma_\mathrm{n}^2 = -20\,\mathrm{dB}$ was selected heuristically, corresponding to a SER, and thus a classification error, of SER $\approx 10^{-3}$ [vBES23].

### 4.5.4 Performance on the LCD Link

#### Comparison of SNN-based Equalizers and Demappers

For the LCD link, we now investigate the impact of neural encoding, recurrent connections, and the regularization. Fig. 4.10 shows the BER and $Z_\mathrm{avg}$ of various SNN-based receivers. For NFE-SNN-structures without recurrent connections, see Fig. 4.10(a), approaches using RFE achieve by far the best performance. Adding recurrent connections leads to a small performance improvement, see Fig. 4.10(b). Approaches using RFE still outperform approaches using QE and TE. For DFE-SNN structures without recurrent connections (Fig. 4.10(c)), the gap between the approaches using different neural encodings is smaller. With recurrent connections, all approaches perform nearly identically, see Fig. 4.10(d). When comparing NFE-$\mathrm{SNN}_\mathrm{RFE,E}$ and DFE-$\mathrm{SNN}_\mathrm{RFE,E}$, both perform approximately alike. However, the structure without decision feedback demonstrates a marginal performance advantage. In terms of BER, regularization has a negligible impact on performance. When comparing $Z_\mathrm{avg}$ for the approaches without regularization, TE and QE achieve a significantly lower $Z_\mathrm{avg}$ than RFE. Using regularization, $Z_\mathrm{avg}$ of all NFE-SNN-approaches is similar, with approximately 20 to 35 spikes per processed sample.

We conclude that when using recurrent connections, QE and TE achieve an improvement in terms of BER at the cost of an increased $Z_\mathrm{avg}$. If regularization is applied, the BER remains approximately the same; however, $Z_\mathrm{avg}$ is greatly decreased. For approaches using RFE, recurrent connections yield only minor gains. When regularization is applied, $Z_\mathrm{avg}$ is significantly reduced, with little to no impact on the BER. For the given parameters of LCD, incorporating decision feedback does not yield major improvements. For approaches using RFE, the opposite is observed: the performance in terms of BER tends to deteriorate slightly when using decision feedback. In terms of BER, we identify the NFE-$\mathrm{SNN}_\mathrm{RFE,R,rec}$ as the best choice. When aiming for a minimal $Z_\mathrm{avg}$ (without major degradation in BER compared to the NFE-$\mathrm{SNN}_\mathrm{RFE,E,R,rec}$), we identify the DFE-$\mathrm{SNN}_\mathrm{TE,E,R,rec}$ and DFE-$\mathrm{SNN}_\mathrm{QE,E,R,rec}$ as the best choice.

#### Significance of the Results

To ensure that the obtained results are representative, we optimized and evaluated 10 SNNs for each structure using recurrent connections and regularization. Fig. 4.11 shows the boxplot of the obtained BER and $Z_\mathrm{avg}$ when evaluated

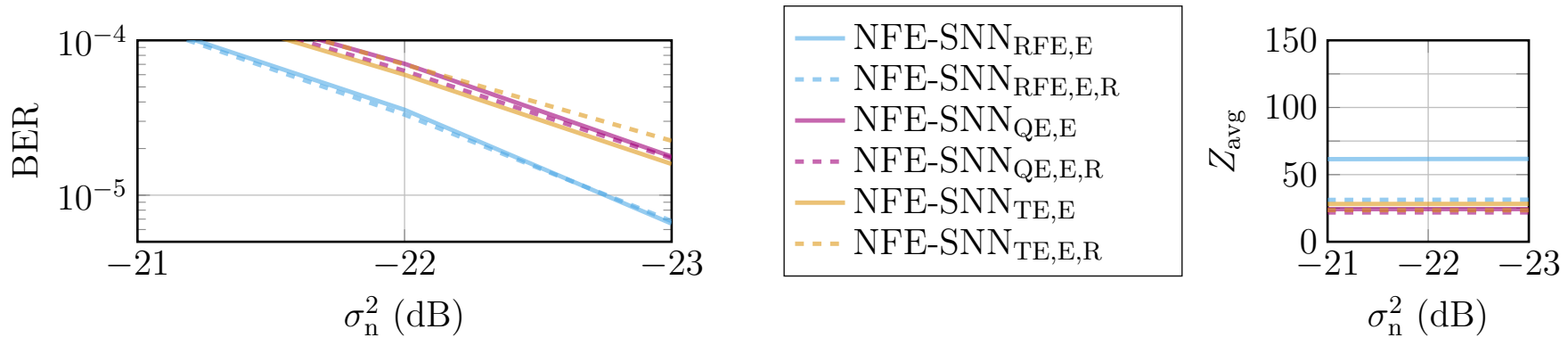

**(a)** BER and $Z_{\text{avg}}$ of NFE-SNN structures without recurrent connections.

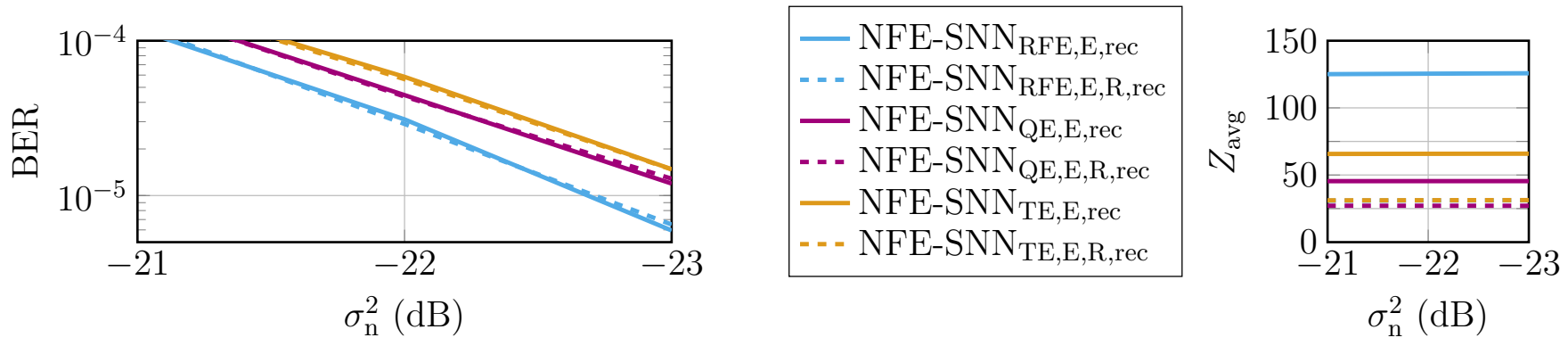

**(b)** BER and $Z_{\text{avg}}$ of NFE-SNN structures with recurrent connections.

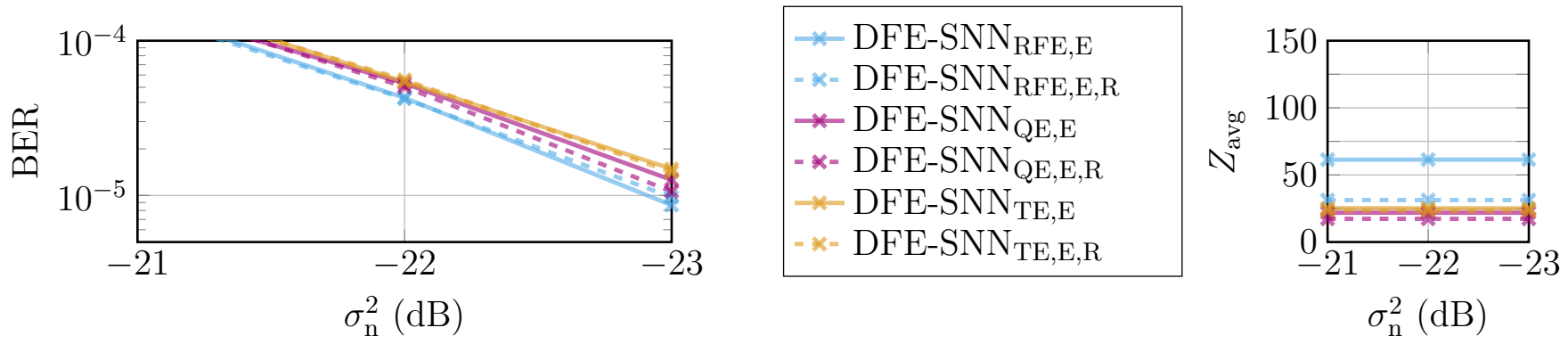

**(c)** BER and $Z_{\text{avg}}$ of DFE-SNN structures without recurrent connections.

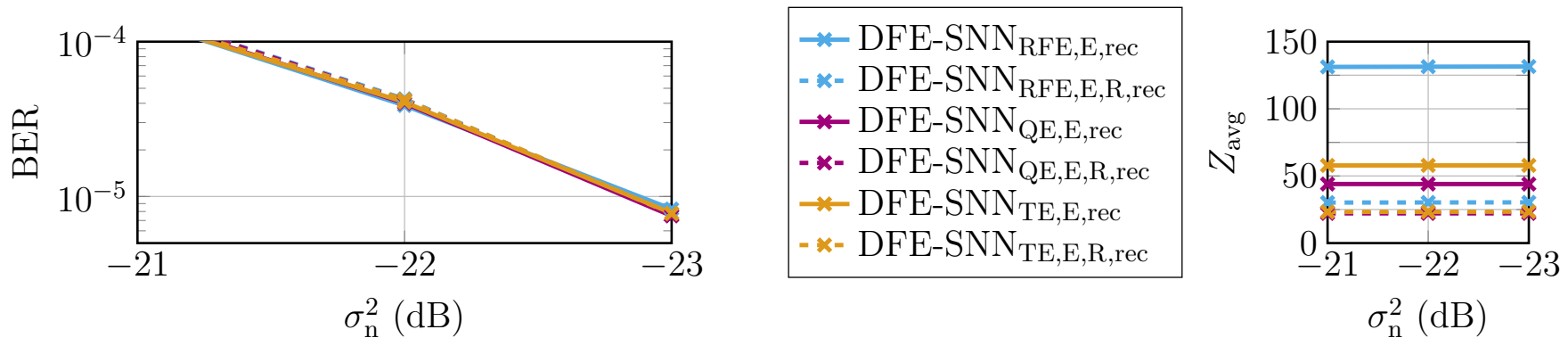

**(d)** BER and $Z_{\text{avg}}$ of DFE-SNN structures with recurrent connections.

**Figure 4.10:** BER and $Z_{\text{avg}}$ of the various SNN-based receivers for the parameters of the LCD link. Different colors indicate different neural encodings, the application of regularization is indicated by dashed lines, and the opacity of the lines indicates the use of recurrent connections. DFE-SNN structures are indicated by a cross-marker.

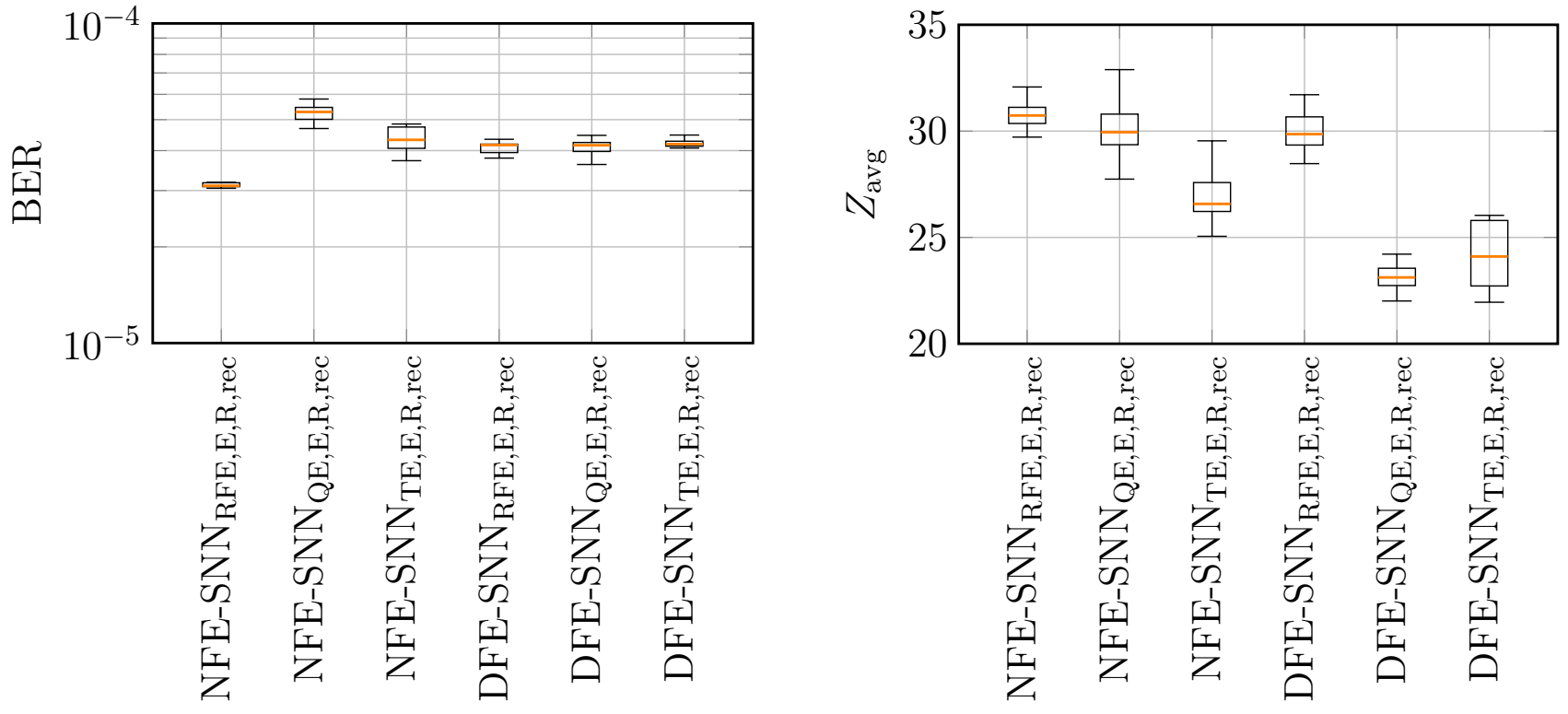

**Figure 4.11:** Boxplot of BER and $Z_{\text{avg}}$ of six different SNN-based equalizers and demappers for the LCD link. To obtain each box, independent initializations of the respective approach were optimized ten times, and evaluated for $\sigma_{\text{n}}^2 = -22\,\text{dB}$. The orange line denotes the median, the box is limited by the lower and upper quantile, and the whiskers are of maximum length $1.5 \cdot \text{IQR}$.

at $\sigma_{\text{n}}^2 = -22\,\text{dB}$. The orange lines denote the median; the boxes are limited by the lower and upper quantiles. The maximum length of the whiskers is $1.5 \cdot \text{IQR}$, where IQR is the interquartile range, i.e., the distance between the upper and lower quartiles. Whiskers are shortened to the nearest actual data point within their range. The plot suggests that different realizations of equal architectures result in nearly identical performance. Thus, we conclude that for each approach of the SNN-based receiver, a single realization is statistically consistent with the ensemble average. Fig. 4.11 consolidates our findings from Fig. 4.10: NFE-SNN$_{\text{RFE,R,rec}}$ achieves the lowest BER, and DFE-SNN$_{\text{TE,E,R,rec}}$ and DFE-SNN$_{\text{QE,E,R,rec}}$ achieve the lowest $Z_{\text{avg}}$.

## Comparison Against Benchmark Equalizers

In Fig. 4.12, we compare the best NFE-SNN, NFE-SNN$_{\text{RFE,R,rec}}$, and the best DFE-SNN, DFE-SNN$_{\text{QE,E,R,rec}}$, with the benchmark receivers introduced in Sec. 4.2. The SNN-based approaches clearly outperform all benchmark equalizers. It is remarkable that the NFE-SNN and DFE-SNN outperform the NFE-ANN$_{\text{class}}$ and DFE-ANN$_{\text{class}}$, which have a similar architecture, solely replacing the ANN with an SNN and neglecting the neural encoding and decoding. This suggests that for the given link, SNN-based receivers are more powerful than their ANN-based counterparts, which is supported by the results of [vBES23, ABS+23]. We furthermore want to emphasize that the NFE-ANN$_{\text{class}}$ achieves a similar BER as the "7 tap ANN" of [ABS+23, Fig. 4], which verifies our implementation. Decision feedback clearly improves performance for ANN-based and FIR filter-based receivers. In contrast, SNN-based receivers show no such improvement As already discussed above, the NFE-SNN$_{RFE,E,R,rec}$ performs best and achieves a slightly smaller BER than DFE-SNN$_{QE,E,R,rec}$. For $\sigma_n^2 > -16$ dB, decision feedback leads to error propagation. For the range of interest, error propagation does not seem to play a major role

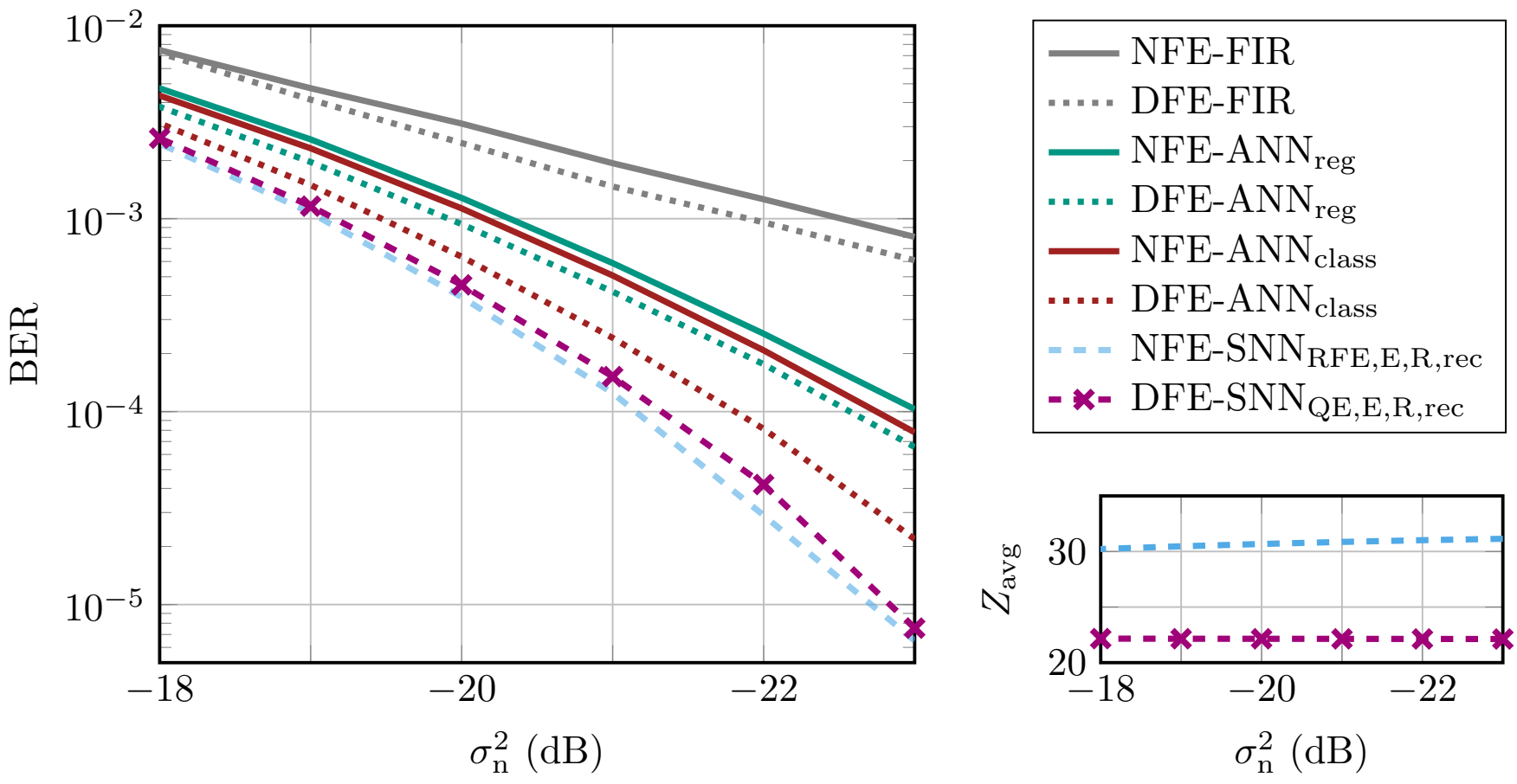

**Figure 4.12:** BER of the best performing NFE-SNN and DFE-SNN and the benchmark receivers for the LCD link. For the SNN-based receivers, the achieved $Z_{avg}$ is also given.

## Further Investigations

It is astonishing that the SNN-based equalizer and demappers significantly outperform their ANN-based counterparts. When optimizing the hyperparameters of the SNN-based equalizers and demappers, we observe that for $n_{tap} > 7$ the performance saturates. Consequently, we fix $n_{tap} = 7$ and consistently applied it across all benchmark equalizers. In Fig. 4.13, the impact of $n_{tap}$ on the BER for the NFE-SNN$_{RFE,R,rec}$, DFE-SNN$_{QE,R,rec}$, and all benchmark equalizers is shown. With increasing $n_{tap}$, the performance improves. For SNN-based approaches, for $n_{tap} \geq 7$ the performance starts to saturate. For the benchmark equalizers, the saturation starts at higher values of $n_{tap}$. For $n_{tap} = 11$, the gap between the SNN-based equalizers and ANN-based equalizers becomes smaller.

Again, for all non-SNN-based equalizers, the approaches with decision feedback outperform the approaches without decision feedback. With increasing $n_{tap}$, the BER saturates, where approaches using decision feedback converge to a lower BER than their counterparts without decision feedback. Both SNN-based receivers converge to the same BER with increasing $n_{tap}$. Note that with increasing $n_{tap}$,

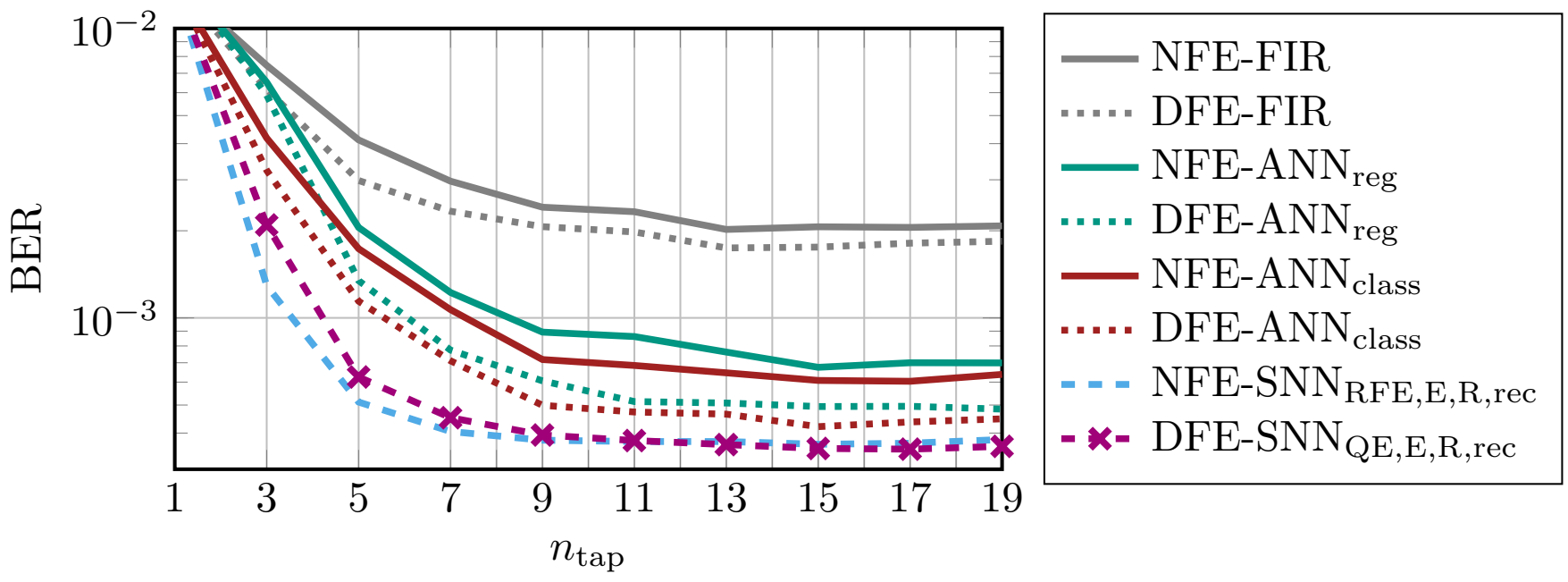

**Figure 4.13:** Impact of $n_{\text{tap}}$ on the BER for NFE-SNN$_{\text{RFE,R,rec}}$, DFE-SNN$_{\text{QE,R,rec}}$ and all benchmark equalizers for the LCD link, when evaluated for $\sigma_n^2 = -20\,\text{dB}$.

the number of input neurons $N_{\text{in}}$, and thus the complexity of the equalizer, increases.

In Fig. 4.14, we study the impact of $N_{\text{enc}}$, $K$, and $N_{\text{hid}}$ on the BER of DFE-SNN$_{\text{QE,E,R,rec}}$. For each value of $N_{\text{enc}}$, $K$, and $N_{\text{hid}}$, a separate SNN is optimized. With increasing values of the three variables, the BER saturates, where $N_{\text{enc}} = 6$, $K = 3$, and $N_{\text{hid}} = 40$ seem to be a good choice. When taking a closer look at the impact of $K$, increasing $K$ from $K = 2$ to $K = 3$ leads to a significant decrease in BER. It is astonishing that even for $K = 1$, the receiver with QE still yields good results. Since QE emits all input spikes in the first time step, this suggests that in the first time step, a large part of the information can be accessed by the SNN. In the hidden layer, in the first time step, a sufficient amount of spikes is generated, which already leads to a low BER. Increasing $K$ allows the SNN to exploit its temporal dynamics and recurrent connections, which further decreases the BER.

It is still peculiar that, contrary to our expectations, the NFE-SNN$_{\text{RFE,E,R,rec}}$ outperforms the DFE-SNN$_{\text{QE,E,R,rec}}$. Since the DFE-SNN$_{\text{QE,E,R,rec}}$ does not achieve any gain in performance when increasing $N_{\text{hid}}$, we conclude that the model using $N_{\text{hid}} = 40$ provides sufficient complexity and does not suffer from underfitting. Since we compare NFE-SNN$_{\text{RFE,E,R,rec}}$ and DFE-SNN$_{\text{QE,E,R,rec}}$, the QE may limit the performance.

To eliminate the difference in neural encoding and hence compare approaches that only differ in decision feedback, we furthermore plot the BER of the NFE-SNN$_{\text{RFE,E,R,rec}}$ and DFE-SNN$_{\text{RFE,E,R,rec}}$ when increasing $N_{\text{hid}}$. However, the performance converges for both approaches with $N_{\text{hid}} \geq 40$, see Fig. 4.14. This suggests that underfitting does not limit the performance of the DFE-SNN$_{\text{RFE,E,R,rec}}$, and that for the LCD link and RFE as neural encoding, using decision feedback does not improve performance. We can summarize that for the LCD link, the NFE-

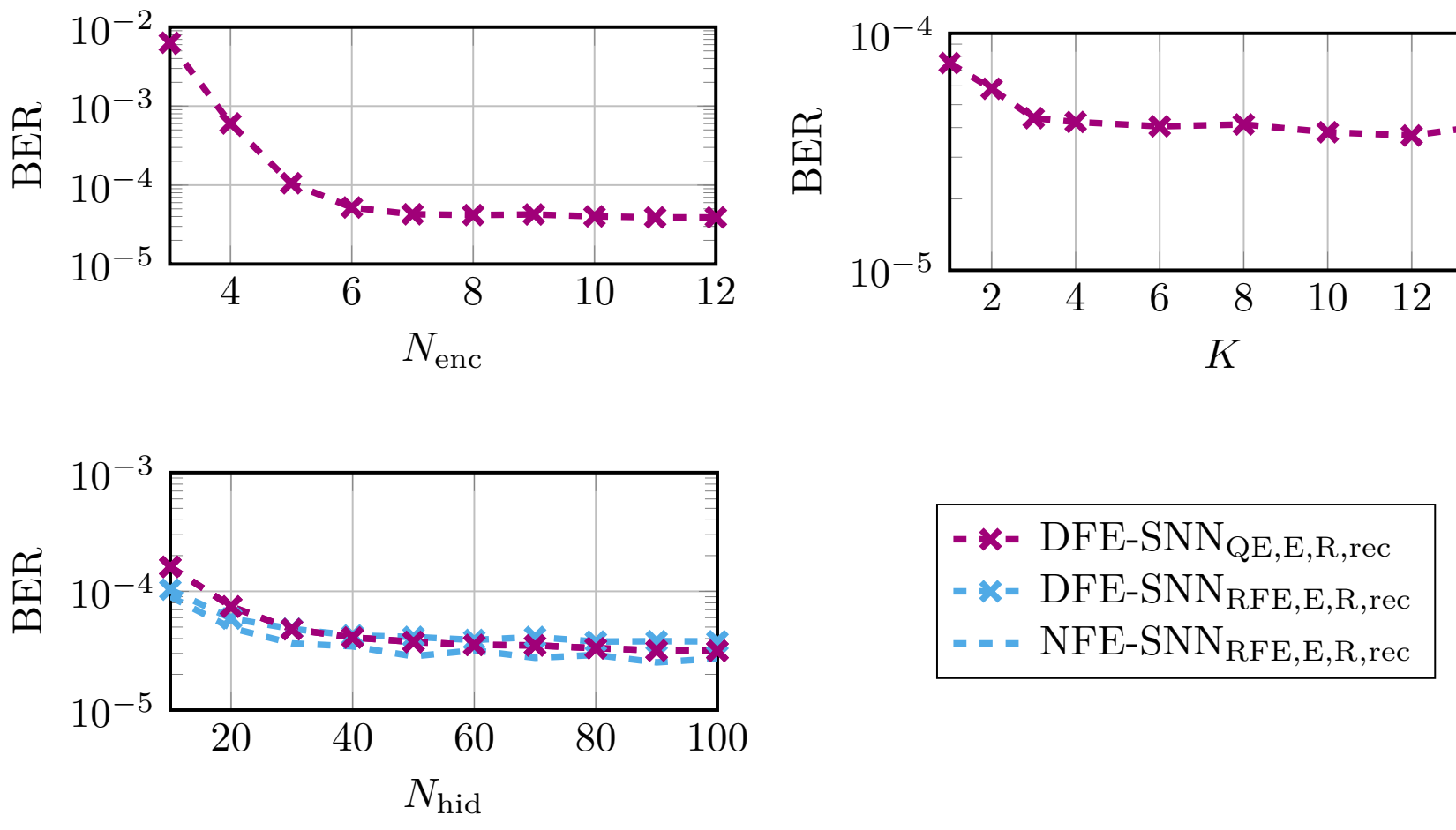

**Figure 4.14:** Impact of $N_{enc}$, $K$ and $N_{hid}$ on the BER of DFE-SNN$_{QE,E,R,rec}$. For DFE-SNN$_{RFE,E,R,rec}$ and NFE-SNN$_{RFE,E,R,rec}$, we also plot the impact of $N_{hid}$. The BER is obtained at $\sigma_n^2 = -22$ dB.

**Table 4.7:** Comparison of number $N_\theta$ of parameters of NFE-SNN$_{RFE,E,R,rec}$ and DFE-SNN$_{RFE,E,R,rec}$.

| | $K$ | $N_{enc}$ | $N_\theta$ |
|---|---|---|---|
| NFE-SNN$_{RFE,E,R,rec}$ | 60 | 10 | 4560 |
| DFE-SNN$_{QE,E,R,rec}$ | 5 | 8 | 3520 |

SNN$_{RFE,E,R,rec}$ is the best performing SNN in terms of BER. DFE-SNN$_{QE,E,R,rec}$ achieves the lowest $Z_{avg}$, with only a small BER performance penalty compared to the NFE-SNN$_{RFE,E,R,rec}$, and hence is also a promising approach.

### Receiver Complexity

Besides the BER and $Z_{avg}$ as a measure of equalization performance and energy consumption, the computational complexity and latency of the receiver must also be considered in practical settings. While the complexity can be measured by the number $N_\theta$ of adaptable parameters [MNHW24], we can measure the latency by the number $K$ of time-discrete steps per processed sample. Tab. 4.7 shows the comparison of latency and complexity for the NFE-SNN$_{RFE,E,R,rec}$ and the DFE-SNN$_{QE,E,R,rec}$. Due to QE with $N_{enc} = 8$ and the decision feedback structure, the DFE-SNN$_{QE,E,R,rec}$ has fewer parameters than the NFE-SNN$_{RFE,E,R,rec}$.

Furthermore, it requires 12 times fewer time steps $K$ during simulation. When looking at Fig. 4.14, we see that $K$ can be further reduced,down to $K = 3$, without expecting major BER penalties.

The difference in $K$ highlights one advantage of QE and TE compared to RFE: Since all spikes are fired at the first time step $\kappa = 0$, the SNN immediately starts processing the input, resulting in a small $K$ and hence fast output. To have a sufficiently high resolution in time when using RFE, a large $K$ is required, see Fig. 3.10.

### 4.5.5 Performance SSMF Link

#### Comparison of SNN-based Equalizers and Demappers

We now optimize and evaluate the various SNN-based equalizers and demappers for the SSMF link. Again, structures with and without decision feedback, different neural encodings, with and without recurrent connections, and with and without regularization are tested. We maintain all parameters as introduced above, and solely modify $N_{\mathrm{hid}} = 60$ and $n_{\mathrm{tap}} = 21$. Fig. 4.15 shows the obtained results. For structures without decision feedback, see Fig 4.15(a) and Fig 4.15(b), the NFE-SNN$_{\mathrm{RFE,E,R,rec}}$ again performs best. While the recurrent connections boost performance, compare NFE-SNN$_{\mathrm{RFE,E,R,rec}}$ and NFE-SNN$_{\mathrm{RFE,E,R}}$, the regularization leads to a low $Z_{\mathrm{avg}}$, compare NFE-SNN$_{\mathrm{RFE,E,R,rec}}$ and NFE-SNN$_{\mathrm{RFE,E,rec}}$. Furthermore, the regularization also results in improved performance. For approaches using QE and TE, similar behavior is observed.

When using decision feedback, see Fig 4.15(c) and Fig 4.15(d), the BER performance of all SNN-based receivers improves significantly. Adding recurrent connections furthermore improves the BER. While DFE-SNN$_{\mathrm{QE,E,R,rec}}$ and DFE-SNN$_{\mathrm{TE,E,R,rec}}$ yield by far the lowest BER, NFE-SNN$_{\mathrm{QE,E,R}}$ yields the smallest $Z_{\mathrm{avg}}$. Since the gap in terms of BER between DFE-SNN$_{\mathrm{QE,E,R,rec}}$ and NFE-SNN$_{\mathrm{QE,E,R}}$ is large, we identify the DFE-SNN$_{\mathrm{QE,E,R,rec}}$ as the best choice for the SSMF link.

#### Comparison Against Benchmark Equalizers

Fig. 4.16 compares the best NFE-SNN, NFE-SNN$_{\mathrm{RFE,E,R,rec}}$, and the best DFE-SNN, DFE-SNN$_{\mathrm{QE,E,R,rec}}$, with the benchmark receivers. Again, the SNN-based receivers clearly outperform their ANN-based counterparts. The DFE-SNN$_{\mathrm{QE,R,rec}}$ achieves the best BER. Except from the DFE-FIR, approaches with decision feedback clearly outperform their counterparts without decision feedback, e.g. NFE-ANN$_{\mathrm{class}}$ and DFE-ANN$_{\mathrm{class}}$. The BER performance of the DFE-FIR is larger than $10^{-1}$ and beyond the range of the plot. We speculate that for the SSMF link, ISI caused by CD is much more severe than for the LCD link, which results in a significant gain when using decision feedback. When comparing NFE-SNN$_{\mathrm{RFE,E,R,rec}}$ and DFE-SNN$_{\mathrm{QE,E,R,rec}}$, the latter achieves both a significantly lower BER and $Z_{\mathrm{avg}}$.

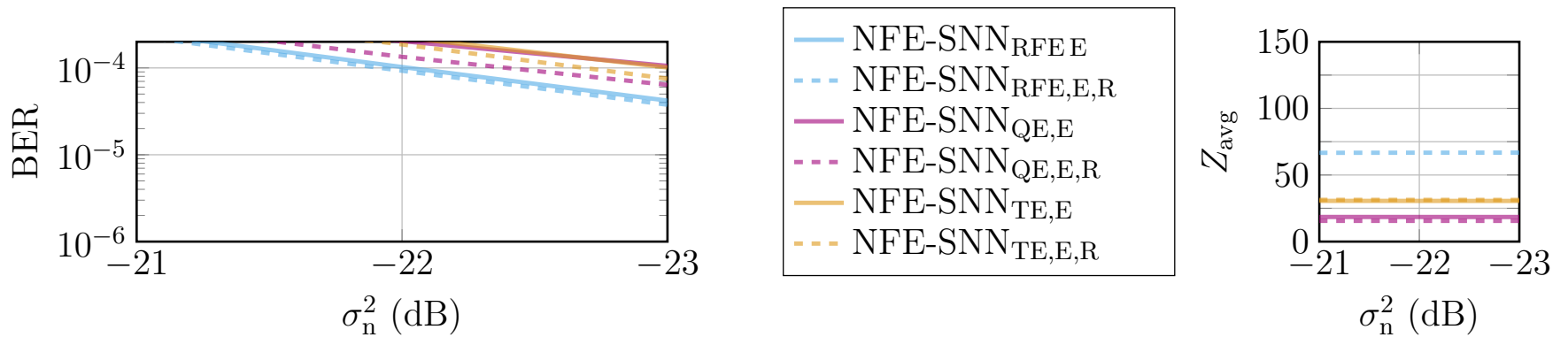

**(a)** BER and $Z_{\mathrm{avg}}$ of NFE-SNN structures without recurrent connections.

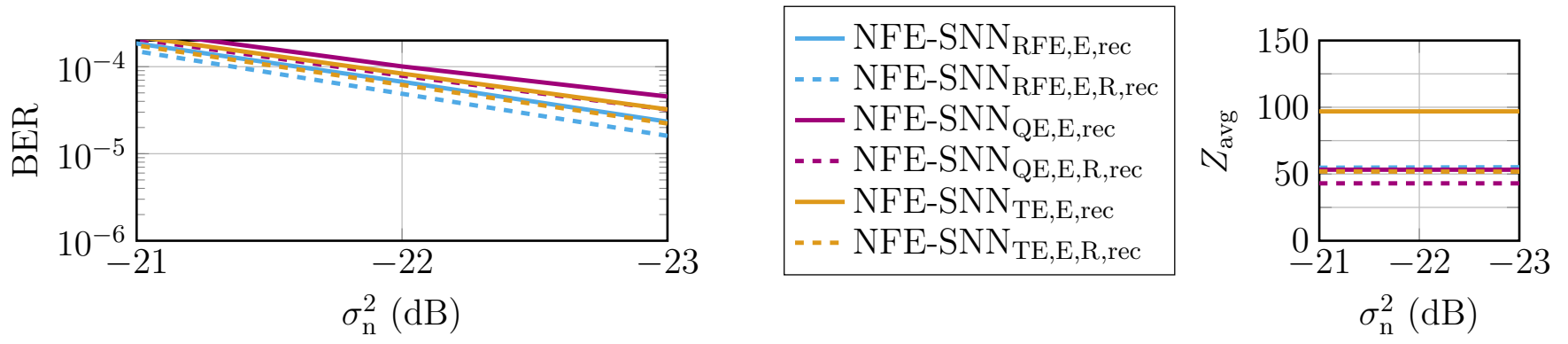

**(b)** BER and $Z_{\mathrm{avg}}$ of NFE-SNN structures with recurrent connections.

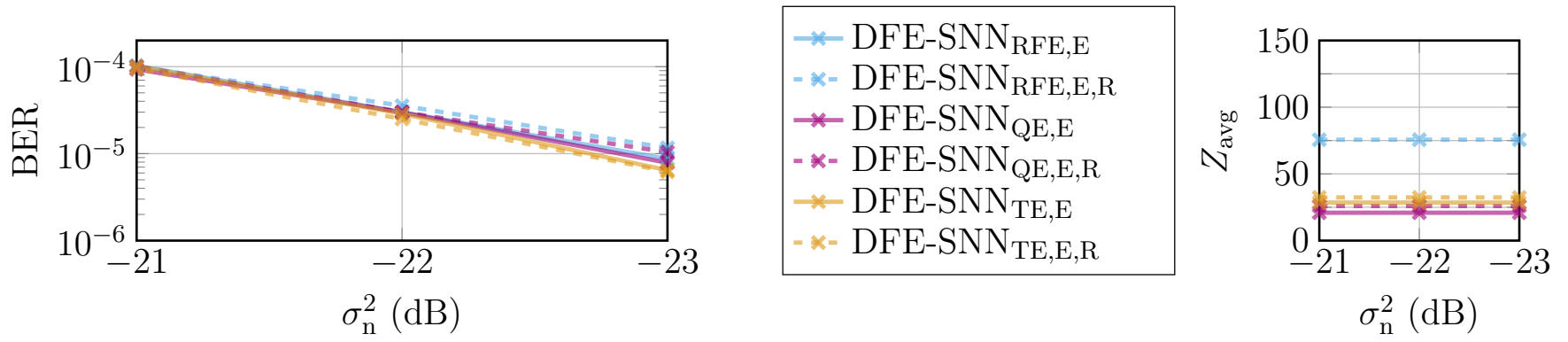

**(c)** BER and $Z_{\mathrm{avg}}$ of DFE-SNN structures without recurrent connections.

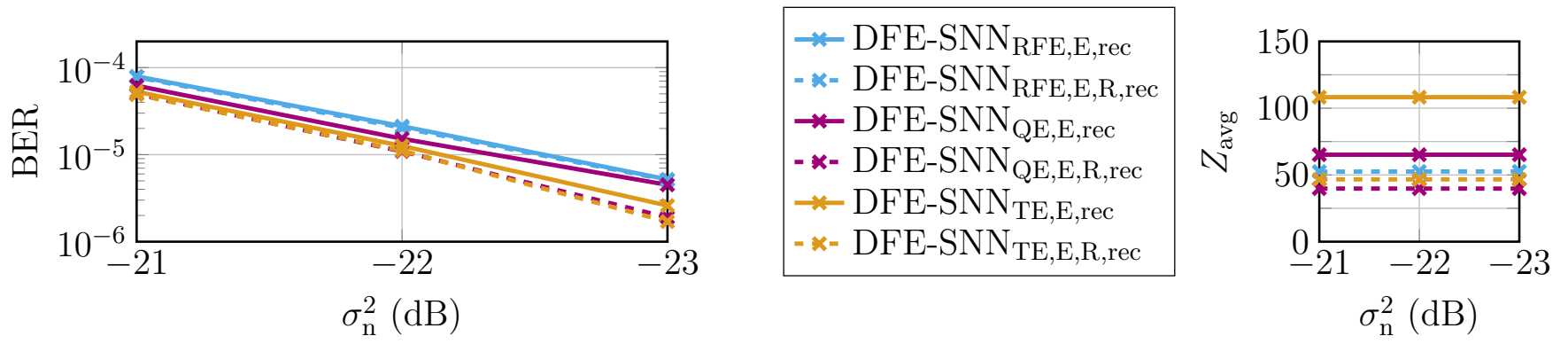

**(d)** BER and $Z_{\mathrm{avg}}$ of DFE-SNN structures with recurrent connections.

**Figure 4.15:** BER and $Z_{\mathrm{avg}}$ of the various SNN-based receivers for the SSMF link. Different colors indicate different neural encodings, the application of regularization is indicated by dashed lines, and lower opacity of the lines indicates the use of recurrent connections. If the DFE-SNN structure is used, it is indicated by a cross-marker.

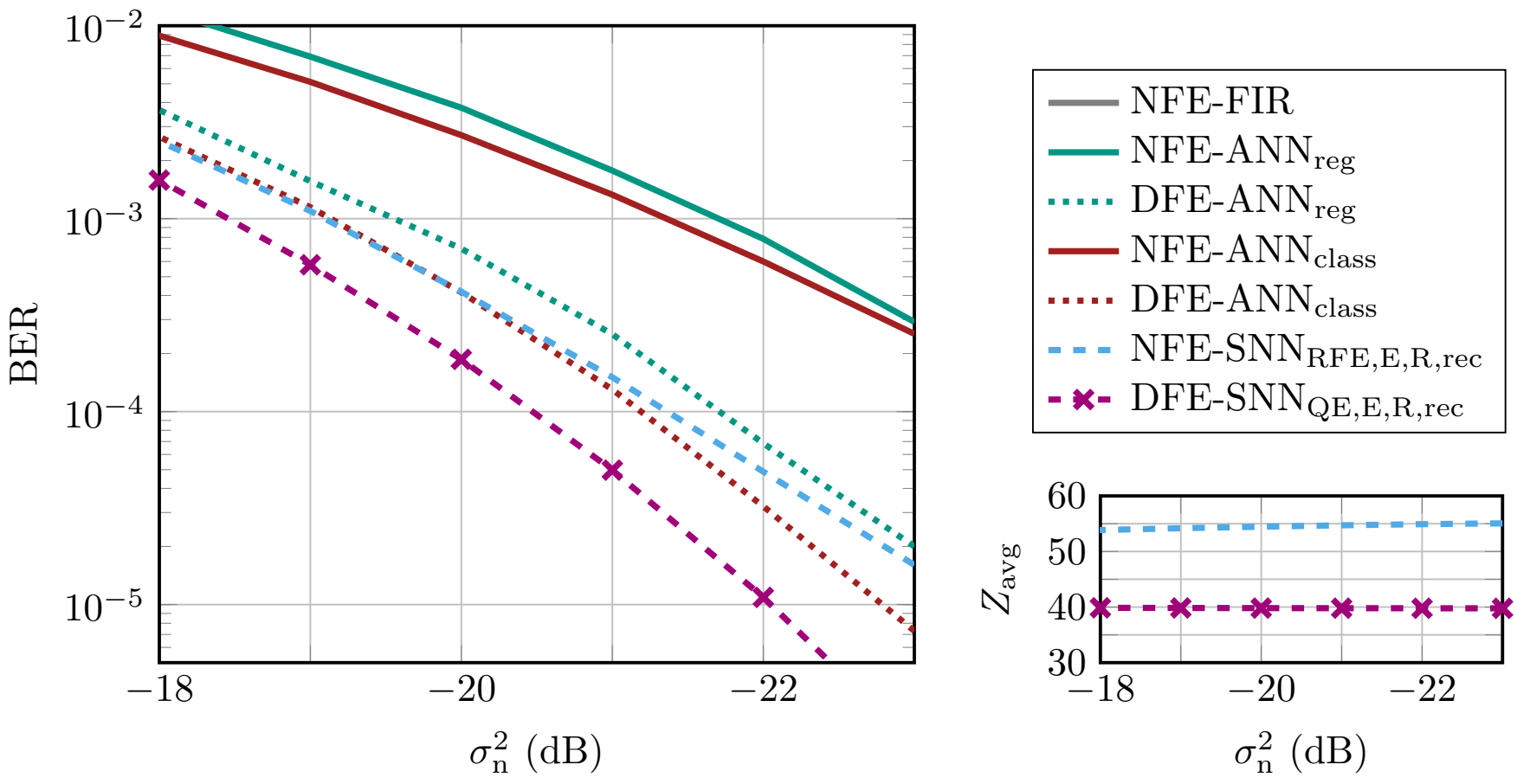

**Figure 4.16:** BER of the best performing NFE-SNN and DFE-SNN and the benchmark receivers for the SSMF link. For the SNN-based receivers, the achieved $Z_{\mathrm{avg}}$ is also given.

**Table 4.8:** Comparison of number $N_\theta$ of parameters of NFE-SNN$_{\mathrm{RFE,E,R,rec}}$ and DFE-SNN$_{\mathrm{RFE,E,R,rec}}$.

| | $K$ | $N_{\mathrm{enc}}$ | $N_\theta$ |
|---|---|---|---|
| NFE-SNN$_{\mathrm{RFE,E,R,rec}}$ | 60 | 10 | 16 440 |
| DFE-SNN$_{\mathrm{QE,E,R,rec}}$ | 5 | 8 | 11 520 |

### Receiver Complexity

Tab. 4.8 displays the latency $K$ and the computational complexity $N_\theta$ of the NFE-SNN$_{\mathrm{RFE,E,R,rec}}$ and DFE-SNN$_{\mathrm{QE,E,R,rec}}$. Due to the encoding, the DFE-SNN$_{\mathrm{QE,E,R,rec}}$ exhibits significantly lower latency than the NFE-SNN$_{\mathrm{RFE,E,R,rec}}$. In terms of complexity, the DFE-SNN$_{\mathrm{QE,E,R,rec}}$ only requires 70% of the parameters needed by the NFE-SNN$_{\mathrm{RFE,E,R,rec}}$.

## 4.6 Key Findings

In this chapter, we introduced and compared two SNN-based equalizers and demappers. While the NFE-SNN exhibits a feed-forward structure, the DFE-SNN incorporates decision feedback. For two different parameterizations of the IM/DD link [AEvB+25], we systematically tested both the NFE-SNN and DFE-SNN. For the LCD link, which is mainly impaired by the nonlinear distortion, both the best performing NFE-SNN and DFE-SNN are able to combat nonlinear impairments, resulting in similar performance. For the SSMF link, the ISI becomes more severe, and the best performing DFE-SNN clearly outperforms the best performing NFE-SNN. When comparing both approaches w.r.t. the achieved average spike rate, latency, and model complexity, the best performing DFE-SNN achieves significantly lower values for all three metrics than the best performing NFE-SNN. When comparing the different neural encoding schemes, we observe that the NFE-SNN performs best when combined with the RFE, and the DFE-SNN performs best when combined with QE. As a result, when comparing the achieved average spike rate $Z_{\text{avg}}$, latency $K$, and model complexity, the best performing DFE-SNN achieves significantly lower values for all three metrics than the best performing NFE-SNN.

We conclude that the DFE-SNN combines multiple advantages: While its internal nonlinear signal processing enables it to combat nonlinear impairments, the decision feedback structure enables it to combat linear impairments, such as ISI. Together with the promise of an energy-efficient implementation [ABS+23, MNHW24], the DFE-SNN is a promising approach for future low-energy, high-performance receivers.

# 5 Reinforcement Learning: Theory and Example Applications

In Ch. 6, we will develop a RL-based update rule for joint optimization of the neural encoding and SNNs [EvBS25]. Beforehand, we provide an introduction to RL. As an illustrative application of RL, we also discuss the resource allocation problem in wireless networks.

## 5.1 Introduction to Reinforcement Learning

The field of RL addresses the problem of determining optimal mappings from situations, termed *states*, to *actions*, with the goal of maximizing a numerical *reward function*. Without explicit supervision, the so-called *agent* must independently discover which actions yield the highest *rewards* through exploration and continuously learning. Often, rewards are delayed: The benefits of an action may not be apparent in the immediate next step, but only after several steps. Learning under such conditions is one of the core challenges of RL [SB18, Sec. 1.1]. Hence, delayed rewards and exploration are two defining characteristics of RL [SB18, Sec. 1.1]. Over the past two decades, RL has emerged as one of the most influential research areas in MaL, significantly contributing to advances in artificial intelligence (AI). Although the process of making sequential decisions, observing outcomes, and adjusting the behavior of the agent accordingly is theoretically guaranteed to converge to an optimal solution, a major challenge remains: the time required for convergence. The agent must thoroughly explore the environment and build an internal model of the system, which can be extremely time consuming [LHG+19].

The key to overcoming this challenge lies in learning an effective approximation from the experience of the agent with a limited subset of situations, enabling generalization to all possible cases. A prominent approach to achieve such generalization is function approximation.Function approximation is rooted in the domain of supervised learning and includes subfields such as NNs. Given example values of a target function, function approximation seeks to estimate the function in its entirety [SB18, p. 195]. Enhancing generalization in RL has been significantly advanced by DL, leading to the emergence of the field of deep reinforcement learning (DRL). DRL harnesses the power of NNs, thereby improving both the performance and the learning efficiency of RL algorithms. This capability has led to its adoption in a wide range of applications [LHG+19]. The effectiveness of DRL has been demonstrated in various domains, including robotics [FTD+16] and computer games [MKS+15, LPM+22], and has even extended to mastering the ancient Chinese game of *Go* [SHM+16]. Due to its effectiveness, DRL has recently emerged as a promising tool for addressing challenges in communications and networking, many of which are surveyed in [LHG+19, PdFC+25]. In par-ticular, DRL has been applied to achieve throughput maximization and energy consumption minimization in networks [LHG+19].

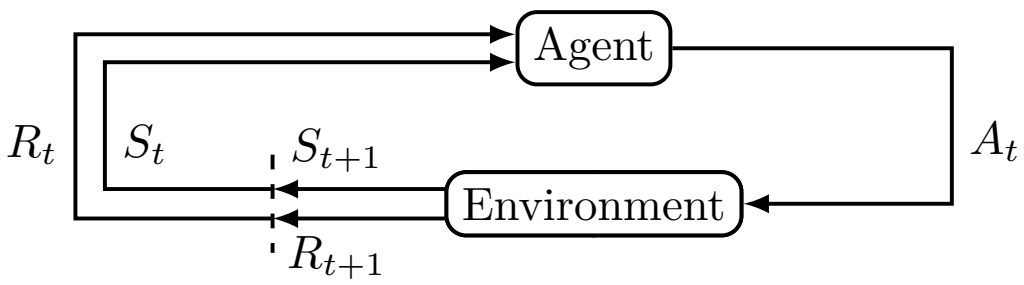

**Figure 5.1:** The agent interacting with the environment in an MDP, adapted from [SB18, Sec. 3.1].

### 5.1.1 The Markov Decision Process

A Markov decision process (MDP) is the mathematically idealized formulation of the RL problem. For an MDP, precise theoretical guarantees can be formulated [SB18, Ch. 3]. In uncertain and stochastic environments, an MDP provides a suitable model for most decision making problems [LHG+19]. Various techniques can be employed to solve MDPs, including dynamic programming [Put14] and the previously mentioned DRL approach [SB18, Sec. 9.7]. In an MDP, the entity responsible for decision-making and learning to achieve a goal is called the *agent*. The agent continuously interacts with an *environment* by selecting *actions* $A_t$ [SB18, Sec. 3.1]. The outcomes of these interactions are influenced partly by the chosen actions $A_t$, and partly by randomness. At environment time $t$, the agent observes the *state* $S_t \in \mathcal{S}$, where $\mathcal{S}$ denotes the set of all possible environment states. Based on $S_t$, an action $A_t \in \mathcal{A}$ is chosen, where $\mathcal{A}$ denotes the set of possible actions. Executing $A_t$ results in the *follow-up state* $S_{t+1} \in \mathcal{S}$, and in receiving the *reward* $R_{t+1} \in \mathcal{R}$, where $\mathcal{R}$ denotes the set of rewards [LHG+19]. Note, that $S_t$, $A_t$ are $R_{t+1}$ are RVs, and that the sets $\mathcal{S}$, $\mathcal{A}$, and $\mathcal{R}$ are assumed to be finite [SB18, Sec. 3.1]. The MDP is defined by the following 3-tuple [LHG+19, Put14]:

$$\big(\mathcal{S},\ \mathcal{A},\ p(S_{t+1}, R_{t+1}|S_t, A_t)\big), \quad R_{t+1} \in \mathcal{R},\, S_t \in \mathcal{S},\, A_t \in \mathcal{A}\,, \tag{5.1}$$

The set of available actions $\mathcal{A}$ may depend on the current state. However, in the following, we assume that the set of actions $\mathcal{A}$ does not depend on the state, and thus contains the same actions for every state $S_t \in \mathcal{S}$. A state $S_t$ satisfies the so-called Markov property if it encapsulates "all aspects of the past agent-environment interaction that make a difference for the future" [SB18, Sec. 3.1]. The dynamics of the MDP are fully captured by the dynamics function $p(S_{t+1}, R_{t+1}|S_t, A_t)$, which specifies the transition probability from state $S_t$ to $S_{t+1}$ after executing action $A_t$. Fig. 5.1 illustrates the interaction between the agent and the environment.

The dynamics function $p$ is a deterministic function that specifies a probability distribution [SB18, Sec. 3.1]:

$$p : \mathcal{S} \times \mathcal{R} \times \mathcal{S} \times \mathcal{A} \to [0,1] \tag{5.2}$$

$$\sum_{S_{t+1} \in \mathcal{S}} \sum_{R_{t+1} \in \mathcal{R}} p(S_{t+1}, R_{t+1} | S_t, A_t) = 1\,, \quad \forall S_t \in \mathcal{S},\, \forall A_t \in \mathcal{A}\,. \tag{5.3}$$

The MDP framework is highly flexible. For instance, time steps do not necessarily have to correspond to fixed real time intervals; instead, they can represent arbitrary sequences of decision making and action-execution steps. Actions may range from low-level control signals to high-level decisions, and the states can take on a wide variety of forms [SB18, Sec. 3.1].

### 5.1.2 Rewards, Goals and Policies

After each step of the environment, a reward $R_t \in \mathcal{R}$ is provided to the agent. The purpose of the reward signal is to formalize the notion of a *goal*, which is a defining characteristic of RL. To evaluate the performance of the agent over time, we introduce the *return* $G_t$, which is defined as a function of the sequence of received rewards. A simple return function would be the finite-horizon undiscounted return

$$G_t := R_{t+1} + R_{t+2} + R_{t+3} + \ldots + R_{T_{\text{env}}}\,, \tag{5.4}$$

where $T_{\text{env}} \in \mathbb{N}$ denotes the number of time steps needed to either achieve a terminal state, or until the MDP is stopped, and hence refers to the final time step of the environment. The agent aims to choose actions $A_t \in \mathcal{A}$ such that the *expected* value of the return $G_t$ is maximized.

Continuous interaction of the agent with the environment in an MDP results in the following sequence:

$$S_0, A_0, R_1, S_1, A_1, R_2, S_2, A_2, R_3, \ldots\,. \tag{5.5}$$

We refer to the period from $t = 0$ to $t = T_{\text{env}}$ as an *episode* (or a *trial*, particularly in applications where it ends in winning or losing a game). A natural terminal state exists only in *episodic tasks.* For many tasks, however, there is no such natural point of termination. These ongoing tasks, which may have long or infinite lifespans, are referred to as *continuing tasks* [SB18, Sec. 3.1]. In this thesis, we first consider the resource allocation task in wireless networks, which is a continuing task. In such cases, we have $T_{\text{env}} = \infty$, which renders the return function in (5.4) ill-defined, as the return $G_t$ being maximized could also become infinite [SB18, Sec. 3.2].

To address this, a *discounted infinite-horizon return* $G_t$ is commonly used

$$G_t := R_{t+1} + \gamma_{\mathrm{G_t}} R_{t+2} + \gamma_{\mathrm{G_t}}^2 R_{t+3} + \ldots = \sum_{k=0}^{\infty} \gamma_{\mathrm{G_t}}^k R_{t+k+1} \,, \tag{5.6}$$

which is the exponential sum over future rewards, discounted by the *discount rate* $\gamma_{\mathrm{G_t}} \in [0,1]$. Each subsequent reward is weighted $\gamma_{\mathrm{G_t}}$ times less than its predecessor. For $\gamma_{\mathrm{G_t}} < 1$, the sum in (5.6) remains finite for any bounded reward sequence $R_k$. Note that for $\gamma_{\mathrm{G_t}} = 0$, the agent is shortsighted, considering only the immediate reward $R_{t+1}$. As $\gamma_{\mathrm{G_t}} \to 1$, the agent becomes increasingly farsighted [SB18, Sec. 3.3]. The recursive relationship between returns in successive time steps is fundamental to many RL algorithms [SB18, Sec. 3.3]:

$$\begin{aligned} G_t &:= R_{t+1} + \gamma_{\mathrm{G_t}} R_{t+2} + \gamma_{\mathrm{G_t}}^2 R_{t+3} + \gamma_{\mathrm{G_t}}^3 R_{t+4} + \ldots \\ &= R_{t+1} + \gamma_{\mathrm{G_t}} (R_{t+2} + \gamma_{\mathrm{G_t}} R_{t+3} + \gamma_{\mathrm{G_t}}^2 R_{t+4} + \ldots) \\ &= R_{t+1} + \gamma_{\mathrm{G_t}} G_{t+1} \,. \end{aligned} \tag{5.7}$$

Any RL algorithm now aims to find a *policy* $\pi$ that maximizes the expected return $\mathbb{E}_\pi\{G_t\}$, where $\pi : \mathcal{S} \to \mathcal{A}$, and $\mathbb{E}_\pi\{\cdot\}$ denotes the expectation when following $\pi$ [1]. The hypothetical optimal policy $\pi^*$ is defined as [LHG$^+$19]

$$\pi^* = \arg\max_{\pi} \mathbb{E}_\pi \left\{ G_t | S_t, \pi(S_t) \right\} \,, \quad \forall\, s \in \mathcal{S} \,, \tag{5.8}$$

where $A_t = \pi(S_t)$ with $A_t \in \mathcal{A}$ denotes the action chosen at state $S_t \in \mathcal{S}$.

RL approximates the optimal policy $\pi^*$ through interaction with the environment: at each time step, the agent observes the current state $S_t$, selects an action $A_t$, and then receives an immediate reward $R_{t+1}$ along with the next state $S_{t+1}$ (see Fig. 5.1). The reward and subsequent state provide information for updating the policy of the agent. This process repeats until the optimal policy is approached.

We furthermore introduce two functions, which indicate the performance of a given policy $\pi$. The first function is the *state-value function* $v_\pi : \mathcal{S} \to \mathbb{R}$. This function maps a state $s \in \mathcal{S}$ to the expected return when following policy $\pi$, formally defined as [SB18, Sec. 3.5]

$$v_\pi(s) := \mathbb{E}_\pi \left\{ G_t \mid S_t = s \right\} = \mathbb{E}_\pi \left\{ \sum_{k=0}^{\infty} \gamma_{\mathrm{G_t}}^k R_{t+k+1} \middle| S_t = s \right\} . \tag{5.9}$$

Secondly, we define $q_\pi : \mathcal{A} \times \mathcal{S} \to \mathbb{R}$ as the *action-value function* when following policy $\pi$, which gives the expected return when following $\pi$ starting from

[1] Since $\pi$ defines the behavior of the action and thus the action probabilities, the expectation is taken w.r.t. $\pi$.

state $s \in \mathcal{S}$ and taking action $a \in \mathcal{A}$:

$$\begin{aligned} q_\pi(s,a) :=& \mathbb{E}_\pi \left\{ G_t \mid S_t = s, A_t = a \right\} \\ =& \mathbb{E}_\pi \left\{ \sum_{k=0}^{\infty} \gamma_{\mathrm{G_t}}^k R_{t+k+1} \middle| S_t = s, A_t = a \right\} . \end{aligned} \tag{5.10}$$

With (5.10), we can reformulate the definition of an optimal policy $\pi^*$ to

$$q^*(s,a) = \arg\max_\pi \, q_\pi(s,a) \,, \quad \forall s \in \mathcal{S}, \, \forall a \in \mathcal{A} \,, \tag{5.11}$$

where $q^*(s,a)$ is the optimal action-value function[SB18, Sec. 3.6]. If $q^*(s,a)$ is known, for each state $s \in \mathcal{S}$, the optimal action $a^* \in \mathcal{A}$ can be selected by

$$a^* = \arg\max_{a \in \mathcal{A}} q^*(s,a) \,. \tag{5.12}$$

One method to approximate $q^*(s,a)$ is the *Q-learning algorithm* [LHG+19].

### 5.1.3 The $Q$-learning Algorithm

The $Q$-learning algorithm tries to iteratively approximate the optimal action-value function $q^*(s,a)$. Thus, it is a so-called *value-based method.* To find $q^*(s,a)$ and hence identify the optimal policy $\pi^*$, we can exploit the recursive property presented in (5.7) [SB18, Sec. 3.5]:

$$q_\pi(s,a) = \mathbb{E}_\pi \left\{ G_t \mid S_t = s, A_t = a \right\} \tag{5.13}$$
$$= \mathbb{E}_\pi \left\{ R_{t+1} + \gamma_{\mathrm{G_t}} G_{t+1} \mid S_t = s, A_t = a \right\} . \tag{5.14}$$

Let $Q(s,a) : \mathcal{S} \times \mathcal{A} \rightarrow \mathbb{R}$ denote the $Q$-function which approximates $q_\pi(s,a)$ and which we try to learn. Using an iterative approach, the optimal $Q$-function $Q^*(s,a)$ can be approximated by updating the $Q$-function according to [SB18, Sec. 6.5]:

$$Q(S_t, A_t) \leftarrow Q(S_t, A_t) + \nu \Big[ R_{t+1} + \gamma_{\mathrm{G_t}} \max_{a \in \mathcal{A}} Q(S_{t+1}, a) - Q(S_t, A_t) \Big] \tag{5.15}$$
$$= (1-\nu) Q(S_t, A_t) + \nu \Big[ R_{t+1} + \gamma_{\mathrm{G_t}} \max_{a \in \mathcal{A}} Q(S_{t+1}, a) \Big], \tag{5.16}$$

where $\nu \in [0,1]$ is the learning rate, and $R_{t+1} + \gamma_{\mathrm{G_t}} \max_{a \in \mathcal{A}} Q(S_{t+1}, a)$ is the *target* value, a more accurate estimate of the state-action value incorporating new reward information. This update reduces the difference between the current estimate $Q_t(S_t, A_t)$ and the target $R_{t+1} + \gamma_{\mathrm{G_t}} \max_{a \in \mathcal{A}} Q(S_{t+1}, a)$, yielding a refined $Q$-function for the given state-action pair. By sampling the tuple $(S_t = s, A_t = a, R_{t+1} = r', S_{t+1} = s')$, where $s' \in \mathcal{S}$ is the follow-up state and $r' \in \mathcal{R}$ the obtained reward, for the given state-action pair $(s, a)$, the $Q$-function can be updated, and a more accurate $Q$-function on the given state-action pair is achieved.

When starting to learn the agent, the action selection is often $\varepsilon$-greedy [SB18, Sec. 6.5]. In an $\varepsilon$-greedy action-selection, a random action $a$ is performed with some probability $\varepsilon \in [0, 1]$, with $a \leftarrow \arg\max_{a \in \mathcal{A}} Q(s, a)$ otherwise. The occasional selection of random actions ensures that previously unexplored states are explored and may thus produce a larger reward in the long run [SB18, Sec. 2.1]. A larger $\varepsilon$ indicates a policy richer in *exploration*, while a smaller $\varepsilon$ indicates *exploitation*. Finding an appropriate trade-off between exploration and exploitation is one of the typical RL challenges [SB18, Sec. 1.1].

### 5.1.4 Policy-gradient Methods

While value-based methods, such as $Q$-learning, exhibit a high sample efficiency and small variance in function estimation, they suffer from not being able to handle large or continuous actions spaces $\mathcal{A} \subseteq \mathbb{R}$ [DDZ+20, Sec. 3.2]. By directly optimizing a parameterized policy, *policy-based methods* overcome this problem. The policy gradient theorem (PGT) provides the mathematical foundation [SMSM99, SLH+14]. In the following, we distinguish between *stochastic* and *deterministic* policies.

#### Stochastic Policy

Let $\boldsymbol{\theta} \in \mathbb{R}^d, d \in \mathbb{N}$ denote the parameters of a parameterized policy $\pi_{\boldsymbol{\theta}}(a)$ with

$$\pi_{\boldsymbol{\theta}}(a) \coloneqq \pi(a|s, \boldsymbol{\theta})\,. \tag{5.17}$$

Given any state $s \in \mathcal{S}$, we can define the probability of taking action $a \in \mathcal{A}$ as

$$p(A_t = a|S_t = s, \boldsymbol{\theta}) \coloneqq \pi_{\boldsymbol{\theta}}(a)\,. \tag{5.18}$$

Thus, $\pi_{\boldsymbol{\theta}}(a) : \mathcal{S} \times \mathcal{A} \to [0, 1]$ is a PDF, and we call the policy a *stochastic* policy [SLH+14]. Now let $J(\boldsymbol{\theta}) \coloneqq v_{\pi_{\boldsymbol{\theta}}(a)}(s)$ be a scalar performance measure, which returns the expected return when starting in state $s \in \mathcal{S}$ and following $\pi_{\boldsymbol{\theta}}(a)$. Using gradient ascent, we can update the policy by [SB18, Ch. 13]

$$\boldsymbol{\theta} \leftarrow \boldsymbol{\theta} + \nu \widehat{\nabla_{\boldsymbol{\theta}} J(\boldsymbol{\theta})}\,, \tag{5.19}$$

where $\widehat{\nabla_{\boldsymbol{\theta}} J(\boldsymbol{\theta})}$ is a stochastic estimate, whose expectation approximates the gradient of $J(\boldsymbol{\theta})$ w.r.t. $\boldsymbol{\theta}$, and $\nu \in \mathbb{R}$ the learning rate. Thus, we can directly optimize the policy, as long as $\nabla_{\boldsymbol{\theta}} J(\boldsymbol{\theta})$ exists and is finite for all $s \in \mathcal{S}$, $a \in \mathcal{A}$.

We can rewrite $v_{\pi_{\boldsymbol{\theta}}(a)}(s)$ to [SB18, Sec. 13.2]

$$v_{\pi_{\boldsymbol{\theta}}(a)}(s) = \int_{a \in \mathcal{A}} \pi_{\boldsymbol{\theta}}(a) q_{\pi_{\boldsymbol{\theta}}}(s,a)\, \mathrm{d}a\,, \quad \forall s \in \mathcal{S}\,, \tag{5.20}$$

which results in

$$J(\boldsymbol{\theta}) = \int_{a \in \mathcal{A}} \pi_{\boldsymbol{\theta}}(a) q_{\pi_{\boldsymbol{\theta}}}(s,a)\, \mathrm{d}a\,, \quad \forall s \in \mathcal{S}\,. \tag{5.21}$$

The PGT for stochastic policies [SMSM99] demonstrates that [SB18, Sec. 13.2]

$$\widehat{\nabla_{\boldsymbol{\theta}} J(\boldsymbol{\theta})} \propto \int_{s \in \mathcal{S}} \mu_{\pi_{\boldsymbol{\theta}}}(s) \int_{a \in \mathcal{A}} q_{\pi_{\boldsymbol{\theta}}}(s,a) \nabla_{\boldsymbol{\theta}} \pi_{\boldsymbol{\theta}}(a)\, \mathrm{d}s\, \mathrm{d}a \tag{5.22}$$

$$= \mathbb{E}_{\pi_{\boldsymbol{\theta}}} \left\{ \int_{a \in \mathcal{A}} q_{\pi_{\boldsymbol{\theta}}}(S_t,a) \nabla_{\boldsymbol{\theta}} \pi_{\boldsymbol{\theta}}(a)\, \mathrm{d}a \right\}\,, \tag{5.23}$$

where $\propto$ denotes proportionality and $\mu_{\pi_{\boldsymbol{\theta}}} : \mathcal{S} \to [0,1]$ is the distribution of states $S_t \in \mathcal{S}$ when following $\pi_{\boldsymbol{\theta}}(a)$. Thus, the computation of the performance gradient is reduced to a simple expectation [SLH+14]. With (5.23), we can rewrite the update (5.19) to [SB18, Sec. 13.2]

$$\boldsymbol{\theta} \leftarrow \boldsymbol{\theta} + \nu \mathbb{E}_{\pi_{\boldsymbol{\theta}}} \left\{ \int_{\mathcal{A}} q_{\pi_{\boldsymbol{\theta}}}(s,a) \nabla_{\boldsymbol{\theta}} \pi_{\boldsymbol{\theta}}(a)\, \mathrm{d}a \right\}\,. \tag{5.24}$$

This update is intuitive: The sign of the gradient $\nabla_{\boldsymbol{\theta}} \pi_{\boldsymbol{\theta}}(a)$ determines the direction of the update, either increasing or decreasing the probability of choosing a certain action. The amplitude of the update is scaled by the expected return $q_{\pi_{\boldsymbol{\theta}}(a)}(S_t, a)$. Hence, actions that yield a high return will experience larger updates, moving the parameters $\boldsymbol{\theta}$ in the directions that favor the choice of these actions. Since $q_{\pi_{\boldsymbol{\theta}}}(s,a)$ is unknown, in practice it is replaced with a learned approximation $\hat{q}_{\pi_{\boldsymbol{\theta}}(a)}$ [SB18, Sec. 13.3].

### Deterministic Policy

We now consider a deterministic policy $\rho_{\boldsymbol{\theta}}(s)$ with $\rho_{\boldsymbol{\theta}} : \mathcal{S} \to \mathcal{A}$, parameterized with $\boldsymbol{\theta} \in \mathbb{R}^d$ $d \in \mathbb{N}$. In contrast to the stochastic policy, where for each state the policy returns a PDF over the actions space $\mathcal{A}$, the deterministic policy maps each state to an action. The deterministic PGT is given by [SLH+14]

$$\nabla_{\boldsymbol{\theta}} J(\boldsymbol{\theta}) \propto \int_{\mathcal{S}} \mu_{\rho_{\boldsymbol{\theta}}}(s)\, \nabla_{\boldsymbol{\theta}} \rho_{\boldsymbol{\theta}}(s) \nabla_a q_{\rho_{\boldsymbol{\theta}}}(s,a) \Big|_{a=\rho_{\boldsymbol{\theta}}(s)} \mathrm{d}s \tag{5.25}$$

$$= \mathbb{E}_{\rho_{\boldsymbol{\theta}}} \left\{ \nabla_{\boldsymbol{\theta}} \rho_{\boldsymbol{\theta}}(S_t)\, \nabla_a q_{\rho_{\boldsymbol{\theta}}}(S_t,a) \Big|_{a=\rho_{\boldsymbol{\theta}}(S_t)} \right\}\,, \tag{5.26}$$

where $\mu_{\rho_{\boldsymbol{\theta}}} : \mathcal{S} \to [0, 1]$ denotes the distribution of states $S_t \in \mathcal{S}$ when following $\rho_{\boldsymbol{\theta}}(s)$. Compared to the stochastic PGT, the deterministic PGT solely integrates over the state space, avoiding the integration over the (possibly large) action space. Furthermore, instead of the action-value function $q_{\pi_{\boldsymbol{\theta}}(a)}(s, a)$, its gradient $\nabla_{\boldsymbol{\theta}} q_{\pi_{\boldsymbol{\theta}}(a)}(s, a)$ is used.

The update step when using a deterministic policy is given by

$$\boldsymbol{\theta} \leftarrow \boldsymbol{\theta} + \nu \mathbb{E}_{\rho_{\boldsymbol{\theta}}} \left\{ \nabla_{\boldsymbol{\theta}} \rho_{\boldsymbol{\theta}}(S_t) \, \nabla_a q_{\rho_{\boldsymbol{\theta}}}(S_t, a) \Big|_{a=\rho_{\boldsymbol{\theta}}(S_t)} \right\} . \tag{5.27}$$

By reverting the chain rule of derivation, we can reformulate to [SLH+14]

$$\boldsymbol{\theta} \leftarrow \boldsymbol{\theta} + \nu \mathbb{E}_{\rho_{\boldsymbol{\theta}}} \left\{ \nabla_{\boldsymbol{\theta}} q_{\rho_{\boldsymbol{\theta}}}(S_t, \rho_{\boldsymbol{\theta}}(S_t)) \right\} . \tag{5.28}$$

The reformulated term is again intuitive: The parameters $\boldsymbol{\theta}$ are moved in the direction of the gradient of $q_{\rho_{\boldsymbol{\theta}}}$, maximizing the outcome of $q_{\rho_{\boldsymbol{\theta}}}$ averaged over all states. However, since $q_{\pi_{\boldsymbol{\theta}}(a)}(S_t, a)$ is unknown, to successfully apply policy-gradient methods, an estimate is required [SLH+14]. Actor-critic methods overcome this issue by learning both the policy and its action-value function.

### 5.1.5 Actor-critic Methods

Actor-critic methods attempt to simultaneously learn both the policy and its action-value function. Hence, they combine the strengths of value-based and policy-based approaches [GBLB12]: The efficient handling of large or continuous action spaces, and lower-variance estimates of expected returns. Actor-critic methods consist of two entities: A parameterized *actor* $\pi_{\boldsymbol{\theta}}(a)$ is defined, where $\boldsymbol{\theta}$ denotes the parameters of the policy. Furthermore, a differentiable *critic* $Q(s, a|\boldsymbol{\phi})$ approximates the true action-value function, i.e., $Q(s, a|\boldsymbol{\phi}) \approx q_\pi(s, a)$, for the current policy $\pi_{\boldsymbol{\theta}}(s)$, where $\boldsymbol{\phi} \in \mathbb{R}^d$ represents the parameters of the critic. While the actor learns the policy, the critic attempts to learn the respective action-value function, which is then used to update the actor via the PGT. Fig. 5.2 displays the interaction of the actor, the critic, and the environment. Both the actor and the critic observe the current state $S_t$. The actor selects an action $A_t$ according to its policy $\pi_{\boldsymbol{\theta}}(a)$. Upon receiving the reward $R_{t+1}$ and next state $S_{t+1}$, the critic updates its estimate $Q(s, a|\boldsymbol{\phi})$. This updated estimate is then used to improve the policy of the actor via a stochastic policy gradient update step, as described in (5.24) [KT99]. For a deterministic policy $\rho_{\boldsymbol{\theta}}$, the actor-critic setup is similarly defined.

### 5.1.6 Deep Reinforcement Learning

Since ANNs define a parametrized mapping $y = f(x, \boldsymbol{\theta})$, RL algorithms can be implemented using multi-layered, so-called deep neural networkss (DNNs) [SB18,

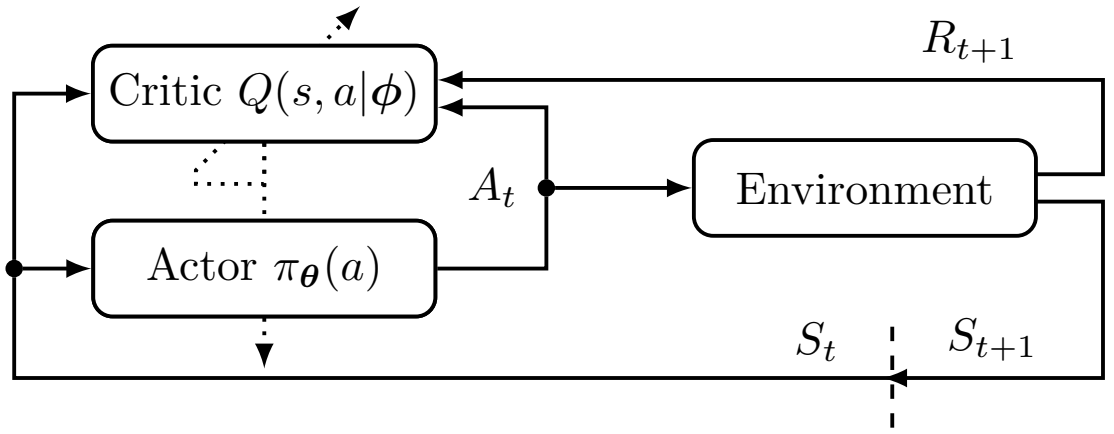

**Figure 5.2:** Schematic overview of an actor-critic algorithm adapted from [GBLB12, Fig. 1]. Dotted lines indicate that the critic is responsible for updating the actor and itself.

Sec. 9.7]. Thus, DRL combines RL and DL, and hence is able to profit from advantages of both DL and RL. It exploits the scalability of multi-layered ANNs to high-dimensional spaces to enable RL even for large (and continuous) action and state spaces. In the following, we briefly discuss deep $Q$-learning (DQL) and the deep deterministic policy gradient (DDPG) algorithm.

### Deep $Q$-Learning

For a small set of possible state-action pairs, tabular $Q$-learning is feasible, storing the expected return for each state-action pair in a table. For larger sets, however, it is limited by its inability to efficiently explore and represent vast state-action spaces. Since a sufficiently large ANN can approximate any continuous function on a closed and bounded subset of $\mathbb{R}^n$ with arbitrarily small error [GBC16, Sec. 6.4.1], DQL uses ANNs to approximate $Q^*(s,a)$ by iteratively updating $Q(s,a)$ [SB18, Sec. 9.7]. The network is called deep $Q$-network (DQN), and its number of output neurons $N_{\text{out}}$ is determined by the number of available actions, i.e., $N_{\text{out}} = |\mathcal{A}|$, where each output neuron corresponds to one action $a \in \mathcal{A}$. Given any state $s \in \mathcal{S}$ as input to the ANN, each output neuron yields the expected return $Q(s,a)$ for taking action $a$ in state $s$. Consequently, by applying $\arg\max_a Q(s,a)$, the action with the highest expected return is selected. Fig. 5.3 illustrates the role of the DQN in a typical RL setting.

To update the parameters of the ANN in order to approach $Q^*(s,a)$, the BP algorithm can be applied, using, e.g., the MSE loss function or Huber loss function between the target value $R_{t+1} + \gamma_{\mathrm{G_t}} \max_{a\in\mathcal{A}} Q(S_{t+1},a)$ and the prediction $Q(S_t, A_t)$. When optimizing $Q(s,a)$, the following issues arise [MKS$^+$15]:

- Small changes to the $Q$-function can cause significant shifts in the policy $\pi_{\boldsymbol{\theta}}$.
- The policy $\pi_{\boldsymbol{\theta}}$ influences the data distribution of the observed sequences, making the input distribution non-stationary.
- Temporal correlations in the observation sequence can further destabilize the learning process.

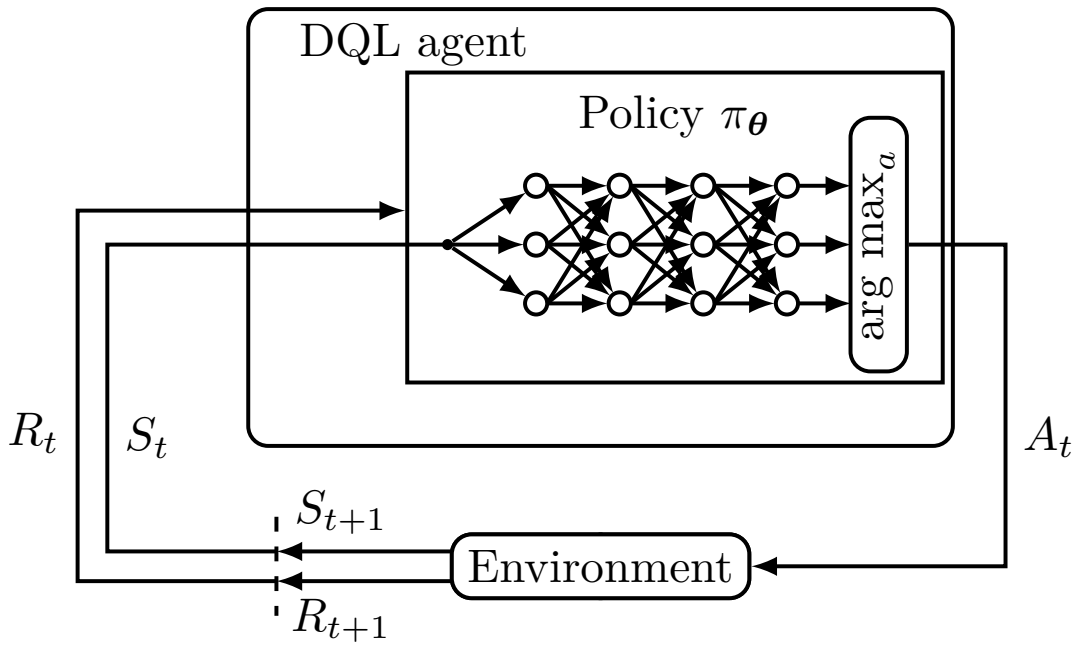

**Figure 5.3:** The DQL agent interacting with the environment, adapted from [LHG+19] and [SB18, Sec. 3.1]. While the DQN approximates $Q^*(s, a)$, the arg max operation returns the action $a \in \mathcal{A}$ yielding the largest return.

- $Q$-values used in both the predictions and target values, i.e., the correlation between $Q(S_t, A_t)$ and $R_{t+1} + \gamma_{\mathrm{G_t}} \max_{a \in \mathcal{A}} Q(S_{t+1}, a)$, introduce additional correlations that may impede convergence.

These concerns can be addressed by two major modifications to the DQL algorithm that reduce correlations and improve training stability [MKS+15]. Instead of updating the DQN after every observed state-action pair, multiple tuples $(S_t, A_t, S_{t+1}, R_{t+1})$ are collected and stored in a *replay memory*. Once the replay memory contains a sufficient number of observations, the backpropagation algorithm is applied on a *mini-batch* of randomly sampled transitions from this memory. This process produces smoother gradient estimates and reduces the correlation among samples. By randomly sampling state-action transitions from the replay memory, the temporal correlation of successive observations is also reduced [SB18, Sec. 16.5]. To further reduce the correlation between the $Q$-values used for prediction and the target values, a separate *target network* $Q_{\boldsymbol{\theta}'}$ is introduced. This target network is a periodically updated copy of the main *policy network* $Q_{\boldsymbol{\theta}}$. Using the target network $Q_{\boldsymbol{\theta}'}$, the $Q$-learning update rule is modified to

$$Q(S_t, A_t) \leftarrow (1 - \nu) Q(S_t, A_t) + \nu \left[ R_{t+1} + \gamma_{\mathrm{G_t}} \max_{a \in \mathcal{A}} Q_{\boldsymbol{\theta}'}(S_{t+1}, a) \right]. \tag{5.29}$$

Alg. 1 provides pseudocode of the DQL algorithm. First, two DQNs to approximate the action-value policy function are initialized. Then, the optimization is carried out for $N_{\mathrm{episodes}} \in \mathbb{N}$ episodes. For each episode, the environment starts with the reception of an initial state $s_0$ and is run for $T_{\mathrm{env}}$ steps. At each time step, an action $a_t$ is chosen $\varepsilon_{\mathrm{Q}}$-greedily. The action $a_t$ is applied to the environment, which returns the follow-up state $s_{t+1}$ and a reward $r_{t+1}$. Afterwards, the tuple $(s_t, a_t, r_{t+1}, s_{t+1})$ is stored in the replay memory. For

**Algorithm 1** DQL with replay memory [MKS+15].

**Require:** Greedy-factor $\varepsilon_{\mathrm{Q}}$, discount factor $\gamma_{\mathrm{Q}}$, empty replay memory $\mathcal{D}$

Initialize action-value policy function $Q_{\boldsymbol{\theta}}(s,a)$ with random weights $\boldsymbol{\theta}$

Initialize action-value target function $Q_{\boldsymbol{\theta}'}(s,a)$, initialize weights as $\boldsymbol{\theta}' = \boldsymbol{\theta}$

**for** $i = 1, \dots, N_{\mathrm{episodes}}$ **do**

  Receive initial state $s_0$

  **for** $t = 0, \dots, T_{\mathrm{env}}$ **do**

    With probability $\varepsilon_{\mathrm{Q}}(t)$: select a random action $a_t \in \mathcal{A}$,
    otherwise: select $a_t = \arg\max_{a\in\mathcal{A}} Q_{\boldsymbol{\theta}}(s_t, a)$

    Send $a_t$ to environment, observe $(r_{t+1}, s_{t+1})$

    Store $(s_t, a_t, r_{t+1}, s_{t+1})$ in $\mathcal{D}$

    Sample a minibatch $\mathcal{B}_{\mathrm{t}}$ of $|\mathcal{B}_{\mathrm{t}}|$ tupels $(s_j, a_j, r_{j+1}, s_{j+1})$ from $\mathcal{D}$

    Set $y_j = r_{j+1} + \gamma_{\mathrm{G_t}} \max_{a\in\mathcal{A}} Q_{\boldsymbol{\theta}'}(s_{j+1}, a)$

    Update $\boldsymbol{\theta}$: Minimize loss $\mathcal{L}$ using gradient descent, with
    $\mathcal{L} = \frac{1}{|\mathcal{B}_{\mathrm{t}}|} \sum_j (Q_{\boldsymbol{\theta}}(s_j, a_j) - y_j)^2$

    $t \leftarrow t + 1$

    Every $T_{\mathrm{target}}$ steps: $\boldsymbol{\theta}' \leftarrow \boldsymbol{\theta}$

  **end for**

  $i \leftarrow i + 1$

**end for**

optimization, a minibatch $\mathcal{B}_{\mathrm{t}}$ is sampled from the replay memory. For each tuple in the minibatch, the corresponding target $R_{t+1} + \gamma_{\mathrm{G_t}} \max_{a\in\mathcal{A}} Q_{\boldsymbol{\theta}'}(S_{t+1}, a)$ is calculated, and used to obtain the loss function $\mathcal{L}$ as the difference of the target $R_{t+1} + \gamma_{\mathrm{G_t}} \max_{a\in\mathcal{A}} Q_{\boldsymbol{\theta}'}(S_{t+1}, a)$ and the estimate of the action-value function $Q_{\boldsymbol{\theta}}$. The parameters $\boldsymbol{\theta}$ are updated in order to minimize $\mathcal{L}$, Afterwards the time $t$ is increased. After $T_{\mathrm{target}} \in \mathbb{N}$ time steps, the parameters of the target network are updated.

### Deep Deterministic Policy Gradient Algorithm

The DDPG algorithm combines the deterministic PGT and ANNs [LHP+16]. It can also be seen as an extension of the DQN algorithm to continuous action spaces [DDZ+20, Sec. 6.1], which uses an actor-critic approach. Thus, DDPG employs two ANN-based entities: First, the critic as the substitute $Q_{\boldsymbol{\phi}}(s,a)$ of the true action-value function, parameterized by $\boldsymbol{\phi}$. Second, the actor, which implements a deterministic policy function $\rho_{\boldsymbol{\theta}}(s)$ with parameters $\boldsymbol{\theta}$.

A drawback of the deterministic policy is that it lacks exploration [SLH+14]. To ensure sufficient exploration during training, the actions need to be chosen from a stochastic policy. However, a deterministic policy is trained. By adding noise $N_t = n$ to the outcome of the deterministic policy $\rho_{\boldsymbol{\theta}}(s)$, a stochastic policy $\rho'_{\boldsymbol{\theta}}(s)$ is obtained [LHP+16]

$$\rho'_{\boldsymbol{\theta}}(s) := \rho_{\boldsymbol{\theta}}(s) + n\,, \tag{5.30}$$

**Algorithm 2** DDPG algorithm, taken from [LHP+16, Alg. 1].

**Require:** Soft update factor $\xi_{\text{up}}$, discount-rate $\gamma_{\text{G}_\text{t}}$, empty replay memory $\mathcal{D}$,

```
Initialize critic network Q_φ(s, a) with parameters φ
Initialize actor network ρ_θ(s) with parameters θ
Initialize target network Q_φ'(s, a) and ρ_θ'(s) with weights φ' ← φ and θ ← θ
for i = 1, ..., N_episodes do
    Initialize a random process N(0, σ_n^2 I) for action exploration
    Receive initial state s_0
    for t = 0, ..., T_env do
        Select action a_t = ρ_θ(s_t) + n_t,  N_t ~ N(0, σ_n^2 I)
        Apply a_t to environment, observe reward r_{t+1} and follow-up state s_{t+1}
        Store transition (s_t, a_t, r_{t+1}, s_{t+1}) in D
        Sample random minibatch of |B_t| transitions (s_j, a_j, r_{j+1}, s_{j+1}) from D
        Set y_j = r_{j+1} + γ_{G_t} Q_φ'(s_{j+1}, ρ_θ'(s_{j+1}))
        Update critic by minimizing the loss L with:
                L = 1/|B_t| Σ_j (Q_θ(s_j, a_j) − y_j)^2
        Update the actor policy using the sampled policy gradient:
                ∇_θ J(θ) ≈ 1/|B_t| Σ_j ∇_a Q_φ(s, a)|_{s=s_j, a=ρ_θ(s_j)} ∇_θ ρ_θ(s)|_{s=s_j}
        Update the target networks:
                θ' ← ξ_up θ + (1 − ξ_up) θ'
                φ' ← ξ_up φ + (1 − ξ_up) φ'
        t ← t + 1
    end for
    i ← i + 1
end for
```

where $N_t$ is sampled from a noise process $N_t \sim \mathcal{N}(0, \sigma_\text{n}^2 \boldsymbol{I})$, matching the dimensionality of the actions. Thus, $\rho'_{\boldsymbol{\theta}}$ outputs a random action with mean $\rho_{\boldsymbol{\theta}}(s)$. This approach is called action-space noise.

Alg. 2 gives the DDPG algorithm as introduced in [LHP+16, Alg. 1]. First, a critic network $Q_{\boldsymbol{\phi}}(s, a)$ and an actor network $\rho_{\boldsymbol{\theta}}$ are initialized. For the same reasons as for the DQL algorithm, a replay memory and target networks are used. For each episode, the environment is initialized, resulting in the reception of the initial state $s_0$. The deterministic policy $\rho_{\boldsymbol{\theta}}$ returns an action, which is superposed by noise sampled from the noise process. Afterwards, the action returned by $\rho'_{\boldsymbol{\theta}}(s)$ is applied to the environment, the reward and follow-up state are received, and the tuple $(s_t, a_t, r_{t+1}, s_{t+1})$ is stored in the replay memory. To update the parameters, a minibatch $\mathcal{B}_\text{t}$ is sampled from the replay memory. Following the same procedure as described in the DQL algorithm, the critic is updated. Afterwardss, the actor is updated according to the deterministic PGT. Then, the target networks are exponentially updated using a soft update factor $\xi_{\text{up}} \in [0, 1]$.

## 5.2 Wireless Resource Allocation Using Reinforcement Learning

One application of DRL is the allocation of resources in a cellular network using multiple carriers, e.g., orthogonal frequency-division multiple access (OFDMA), where each subcarrier of the multi-carrier system represents a physical resource block (PRB) to be allocated to a user [WXH+19]. OFDMA resource allocation algorithms are typically classified by the objective of the underlying optimization problem [SBL14]. Margin-adaptive schemes aim at minimizing the transmit power consumption while complying with a set of fixed user requirements, such as the guaranteed bit rate (GBR). In contrast, rate-adaptive schemes aim at maximizing the total sum data rate over all user equipments (UEs) while satisfying power consumption constraints [SBL14]. The following section focuses on the rate-adaptive scheme.

In today's cellular networks, different UEs inside a cell can be assigned to different quality of service (QoS) classes, e.g., conversational voice, conversational video or web browsing. The QoS classes differ in their GBR and packet delay budget (PDB) [Val20]. For UEs belonging to a single QoS class with constant data rate requirements, the OFDMA resource allocation can be modeled as a convex optimization problem aiming to maximize the overall data rate of the system. By adding constraints to the problem, which, e.g., describe the fairness among UEs by the GBR of each UE, a trade-off between maximizing total throughput and ensuring fairness can be achieved [SBL14]. Various approaches for solving the resource allocation problem exist. One example is the water filling algorithm [CV93], which maximizes the throughput but is unfair, since UEs with bad channel conditions may not get any resources. While the max-min algorithm [RC00] maximizes fairness at the cost of data rate, the proportional fairness algorithm [Kel97] aims at a trade-off of both. However, as the number of UEs with diverse data rate and latency demands grows, i.e., UEs differ in their QoS class, resource allocation becomes increasingly complex. Meeting all these heterogeneous requirements results in a non-convex optimization problem when maximizing the overall system data rate while maintaining fairness among UEs [SBL14]. To still enable optimization, the allocation problem can be modeled as an MDP and solved using DRL [WXH+19].

In this section, we briefly discuss two DRL-based approaches for resource allocation, which we proposed in [BES21] and [BKS22]. The DRL agent architecture in [BES21] is designed for resource allocation in OFDMA systems. Based on the channel quality, size and age of packets inside the buffer waiting for transmission, as well as other UE-specific parameters, a DRL agent is optimized. The agent assigns the PRB to one out of $K_{\mathrm{UE}}$ UEs, with $K_{\mathrm{UE}} \in \mathbb{N}$. Thus, the cardinality of the action set $\mathcal{A}$ is limited by $K_{\mathrm{UE}}$. Based on the DQL algorithm, we designed an agent that clearly outperformed given benchmark agents.

The novelty of the proposed agent lies in the incorporation of the size and age of packets waiting inside the buffer, which we refer to as buffer state information (BSI).

In, e.g., [WXH+19], a DRL agent for a downlink OFDMA resource allocation scenario is optimized. For each time step of the environment, this agent allocates all PRBs to a single UE. The agent knows the instantaneous data rate and average data rate for each UE, while all UEs have the same QoS and the base station (BS) buffer is always filled with packets to transmit. In [ACR20] and [XWY+20], a DRL agent which also allocates the PRBs of an OFDMA system to different UEs is proposed. By randomly generating packets, time-varying data rates of UE are simulated by unoccupied BS buffer slots. While the approach of [ACR20] includes BSI, solely indicating whether a packet for a certain UE is stored inside the buffer and waiting for transmission, [XWY+20] extends BSI to the waiting time of the oldest packet for each UE, as well as the spare space in the buffer. Both outperform proportional fair scheduling, [ACR20] for 4 and 8 UEs with alike QoS and [XWY+20] for 5 UEs with alike QoS. Our agent proposed in [BES21] incorporates more detailed BSI, and we demonstrated that it can handle up to 32 UEs belonging to different QoS classes.

In [BKS22], the multi-carrier system is extended by non-orthogonal multiple access (NOMA). Since NOMA allows multiple UEs to superpose their signals on the same carrier, each carrier now provides $N_{\text{NOMA}}$ resources, where $N_{\text{NOMA}} \in \mathbb{N}$ is the number of UEs that can access a carrier simultaneously. Thus, for each subcarrier, $K_{\text{UE}}^{N_{\text{NOMA}}}$ variations on how to allocate the UEs to the carrier exist, leading to an extremely large action space $\mathcal{A}$. Due to the large action space, the DQL-based agent from [BES21] fails to converge. In [WLN+21], an DDPG-based approach for a downlink multi-carrier NOMA (MC-NOMA) scenario is proposed, which outperforms various benchmarks. However, BSI is neglected. In [XYHZ20], an uplink MC-NOMA system is investigated. For time-varying data and two QoS classes, the DDPG-based approach outperforms a $Q$-learning based reference. However, BSI is reduced to the buffer queue length. In [BKS22], we designed an agent based on the DDPG-algorithm. Again, the novelty of the agent lies in the incorporation of more detailed BSI, i.e., the size and age of packets inside the buffer waiting for transmission.

In the following, we discuss the setup of the agents, the setup of the used environments, and the obtained results in more detail.

### 5.2.1 Multicarrier Resource Allocation

#### The Time-frequency Resource Allocation Environment

Let $\mathcal{N}_{\text{PRB}} = \{1, \ldots, N_{\text{PRB}}\}$ represent the set of $N_{\text{PRB}} \in \mathbb{N}$ subcarriers, and let $\mathcal{K}_{\text{UE}} = \{1, \ldots, K_{\text{UE}}\}$ denote the set of $K_{\text{UE}} \in \mathbb{N}$ UEs. Furthermore, we define $\mathcal{N}_{k_{\text{UE}}} \subseteq \mathcal{N}$ as the subset of subcarriers that are assigned to UE $k_{\text{UE}} \in \mathcal{K}_{\text{UE}}$. Since each subcarrier can only be assigned to one UE, the subsets $\mathcal{N}_{k_{\text{UE}}}$ are pairwise disjoint. We further define $r_{k_{\text{UE}},n_{\text{PRB}}}$ as the data rate that the $k_{\text{UE}}$th UE can achieve over the $n_{\text{PRB}}$th subcarrier with $n_{\text{PRB}} \in \mathcal{N}_{k_{\text{UE}}}$, and $P_{\text{Tx},k_{\text{UE}},n_{\text{PRB}}}$ as the transmit power of the $k_{\text{UE}}$th UE on the $n_{\text{PRB}}$th PRB. Finally, $\underline{r}_{k_{\text{UE}}}$ denotes the target data rate of the $k_{\text{UE}}$th UE, with the individual maximum total trans-mit power constraint $P_{\mathrm{Tx,max},k_{\mathrm{UE}}}$ per UE $k_{\mathrm{UE}}$. Then, the rate-adaptive OFDMA optimization problem is given by [SBL14]

$$\underset{\boldsymbol{P}_{\mathrm{Tx}},\,\mathcal{N}_{k_{\mathrm{UE}}}}{\text{maximize}} \quad \sum_{k_{\mathrm{UE}}\in\mathcal{K}_{\mathrm{UE}}} \sum_{n_{\mathrm{PRB}}\in\mathcal{N}_{k_{\mathrm{UE}}}} r_{k_{\mathrm{UE}},n_{\mathrm{PRB}}} \tag{5.31}$$

$$\text{subject to} \quad \sum_{n_{\mathrm{PRB}}\in\mathcal{N}_{k_{\mathrm{UE}}}} r_{k_{\mathrm{UE}},n_{\mathrm{PRB}}} \geq \underline{r}_{k_{\mathrm{UE}}} \qquad \forall\, k_{\mathrm{UE}} \in \mathcal{K}_{\mathrm{UE}} \tag{5.32}$$

$$\sum_{n_{\mathrm{PRB}}\in\mathcal{N}_{k_{\mathrm{UE}}}} P_{\mathrm{Tx},k_{\mathrm{UE}},n_{\mathrm{PRB}}} \leq P_{\mathrm{Tx,max},k_{\mathrm{UE}}} \qquad \forall\, k_{\mathrm{UE}} \in \mathcal{K}_{\mathrm{UE}} \tag{5.33}$$

$$\mathcal{N}_{k_{\mathrm{UE}}} \cap \mathcal{N}_{j_{\mathrm{UE}}} = \emptyset \qquad \forall\, k_{\mathrm{UE}},\, j_{\mathrm{UE}} \in \mathcal{K}_{\mathrm{UE}},\, k_{\mathrm{UE}} \neq j_{\mathrm{UE}} \tag{5.34}$$

$$\bigcup_{k_{\mathrm{UE}}=1}^{K_{\mathrm{UE}}} \mathcal{N}_{k_{\mathrm{UE}}} = \mathcal{N}_{\mathrm{PRB}} \qquad . \tag{5.35}$$

The goal is to maximize the sum data rate of the system, see (5.31). However, four constraints apply: Each UE must achieve at least its target data rate, see (5.32). Furthermore, for each UE the sum of the transmit power distributed to its assigned carriers must not exceed the predefined maximum transmit power, see (5.33). Each subcarrier can only be assigned to one UE, see (5.34), and all subcarriers need to be assigned, see (5.35).

An implementation of the rate-adaptive resource allocation problem is found within Nokia's "Wireless Suite" problem collection [Nok20]. This platform enhances the comparability and reproducibility of research outcomes by offering a standardized set of environments for benchmarking purposes. The time-frequency resource allocation (TFRA) environment simulates a resource allocation task in an OFDMA downlink setting. An RL agent acts as the scheduler, assigning a limited number of frequency resources, bundled into PRBs, to numerous UEs. During each allocation step, the agent designates one PRB to a UE. All available PRBs are assigned consecutively. Once the entire set of PRBs has been assigned, one time step of 1 ms, is completed. The reward is the accumulation of penalties associated with failing to meet the traffic requirements of the UEs, and as a result, it is always negative. These requirements differ in terms of GBR and PDB, based on the QoS class of the UEs [Val20]. The TFRA environment consists of four distinct QoS classes, which are distinguished by their QoS identifier (QI) $\mathfrak{q}_{k_{\mathrm{UE}}} \in \{1, 2, 3, 4\}$. When the environment is initialized, $K_{\mathrm{UE}} \in \mathbb{N}^{+}$, $4 \mid K_{\mathrm{UE}}$, UEs are randomly distributed throughout a $1\,\mathrm{km}^2$ squared area. This area represents an empty Euclidean space with a transceiver BS positioned at its center. The $K_{\mathrm{UE}}$ UEs roam around the square at constant speeds, which are selected to mimic the speeds of pedestrians [CB13]. The UEs follow random linear paths, reflecting off the edges of the square at specular angles. The scheduler possesses information regarding the age $e_{k_{\mathrm{UE}},l}$ and size $s_{k_{\mathrm{UE}},l}$ of all packets stored in the BS buffer awaiting transmission, where $e_{k_{\mathrm{UE}},l}$ and $s_{k_{\mathrm{UE}},l}$ denote the age and size of the $l$th packet, $l \in \mathbb{N}$, waiting inside the buffer to be transmitted to the $k_{\mathrm{UE}}$th UE. In addition, the channel quality indicator (CQI) $c_{k_{\mathrm{UE}}} \in \{0, \ldots, 15\}$ for UEs $k_{\mathrm{UE}}$, which reflects the channel condition of a UE [ZLYZ19], along with details about the QoS class for each UE, is available to the scheduler.

### Setup of the Agent

The agent introduced in [BES21] is based on DQL. The state of the TFRA environment is characterized by the age and size of all packets held inside the buffer, the CQI and QoS of each UE, along with the index $n_{\mathrm{PRB}}$ of the PRB that is to be allocated. The action space of the agent is defined by $\mathcal{A} = \mathcal{K}_{\mathrm{UE}}$. Consequently, the actions specify which UE gets assigned the current PRB. To facilitate quick and successful optimization of the DQL-based agent, we enhance the DQN and its updating algorithm with several features discussed in what follows.

#### Encoder Neural Networks

To improve the agent, complete BSI can be utilized. Nevertheless, the dimensionality of the state vector increases rapidly, rendering the learning process difficult. Dimensionality reduction can capture the essence of the data while eliminating non-essential features [Mur12]. Building on the concept of autoencoders [GBC16, p. 502], which are NNs trained to reconstruct their input through a compressed hidden-layer representation, we propose encoder neural networks (ENNs). All accessible state information of a UE is input into an ENN to acquire a condensed representation of it. This state information encompasses the CQI $c_{k_{\mathrm{UE}}}$ of UE $k_{\mathrm{UE}}$ and its mean $c_{k_{\mathrm{UE}},\mathrm{mean}}$, along with the age $e_{k_{\mathrm{UE}},l}$ and size $s_{k_{\mathrm{UE}},l}$ of packets stored within the buffer of the BS at slot $l$, $l \in \{1, 2, \ldots L_{\mathrm{BS}}\}$, where $L_{\mathrm{BS}} \in \mathbb{N}$ indicates the total buffer length. To optimize the training of the ENNs for extensive sets of UEs, we instantiate four ENNs, with one ENN having parameters $\boldsymbol{\theta}(\mathfrak{q})$ shared among all UEs possessing the same QI. Fig. 5.4 illustrates the configuration of an ENN. For a UE within the TFRA environment, we reduce dimensionality from $(2L_{\mathrm{BS}} + 2)$ DQN input features to 3.

#### Reward Design

The effectiveness of learning a policy heavily depends on how well the reward signal represents the goal of the application [SB18, Sec. 17.4]. To train our resource allocation agent, we use the reward $r^{(\mathrm{TFRA})}$ that has been pre-defined by the TFRA environment [Val20]. However, to accelerate and kickstart training process, we introduce an additional reward term. Motivated by kick-starting DRL training through the assistance of expert agents [SvHHS18] and expert learning [WXH+19], we present mimicking learning (MICKI). A prerequisite for MICKI is that we already have a (suboptimal) solution approach from which we can initially learn. We refer to this solution approach as the expert agent. We now adjust the reward as

$$r_t = r_t^{(\mathrm{TRFA})} + r_t^{(\mathrm{MICKI})}$$

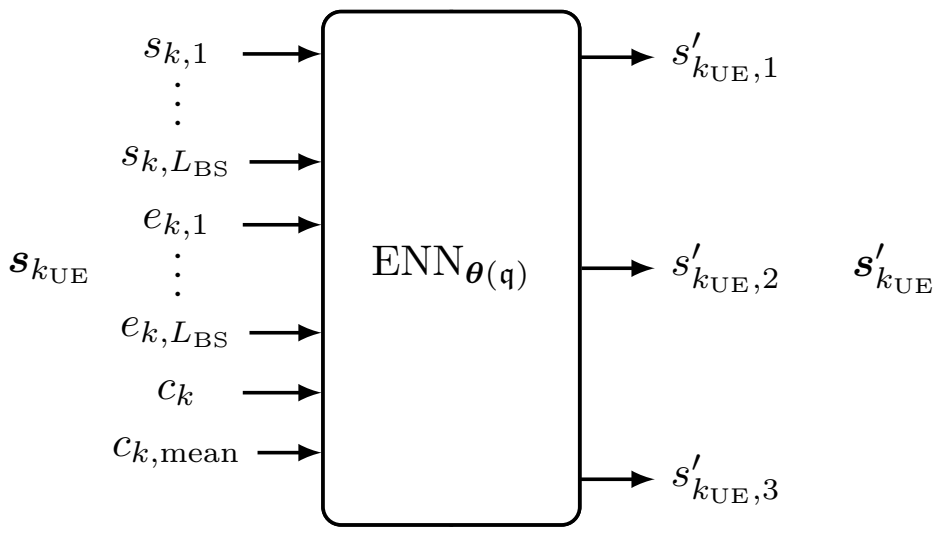

**Figure 5.4:** Configuration of an ENN, adapted from [BES21, Fig. 2].

with

$$r_t^{(\mathrm{MICKI})} = \begin{cases} \mu_{\mathrm{MICKI}}(i), & \text{if } a_{t-1}^{(\mathrm{DQN})} = a_{t-1}^{(\mathrm{expert})} \\ 0, & \text{otherwise,} \end{cases}$$

where $\mu_{\mathrm{MICKI}} : \mathbb{N} \rightarrow \mathbb{R}^+$ denotes a monotonously decreasing function characterized by $\lim_{i\rightarrow\infty} \mu(i) = 0$, with $i = 1, \ldots, N_{\mathrm{episodes}}$. We compare the action $a_{t-1}^{(\mathrm{DQN})}$ of our agent against the action choice $a_{t-1}^{(\mathrm{expert})}$ made by the expert agent that operates concurrently in the same state $s_t$. If our agent selects the same action as the expert agent, it earns a bonus reward. The value of the bonus reward decreases with the number of passed episodes, motivating our agent to discover a policy that outperforms the expert agent in performance. MICKI can be easily integrated into any RL framework where an expert agent is available. In our implementation, we select $\mu_{\mathrm{MICKI}}(t)$ to be constant throughout a training episode, and then exponentially decaying with increasing episodes.

**Packet Shuffling**

We noted that during training, the BS buffer seldomly reaches full capacity, and certain buffer slots rarely contain a packet. This discrepancy in packet placement within the BS buffer results in inadequate training of the ENN, as input neurons linked to infrequently utilized buffer slots generally receive zero valued input throughout training. To enable generalization for filled buffers, we propose packet shuffling, which involves rearranging the packet positions within the BS buffer during the training of the ENNs. In the case of random packet shuffling (RPS), we shuffle the packet positions randomly to achieve a uniform distribution of packets across the buffer slots. However, this random shuffling results in the ENN being unable to infer the order of packets based on their positions. Consequently, we also present sorted packet shuffling (SPS), where we randomly shift the position of the sequence of packets inside the buffer, which maintains the order of the packets.

**User Equipment Shuffling**

During the training of the agent, we furthermore noted that the agent frequently converges to a behavior where the same action is taken at every step. This indicates that a UE bias infiltrates the network, as the agent consistently selects the same UE for PRB allocation. Consequently, the content of the replay memory is biased which hinders the agent to learn to select UEs other than the one it is biased towards. To address this issue, we randomly shuffle the order of all UEs in the dataset prior to feeding it into the DQN, and we reverse the shuffling at the output of the network. We define the shuffling operation as

$$\mathbf{Q} = \mathbf{P}_{\text{rand}}^{-1} \cdot \text{DQN}(\mathbf{P}_{\text{rand}} \boldsymbol{s}'_{\text{UEs}})$$

where $\boldsymbol{s}'_{\text{UEs}} = \left(\boldsymbol{s}'_1, \dots, \boldsymbol{s}'_{K_{\text{UE}}}\right)^{\text{T}}$ represents the vector containing all data for the $k_{\text{UE}}$th UE, $\text{DQN} : \mathcal{S} \times \mathcal{A} \rightarrow \mathbb{R}$ the application of the DQN, and $\mathbf{Q} \in \mathbb{R}^{K_{\text{UE}}}$ the obtained $Q$-values after shuffling is reverted. $\mathbf{P}_{\text{rand}} \in \{0,1\}^{K_{\text{UE}} \times K_{\text{UE}}}$ denotes a random permutation matrix that shuffles the UEs.

**Age Capping**

During training, the environment is simulated for $T_{\text{env}} \in \mathbb{N}$ time steps. Hence, the age $e_{k_{\text{UE}},l}$ of a packet belonging to UE $k_{\text{UE}}$ and held in buffer slot $l$, $l \in \{1, \dots, L_{\text{BS}}\}$ is limited by $T_{\text{env}}$. Conversely, during deployment and validation, packets may exhibit an arbitrary age if they are not transmitted in a timely manner. As gradient descent optimizes the DQN parameters to minimize the loss on the training data only, the network struggles to generalize to data that is not covered by the training set. Thus, unpredictable behavior and limited robustness might occur in situations with large packet ages that have not been covered during training. We tackle this concern by setting a limit on packet ages by modifying ("$\leftarrow$") the ages of the packets $e_{k_{\text{UE}},l}$ as

$$e_{k_{\text{UE}},l} \leftarrow \min\left(e_{k_{\text{UE}},l}\,,\ \text{PDB}(\mathfrak{q}_{k_{\text{UE}}}) + 1\right),$$

where $\text{PDB}(\mathfrak{q}_{k_{\text{UE}}})$ denotes the PDB of the UEs. We hypothesize that the packet age data is primarily relevant for assessing when the PDB of the packet will be depleted. Restricting packet ages to the PDB of the QoS class guarantees that the DQN is aware of which packets have already depleted their corresponding PDB.

**Embedding**

During each allocation step, categorical data $n_{\text{PRB}} \in \{1, \dots, N_{\text{PRB}}\}$ that indicates the PRB to be allocated in the upcoming step is provided to the agent. To enhance the significance of $n_{\text{PRB}}$ for the DQN, it is converted into an $n_{\text{emb}}$-dimensional vector utilizing learnable embeddings, which are specified as parameterized lookup tables. We select an embedding dimension $n_{\text{emb}} = \left\lceil \sqrt[4]{N_{\text{PRB}}} \right\rceil$ [Ten17]. For the TFRA environment with $N_{\text{PRB}} = 25$, we obtain $n_{\text{emb}} = 3$.

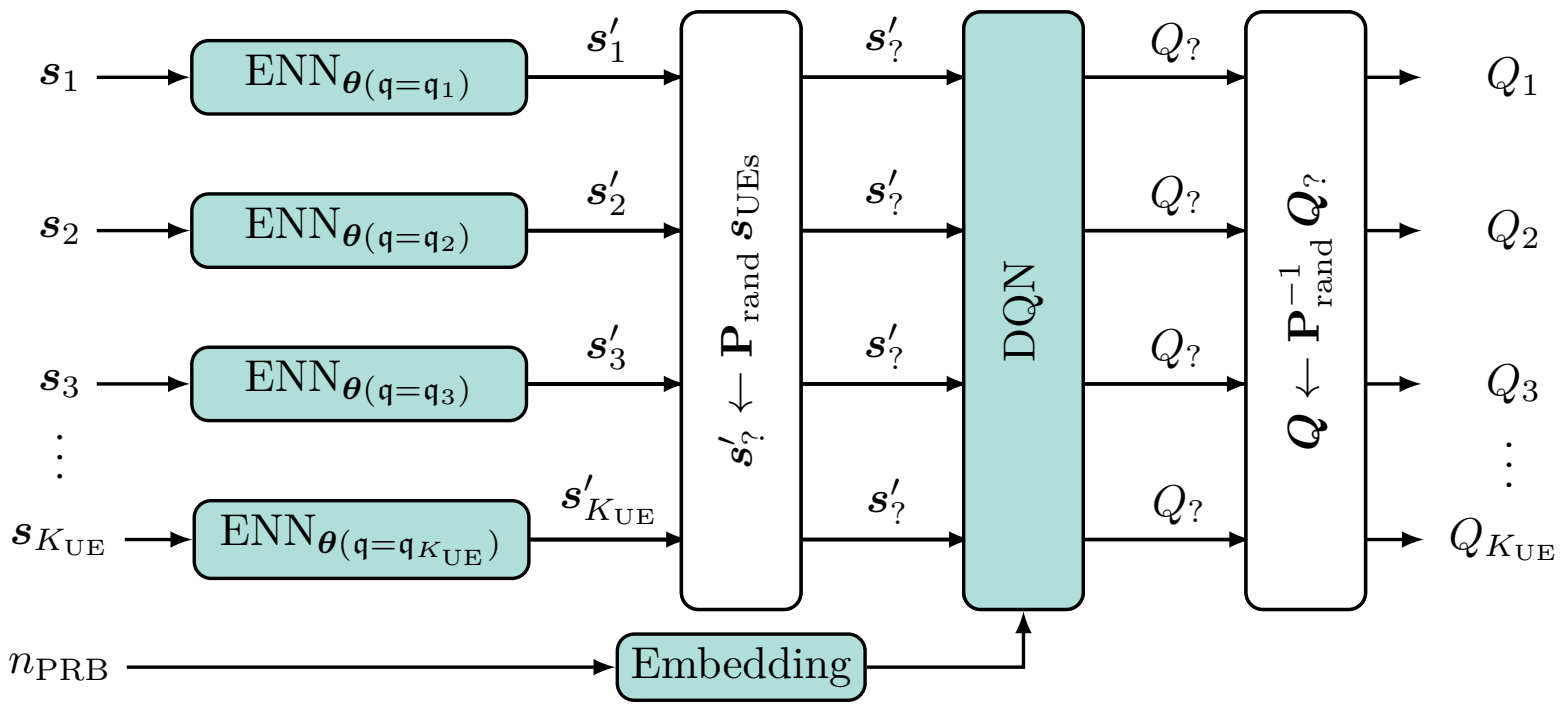

**Figure 5.5:** Architecture of the proposed DQN-based agent, adapted from [BES21, Fig. 3]. All segments with adaptable parameters are colorized.

### Resulting Setup

Fig. 5.5 displays the resulting setup. First, the state $\boldsymbol{s}_{k_{\mathrm{UE}}}$ of each UE is compressed into $\boldsymbol{s}'_{k_{\mathrm{UE}}}$ using ENNs. Afterwards, UE shuffling is applied. Together with $s_{\mathrm{PRB}}$ as indicator of the PRB to be allocated, the permuted and compressed states of all UEs are input to the DQN. At the output of the DQN, the $Q$-values are obtained. By reverting the shuffling, the input order of the UEs is restored. Note, that the packet shuffling executed in certain versions of the ENN is not illustrated.

### Experimental Setup

For validation purposes, the TFRA environment [Nok20] is utilized. We initialize $K_{\mathrm{UE}} = 32$ UEs with 8 UEs per QoS class. Prior to advancing the TFRA environment by a physical time step, $N_{\mathrm{PRB}} = 25$ PRBs are distributed to the UEs. For each UE, the buffer contains $L_{\mathrm{BS}} = 32$ buffer slots.

We optimize and evaluate four agents, all of which implement MICKI, UE shuffling, and full BS buffer state information compressed by ENNs. The ENN agent does not incorporate any additional techniques, whereas the no packet shuffling (NPS) agent additionally employs age capping. The RPS and SPS agents both utilize age capping and either shuffle their packets randomly or in a sorted manner. Tab. 5.1 presents the methods implemented by the agents, and Tab. 5.2 displays the dimensions of the ANNs.

Each of the four approaches is evaluated by independently training seven agents, each initialized with different environment seeds. Let $\mathfrak{S}$ denote the set of distinct initializations of the environment. We create seven distinct training sets $\mathfrak{S}_{\mathrm{train},m}$ with $m = 1, \dots, 7$ and $\mathfrak{S}_{\mathrm{train},m} \cap \mathfrak{S}_{\mathrm{train},j} = \emptyset$, $j, m = 1, \dots, 7$, $j \neq m$, where each set has seven environment realizations, i.e., $|\mathfrak{S}_{\mathrm{train},m}| = 7$. Thus, the set of environments used during training is given by $\mathfrak{S}_{\mathrm{train}} = \cup_{k=1}^{7} \mathfrak{S}_{\mathrm{train},m}$. We run te training for 800 episodes, where for each episode, we conduct 17 500 allocation steps, which aligns with a runtime of the environment of $T_{\text{env}} = 700\,\text{ms}$. After every 10 training episodes, we evaluate the trained agents on an evaluation set $\mathfrak{S}_{\text{eval}} \subset \mathfrak{S}$, $|\mathfrak{S}_{\text{eval}}| = 8$, $\mathfrak{S}_{\text{train}} \cap \mathfrak{S}_{\text{eval}} = \emptyset$. For the final evaluation, we select a test set $\mathfrak{S}_{\text{test}} \in \mathfrak{S}$ with $\mathfrak{S}_{\text{test}} \cap \mathfrak{S}_{\text{train}} = \emptyset$, $\mathfrak{S}_{\text{test}} \cap \mathfrak{S}_{\text{eval}} = \emptyset$ and $|\mathfrak{S}_{\text{test}}| = 300$.

**Table 5.1:** Comparison of methods applied to the agents.

| | ENN | NPS | RPS | SPS |
|---|---|---|---|---|
| MICKI | ✓ | ✓ | ✓ | ✓ |
| UE shuffling | ✓ | ✓ | ✓ | ✓ |
| ENNs | ✓ | ✓ | ✓ | ✓ |
| Age capping | — | ✓ | ✓ | ✓ |
| Packet shuffling | — | — | random | sorted |

**Table 5.2:** Parameters of the ANNs used within the agent.

| Parameter | Embedding | ENN | Main DQN |
|---|---|---|---|
| $N_{\text{in}}$ | 25 | 66 | 99 |
| $N_{\text{out}}$ | 3 | 3 | 32 |
| Number hid. layers | – | 2 | 2 |
| $N_{\text{hid}}$ | – | (16,8) | (79,79) |
| Activation functions | linear | ReLU | ReLU, linear output |

We conduct benchmarking against the agents provided by [Nok20]: the round-robin if traffic (RRiT), proportional fair-channel aware (PFCA), and the knapsack agent. For benchmarking, we initialize $\mathfrak{S}_{\text{test}}$ and evaluate our agents in comparison with the benchmark agents for 65 536 allocation steps, which corresponds to an environment runtime of $T_{\text{env}} = 2621\,\text{ms}$. This odd number is the maximum simulation time predefined by [Nok20].

#### Results

Fig. 5.6 displays the evaluation reward achieved during training of the seven agents for NPS, RPS, and SPS. The dark green line represents the median evaluation value for all training runs. The upper limit of the dark shaded region is calculated by averaging the second and third highest values for each episode, whereas the lower limit is determined by averaging the second and third lowest values. For each episode, the light shaded area encompasses the highest and lowest evaluation values. The performance of the knapsack, PFCA, and RRiT agents for $N_{\text{PRB}} = 25$ and $K_{\text{UE}} = 32$, is represented by the dashed and dotted lines. With respect to the average obtained reward, all three approaches are able to outperform the knapsack agent, which is the best performing benchmark agent. However, NPS and SPS exhibit large fluctuations. The graph clearly illustrates the improving reward across episodes.

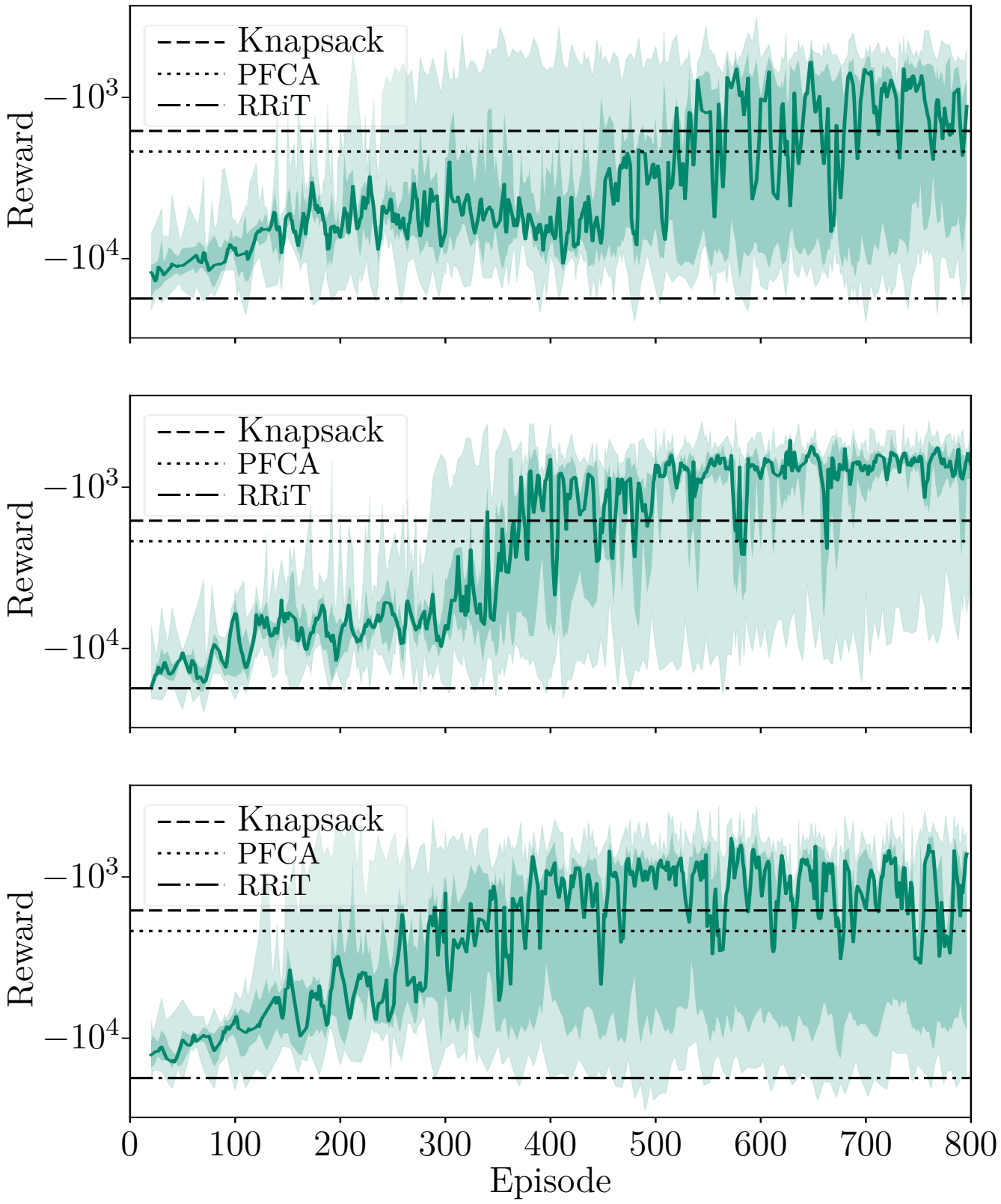

**Figure 5.6:** Evaluation reward of the seven training runs of different agents: NPS (upper plot), RPS (middle plot), and SPS agent (bottom plot).

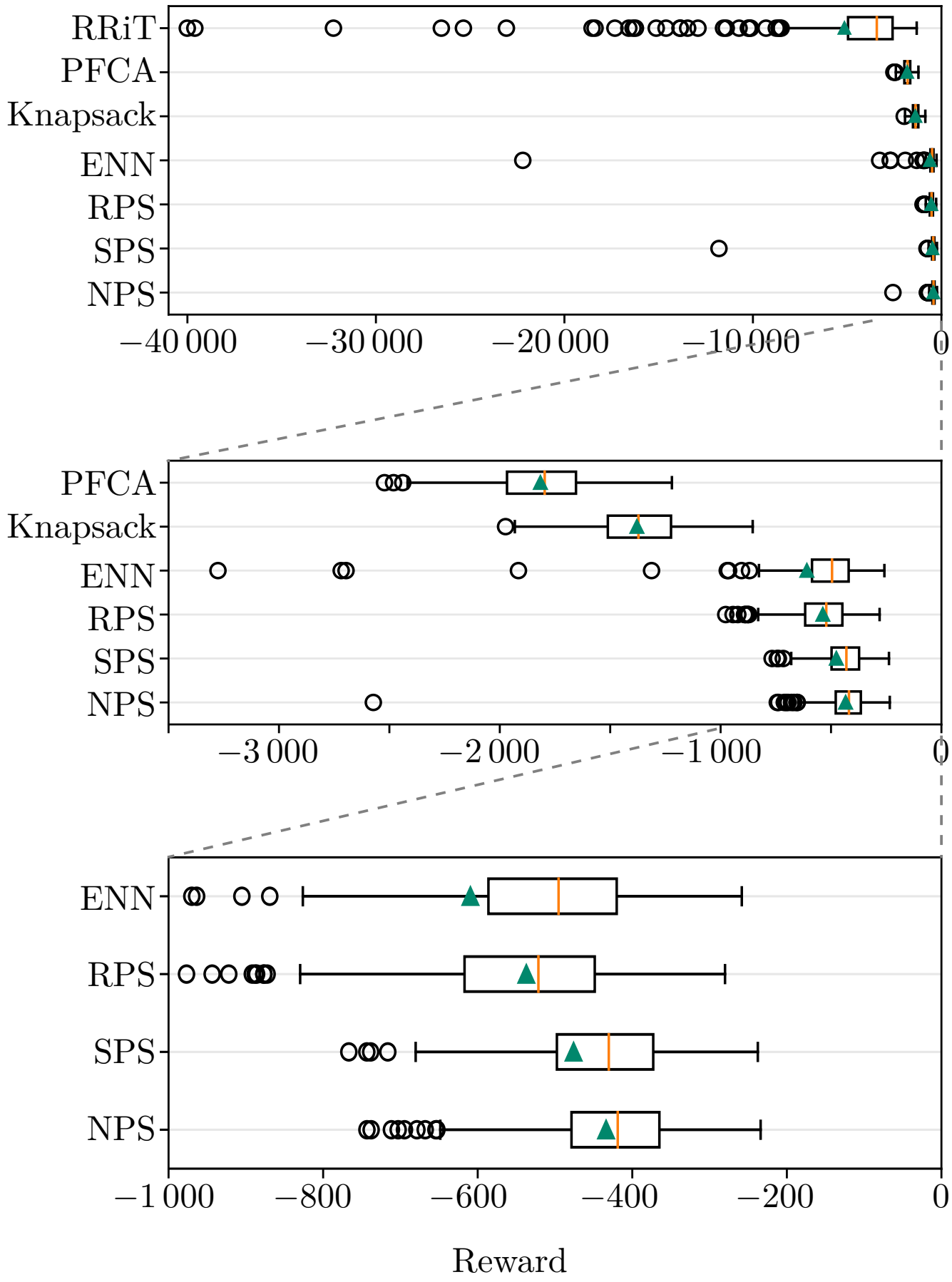

**Figure 5.7:** Boxplots of achieved rewards for our agents and benchmark agents during testing. The plots show the same data on different scales.

Next, we test the different agents. Therefore, for each approach (ENN, NPS, SPS, RPS), out of the seven optimized agents the best performing one is chosen. For each of the 300 environment initializations of $\mathfrak{S}_{\text{test}}$, we calculate the rewards achieved by the agents. Fig. 5.7 displays the distribution of test rewards using boxplots. Note that the three subplots display the same data, but use different scales. Triangles indicate the mean of the reward, and vertical lines inside the box, which is bounded by the lower and upper quartile, depict the median reward over all environments. The maximum length of the whiskers is $1.5 \cdot \text{IQR}$, where IQR is the interquartile range. Outliers are marked by circles.

All our agents significantly outperform the reference agents. When we analyze the median and mean performance of our agents in comparison to the knapsack and PFCA agent, our agents reduce the reward by a factor of three. In the absence of ENNs, we were unable to develop an agent that exceeded the performance of the RRiT. Since every agent shown in Fig. 5.7 employs ENNs, we can conclude that compressing the state information and thereby minimizing the dimensionality of the input to the DQN facilitates optimization. In our comparative analysis of the agents, we find that the ENN agent has a considerable number of significant outliers, while the NPS and SPS agents display only one significant outlier each; the RPS agent shows no significant outliers. Given that the ENN agent is the only one not utilizing age capping, we deduce that age capping contributes to the generalization. Moreover, since the RPS is the only agent absent of significant outliers, we conclude that RPS provides the best generalization, since all buffer slots are equally likely incorporated into the training. Nevertheless, the SPS and NPS agents achieve superior mean and median rewards compared to the RPS agent. This suggests that through the application of packet shuffling techniques, we can trade generalization with performance. Without packet shuffling, the best mean and median performance is achieved, which we contribute to the fact that the first buffer slots, which contain the most packets, are excessively incorporated during training.

### 5.2.2 Multicarrier-NOMA Resource Allocation

For future mobile communication networks, NOMA is identified as a key enabling technology to enhance the spectral efficiency [YLZ+21]. By superimposing UEs in the power domain at the transmitter and applying successive interference cancellation (SIC) at the receiver, NOMA allows multiple UEs to share the same PRB [JHHS21, HFND15]. Distinguishing the superposed messages from various UEs can be achieved through either transmission power control of the UEs, or by aggregating UEs that exhibit sufficiently distinct channel gains. Due to the high robustness of OFDMA to frequency-selective fading [SBL14], MC-NOMA, which merges NOMA and OFDMA [WLN+21], is viewed as a solution to the challenges faced by next-generation communication networks [JHHS21].

Below, we expand the TFRA environment from Sec. 5.2.1 by NOMA. We denote the extended environment as NOMA-TFRA. Once more, the system model is represented by Nokia's “Wireless Suite” problem collection [Nok20]. In the model, all UEs have alike transmit power, and we have to ensure discriminability by learning an agent, which combines UEs with sufficiently distinct channel gains.

Since NOMA increases the number of possible combinations when allocating resources, a DRL algorithm that can handle a large action space is necessary. In [LHP+16], the DDPG algorithm is presented, which can handle a continuous set of DRL actions. Using the DDPG-algorithm, an agent for resource allocation in a downlink MC-NOMA scenario is optimized in [WLN+21], outperforming all benchmarks. In the subsequent section, we build upon the agent from Sec. 5.2.1 to design a DDPG-based agent specifically for uplink MC-NOMA resource allocation with full BSI, i.e., the size and age of packets inside the buffer. The proposed methods and results are based on [BKS22].

### 5.2.3 The NOMA-TFRA Environment

The setup of the NOMA-TFRA environment is smiliar to the TFRA environment used in Sec. 5.2.1. Again, $\mathcal{K}_{\mathrm{UE}}$ denotes the set of $K_{\mathrm{UE}}$ UEs, $\mathcal{N}_{\mathrm{PRB}}$ the set of $N_{\mathrm{PRB}}$ PRBs, and $L_{\mathrm{BS}} \in \mathbb{N}$ the length of the buffer of each UE. Upon initialization of an environment, the UEs are randomly spread over a squared area of dimensions $1\,\mathrm{km} \times 1\,\mathrm{km}$ and assigned a QI $\mathfrak{q}_{k_{\mathrm{UE}}} \in \{1, 2, 3, 4\}$, indicating their QoS class. The area is an empty Euclidean space with a transceiver BS at its center. The UEs roam around at rectilinear trajectories with random speeds [Val20]. At the egdes of the area, the UEs bounce off at specular angles [Val20]. In contrast to the TFRA environment, where each PRB can only serve one UE, one PRB can serve up to $M_{\mathrm{NOMA}}$ UEs, where $M_{\mathrm{NOMA}} \in \mathbb{N}$ defines the number of UEs that can be combined on the same PRB. Since multiple UEs are assigned the same PRB, the achievable data rate depends on the interference caused by other UEs occupying the same PRB. For each UE allocated to a PRB, the achievable data rate $R_{\mathrm{bit}}$ is given by

$$R_{\mathrm{bit}} = B_{\mathrm{PRB}} \log_2(1 + \mathrm{SINR}) \,, \tag{5.36}$$

where $B_{\mathrm{PRB}}$ is the bandwidth of the PRB, and SINR the signal-to-inference-plus-noise-ratio (SINR). The latter is calculated by

$$\mathrm{SINR} = \frac{P_{\mathrm{RX}}}{\sigma_{\mathrm{n}}^2 + \zeta + \xi} \,, \tag{5.37}$$

where $P_{\mathrm{RX}}$ denotes the power of the UE received at the BS, $\sigma_{\mathrm{n}}^2$ the power of AWGN, $\zeta = -105\,\mathrm{dBm}$ a constant interference power throughout the coverage area, and $\xi \in \mathbb{R}^+$ the interference power caused by other UEs occupying the current PRB. The received power $P_{\mathrm{RX}}$ is determined by the distance between the UE and the BS, applying the free-space path loss to the transmit power $P_{\mathrm{TX}} = 13$ dBm, which remains constant across all UEs. Once each PRB is allocated and the transmitted bits are removed from the buffers, it is assessed whether the leftover packets surpass their latency requirements, as defined by their PDB, which depends on the QoS class of the UEs. If there are packets that exceed their PDB, the environment returns a negative reward (penalty) by aggregating the bits of packets that have surpassed their PDB [Val20]. Subsequently, the UEs move and new packets are created. While SIC demodulates the signals from UEs in order of diminishing received power and progressively removes the interfering waveforms one at a time [PS08, Sec. 16.3-4], the NOMA-TFRA environment simulates SIC by incrementally adding interference.

Another difference of the TFRA and NOMA-TFRA environment is given when comparing their action spaces. For NOMA-TFRA, the action space is given by $\mathcal{A} = \{\mathcal{K}_{\mathrm{UE}}, \mathrm{empty}\}$. Either one of the $K_{\mathrm{UE}}$ UEs is chosen, or the resource block is kept empty. When leaving a NOMA resource empty, the SINR and hence the achieved data rate of the UEs already assigned to the PRB is increased, compare (5.36) and (5.37). Thus, adding an action that leaves a NOMA resource empty can be beneficial.

### Setup of the Agent

#### Sequential NOMA Allocation

While the general DRL framework introduced in Sec. 5.1.1 receives a state $S_t$, takes an action $A_t$, and immediately receives the reward $R_{t+1}$ and follow-up state $S_{t+1}$, the uplink MC-NOMA scenario exhibits a significant challenge: The BSI and, consequently the state $S_t$, is modified after the allocation of all NOMA resources of a PRB has been completed. As a result, multiple sequential allocation actions, which involve distributing the NOMA resources of a PRB, must occur without updated state information. For instance, if UE $k_{\mathrm{UE}}$ is permitted to utilize the first NOMA resource and successfully transmits all packets inside its buffer, allocating the second NOMA resource to the same UE becomes inefficient. However, the state that holds information regarding the status of all buffers remains unchanged and continues to indicate that UE $k_{\mathrm{UE}}$ has packets awaiting transmission. To address this challenge, we propose a sequential decision-making structure illustrated in Fig. 5.8, which is inspired by [ZVS+18]. Let $\boldsymbol{x}_{\mathrm{a},m_{\mathrm{NOMA}}} \in [0,1]^{(K_{\mathrm{UE}}+1)}$, $m_{\mathrm{NOMA}} = 1, \dots M_{\mathrm{NOMA}}$, denote a vector, whose elements represent the probabilities of the $K_{\mathrm{UE}}+1$ possible actions that can be selected for the $m_{\mathrm{NOMA}}$th NOMA resource of a PRB. We further define $\boldsymbol{X}_{\mathrm{a}} = (\boldsymbol{x}_{\mathrm{a},1} \dots \boldsymbol{x}_{\mathrm{a},M_{\mathrm{NOMA}}})^{\mathrm{T}}$, $\boldsymbol{X}_{\mathrm{a}} \in [0,1]^{M_{\mathrm{NOMA}} \times (K_{\mathrm{UE}}+1)}$, as a matrix which concatenates the probability vectors $\boldsymbol{x}_{\mathrm{a},m_{\mathrm{NOMA}}}$ of all NOMA resources of a PRB. Thus, the $m_{\mathrm{NOMA}}$th row of $\boldsymbol{X}_{\mathrm{a}}$ contains the probabilities of how to allocate the $m_{\mathrm{NOMA}}$th NOMA resource.

When running the sequential decision-making, we initialize $\boldsymbol{X}_{\mathrm{a}} = \mathbf{0}$, and make iterative decisions for allocation, substituting the $m_{\mathrm{NOMA}}$th row of $\boldsymbol{X}_{\mathrm{a}}$ with the output $\boldsymbol{x}_{\mathrm{a},m_{\mathrm{NOMA}}}$ of the actor. The updated $\boldsymbol{X}_{\mathrm{a}}$ is used to enable prediction of the changes applied to the state, such as identifying which UE might have been allocated to previous NOMA resources, potentially emptying the respective buffer. Once the actor returns $\boldsymbol{x}_{\mathrm{a},M_{\mathrm{NOMA}}}$, the actions $\boldsymbol{a} = (a_1, \dots, a_{M_{\mathrm{NOMA}}})$ are drawn from the probability distributions given by $\boldsymbol{X}_a$.

#### Traffic-based Masking

In order to prevent the actor from selecting UEs that have empty buffers, we implement traffic-based masking. A Boolean mask $\boldsymbol{h} \in \{0,1\}^{(K+1)}$ is defined, which indicates whether a UE has packets available in its buffer. The probability vector $\boldsymbol{x}_{\mathrm{a},m_{\mathrm{NOMA}}}$ is then modified by an elementwise multiplication $\boldsymbol{x}_{\mathrm{a},m_{\mathrm{NOMA}}} \leftarrow \boldsymbol{x}_{\mathrm{a},m_{\mathrm{NOMA}}} \odot \boldsymbol{h}$. To guarantee that $\boldsymbol{x}_{\mathrm{a},m_{\mathrm{NOMA}}}$ results in a valid probability distribution after masking, we perform normalization.

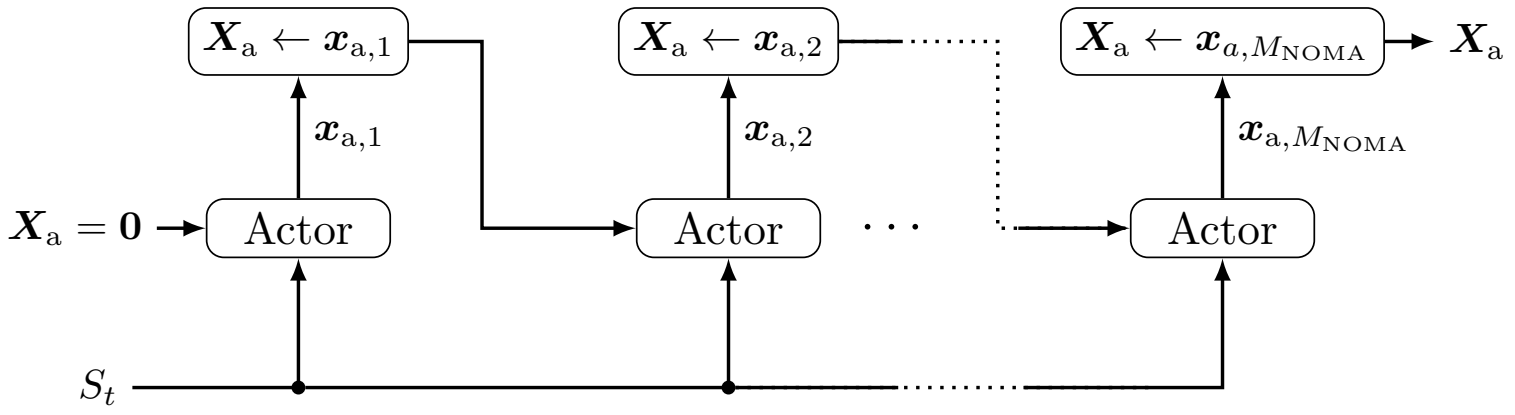

**Figure 5.8:** Block diagram of the sequential decision-making structure for MC-NOMA resource allocation, adapted from [BKS22, Fig. 2].

### Resulting Setup

The resulting setup of the actor network is displayed in Fig. 5.9, which can be seen as an extension of Fig. 5.5. ENNs compress the BSI and CQI of each UE, see Fig. 5.4. Afterwards, UE shuffling is applied. Together with $s_\mathrm{PRB}$ as indicator of the PRB to be allocated and $\boldsymbol{X}_\mathrm{a}$, the permuted and compressed states of all UEs are input to the DQN. A softmax function ensures that the vector generated by the DNN satisfies the conditions of a probability distribution. For the outputs associated with the allocation of UEs, the permutation is reverted and the order of the UEs is restored. Afterwards, the probabilities $x_a \in [0, 1]$ of all actions, including the empty action, are fed to the traffic-based masking.

To mitigate overfitting in the DNN, dropout layers are utilized. Additionally, we implement age capping to manage packets that have surpassed their PDB. The architecture of the critic is similar in terms of input information processing; however, the output layer of the DNN consists of a single neuron, since solely the expected return for executing actions $\boldsymbol{X}_\mathrm{a}$ given state $(\boldsymbol{s}_1, \dots \boldsymbol{s}_{K_\mathrm{UE}}, n_\mathrm{PRB})$ needs to be estimated.

In the beginning of the training procedure, we observed that the untrained agent selects suboptimal actions, which results in full UE buffers and, hence, to situations without a chance for selecting beneficial actions. If the training under such an ill-conditioned situation is continued, the training is prolonged, and the replay memory is filled with samples, which are not statistically significant. We assume that a well-trained agent is able to avoid such ill-conditioned situations, and terminate a training episode early if the training reward $R_{t+1} < R_\mathrm{cap}$, where $R_\mathrm{cap} \in \mathbb{R}$, $R_\mathrm{cap} < 0$ is an empirically determined value.

### Results

We optimized agents for two different parameterizations of the NOMA-TFRA environment: For $K_\mathrm{UE} = 20$ UEs, $N_\mathrm{PRB} = 10$ PRBs, and $R_\mathrm{cap} = -80\,000$, which we refer to as small environment (SE), we demonstrate the effectiveness of masking. For $K_\mathrm{UE} = 32$ UEs, $N_\mathrm{PRB} = 25$ PRBs, and $R_\mathrm{cap} = -150\,000$, which we denote as large environment (LE), we demonstrate the importance of dropout layers. All configurations include $L_\mathrm{BS} = 8$ buffer slots per UE and $M_\mathrm{NOMA} = 2$. The parameters of the embedding and ENNs are presented in Tab. 5.2. Depending on the used environment, the DNN architecture varies as indicated in Tab. 5.3. The various agents that we train and evaluate are outlined in Tab. 5.4.

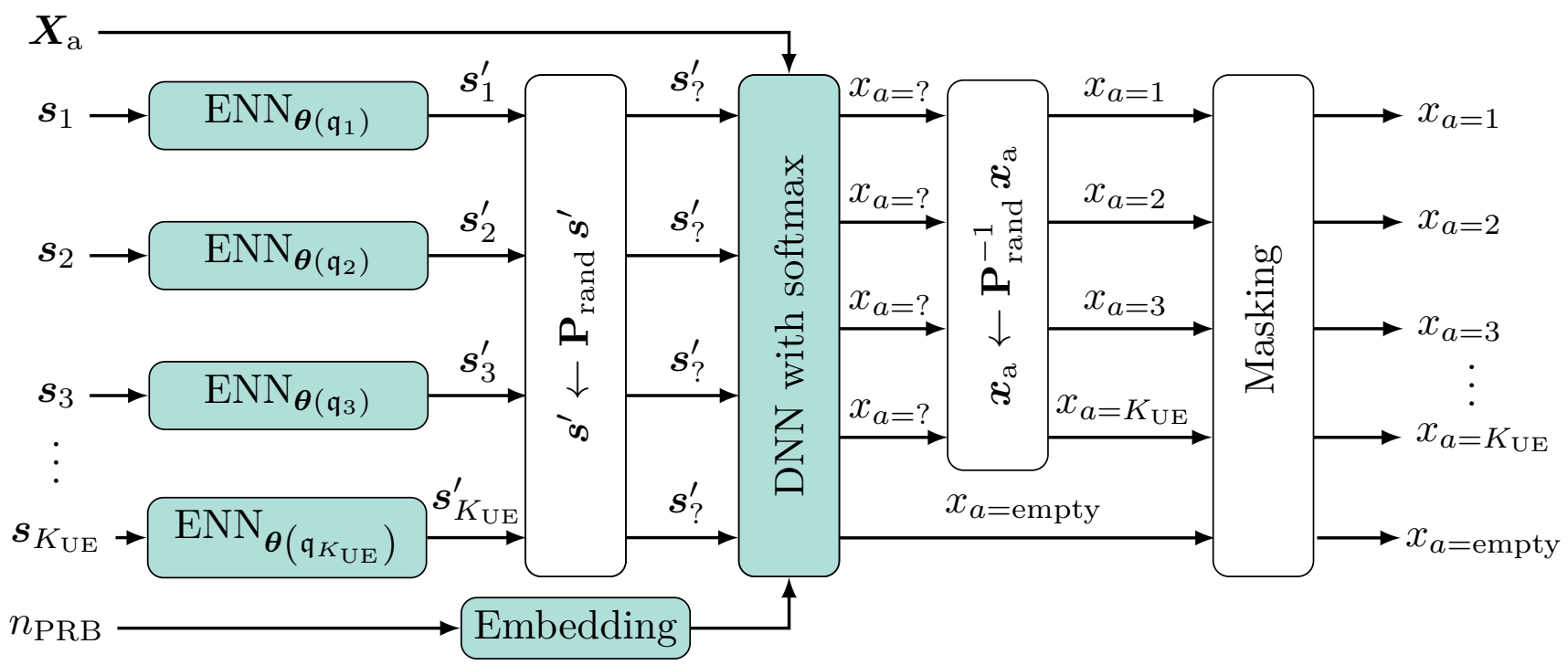

**Figure 5.9:** Architecture of the proposed agent for resource allocation in uplink NOMA scenarios, adapted from [BKS22, Fig. 3]. All segments with adaptable parameters are colorized.

**Table 5.3:** Parameters of the DNNs for the SE and the LE.

| | $N_{in}$ | $N_{hid}$ | $N_{out}$ | Number hid. layers | Activation function |
|---|---|---|---|---|---|
| SE | 105 | 603 | 21 | 3 | ReLU |
| LE | 165 | 963 | 33 | 3 | ReLU |

We conduct benchmarking against the *NOMA-uplink-PFCA (NPFCA)* agent, which is supplied by [Nok20]. Given the infinite set of environment realizations $\mathfrak{S}$, we select $|\mathfrak{S}_{eval}| = 4$ realizations for evaluation and $|\mathfrak{S}_{test}| = 100$ for the test set, with $\mathfrak{S}_{eval} \cap \mathfrak{S}_{test} = \emptyset$. Additionally, $\mathfrak{S}_{train} \subset \mathfrak{S} \setminus (\mathfrak{S}_{eval} \cup \mathfrak{S}_{test})$ represents the training set, from which the environment initializations utilized for training are sampled. During the training phase, an environment is stopped either after $T_{env} = 600\,\mathrm{ms}$, or by an early termination due to $R_{t+1} < R_{cap}$. The agent undergoes evaluation following every five training episodes. To save computing time for the LE, agents are only evaluated if they achieve a training reward surpassing the average reward of the benchmark agent, which results in $R_{cap} = -1552$. The evaluation and testing phases are restricted to 65 536 allocation steps, leading to an environment runtime of $T_{env} = 6553\,\mathrm{ms}$ for the SE, and $T_{env} = 2621\,\mathrm{ms}$ for the LE.

Fig. 5.10(a) illustrates the training performance of the different agents. When evaluating an agent, its average evaluation reward over $\mathfrak{S}_{eval}$ is displayed as a function of the training episode. The S-def agent requires approximately 1020

**Table 5.4:** Overview of agents for the NOMA-TFRA environment.

| Agent | Environment | $K_{\mathrm{UE}}$ | $N_{\mathrm{PRB}}$ | Masking | Dropout |
|---|---|---|---|---|---|
| S-def | SE | 20 | 10 | — | — |
| S-mask | SE | 20 | 10 | ✓ | — |
| L-mask | LE | 32 | 25 | ✓ | — |
| L-drop | LE | 32 | 25 | ✓ | ✓ |

episodes to achieve a training performance which is sufficient for evaluation. After another 180 episodes, the performance of the S-def agent decreases, resulting in large negative rewards. Thus, the training of S-def is unstable. In comparison, with progressing training, the S-mask agent shows a consistent improvement. We conclude that traffic-based masking reduces the demands on the agent, and hence stabilizes and improves training significantly. For the LE, the L-mask agent produces unstable evaluation rewards, and its performance deteriorates after around 650 training episodes. We attribute this to overfitting of the DNN. Introducing dropout layers with an outage probability of $p_{\mathrm{drop}} = 0.2$ helps to stabilize the training, as evidenced by the performance of the L-drop agent. Due to sufficient evaluation performance and high computation effort, training of the L-drop agent was stopped after 440 training episodes. Fig. 5.10(b) illustrates the effect of early termination on the training of the L-drop agent. As training progresses, the runtime $T_{\mathrm{env}}$ of the environment realizations also increases. Notably, in the initial phase of training, early termination considerably accelerates the training process by stopping ill-conditioned environments early. The L-drop agent experiences a performance boost around 200 episodes, marking the beginning of its evaluation as seen in Fig. 5.10.

The improvement of our proposed agents in comparison to the benchmark agent is illustrated in Fig. 5.11. The agents are tested using identical $|\mathfrak{S}_{\mathrm{test}}| = 100$ environment realizations, and we show the obtained rewards. During the training phase, we record the parameters of each agent after every episode. For the testing phase, we employ the network parameters that yield the best training outcome for each agent; for instance, in the case of the L-mask agent, we utilized the parameters obtained after episode 342. Once more, green triangles represent the average evaluation reward, and the vertical orange lines within the box, constrained by the lower and upper quartile, illustrate the median reward across all environments. Outliers are indicated by circles. Our agents exceed the performance of the benchmark agent by receiving only 37% and 27% of the penalties imposed by the benchmarks, for both the SE and the LE respectively.

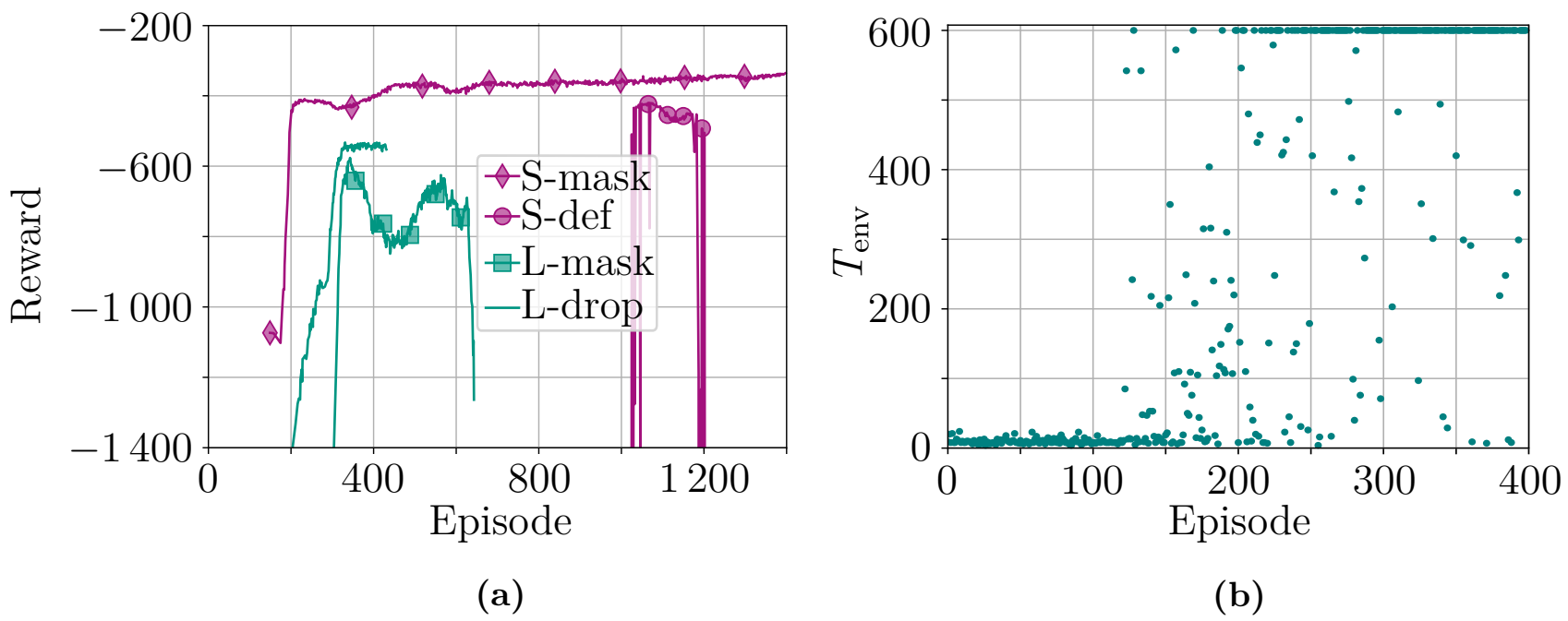

**Figure 5.10:** (a): Training performance in the NOMA-TFRA environment. (b): Runtime $T_{\text{env}}$ of the environment in dependence of the episode for the L-drop agent.

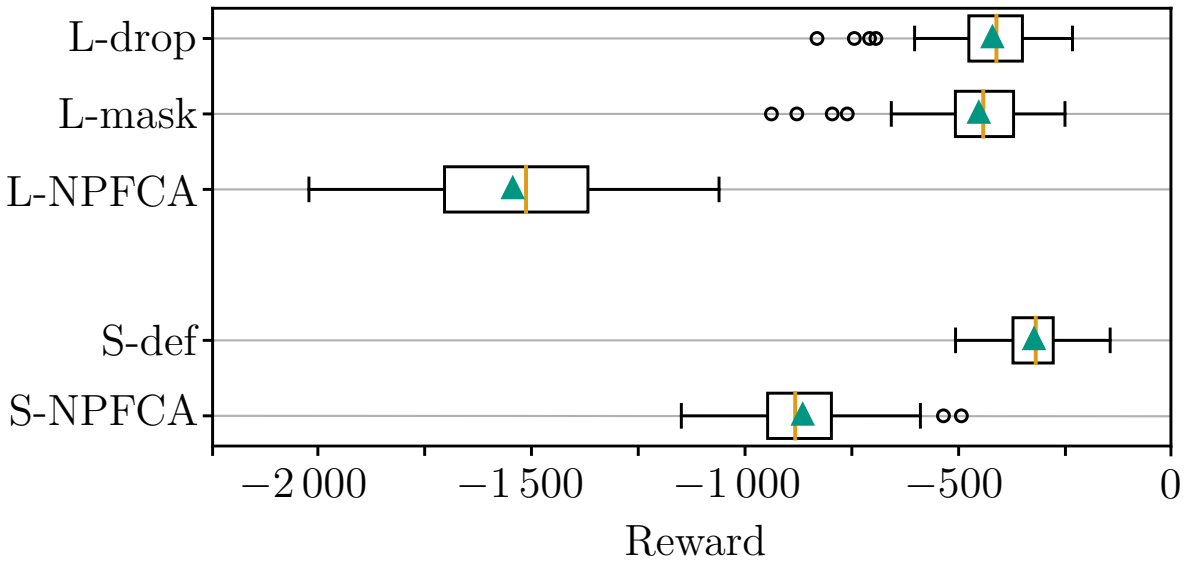

**Figure 5.11:** Boxplot of achieved rewards of the agents introduced in Tab. 5.4 and the benchmark agents. The lower two agents are specialized to the SE, the upper three agents to the LE.

## 5.3 Key Findings

In this chapter, we provided two DRL-based agents for resource allocation in wireless multi-carrier scenarios. Both agents incorporate BSI, i.e., the size and age of packets inside the buffer awaiting transmission. To downscale the input to the decision-making DNN, we introduce ENNs, which compress the BSI to a vector of lower dimensionality. For the TFRA environment, we outperform the benchmarks using ENNs and compressed BSI. When using age capping, a technique that limits the age of packets to their PDB, we can further generalize and hence improve the agent. If NOMA capabilities are considered, multiple UEs can be assigned to a single PRB. Without an update of the state, multiple actions need to be selected per PRB. We propose to use sequential decision-making, which takes the probabilities of past actions on the same PRB into account. Furthermore, we propose to use traffic-based masking to avoid the allocation of resources to UEs with empty buffers. We demonstrate that our proposed setup outperforms the benchmark agents supplied by [Nok20].

We conclude that the incorporation of BSI enhances the performance of DRL-based agents for resource allocation in multi-carrier systems. However, to stabilize training and achieve generalization, techniques such as ENNs, age capping, and traffic-based masking are necessary.

# 6 Optimization of Spiking Neural Networks Using Reinforcement Learning

In Ch. 4, we demonstrated the successful application of SNNs for equalization and demapping. To convert the observation $y \in \mathbb{R}$ at the channel output into spike signals, different neural encodings were applied. However, the parameters of the neural encodings, e.g., the centers and widths of the linear RFE, were determined heuristically. The heuristic optimization of these parameters is time-consuming and can easily result in non-optimal parameters. Hence, it is desirable to have a joint optimization of the parameters of the encoding and the parameters of the SNN. For the linear RFE, the non-differentiable conversion from spike timings $\kappa_{\mathrm{enc,z},j}[\kappa] \in \mathbb{N}_0$ into spike signals $z_{\mathrm{enc}}[\kappa] \in \{0,1\}$ (see (2.20)) prohibits the application of the BP algorithm. To enable the optimization of the parameters of the encoding, we can apply RL.

In this chapter, we exploit the PGT to update the parameters of the linear RFE [EvBS25]. First, we derive a policy gradient-based update rule, which we hereafter refer to as policy gradient-based update (PGU). Using a test function, we investigate the PGU and propose some modifications that stabilize and speed up training. By comparing with the results obtained in Ch. 4, we demonstrate that we can significantly reduce the dimensionality of the output of the linear RFE, while achieving similar performance. Furthermore, we demonstrate that the proposed approach can be used to directly optimize the parameters of the SNN.

## 6.1 Derivation of the Policy Gradient-based Update Rule

The following derivation is motivated by the problem setting and solution given in [AH18], which successfully optimizes a parameterized transmitter prior to a non-differentiable channel using RL-based methods.

Let $\boldsymbol{\theta} \in \mathbb{R}^d$, $d \in \mathbb{N}$, denote a parameter vector of a system, which we aim to optimize in order to fulfill a given task. Furthermore, let $\ell : \mathbb{R}^d \to \mathbb{R}$ denote a performance metric, where $\ell(\boldsymbol{\theta})$ indicates how well $\boldsymbol{\theta}$ performs on the given task. We now aim to find the optimal parameter vector $\boldsymbol{\theta}^*$ such that

$$\boldsymbol{\theta}^* = \underset{\boldsymbol{\theta} \in \mathbb{R}^d}{\arg\min}\ \ell(\boldsymbol{\theta})\,. \tag{6.1}$$

To approximate $\boldsymbol{\theta}^*$, we iteratively update and refine $\boldsymbol{\theta}$ by exploring its vicinity. Therefore, we define a Gaussian policy $\pi_{\tilde{\boldsymbol{\Theta}}|\boldsymbol{\theta}} : \mathbb{R}^d \to \mathbb{R}^+$, representing a Gaussian

PDF with mean $\boldsymbol{\theta}$. It is given by

$$\pi_{\tilde{\boldsymbol{\Theta}}|\boldsymbol{\theta}}\left(\tilde{\boldsymbol{\theta}}|\boldsymbol{\theta}\right) = \frac{1}{(2\pi\sigma_{\pi}^2)^{\frac{d}{2}}} \exp\left(-\frac{||\tilde{\boldsymbol{\theta}} - \boldsymbol{\theta}||_2^2}{2\sigma_{\pi}^2}\right) , \tag{6.2}$$

where $\sigma_{\pi}^2$ denotes the variance, and $\tilde{\boldsymbol{\Theta}} \in \mathbb{R}^d$ denotes a RV modeling the variations $\tilde{\boldsymbol{\theta}}$, which we sample from the Gaussian policy. Similarly to $\ell(\boldsymbol{\theta})$, the performance metric $\ell(\tilde{\boldsymbol{\theta}})$ indicates the performance of $\tilde{\boldsymbol{\theta}}$ w.r.t. the given task.

In analogy to the objective function of the RL problem (5.21), we define the objective function $J(\boldsymbol{\theta})$ of the given problem as [AH18]

$$J\left(\boldsymbol{\theta}\right) := \int_{\mathbb{R}^d} \ell\left(\tilde{\boldsymbol{\theta}}\right) \cdot \pi_{\tilde{\boldsymbol{\Theta}}|\boldsymbol{\theta}}\left(\tilde{\boldsymbol{\theta}}|\boldsymbol{\theta}\right) \, \mathrm{d}\tilde{\boldsymbol{\theta}} \tag{6.3}$$

$$= \mathbb{E}_{\tilde{\boldsymbol{\Theta}}|\boldsymbol{\theta}}\left\{\ell\left(\tilde{\boldsymbol{\Theta}}\right)\middle|\boldsymbol{\theta}\right\} , \tag{6.4}$$

which is the expected performance when following $\pi_{\tilde{\boldsymbol{\Theta}}|\boldsymbol{\theta}}\left(\tilde{\boldsymbol{\theta}}|\boldsymbol{\theta}\right)$. Thus, we can rewrite (6.1) as

$$\begin{aligned}\boldsymbol{\theta}^* &= \underset{\boldsymbol{\theta}\in\mathbb{R}^d}{\arg\min}\; J\left(\boldsymbol{\theta}\right)\\ &= \underset{\boldsymbol{\theta}\in\mathbb{R}^d}{\arg\min}\; \mathbb{E}_{\tilde{\boldsymbol{\Theta}}|\boldsymbol{\theta}}\left\{\ell\left(\tilde{\boldsymbol{\Theta}}\right)\middle|\boldsymbol{\theta}\right\} .\end{aligned}$$

Remember that $\boldsymbol{\theta}$ constitutes the mean value of the policy $\pi_{\tilde{\boldsymbol{\Theta}}|\boldsymbol{\theta}}(\tilde{\boldsymbol{\theta}}|\boldsymbol{\theta})$. Thus, we now aim to find the policy (and hence the PDF) that minimizes $J(\boldsymbol{\theta})$, which we expect to be centered around $\boldsymbol{\theta}^*$. We can update $\boldsymbol{\theta}$ (and thus the policy $\pi_{\tilde{\boldsymbol{\Theta}}|\boldsymbol{\theta}}(\tilde{\boldsymbol{\theta}}|\boldsymbol{\theta})$) using gradient descent

$$\boldsymbol{\theta} \leftarrow \boldsymbol{\theta} - \nu \cdot \nabla_{\boldsymbol{\theta}} J\left(\boldsymbol{\theta}\right) , \tag{6.5}$$

where $\nu \in \mathbb{R}^+$, is the learning rate. To determine the gradient, we can apply the PGT to (6.3) [AH18]:

$$\nabla_{\boldsymbol{\theta}} J\left(\boldsymbol{\theta}\right) = \int_{\mathbb{R}^d} \ell\left(\tilde{\boldsymbol{\theta}}\right) \; \nabla_{\boldsymbol{\theta}} \pi_{\tilde{\boldsymbol{\Theta}}|\boldsymbol{\theta}}\left(\tilde{\boldsymbol{\theta}}|\boldsymbol{\theta}\right) \, \mathrm{d}\tilde{\boldsymbol{\theta}} \tag{6.6}$$

$$= \int_{\mathbb{R}^d} \ell\left(\tilde{\boldsymbol{\theta}}\right) \; \pi_{\tilde{\boldsymbol{\Theta}}|\boldsymbol{\theta}}\left(\tilde{\boldsymbol{\theta}}|\boldsymbol{\theta}\right) \; \nabla_{\boldsymbol{\theta}} \ln\left(\pi_{\tilde{\boldsymbol{\Theta}}|\boldsymbol{\theta}}\left(\tilde{\boldsymbol{\theta}}|\boldsymbol{\theta}\right)\right) \, \mathrm{d}\tilde{\boldsymbol{\theta}} \tag{6.7}$$

$$= \mathbb{E}_{\tilde{\boldsymbol{\Theta}}|\boldsymbol{\theta}}\left\{\ell\left(\boldsymbol{\Theta}\right) \; \nabla_{\boldsymbol{\theta}} \ln\left(\pi_{\tilde{\boldsymbol{\Theta}}|\boldsymbol{\theta}}\left(\boldsymbol{\Theta}|\boldsymbol{\theta}\right)\right)\right\} , \tag{6.8}$$

where we exploit $\nabla_{\boldsymbol{x}} \ln(f(\boldsymbol{x})) = \frac{\nabla_{\boldsymbol{x}} f(\boldsymbol{x})}{f(\boldsymbol{x})}$. Furthermore, the proportionality of the PGT, see (6.3), can be replaced by an equality, since $\ell(\tilde{\boldsymbol{\theta}})$ is independent of $\boldsymbol{\theta}$.

We now calculate the gradient $\nabla_{\boldsymbol{\theta}} \ln \pi_{\tilde{\boldsymbol{\Theta}}|\tilde{\boldsymbol{\Theta}}} \left(\tilde{\boldsymbol{\theta}}|\boldsymbol{\theta}\right)$:

$$\begin{aligned}
\nabla_{\boldsymbol{\theta}} \ln \left(\pi_{\tilde{\boldsymbol{\Theta}}|\boldsymbol{\theta}} \left(\tilde{\boldsymbol{\theta}}\middle|\boldsymbol{\theta}\right)\right) &= \nabla_{\boldsymbol{\theta}} \ln \left(\frac{1}{(2\pi\sigma_\pi^2)^{\frac{d}{2}}} \exp\left(-\frac{\|\tilde{\boldsymbol{\theta}} - \boldsymbol{\theta}\|_2^2}{2\sigma_\pi^2}\right)\right) && (6.9)\\
&= \frac{(2\pi\sigma_\pi^2)^{\frac{d}{2}}}{\exp\left(-\frac{\|\tilde{\boldsymbol{\theta}}-\boldsymbol{\theta}\|_2^2}{2\sigma_\pi^2}\right)} \cdot \nabla_{\boldsymbol{\theta}} \left(\frac{1}{(2\pi\sigma_\pi^2)^{\frac{d}{2}}} \exp\left(-\frac{\|\tilde{\boldsymbol{\theta}} - \boldsymbol{\theta}\|_2^2}{2\sigma_\pi^2}\right)\right)\\
&= \nabla_{\boldsymbol{\theta}} \left(-\frac{\|\tilde{\boldsymbol{\theta}} - \boldsymbol{\theta}\|_2^2}{2\sigma_\pi^2}\right)\\
&= -\frac{1}{2\sigma_\pi^2} \nabla_{\boldsymbol{\theta}} \left(\left(\tilde{\boldsymbol{\theta}} - \boldsymbol{\theta}\right)^{\mathrm{T}} \left(\tilde{\boldsymbol{\theta}} - \boldsymbol{\theta}\right)\right)\\
&= \frac{1}{\sigma_\pi^2} \left(\tilde{\boldsymbol{\theta}} - \boldsymbol{\theta}\right) . && (6.10)
\end{aligned}$$

Consequently, we can rewrite the update step to

$$\boldsymbol{\theta} \leftarrow \boldsymbol{\theta} - \nu \frac{1}{\sigma_\pi^2} \cdot \mathbb{E}_{\tilde{\boldsymbol{\Theta}}|\boldsymbol{\theta}} \left\{\ell\left(\tilde{\boldsymbol{\Theta}}\right) \cdot \left(\tilde{\boldsymbol{\Theta}} - \boldsymbol{\theta}\right)\right\} . \tag{6.11}$$

The update is quite intuitive. For each $\tilde{\boldsymbol{\theta}}$, the difference $\tilde{\boldsymbol{\theta}} - \boldsymbol{\theta}$, which is a $d$-dimensional vector indicating the direction from $\boldsymbol{\theta}$ to $\tilde{\boldsymbol{\theta}}$, is weighted by $\ell(\tilde{\boldsymbol{\theta}})$. Hence, if $\ell(\tilde{\boldsymbol{\theta}})$ is large, the respective update $\tilde{\boldsymbol{\theta}} - \boldsymbol{\theta}$ will significantly contribute to the update vector, moving $\boldsymbol{\theta}$ towards $\tilde{\boldsymbol{\theta}}$. In contrast, if $\ell(\tilde{\boldsymbol{\theta}})$ is small, the scaled vector will not impact the update vector.

## 6.2 Modifications

If an analytical expression for $\ell(\cdot)$ is unavailable, the expectation in (6.11) cannot be computed directly. However, it can be approximated by leveraging the law of large numbers. Therefore, we draw $|\mathcal{B}_\mathrm{t}|$ samples from $\pi_{\tilde{\boldsymbol{\Theta}}|\boldsymbol{\theta}}(\tilde{\boldsymbol{\theta}}|\boldsymbol{\theta})$, which form the update batch $\mathcal{B}_\mathrm{t} = \{\tilde{\boldsymbol{\theta}}_1, \ldots, \tilde{\boldsymbol{\theta}}_{|\mathcal{B}_\mathrm{t}|}\}$. We now approximate the expectation by

$$\boldsymbol{\theta} \leftarrow \boldsymbol{\theta} - \nu \cdot \frac{1}{\sigma_\pi^2 |\mathcal{B}_\mathrm{t}|} \sum_{\tilde{\boldsymbol{\theta}} \in \mathcal{B}_\mathrm{t}} \ell\left(\tilde{\boldsymbol{\theta}}\right) \cdot \left(\tilde{\boldsymbol{\theta}} - \boldsymbol{\theta}\right) . \tag{6.12}$$

Furthermore, we absorb the constant factor $\frac{1}{\sigma_\pi^2 |\mathcal{B}_\mathrm{t}|}$ into the learning rate $\nu$

$$\boldsymbol{\theta} \leftarrow \boldsymbol{\theta} - \nu \sum_{\tilde{\boldsymbol{\theta}} \in \mathcal{B}_\mathrm{t}} \ell\left(\tilde{\boldsymbol{\theta}}\right) \cdot \left(\tilde{\boldsymbol{\theta}} - \boldsymbol{\theta}\right) . \tag{6.13}$$

We now test the update algorithm using the Rosenbruck function [Ros60], which is a simple function $f : \mathbb{R}^2 \rightarrow \mathbb{R}$ with two parameters $\theta_1$ and $\theta_2$, and is commonly used for testing optimization algorithms. It is defined by

$$f(\theta_1, \theta_2) = 100\left(\theta_2 - \theta_1^2\right)^2 + (1 - \theta_1)^2 , \tag{6.14}$$

and has a minimum value at $\theta_1 = \theta_2 = 1$. The Rosenbruck function is a non-convex function, where the global minimum is inside a long flat valley with steep slopes. We define our parameter vector as $\boldsymbol{\theta} = (\theta_1, \theta_2)$, and the performance measure as $\ell(\boldsymbol{\theta}) := f(\theta_1, \theta_2)$.

When updating $\boldsymbol{\theta}$ according to (6.13), we observe that the optimization algorithm updates $\boldsymbol{\theta}$ in a direction that actually increases $f(\theta_1, \theta_2)$, thereby moving it away from the global minimum. This is not surprising, since a small $\ell(\tilde{\boldsymbol{\theta}})$ (and thus well performing $\tilde{\boldsymbol{\theta}}$) has only a minor effect on the update. Far worse, a large $\ell(\tilde{\boldsymbol{\theta}})$ (and hence poor performing $\tilde{\boldsymbol{\theta}}$) strongly affects the update, however, in the opposite direction of $\tilde{\boldsymbol{\theta}} - \boldsymbol{\theta}$. The outcome of the resulting update becomes unpredictable.

To remedy this issue, we add $\ell(\boldsymbol{\theta})$ as a baseline to the update step:

$$\boldsymbol{\theta} \leftarrow \boldsymbol{\theta} - \nu \sum_{\tilde{\boldsymbol{\theta}} \in \mathcal{B}_\mathrm{t}} \left(\ell\left(\tilde{\boldsymbol{\theta}}\right) - \ell\left(\boldsymbol{\theta}\right)\right) \cdot \left(\tilde{\boldsymbol{\theta}} - \boldsymbol{\theta}\right) . \tag{6.15}$$

Three different cases exist: The first is $\ell(\tilde{\boldsymbol{\theta}}) - \ell(\boldsymbol{\theta}) < 0$, which is equal to $\ell(\tilde{\boldsymbol{\theta}}) < \ell(\boldsymbol{\theta})$. Thus, $\tilde{\boldsymbol{\theta}}$ achieves better performance than $\boldsymbol{\theta}$, and we want to update $\boldsymbol{\theta}$ in the direction of $\tilde{\boldsymbol{\theta}}$. Since we use gradient descent and $\ell(\tilde{\boldsymbol{\theta}}) - \ell(\boldsymbol{\theta}) < 0$, $\boldsymbol{\theta}$ is updated in the direction of $\tilde{\boldsymbol{\theta}}$. The second case is $\ell(\tilde{\boldsymbol{\theta}}) - \ell(\boldsymbol{\theta}) > 0$, which is equal to $\ell(\tilde{\boldsymbol{\theta}}) > \ell(\boldsymbol{\theta})$. Thus, $\tilde{\boldsymbol{\theta}}$ achieves worse performance than $\boldsymbol{\theta}$, and we do not want to update $\boldsymbol{\theta}$ in the direction of $\tilde{\boldsymbol{\theta}}$. Since we use gradient descent and $\ell(\tilde{\boldsymbol{\theta}}) - \ell(\boldsymbol{\theta}) > 0$, the parameter vector $\boldsymbol{\theta}$ is updated in the opposite direction of $\tilde{\boldsymbol{\theta}}$. Note that we update in a direction that we have not explored, hoping to find a better solution there. The third case is $\ell(\tilde{\boldsymbol{\theta}}) - \ell(\boldsymbol{\theta}) \approx 0$, where $\tilde{\boldsymbol{\theta}}$ does not contribute to the update. As long as all $\tilde{\boldsymbol{\theta}}$ used within an update step experience the same baseline $\ell(\boldsymbol{\theta})$, the baseline acts as a constant factor, and the derivation of the update algorithm remains valid [SB18, p. 329].

We furthermore observed that during optimization, $\ell(\tilde{\boldsymbol{\theta}}) - \ell(\boldsymbol{\theta})$ can become small, and the optimization gets stuck. In contrast, at the beginning of the optimization, $\ell(\tilde{\boldsymbol{\theta}}) - \ell(\boldsymbol{\theta})$ may be extremely large. To avoid both extreme cases, we introduce a normalization by $\ell(\boldsymbol{\theta})$, and further modify the update step to

$$\boldsymbol{\theta} \leftarrow \boldsymbol{\theta} - \nu \sum_{\tilde{\boldsymbol{\theta}} \in \mathcal{B}_\mathrm{t}} \frac{\ell\left(\tilde{\boldsymbol{\theta}}\right) - \ell\left(\boldsymbol{\theta}\right)}{\ell\left(\boldsymbol{\theta}\right)} \cdot \left(\tilde{\boldsymbol{\theta}} - \boldsymbol{\theta}\right) . \tag{6.16}$$

As before, the normalization can be treated as a constant; hence, the derivation of the update algorithm remains valid.

Fig. 6.1 displays different variations of the update algorithm when optimizing the Rosenbruck function with a batch size of $|\mathcal{B}_\mathrm{t}| = 4$, a policy variance of $\sigma_\pi^2 = 5 \cdot 10^{-4}$, and a learning rate of $\nu = 0.5$. All optimization approaches share the starting point $(\theta_1, \theta_2) = (1.8, 1.9)$, which is indicated by a gray cross marker. The global minimum at $(\theta_1, \theta_2) = (1, 1)$ is indicated by a black cross marker. While the left column displays $\boldsymbol{\theta}$ with progressing optimization, the right column displays the achieved function value $f(\theta_1, \theta_2)$.

Fig. 6.1(a) shows the optimization when updating following (6.15), thus no normalization is used. As the training progresses, $f(\theta_1, \theta_2)$ decreases. However, with decreasing $f(\theta_1, \theta_2)$, the amplitude of $(\ell(\tilde{\boldsymbol{\theta}}) - \ell(\boldsymbol{\theta}))$ also decreases, resulting in vanishing update steps. In Fig. 6.1(b), the update is carried out following (6.16), thus with the normalization of $(\ell(\tilde{\boldsymbol{\theta}}) - \ell(\boldsymbol{\theta}))$. The normalization is able to combat the vanishing update steps, see the first 25 000 epochs. Afterwards, the update oscillates along the valley around the global minimum.

To combat the oscillations, we can either decrease the learning rate $\nu$ or the variance $\sigma_\pi^2$ of the policy. When decreasing $\sigma_\pi^2$, the search radius shrinks, which is preferable as we get closer to the minimum. Hence, we couple the variance of our policy to the current state of our optimization by $\sigma_{\pi,\mathrm{c}}^2 := \sigma_\pi^2 \cdot \max(\ell(\boldsymbol{\theta})), 1)$. For values $\ell(\boldsymbol{\theta}) > 1$, the max-operation avoids an exploding variance. We modify the Gaussian policy $\pi_{\tilde{\boldsymbol{\Theta}}|\boldsymbol{\theta}}(\tilde{\boldsymbol{\theta}}|\boldsymbol{\theta})$ to

$$\pi_{\tilde{\boldsymbol{\Theta}}|\boldsymbol{\theta}}\left(\tilde{\boldsymbol{\theta}}|\boldsymbol{\theta}\right) = \frac{1}{(2\pi\sigma_{\pi,\mathrm{c}}^2)^{\frac{d}{2}}} \cdot \exp\left(-\frac{||\tilde{\boldsymbol{\theta}} - \boldsymbol{\theta}||_2^2}{2\sigma_{\pi,\mathrm{c}}^2}\right). \qquad (6.17)$$

Fig. 6.1(c) displays the optimization when updating according to (6.15) and the modified policy (6.17). The performance measure decreases continuously to very small values. However, after 100 000 epochs, the algorithm still has not converged.

To decrease the number of epochs, we can increase $\sigma_\pi^2$. This results in a larger search region at the beginning, increasing the update steps initially and hence the convergence speed. In Fig. 6.2, we investigate the impact of an increased $\sigma_\pi^2$. Note that the number of epochs is reduced by a factor of 10, hence to $10,000$ epochs. Still, $\nu = 0.5$ and $|\mathcal{B}_\mathrm{t}| = 4$ is used.

Fig. 6.2(a) displays the results for $\sigma_\pi^2 = 10^{-3}$. Initially, $f(\theta_1, \theta_2)$ declines rapidly. After roughly 1000 epochs, the performance starts oscillating. When approaching the valley of the Rosenbruck function, $\boldsymbol{\theta}$ oscillates between the steep slopes of the valley. Due to the very steep slopes, the contribution of variations $\tilde{\boldsymbol{\theta}}$ with $\ell(\tilde{\boldsymbol{\theta}}) - \ell(\boldsymbol{\theta}) > 0$ (variations that achieve a worse performance measure) to the overall update has a much larger amplitude than the contribution of variations $\tilde{\boldsymbol{\theta}}$ with $\ell(\tilde{\boldsymbol{\theta}}) - \ell(\boldsymbol{\theta}) < 0$ (variations that achieve a better performance). Hence, instead of moving towards the global minimum, $\boldsymbol{\theta}$ oscillates perpendicular to

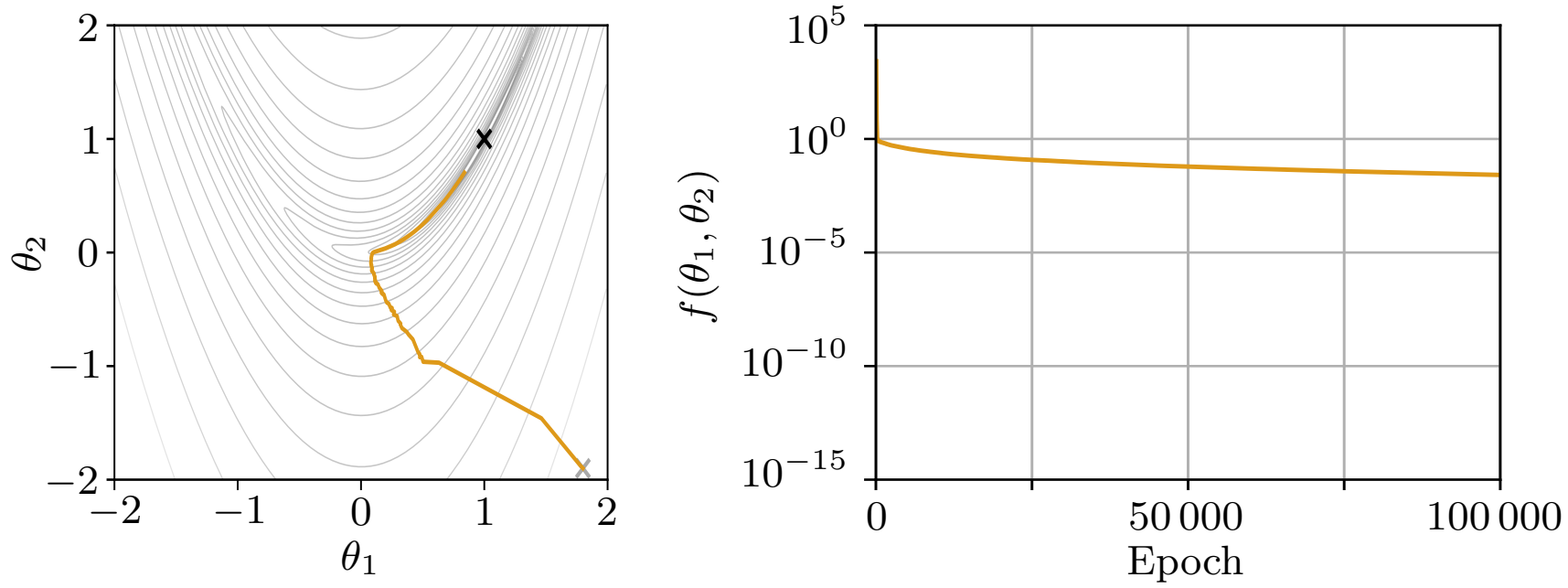

**(a)** Optimization progress when following (6.15).

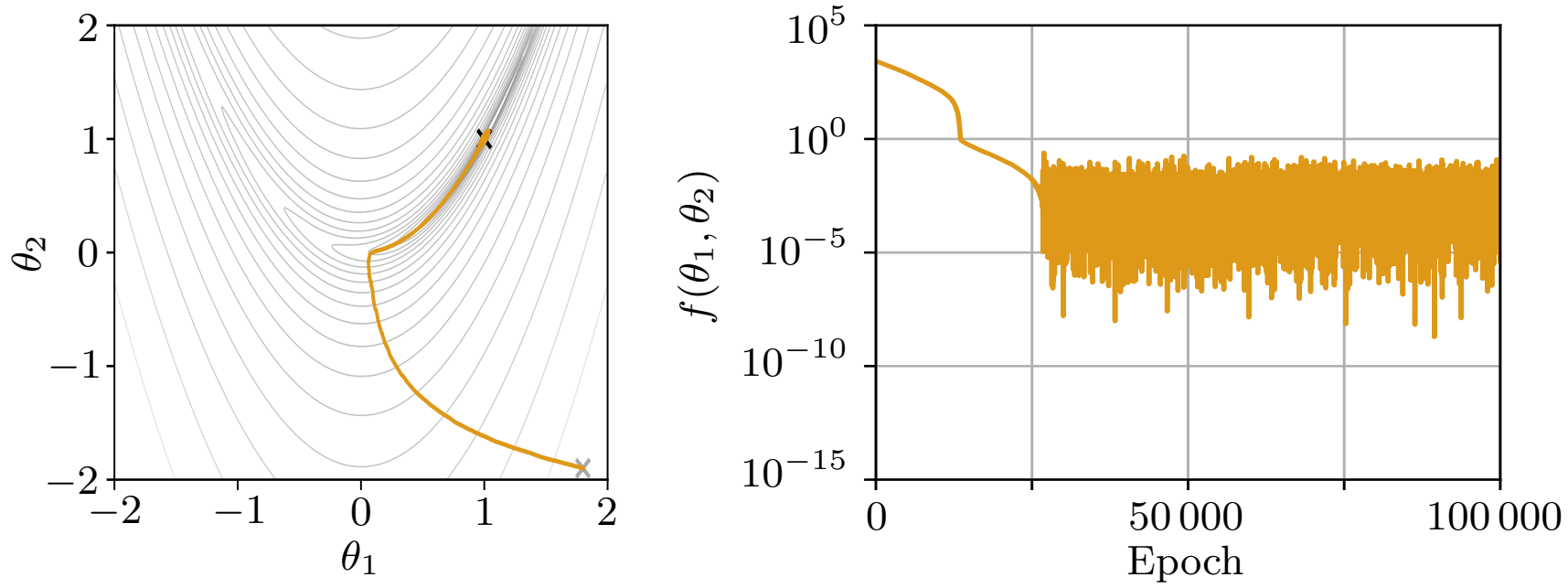

**(b)** Optimization progress when following (6.16).

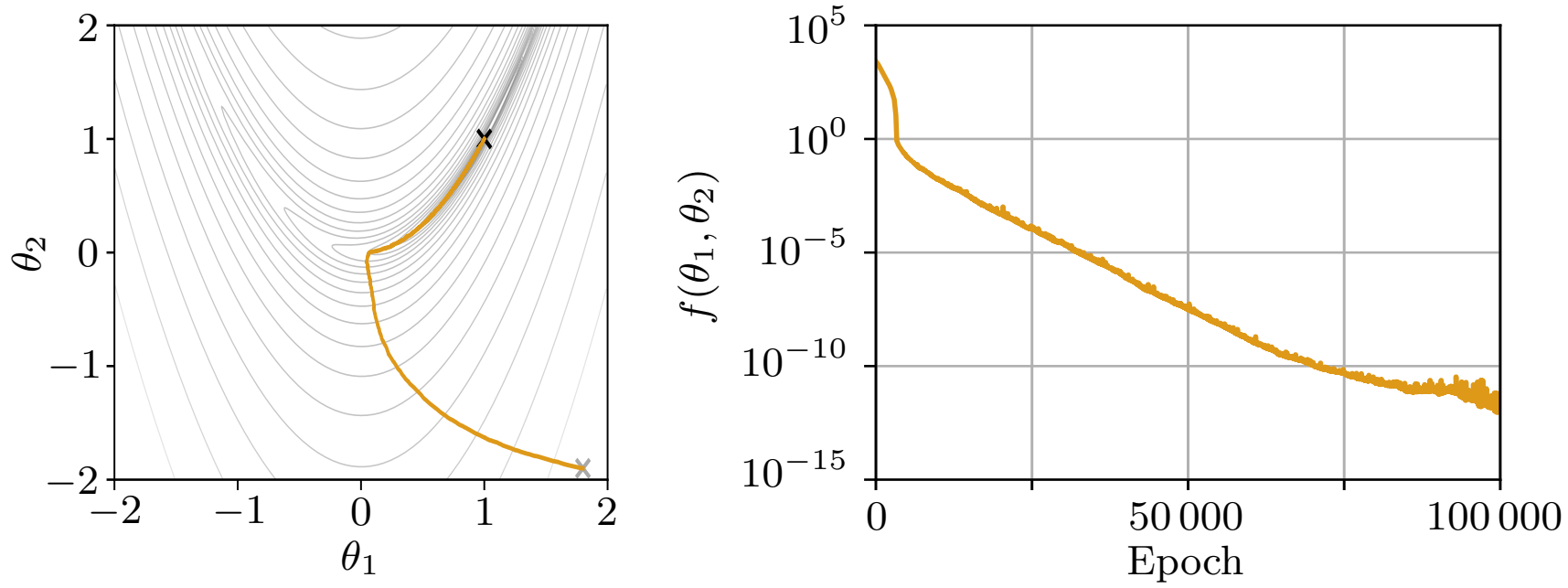

**(c)** Optimization progress when following (6.16) and modified policy (6.17).

**Figure 6.1:** Optimization progress on the Rosenbruck function. While the starting point is indicated by a gray marker, the global minimum is indicated by a black marker.

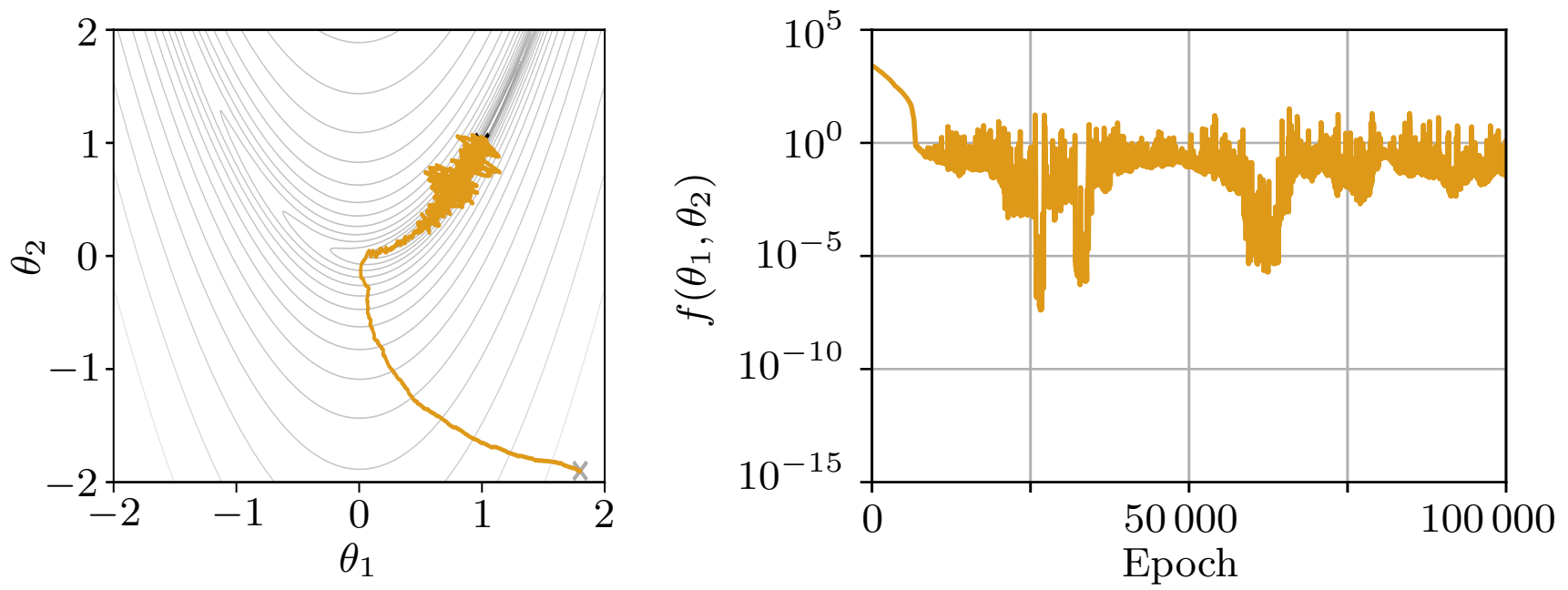

**(a)** Optimization progress when following (6.16) with $\sigma_\pi^2 = 10^{-1}$.

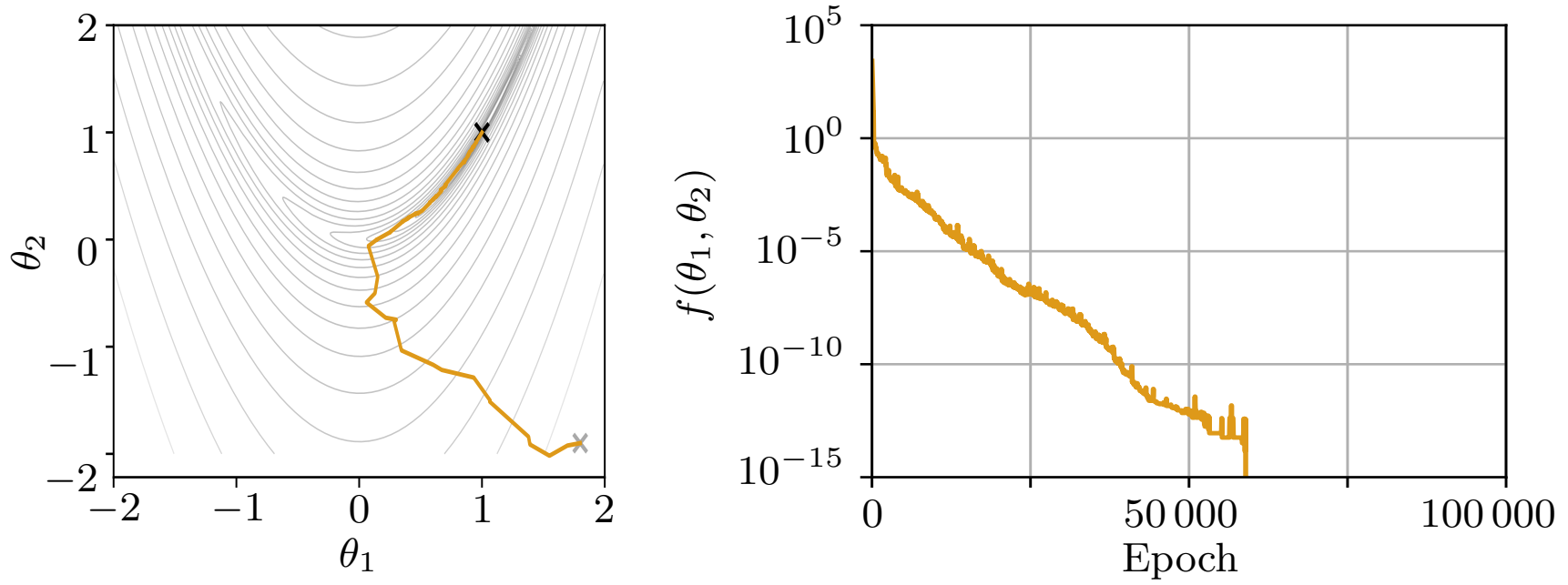

**(b)** Optimization progress when following (6.18) with $\sigma_\pi^2 = 10^{-3}$ and $\alpha_{\text{damp}} = 10^{-4}$.

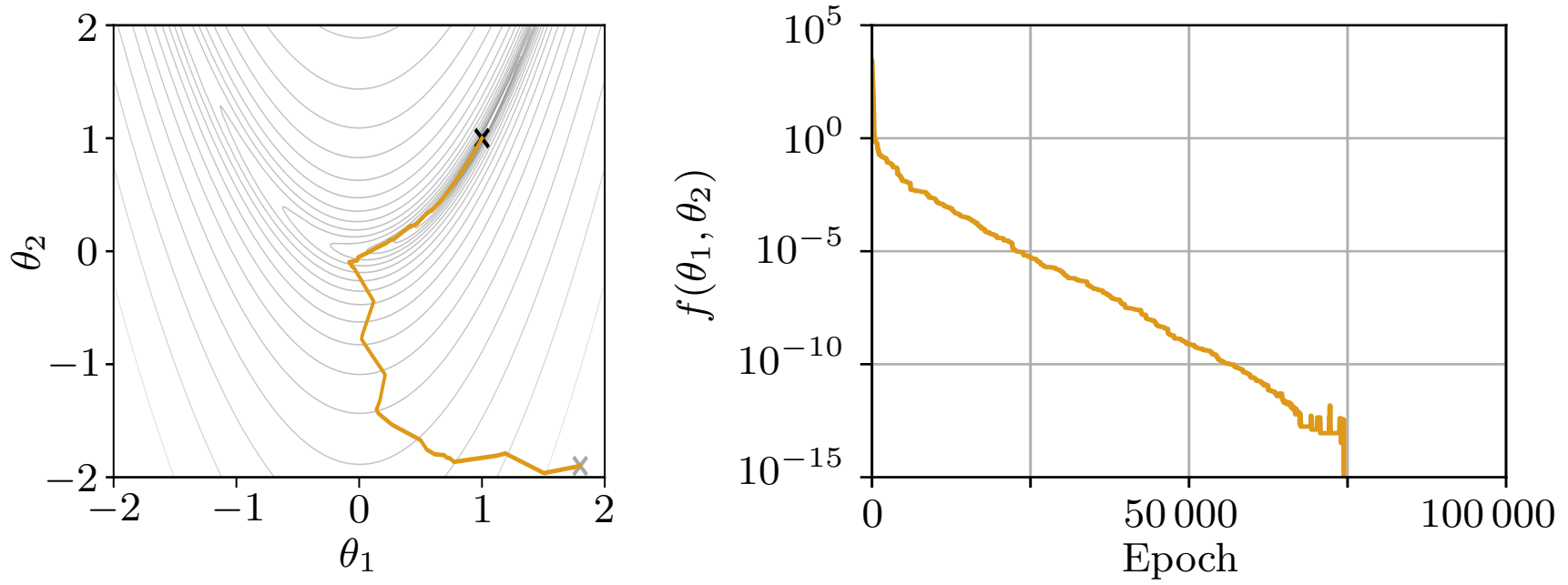

**(c)** Optimization progress when following (6.18) with $\sigma_\pi^2 = 10^{-3}$ and $\alpha_{\text{damp}} = 10^{-6}$.

**Figure 6.2:** Optimization progress on the Rosenbruck function. While the starting point is indicated by a gray marker, the global minimum is indicated by a black marker.

the valley. To avoid these oscillations, we propose to dampen positive values of $\ell(\tilde{\boldsymbol{\theta}}) - \ell(\boldsymbol{\theta})$. We redefine the update rule to

$$\boldsymbol{\theta} \leftarrow \boldsymbol{\theta} - \nu \sum_{\tilde{\boldsymbol{\theta}} \in \mathcal{B}_\mathrm{t}} \frac{\ell_\mathrm{diff}(\boldsymbol{\theta}, \tilde{\boldsymbol{\theta}})}{\ell(\boldsymbol{\theta})} \cdot \left(\tilde{\boldsymbol{\theta}} - \boldsymbol{\theta}\right) , \tag{6.18}$$

where $\ell_\mathrm{diff} : \mathbb{R}^{2d} \to \mathbb{R}$ is given by

$$\ell_\mathrm{diff}(\boldsymbol{\theta}, \tilde{\boldsymbol{\theta}}) = \begin{cases} \ell\left(\tilde{\boldsymbol{\theta}}\right) - \ell(\boldsymbol{\theta}) , & \text{if } \ell\left(\tilde{\boldsymbol{\theta}}\right) - \ell(\boldsymbol{\theta}) > 0 , \\ \left(\ell\left(\tilde{\boldsymbol{\theta}}\right) - \ell(\boldsymbol{\theta})\right) \cdot \alpha_\mathrm{damp} , & \text{else} , \end{cases} \tag{6.19}$$

with $\alpha_\mathrm{damp} \in [0, 1]$ as the damping factor.

Fig. 6.2(b) displays the optimization when updating according to (6.18) with $\sigma_\pi^2 = 0.1$ and $\alpha_\mathrm{damp} = 10^{-4}$. The further increase of $\sigma_\pi^2$ leads to very fast movement at the beginning of the optimization. Using the damping factor $\alpha_\mathrm{damp}$, the oscillations perpendicular to the valley are avoided. After approximately 6000 epochs, the performance measure drops to a value below $f(\theta_1, \theta_2) < 10^{-15}$, indicating that the optimization is very close to the global minimum.

Note the effects of the initialization of $\sigma_\pi^2$ and $\alpha_\mathrm{damp}$. When increasing $\sigma_\pi^2$, the optimization accelerates, potentially resulting in oscillations if the minimum is inside a long valley with steep walls. Damping positive values of $\ell_\mathrm{diff}(\boldsymbol{\theta}, \tilde{\boldsymbol{\theta}})$ can avoid the oscillations. However, $\alpha_\mathrm{damp}$ needs to be chosen carefully: If $\alpha_\mathrm{damp}$ is chosen too large, its effect diminishes, and oscillations occur. If $\alpha_\mathrm{damp}$ is chosen too small, the contribution of variations with $\ell_\mathrm{diff}(\boldsymbol{\theta}, \tilde{\boldsymbol{\theta}}) < 0$ is neglected, slowing down the optimization. If the function we want to optimize exhibits plateaus or local minima, choosing $\alpha_\mathrm{damp}$ too small can result in the algorithm getting stuck in local minima. Fig. 6.2(c) shows the optimization for $\sigma_\pi^2 = 10^{-1}$ and $\alpha_\mathrm{damp} = 10^{-6}$. Compared to Fig. 6.2(b), the optimization requires more epochs to converge.

So far, we have used $|\mathcal{B}_\mathrm{t}| = 4$ variations to explore the vicinity of $\boldsymbol{\theta}$. When increasing $|\mathcal{B}_\mathrm{t}|$, we expect a more detailed exploration and, hence, faster convergence. Fig. 6.3 shows the optimization for $|\mathcal{B}_\mathrm{t}| = 40$, $\sigma_\pi^2 = 10^{-1}$, $\alpha_\mathrm{damp} = 10^{-4}$, and $\nu = 0.5$. The number of epochs is further reduced to 2 000. Five different starting points are chosen, one of which is the previously used point; see the orange line. The remaining four are chosen randomly.

When comparing Fig. 6.3 and Fig. 6.2(b), whose setups differ only in $|\mathcal{B}_\mathrm{t}|$, increasing $|\mathcal{B}_\mathrm{t}|$ massively speeds up convergence. Furthermore, all curves in Fig. 6.3 converge close to the global minimum.

We conclude that the modifications we applied to the PGU stabilize the algorithm and speed up training. The final PGU algorithm is given by Alg. 3.

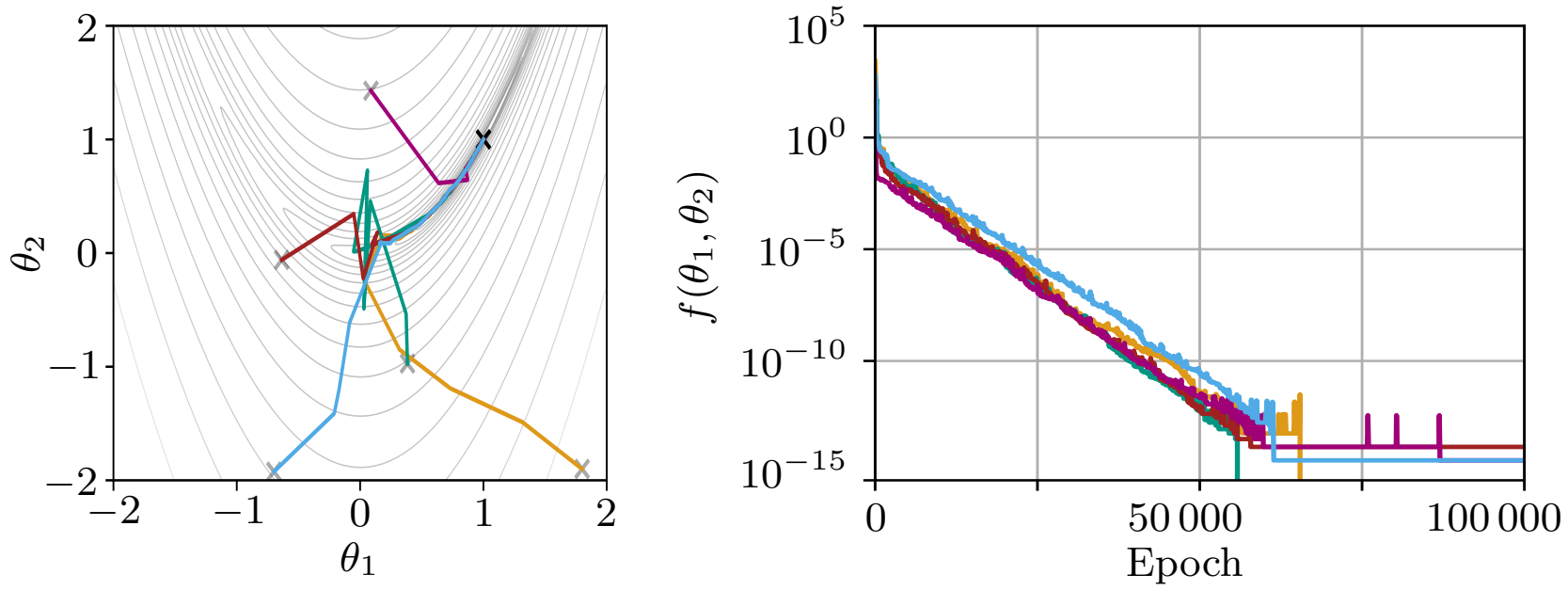

**Figure 6.3:** Optimization progress when following (6.16) with $|\mathcal{B}_\mathrm{t}| = 40$, $\nu = 0.5$, $\sigma_\pi^2 = 10^{-1}$, and $\alpha_\mathrm{damp} = 10^{-4}$.

## 6.3 SNN-based Detector for the AWGN Channel

We can apply the PGU as described by Alg. 3 to optimize the parameters of an SNN. Based on the problem statement of Ch. 3, we implement an SNN-based detector for the AWGN-channel and a 16-QAM. Its setup is given in Fig. 3.2. As in Sec. 3.4, we optimize an SNN with a single hidden layer with $N_\mathrm{hid} = 16$ hidden neurons. As neural encoding, we use QE with $N_\mathrm{enc} = 2$, $y_\mathrm{enc,max} = 4c$, and $K_\mathrm{enc} = 5$ time steps. This setup results in $N_\mathrm{hid} \cdot (2N_\mathrm{enc} + N_\mathrm{out}) = 320$ adaptable weights. Thus, the dimension of the parameter vector is given by $\boldsymbol{\theta} \in \mathbb{R}^{320}$. For the given setup, the SNN-based detector updated with BPTT and SGs achieves near-ML performance, see Fig. 3.13.

As performance measure $\ell(\cdot)$, we choose the CE between the true label and the values returned by EOTM decoding. To obtain probability distributions, the output of the EOTM is passed through the softmax function, while the true label is represented using one-hot encoding. To obtain a statistically significant value of the CE, for each $\boldsymbol{\theta}$ (and $\tilde{\boldsymbol{\theta}}$), $|\mathcal{B}_\mathrm{pol}| = 10\,000$ samples disturbed by AWGN are processed. Thus, we introduce a second batch size $|\mathcal{B}_\mathrm{pol}|$ that defines how many samples are used to evaluate a single parameter vector $\boldsymbol{\theta}$ (and $\tilde{\boldsymbol{\theta}}$). During an update step, $\ell\left(\boldsymbol{\theta}\right)$ and all $\ell\left(\tilde{\boldsymbol{\theta}}\right)$ are obtained using the same 10 000 samples. After the update, new samples are created. In Alg. 3, new samples are created at the beginning of the for-loop for the evaluation of the CE. This ensures that all $\ell(\cdot)$ contributing to an update step are evaluated using the same samples.

We furthermore choose $\sigma_\pi^2 = 0.1$, $\nu = 1$, $|\mathcal{B}_\mathrm{t}| = 50$, and $\alpha_\mathrm{damp} = 10^{-1}$. The optimization is carried out at an $E_\mathrm{b}/N_0 = 10\,\mathrm{dB}$ for 1 000 epochs, and the initial parameter vector $\boldsymbol{\theta}$ is drawn from a normal distribution. Fig. 6.4(a) shows the results, where the left plot displays the BER, and the right plot displays the CE

**Algorithm 3** PGU algorithm.

**Require:** Performance measure $\ell(\cdot)$, policy variance $\sigma_\pi^2$, learning rate $\nu$, batch size $|\mathcal{B}_\mathrm{t}|$, damping factor $\alpha_\mathrm{damp}$

Randomly initialize $\boldsymbol{\theta} \in \mathbb{R}^d$
**for** $i = 1, \ldots, N_\mathrm{epochs}$ **do**
  Acquire $\ell(\boldsymbol{\theta})$
  Update $\sigma_{\pi,\mathrm{c}}^2 \coloneqq \sigma_\pi^2 \cdot \max\left(\ell\left(\boldsymbol{\theta}\right), 1\right)$
  Draw $|\mathcal{B}_\mathrm{t}|$ samples $\tilde{\boldsymbol{\theta}}$, $b = 1, \ldots, |\mathcal{B}_\mathrm{t}|$ from Gaussian policy $\pi_{\tilde{\boldsymbol{\Theta}}|\boldsymbol{\theta}}(\tilde{\boldsymbol{\theta}}|\boldsymbol{\theta})$ with

$$\pi_{\tilde{\boldsymbol{\Theta}}|\boldsymbol{\theta}}\left(\tilde{\boldsymbol{\theta}}|\boldsymbol{\theta}\right) = \frac{1}{(2\pi\sigma_{\pi,\mathrm{c}}^2)^{\frac{d}{2}}} \cdot \exp\left(-\frac{||\tilde{\boldsymbol{\theta}} - \boldsymbol{\theta}||_2^2}{2\sigma_{\pi,\mathrm{c}}^2}\right)$$

  Obtain $\ell\left(\tilde{\boldsymbol{\theta}}\right)$ as performance of $\tilde{\boldsymbol{\theta}}$
  Calculate relative performance $\ell_\mathrm{diff}(\boldsymbol{\theta}, \tilde{\boldsymbol{\theta}})$ and dampen

$$\ell_\mathrm{diff}(\boldsymbol{\theta}, \tilde{\boldsymbol{\theta}}) = \begin{cases} \ell\left(\tilde{\boldsymbol{\theta}}\right) - \ell\left(\boldsymbol{\theta}\right), & \text{if } \ell\left(\tilde{\boldsymbol{\theta}}\right) - \ell\left(\boldsymbol{\theta}\right) < 0 \\ \left(\ell\left(\tilde{\boldsymbol{\theta}}\right) - \ell\left(\boldsymbol{\theta}\right)\right) \cdot \alpha_\mathrm{damp}, & \text{else} \end{cases}$$

  Update

$$\boldsymbol{\theta} \leftarrow \boldsymbol{\theta} - \nu \sum_{\tilde{\boldsymbol{\theta}} \in \mathcal{B}_\mathrm{t}} \frac{\ell_\mathrm{diff}(\boldsymbol{\theta}, \tilde{\boldsymbol{\theta}})}{\ell\left(\boldsymbol{\theta}\right)} \left(\tilde{\boldsymbol{\theta}} - \boldsymbol{\theta}\right)$$

  $i \leftarrow i + 1$
**end for**

and SER during optimization. Since the BER curve matches the BER curve of the ML-detector, we conclude that the optimization using the PGU is successful.

In Fig. 6.4(b), we furthermore apply the optimization with similar parameters, but instead of the CE the SER is used as performance measure $\ell(\cdot)$. Note that the SER constitutes a hard-decision metric, whereas the CE represents a soft metric. The SER during optimization consists of flat plateaus with a few steep declines. The values of the plateaus in terms of the SER are at multiples of $\frac{1}{16}$, which is obvious since each symbol class is either classified correctly or incorrectly. The exploration of the PGU is too limited to find parameters $\tilde{\boldsymbol{\theta}}$ that significantly reduce the SER, resulting in vanishing updates and the observed plateaus. For varying parameters $\sigma_\pi^2$, $\nu$, and $\alpha_\mathrm{damp}$, we always experienced the plateaus and did not achieve ML performance. We conclude that using the SER as a performance measure is more likely to experience flat plateaus during optimization, which may result in the optimization getting stuck.

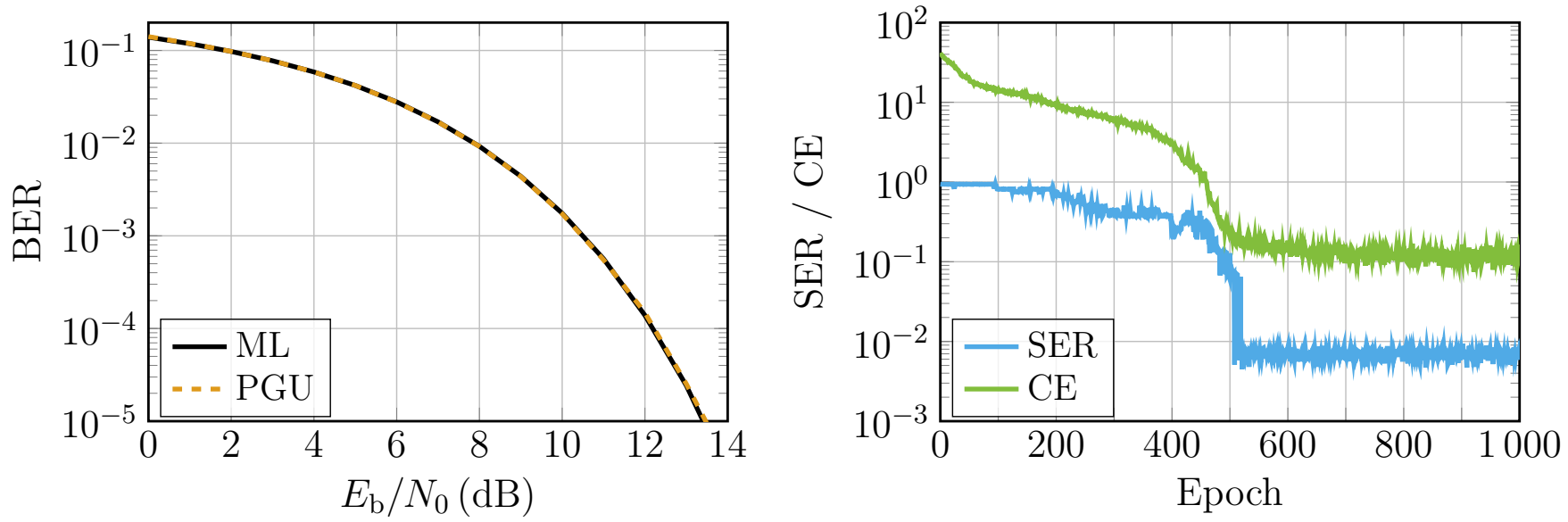

**(a)** CE as performance measure.

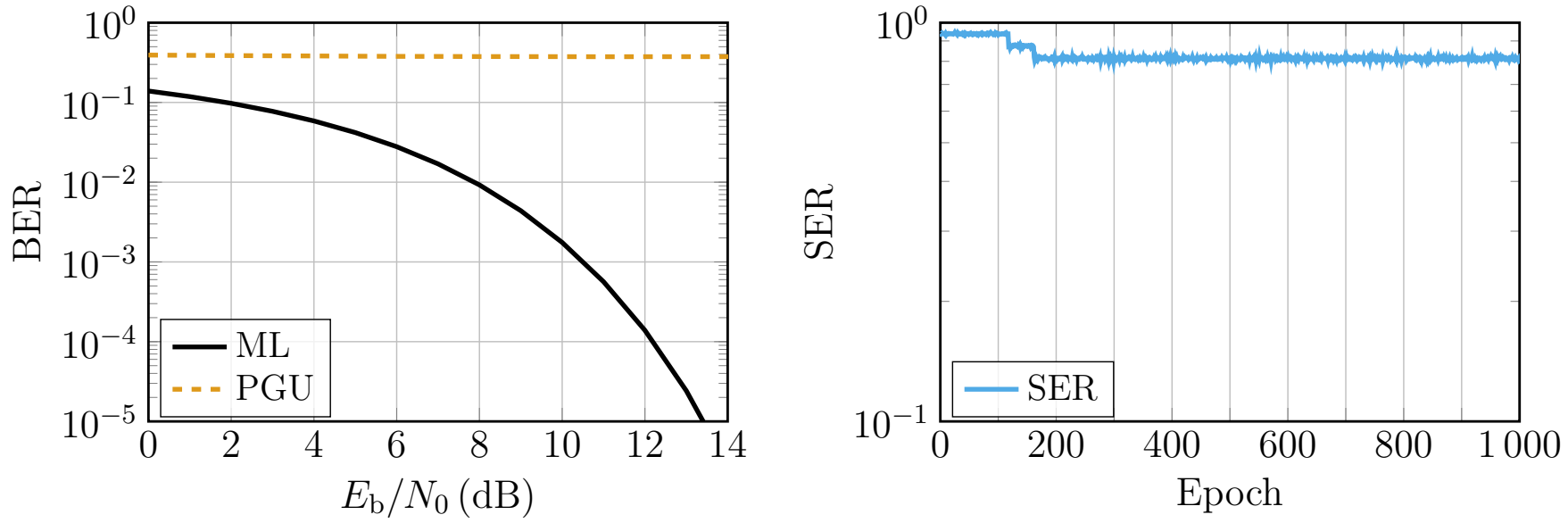

**(b)** SER as performance measure.

**Figure 6.4:** SNN-based detector updated using PGU. As performance measure $\ell(\cdot)$ the CE and SER is applied.

## 6.4 SNN-based Equalizer and Demapper

Next we investigate the behavior of PGU for a search space with a larger dimensionality $d \in \mathbb{N}$. In Ch. 4, the BPTT with SG delivered promising results for SNN-based equalizers and demappers. We now try to reproduce these results using the PGU as the update algorithm.

Remember that each PGU update step requires the simulation of $|\mathcal{B}_t|$ variations. Thus, to update the SNN-based equalizer and demapper, the simulation of $|\mathcal{B}_t|$ SNNs is required for each update step. The simulation of the different SNNs for an update step can be executed either in parallel or in a time-serial manner. When executing the SNNs in parallel, the memory of the processing unit, e.g., a GPU, constrains $|\mathcal{B}_t|$. When executing the SNNs in a time-serial manner, we observed that the runtime of the optimization rapidly grows with $|\mathcal{B}_t|$.

To accelerate the optimization of the SNN-based equalizer and demapper, we choose to update the NFE-SNN$_{\mathrm{QE,E}}$, which is the equalizer without decision feedback, QE as neural encoding, and EOTM as neural decoding. Compared to RFE, QE exhibits a lower number of discrete-time simulation steps $K$, which reduces the runtime of each SNN and hence of the optimization. The parameter vector $\boldsymbol{\theta} \in \mathbb{R}^d$ contains the weights of the linear layers connecting the input to the hidden layer, and the hidden to the output layer. Thus, the dimensionality of $\boldsymbol{\theta}$ is given by $d = N_{\mathrm{hid}} \cdot (N_{\mathrm{in}} + N_{\mathrm{out}}) = N_{\mathrm{hid}} \cdot (n_{\mathrm{tap}} \cdot N_{\mathrm{enc}} + N_{\mathrm{out}}) = 2400$, see Tab. 4.4 and Tab. 4.6.

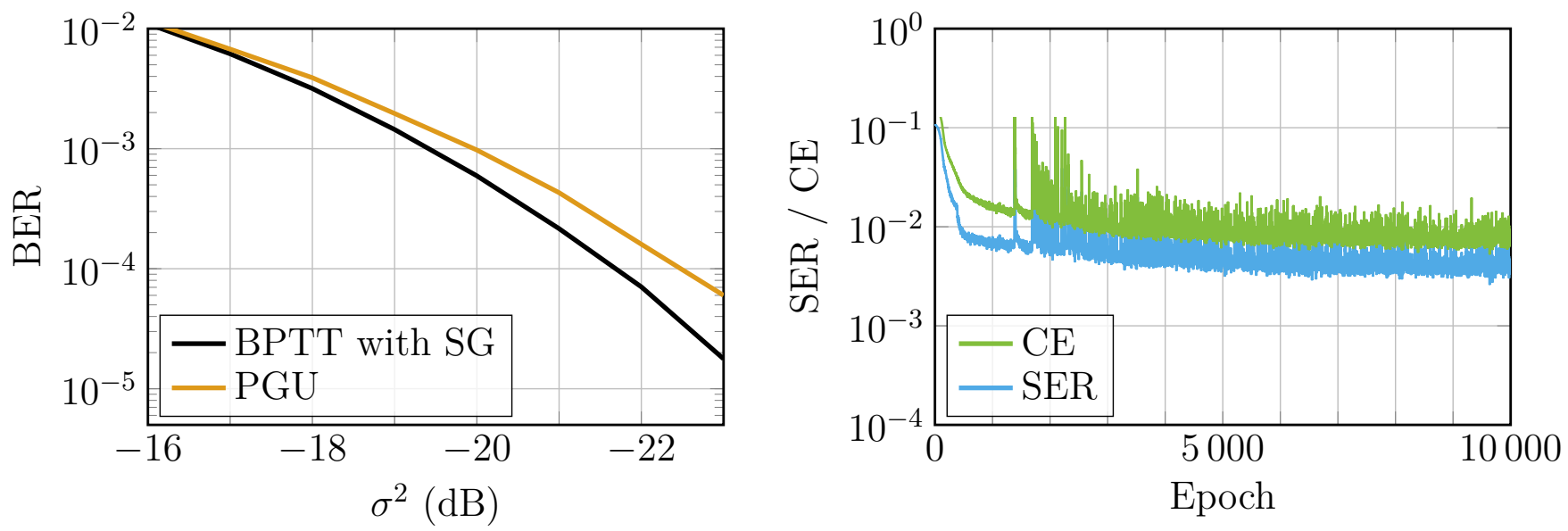

**Figure 6.5:** Performance of the NFE-SNN$_{\mathrm{QE,E}}$ when updated using BPTT with SG, and PGU.

To update the NFE-SNN$_{\mathrm{QE,E}}$ using PGU, we choose $\sigma_\pi^2 = 10^{-3}$, $\nu = 0.25$, $|\mathcal{B}_{\mathrm{t}}| = 2\,000$, $|\mathcal{B}_{\mathrm{pol}}| = 20\,000$, and $\alpha_{\mathrm{damp}} = 10^{-1}$. The optimization is carried out at $\sigma_{\mathrm{n}}^2 = -20\,\mathrm{dB}$ using 10 000 epochs and the CE as the performance measure $\ell(\cdot)$. The initial parameter vector $\boldsymbol{\theta}$ is drawn from a normal distribution. Fig. 6.5 displays the CE and SER during optimization, as well as the results when comparing the PGU with the update carried out using BPTT with SG. The NFE-SNN$_{\mathrm{QE,E}}$ updated using the PGU achieves a performance worse than the performance of the NFE-SNN$_{\mathrm{QE,E}}$ updated with BPTT with SG. When investigating the course of the CE during optimization, we notice heavy fluctuations. At the start of the optimization, the CE declines rapidly; afterwards, fluctuations occur. Unfortunately, the source of the fluctuations could not be identified. When decreasing $\sigma_\pi^2$, the fluctuations diminish; however, the optimization process slows down significantly, making it impractical.

At this point, we would like to emphasize once again that the PGU is a trial-and-error method. With growing network size, the dimensionality $d \in \mathbb{N}$ of the parameter vector $\boldsymbol{\theta} \in \mathbb{R}^d$, and hence the search space, also grows. We failed to reproduce the results achieved using BPTT with SG using the NFE-SNN$_{\mathrm{QE,E}}$ updated with the PGU. Thus, we conclude that the PGU scales poorly for larger search spaces. Consequently, we suggest carrying out the updates of SNNs using BPTT with SG.

## 6.5 Optimization of Neural Encoding

Since the PGU does not require differentiable functions, we can exploit the PGU to optimize parts of a system that cannot be updated with the BP algorithm due to non-differentiability. In Ch. 4, several SNN-based equalizers and demappers were tested, some of them combined with RFE. The parameters of the RFE were taken from [ABS+23, AEvB+25], with $K = 60$ encoding time steps and $N_{\text{enc}} = 10$ encoding spike signals. In contrast, the SNN-based equalizers and demappers using QE and TE as neural encoding use $K = 5$ and $N_{\text{enc}} = 8$. Thus, using RFE, the implementation requires 12 times more time steps than implementations using QE and TE. Furthermore, when using RFE as neural encoding, the higher number $N_{\text{enc}}$ of encoding spike signals results in a larger input layer, and thus a larger model complexity.

We can reduce $N_{\text{enc}}$ or $K$, however, the field centers $\mu_j$, $j = 1, \dots, N_{\text{enc}}$ and the field width $\Delta_j$ of the RFE are chosen to fit the setup with $K = 60$ and $N_{\text{enc}} = 10$, and hence yield poor results. Consequently, the parameters $\mu_j$ and $\Delta_j$ need to be updated. Suitable parameters can be determined manually by brute-force testing of different values. A more elegant method is Bayesian optimization, which builds a probabilistic model based on previous results and selectively proposes promising new parameters. However, both methods require training and testing multiple realizations and are therefore very time-consuming.

Instead, we can apply the PGU to update the encoding parameters $\mu_j$ and $\Delta_j$ [EvBS25]. First, we fix $N_{\text{enc}}$ and $K$ and initialize $\mu_j$ and $\Delta_j$ with reasonable starting points. Based on the findings of Sec. 6.4, we propose to update the SNN using BPTT with SG, and to update the parameters of the RFE using PGU. Hence, we alternately update the parameters of the SNN and the parameters of the encoding. As a result, only a single SNN-based equalizer and demapper needs to be trained.

In the following, we revisit the problem statement from Ch. 4. For the LCD link, see Sec. 4.5.1, we implement the NFE-$\text{SNN}_{\text{RFE,E}}$ with and without recurrent connections, and with and without regularization. Using the PGU, we opt to reduce $K$ and $N_{\text{enc}}$ of the RFE, while preserving performance.

### 6.5.1 Parameters

The parameters of the SNN are identical to those employed in Tab. 4.5 and Tab. 4.6. For the RFE, we initialize the $j$th field center $\mu_j$ by $\mu_j = j \cdot \frac{y_{\text{enc,max}}}{N_{\text{enc}}}$, with $j = 1, \dots, N_{\text{enc}}$, and the respective field width by $\sigma_j = 2$. Note that in Ch. 4, all fields shared the same width, i.e., $\sigma_j = \sigma$. We now assign each receptive field its own field width $\sigma_j$, which can be updated independently. Consequently, the parameter vector $\boldsymbol{\theta}$ to be optimized using PGU is defined by $\boldsymbol{\theta} = (\mu_1, \dots, \mu_{N_{\text{enc}}}, \sigma_1, \dots, \sigma_{N_{\text{enc}}})$, with $\boldsymbol{\theta} \in \mathbb{R}^{2N_{\text{enc}}}$.

For optimization, we alternately update the parameters $\boldsymbol{\theta}$ of the neural encoding using PGU, and the parameters of the SNN using BPTT with SG. Overall, we update using 20 000 epochs. For the first 10 000 epochs, for each epoch, first the RFE and then the SNN are updated once. Afterwards, the parameters $\boldsymbol{\theta}$ of the RFE are fixed, and solely the SNN is further updated. The PGU is executed using a batch size of $|\mathcal{B}_\mathrm{t}| = 20$ variations, a policy variance of $\sigma_\pi^2 = 10^{-4}$, a learning rate of $\nu = 10^{-3}$, and $|\mathcal{B}_\mathrm{pol}| = 10^5$ samples per evaluated parameter vector $\boldsymbol{\theta}$ (and $\tilde{\boldsymbol{\theta}}$). As performance measure $\ell(\cdot)$, the CE loss is used. The SNN is updated using the same parameters as given in Ch. 4: a batch size of $|\mathcal{B}_\mathrm{t}|_\mathrm{BPTT} = 2 \cdot 10^4$ and a learning rate of $\nu_\mathrm{BPTT} = 10^{-3}$. The optimization is carried out at a fixed noise power $\sigma_\mathrm{n}^2 = -20\,\mathrm{dB}$.

In addition to the BER and the average number $Z_\mathrm{avg}$ of hidden-layer spikes, we introduce the number of required MAC operations per processed sample, denoted by $\#\mathrm{MAC} \in \mathbb{N}_0$, as a third metric. It provides a measure of the computational complexity required for implementing SNNs in digital hardware (compare (2.5)) and is obtained by $\#\mathrm{MAC} = N_\theta \cdot K$ [MNHW24], where we defined $N_\theta$ as the number of parameters of the linear layers, see (4.9) and (4.10).

### 6.5.2 Results

We optimize and compare four different approaches for $N_\mathrm{enc} = \{4, 6, 8, 10\}$ and $K = \{4, 6, 8, 10, 60\}$: NFE-SNN$_\mathrm{RFE,E}$, NFE-SNN$_\mathrm{RFE,E,R}$, NFE-SNN$_\mathrm{RFE,E,rec}$, and NFE-SNN$_\mathrm{RFE,ER,rec}$. In Fig. 6.6, we compare the BER, $Z_\mathrm{avg}$, and #MAC at a noise power of $\sigma_\mathrm{n}^2 = -20\,\mathrm{dB}$. In general, with decreasing $K$ and $N_\mathrm{enc}$ the performance of all approaches degrades. The level of degradation heavily depends on the chosen combination of $N_\mathrm{enc}$ and $K$, and the variation of the NFE-SNN. When comparing all approaches using $K = 60$, they perform similarly, with the NFE-SNN$_\mathrm{RFE,E,rec}$ performing best. Nonetheless, this comes at the cost of a significantly higher $Z_\mathrm{avg}$. In terms of $Z_\mathrm{avg}$, both variations using regularization perform best. In terms of #MAC, the approaches without recurrent connections exhibit fewer required MAC operations than the approaches with recurrent connections.

When comparing the approaches with $N_\mathrm{enc} = 10$ and $K = 4$ to those with $N_\mathrm{enc} = 10$ and $K = 60$, only a small performance gap is observed in terms of the BER. However, the number of simulation steps $K$ is reduced to $K = 4$, which corresponds to a 15-fold reduction. For illustration, we conduct a comparison of the three metrics (BER, $Z_\mathrm{avg}$, and #MAC) using the NFE-SNN$_\mathrm{RFE,E}$ with fixed $N_\mathrm{enc} = 10$. Tab. 6.1 shows the obtained metrics for various $K$. When comparing, e.g., $K = 10$ and $K = 60$, we can reduce $K$ by 83%, $Z_\mathrm{avg}$ by 30%, and #MAC by 83%, at a BER penalty of approximately 5.8%.

Besides $K$, we can also decrease $N_\mathrm{enc}$. For $N_\mathrm{enc} = 8$, the gap between the approaches using $K = 4, 5, 8, 10$ and the approach using $K = 60$ is significant. When further decreasing $N_\mathrm{enc}$, the performance degrades significantly.

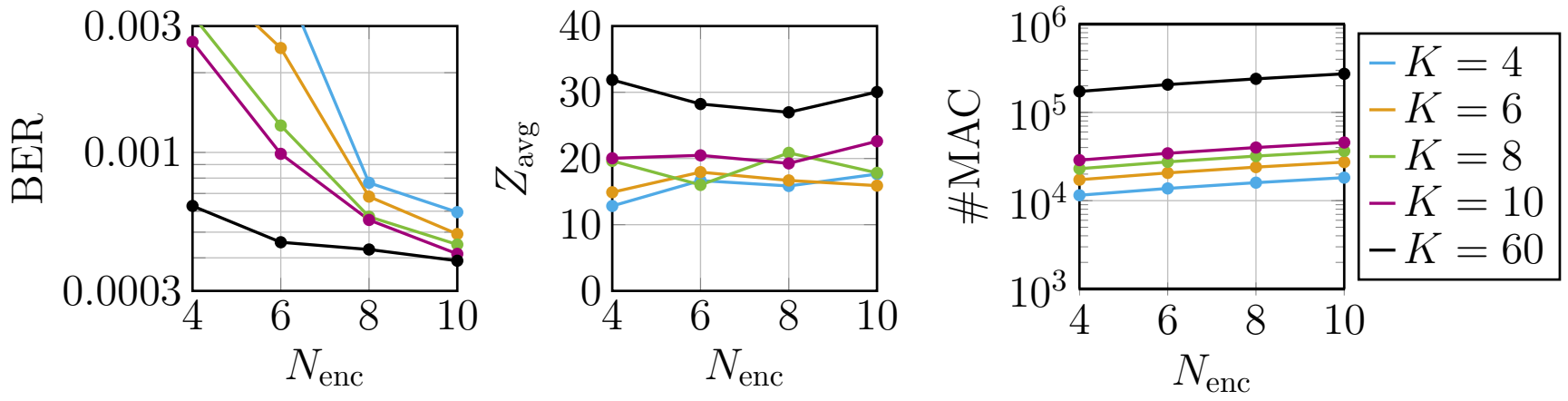

**(a)** NFE-SNN$_{\mathrm{RFE,E,R,rec}}$

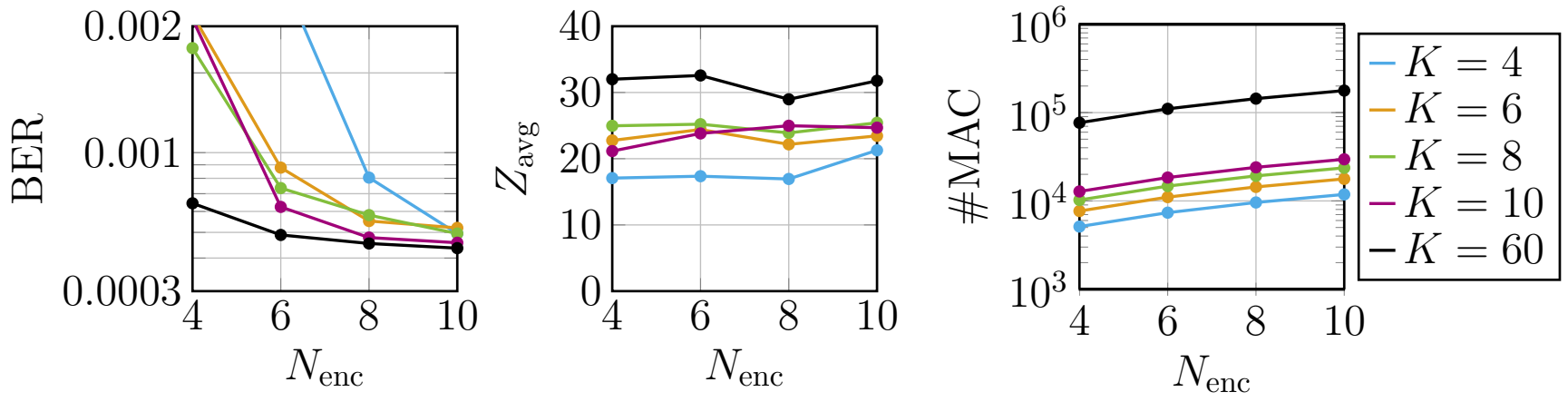

**(b)** NFE-SNN$_{\mathrm{RFE,E,R}}$

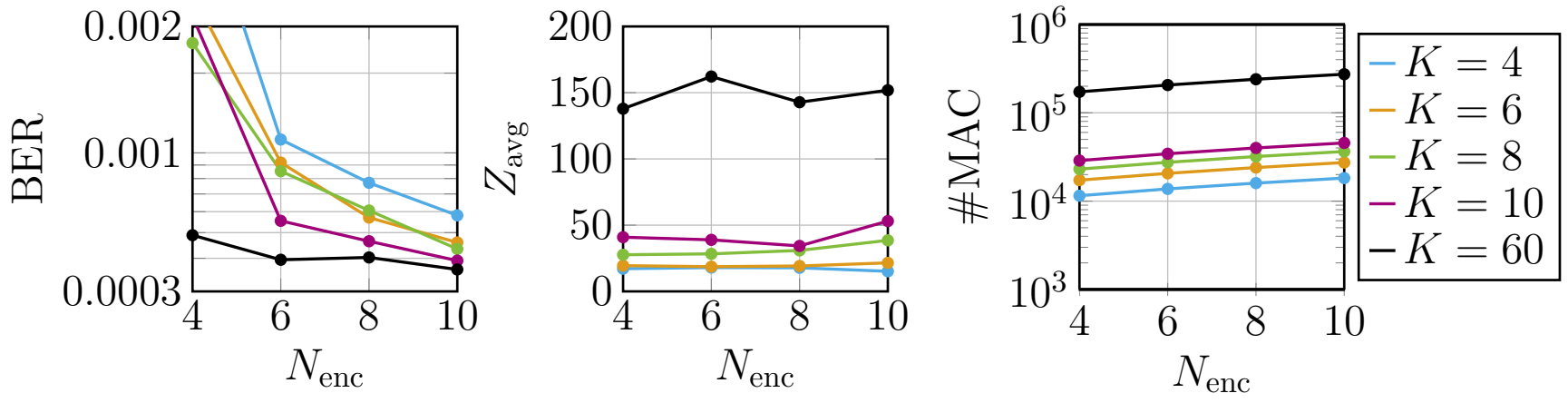

**(c)** NFE-SNN$_{\mathrm{RFE,E,rec}}$

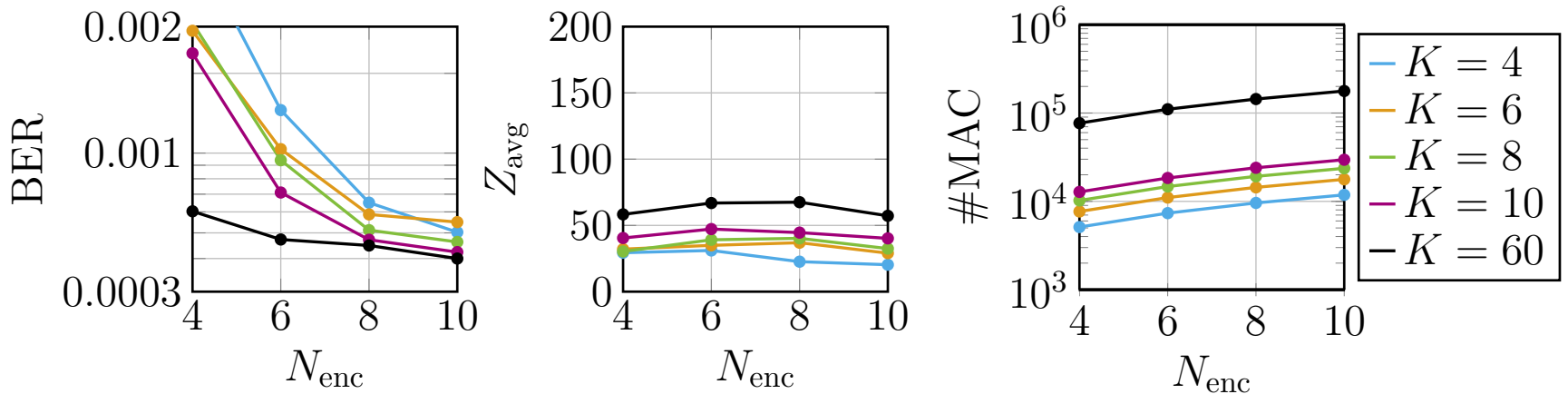

**(d)** NFE-SNN$_{\mathrm{RFE,E}}$

**Figure 6.6:** Performance of four variations of the NFE-SNN$_{\mathrm{RFE,E}}$ and varying parameters of the RFE. All curves are obtained for $\sigma_{\mathrm{n}}^2 = -20\,\mathrm{dB}$.

**Table 6.1:** Performance of NFE-SNN$_{\text{RFE,E}}$ for $N_{\text{enc}} = 10$.

| | BER | $Z_{\text{avg}}$ | #MAC |
|---|---|---|---|
| $K = 4$ | $5.03 \cdot 10^{-4}$ | 20.3 | 11 840 |
| $K = 6$ | $5.49 \cdot 10^{-4}$ | 29.1 | 17 760 |
| $K = 8$ | $4.62 \cdot 10^{-4}$ | 32.5 | 23 680 |
| $K = 10$ | $4.23 \cdot 10^{-4}$ | 40.2 | 29 600 |
| $K = 60$ | $4 \cdot 10^{-4}$ | 57.2 | 177 600 |

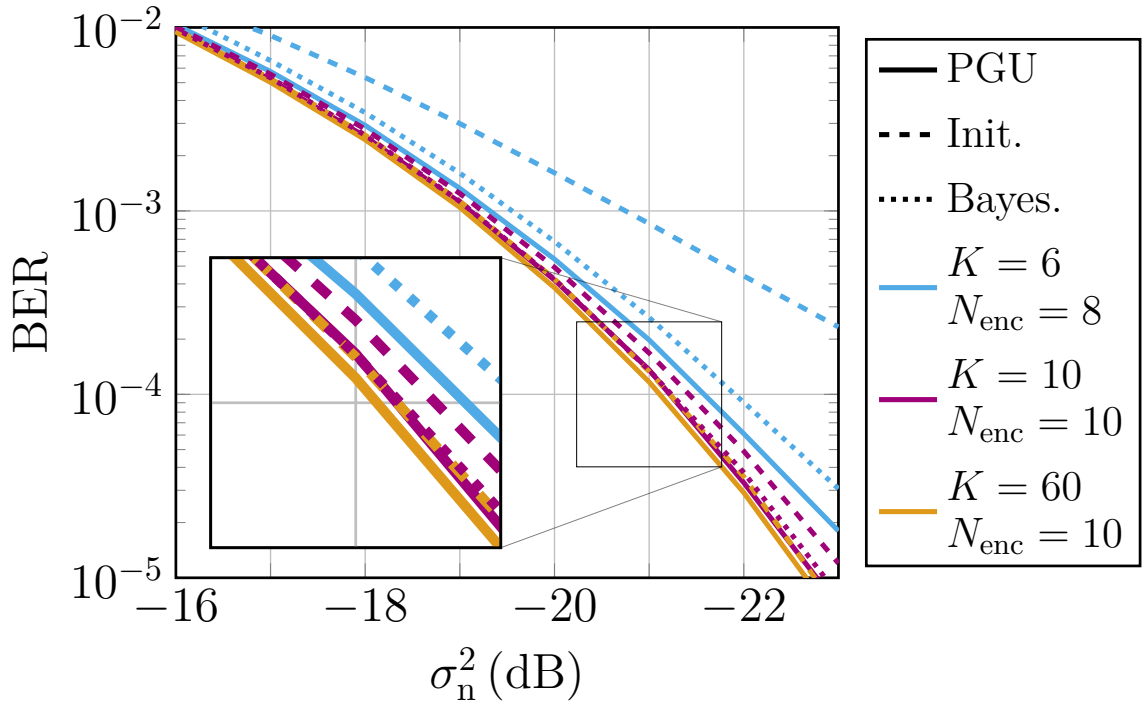

**Figure 6.7:** BER of NFE-SNN$_{\text{RFE,E}}$ for different update methods of the RFE.

In summary, we observe that we can reduce $K$ and $N_{\text{enc}}$ to a certain degree. With a reduction of $K$, we can reduce $Z_{\text{avg}}$ as well as #MAC. Reducing $N_{\text{enc}}$ does not have any observable impact on $Z_{\text{avg}}$. However, the dimensionality $N_{\text{in}}$ of the input layer decreases, and hence the model complexity $N_\theta$ and thus #MAC.

To demonstrate that the PGU significantly contributes to the above results, we plot the BER obtained for the NFE-SNN$_{\text{RFE,R}}$ when applying the PGU and when not applying the PGU in Fig. 6.7. Since the latter uses the initial parameters of the RFE, it is indicated by "Init.". For some combinations of $N_{\text{enc}}$ and $K$, we furthermore update the parameters of the RFE through Bayesian optimization, based on 100 evaluations. The Bayesian optimization is initialized in the same way as the PGU, and is executed using the MaL experiment tracking tool `wandb` [Bie20].

For $K = 60$ and $N_{\text{enc}} = 10$, the orange dashed curve (Init., $K = 60$, $N_{\text{enc}} = 10$) corresponds to the BER curve of the NFE-SNN$_{\text{RFE,E}}$ of Ch. 4. Using the PGU, we observe a small gain. We conclude that the initial parameters already perform very well. For $K = 10$ and $N_{\text{enc}} = 10$, both PGU and Bayesian optimization can improve the initial parameters, resulting in a BER improvement by a good margin. The PGU slightly outperforms Bayesian optimization. Lastly, for $K = 6$ and $N_{\text{enc}} = 8$, the initial parameters of the RFE perform poorly. The application of PGU and Bayesian optimization both significantly improve the BER. The application of PGU results in a lower BER.

We can emphasize three findings: For a poorly initialized RFE, the application of PGU results in parameters enhancing the overall BER performance. When comparing PGU with Bayesian optimization, PGU achieves typically better results and is faster since it requires the optimization of a single SNN, instead of the sequential optimization and evaluation of multiple SNNs with varying encoding parameters $\boldsymbol{\theta}$.

## 6.6 Key Findings

In this chapter, we derived the PGU, an RL-based update algorithm for directly updating the parameters of a model [EvBS25]. To improve its convergence behavior, we applied heuristic modifications using the Rosenbrock test function. Using the modified PGU, we demonstrated the successful optimization of the parameters of an SNN-based detector for AWGN. Afterwards, we investigated the optimization of a higher-dimensional model, the SNN-based equalizer and demapper for the LCD link. Although the optimization converged, the resulting performance remains substantially inferior to that achieved through optimization with BPTT and SG. Since PGU is a trial-and-error method, we suspect that the PGU scales poorly for larger search spaces.

For the SNN-based equalizer and demapper combined with RFE, the RFE applies a non-differentiable function, converting spike times into spike signals. As a result, the parameters of the RFE cannot be updated using the BP algorithm and need to be updated heuristically, which can be done manually or using, for example, Bayesian optimization. Both methods rely on the optimization of multiple SNNs, where the encoding parameters differ between runs. Instead, we can exploit the PGU to update the encoding parameters. Both PGU and BPTT with SG are applied in an alternating manner, the first updating the parameters of the RFE, and the latter updating the parameters of the SNN [EvBS25]. Using this approach, we demonstrated that we can significantly reduce the dimensionality of the RFE, significantly reducing the simulation time $K$, the number of generated spikes $Z_{\text{avg}}$, and the number of required MAC operations #MAC, while only observing a negligibly small BER degradation. Moreover, the PGU delivers superior results compared to Bayesian optimization while requiring substantially less optimization time.

Consequently, utilizing PGU for optimization is particularly suitable in cases where the model exhibits moderate dimensionality, and the BP algorithm cannot be used due to, e.g., non-differentiability.

# 7 Summary and Outlook

## Summary

Recent advances in communications engineering have enabled the development of highly capable transmitters and receivers that approach channel capacity across a range of transmission scenarios. A key factor behind this progress is the use of machine learning (MaL) techniques, particularly artificial neural networks (ANNs), which can effectively compensate for device and channel impairments and address complex tasks that are challenging with conventional methods. When focusing on the problem of channel equalization, MaL and ANNs are promising. However, the performance of ANN-based equalizers is often tied to their computational complexity, which can make them unsuitable for ultra-low-complexity systems. Moreover, higher complexity typically translates to higher energy consumption, resulting in power-intensive receivers when implemented using digital hardware.

In contrast, the human brain only consumes tens of Watts of power on average to execute complex tasks. By mimicking the human brain more closely, spiking neural networks (SNNs) aim to bridge the gap between ANNs and the human brain, possibly achieving low-power artificial intelligence. Unlike ANNs, SNNs feature time-dependent neuron and synapse dynamics and communicate information through discrete binary events known as spikes. Because spikes are transmitted only when information is actively processed, SNNs can achieve significantly lower energy consumption compared to conventional networks. When deployed on neuromorphic hardware, SNNs enable efficient, high-speed, and low-power signal processing.

However, major issues persist when optimizing SNNs: First, it is still not fully understood how the human brain learns. The literature describes several update rules, which vary in their degree of biological plausibility. Furthermore, prior to an SNN, the neural encoding, which converts real-world signals into spike signals, needs to be applied. Similar to the update rule, numerous methods have been proposed, yet no commonly accepted standard has emerged to date.

In this thesis, we have explored the design of SNN-based receivers with the goal of achieving energy-efficient communication systems. First, we investigated the design and optimization of an SNN-based detector for the additive white Gaussian noise (AWGN) channel. Based on the findings, we developed an SNN-based equalizer and demapper with decision feedback. Using methods of reinforcement learning (RL), we developed policy gradient-based update (PGU), a method that enables the joint optimization of the neural encoding and the SNN.

### SNN-based Detector

Based on the transmission of a 16-quadrature amplitude modulation (QAM) signal over an AWGN channel, we designed an SNN-based detector. In this context, we investigated three different update rules and various neural encoding schemes. First, we investigated spike-timing-dependent plasticity (STDP), which is a biologically plausible update rule, but it is limited to a local learning signal. Due to the locality of the learning signal, we initially failed to optimize an SNN with a single hidden layer. By fixing the input weights and solely updating the output weights, we demonstrated the successful optimization of the SNN-based detector using STDP. As a second update rule, we investigated backpropagation through time (BPTT) with surrogate gradients (SGs). SGs enable backpropagation, circumventing the non-differentiability of spiking neurons by approximating the true gradient. Using BPTT with SGs, we also demonstrated the successful optimization of the SNN-based detector. The third update rule is based on probabilistic SNNs. Instead of neuron models with deterministic spiking behavior, the neuron model assumes a stochastic nature of the spike behavior. The spikes output by the SNN are modeled by a jointly distributed random process, and the parameters of the probabilistic SNN are determined by maximizing the probability of observing a desired target realization. We developed a strategy to convert the probabilistic SNNs into a deterministic SNN, and also demonstrated the successful optimization of the SNN-based detector using probabilistic SNNs. However, we experienced that the optimization performance is highly sensitive to the formulation of the target realization at the SNN output. From the findings, we identified BPTT with SG as the most promising update rule, since it can be easily upscaled to multilayer SNNs, directly optimizes deterministic neuron models, such as the spiking leaky integrate-and-fire (LIF) neuron model, and enables the application of well-studied MaL techniques.

Furthermore, we investigated various neural encodings. For the given task of the SNN-based detector, a complex-valued sample needs to be converted into (multiple) spike signals. Building on Bernoulli encoding (BE) and time-to-first-spike encoding (TTFS) as the core component, we developed multiple encoding schemes. We observed that the interconnection of these blocks plays a crucial role in the success of the encoding and, in some cases, requires substantial prior knowledge about the decision boundaries of the corresponding maximum likelihood (ML)-based detector. Based on quantization and the consecutive conversion of the binary signals into spike signals, we developed ternary encoding (TE) and quantization encoding (QE). Furthermore, we examined receptive field encoding (RFE). All three encoding schemes were successfully applied without requiring any prior knowledge of the data and followed a straightforward implementation. Consequently, we identified RFE, TE and QE as particularly promising candidates.

**SNN-based Equalizer and Demapper**

Next, we introduced various SNN-based equalizers and demappers. The various approaches differ in their use of feedback from already decided symbols, application of recurrent connections, neural encoding, neural decoding, and regularization. For two different parameterizations of the intensity modulation with direct detection (IM/DD) link, we systematically compared all approaches in terms of their bit error rate (BER) and average number $Z_{\mathrm{avg}}$ of generated spikes per symbol .

The IM/DD link suffers from chromatic dispersion (CD), non-linear distortion, and AWGN. We observed that max-of-time membrane potential (MOTM) and end-of-time membrane-potential (EOTM) decoding achieve similar performance, with EOTM emitting fewer spikes. Furthermore, adding recurrent connections improves performance, at the expense of increased spike activity. Regularization can combat the increased spike activity, resulting in both good performance and low spike activity.

With respect to decision feedback and neural encoding, the SNN without feedback combined with RFE, as well as the feedback-based SNN with QE, were identified as promising candidates. When non-linearity is the dominant impairment, both architectures achieve comparable performance. However, under stronger influence of CD, the feedback-based SNN with QE clearly outperforms its counterpart without feedback and RFE. Due to the neural encoding, the feedback-based SNN with QE achieves a 12 times smaller latency than the SNN without feedback and RFE. Furthermore, depending on the parameterization of the IM/DD link, its complexity amounts to 77% or 70% of that of the SNN without feedback and RFE, respectively.

In conclusion, the SNN-based equalizer and demapper, incorporating decision feedback, recurrent connections, QE, and regularization, proves effective in addressing both CD and non-linear impairments. It achieves superior performance alongside low spike activity, latency, and computational complexity. Furthermore, it significantly outperforms all benchmark equalizers and demappers, even ANN-based approaches.

**Optimization of Neural Encoding**

So far we have determined the parameters of RFE heuristically. Depending on the parameters, the performance of the overall system may suffer. Due to the non-differentiable spike-conversion in the RFE, the joint optimization of the RFE and SNN using BPTT is not feasible. We developed PGU, an RL-based update algorithm, and successfully applied it to update the parameters of the SNN-based detector for the AWGN channel. Since PGU relies on an exploration-and-learning approach and the SNN-based equalizer and demapper contain substantially more parameters than the SNN-based detector, we observed that PGU scales poorly. Consequently, we updated the SNN using BPTT with SG, and used the PGU to update the parameters of the RFE. We demonstrated that we can significantly reduce the dimensions of the RFE without major BER penalty. As a result, the latency and complexity of the SNN-based equalizer and demapper are also significantly decreased.

In summary, this thesis developed SNN-based receivers for the AWGN channel and non-linear frequency-selective channels. It discussed caveats and pitfalls when optimizing SNN-based receivers, and identified update algorithms, neural encodings, and SNN architectures that result in well-performing SNN-based receivers.

## Outlook

ANNs have already led to solutions to many problems in communications engineering that are challenging for traditional approaches. It is reasonable to expect that a low-power SNN-based counterpart exists for each ANN-based algorithm. However, as demonstrated in this work, the conversion of ANN-based algorithms into SNN-based algorithms is not straightforward. Leveraging the foundation established in this thesis, this conversion process should be simplified, potentially unlocking the (hopefully) vast field of low-power SNN-based solutions for communication engineering.

One example of further SNN-based algorithms is the enlarge-likelihood-each-notable-amplitude-SNN (ELENA-SNN) decoder, which is a novel SNN-based decoding algorithm for low-density parity-check (LDPC) codes [vBEM+25]. Typically, LDPC codes are decoded using the sum-product algorithm (SPA), which in combination can achieve performance close to channel capacity. However, the SPA involves computationally complex check node updates. ELENA-SNN approximates these updates using one SNNs per check node. For the (273, 191) and (1023, 781) finite-geometry LDPC code, ELENA-SNN outperforms the sum-product decoder at high signal-to-noise ratios. For LDPC codes with small variable node degrees, e.g., the $(38400, 30720)$ regular LDPC code, ELENA-SNN yields poor performance due to a limited dynamic range of the single SNN. To increase the dynamic range, we introduced multi-level ELENA-SNN (ML-ELENA-SNN), which applies multiple SNNs in parallel per check node [vBEMS25]. For the $(38400, 30720)$ regular LDPC code, ML-ELENA-SNN achieves results comparable to the SPA algorithm. Due to non-differentiabilities in the decoding process, the parameters of the SNNs of ELENA-SNN and ML-ELENA-SNN are determined heuristically by consecutive line searches. Currently, we investigate the application of PGU to determine these parameters, as well as the efficient implementation on field programmable gate arrays (FPGAs).

# A Appendix

## A.1 Update Algorithm for Fully Observed Probabilistic SNNs

To obtain the gradient $\nabla_{\boldsymbol{\theta}}\mathcal{L}_{\leq K-1}$, we exploit that the probability of observing $\boldsymbol{Z}_{\leq K-1}$ can be expressed as the product of the individual probabilities $p_{\boldsymbol{\theta}_j}(Z_j[\kappa]|v_j[\kappa])$ of all neurons $j \in \mathcal{Z}$ at all time steps $\kappa$. We can simplify the gradient of the log-likelihood $\mathcal{L}_{\leq K-1}$ to [JSGG19]

$$\nabla_{\boldsymbol{\theta}}\mathcal{L}_{\leq K-1} = \nabla_{\boldsymbol{\theta}} \ln\left(p_{\boldsymbol{\theta}}(\boldsymbol{Z}_{\leq K-1})\right) \tag{A.1}$$

$$= \sum_{\kappa=0}^{K-1}\sum_{j\in\mathcal{Z}} \nabla_{\boldsymbol{\theta}_j} \ln\left(p_{\boldsymbol{\theta}_j}\left(Z_j[\kappa]\middle|v_j[\kappa]\right)\right) . \tag{A.2}$$

The spiking probability per neuron $p_{\boldsymbol{\theta}_j}\left(Z_j[\kappa]\middle|v_j[\kappa]\right)$ can be expressed using the sigmoid-function of (2.12)

$$p_{\boldsymbol{\theta}_j}\left(Z_j[\kappa]\middle|v_j[\kappa]\right) = \sigma(v_j[\kappa])^{Z_j[\kappa]} \cdot (1-\sigma(v_j[\kappa]))^{1-Z_j[\kappa]} , \tag{A.3}$$

which is consistent with the spike response model (SRM) as introduced in Sec. 2.1.2. Using (A.3), we can rewrite (A.2)

$$\begin{aligned}
\nabla_{\boldsymbol{\theta}}\mathcal{L}_{\leq K-1}(\boldsymbol{\theta}) &= \sum_{\kappa=0}^{K-1}\sum_{j\in\mathcal{Z}} \nabla_{\boldsymbol{\theta}_j} \ln\left(\sigma(v_j[\kappa])^{Z_j[\kappa]} \cdot (1-\sigma(v_j[\kappa]))^{1-Z_j[\kappa]}\right) \\
&= \sum_{\kappa=0}^{K-1}\sum_{j\in\mathcal{Z}} \Big(\nabla_{\boldsymbol{\theta}_j} Z_j[\kappa] \ln \sigma(v_j[\kappa]) \\
&\quad + \nabla_{\boldsymbol{\theta}_j}(1-Z_j[\kappa]) \ln(1-\sigma(v_j[\kappa]))\Big) \\
&= \sum_{\kappa=0}^{K-1}\sum_{j\in\mathcal{Z}} \left(Z_j[\kappa]\frac{\nabla_{\boldsymbol{\theta}_j}\sigma(v_j[\kappa])}{\sigma(v_j[\kappa])} + (1-Z_j[\kappa])\frac{\nabla_{\boldsymbol{\theta}_j}(1-\sigma(v_j[\kappa]))}{1-\sigma(v_j[\kappa])}\right)
\end{aligned}$$

$$= \sum_{\kappa=0}^{K-1}\sum_{j\in\mathcal{Z}} \Bigg(Z_j[\kappa]\frac{\sigma(v_j[\kappa])\,(1-\sigma(v_j[\kappa]))}{\sigma(v_j[\kappa])}\nabla_{\boldsymbol{\theta}_j} v_j[\kappa]$$

$$
\begin{aligned}
&\quad + (1 - Z_j[\kappa]) \frac{-\sigma(v_j[\kappa])(1-\sigma(v_j[\kappa]))}{1-\sigma(v_j[\kappa])} \nabla_{\boldsymbol{\theta}_j} v_j[\kappa] \Bigg) \\
&= \sum_{\kappa=0}^{K-1} \sum_{j \in \mathcal{Z}} \Bigg( Z_j[\kappa] \cdot (1 - \sigma(v_j[\kappa])) \\
&\quad - \sigma(v_j[\kappa]) + Z_j[\kappa]\sigma(v_j[\kappa]) \Bigg) \cdot \nabla_{\boldsymbol{\theta}_j} v_j[\kappa] \\
&= \sum_{\kappa=0}^{K-1} \sum_{j \in \mathcal{Z}} (Z_j[\kappa] - \sigma(v_j[\kappa])) \cdot \nabla_{\boldsymbol{\theta}_j} v_j[\kappa] \,,
\end{aligned} \tag{A.4}
$$

where we exploited $\nabla_x \sigma(f(x)) = \sigma(f(x)) \cdot (1 - \sigma(f(x))) \cdot \nabla_x f(x)$.

We can now separate the parameters $\boldsymbol{\theta}_j$ of the $j$th neuron into the feedforward weights $\boldsymbol{\theta}_j^{(\mathrm{f})}$ and its recurrent weight $\theta_j^{(\mathrm{r})}$. We derive (2.10) and obtain

$$
\nabla_{\boldsymbol{\theta}_j^{(\mathrm{f})}} v_j[\kappa] = (\boldsymbol{z}_{\mathrm{in},j} * \alpha)[\kappa] \,, \tag{A.5}
$$

$$
\nabla_{\theta_j^{(\mathrm{r})}} v_j[\kappa] = (z_{\mathrm{out},j} * \beta)[\kappa - 1] \,, \tag{A.6}
$$

where $\boldsymbol{z}_{\mathrm{in},j}[\kappa] \in \{0,1\}^J$, $J \in \mathbb{N}^+$ is a vector containing the input signals received by the $J$ presynaptic neurons of neuron $j$, and $z_{\mathrm{out},j}[\kappa] \in \{0,1\}$ the spike signal fed back from neuron $j$ itself. The gradient of each neuron with respect to (w.r.t.) $\boldsymbol{\theta}_j^{(\mathrm{f})}$ and $\theta_j^{(\mathrm{r})}$ is then obtained by

$$
\nabla_{\boldsymbol{\theta}_j^{(\mathrm{f})}} \ln\left(p_{\boldsymbol{\theta}_j}\left(Z_j[\kappa] \middle| v_j[\kappa]\right)\right) = (\boldsymbol{z}_{\mathrm{in},j} * \alpha)[\kappa] \cdot (Z_j[\kappa] - \sigma(v_j[\kappa])) \,, \tag{A.7}
$$

$$
\nabla_{\theta_j^{(\mathrm{r})}} \ln\left(p_{\boldsymbol{\theta}_j}\left(Z_j[\kappa] \middle| v_j[\kappa]\right)\right) = (z_{\mathrm{out},j} * \beta)[\kappa - 1] \cdot (Z_j[\kappa] - \sigma(v_j[\kappa])) \,. \tag{A.8}
$$

## A.2 Update Algorithm for Partially Observed Probabilistic SNNs

In the following, a detailed derivation of the update algorithm for partially observed probabilistic SNNs is provided. The derivation closely follows the steps of [JSGG19]. For better understanding, the derivation is extended using [Sim18, Ch. 8], [MG14], and [BB24, Sec. 15.4].

We use the notation as introduced in Sec. 2.5.3. For partially observed SNNs, the set of hidden neurons is not empty, i.e., $\mathcal{H} \neq \{\emptyset\}$. Again, we can formulate the learning problem as maximizing the probability of observing the desired output spike pattern $\boldsymbol{Z}_{\leq K-1}$, see (2.44). The gradient of (A.1) w.r.t. $\boldsymbol{\theta}$ is obtained by

$$\begin{aligned}
\nabla_{\boldsymbol{\theta}} \mathcal{L}_{\leq K-1}(\boldsymbol{\theta}) &= \nabla_{\boldsymbol{\theta}} \ln p_{\boldsymbol{\theta}}(\boldsymbol{Z}_{\leq K-1}) && \text{(A.9)} \\
&= \frac{1}{p_{\boldsymbol{\theta}}(\boldsymbol{Z}_{\leq K-1})} \nabla_{\boldsymbol{\theta}} p_{\boldsymbol{\theta}}(\boldsymbol{Z}_{\leq K-1}) \\
&\overset{\text{(a)}}{=} \frac{1}{p_{\boldsymbol{\theta}}(\boldsymbol{Z}_{\leq K-1})} \nabla_{\boldsymbol{\theta}} \sum_{\boldsymbol{H}_{\leq K-1}} p_{\boldsymbol{\theta}}(\boldsymbol{Z}_{\leq K-1}, \boldsymbol{H}_{\leq K-1}) \\
&= \frac{1}{p_{\boldsymbol{\theta}}(\boldsymbol{Z}_{\leq K-1})} \sum_{\boldsymbol{H}_{\leq K-1}} \nabla_{\boldsymbol{\theta}} p_{\boldsymbol{\theta}}(\boldsymbol{Z}_{\leq K-1}, \boldsymbol{H}_{\leq K-1}) \\
&= \sum_{\boldsymbol{H}_{\leq K-1}} \frac{1}{p_{\boldsymbol{\theta}}(\boldsymbol{Z}_{\leq K-1})} \frac{p_{\boldsymbol{\theta}}(\boldsymbol{Z}_{\leq K-1}, \boldsymbol{H}_{\leq K-1})}{p_{\boldsymbol{\theta}}(\boldsymbol{Z}_{\leq K-1}, \boldsymbol{H}_{\leq K-1})} \nabla_{\boldsymbol{\theta}} p_{\boldsymbol{\theta}}(\boldsymbol{Z}_{\leq K-1}, \boldsymbol{H}_{\leq K-1}) \\
&\overset{\text{(b)}}{=} \sum_{\boldsymbol{H}_{\leq K-1}} p_{\boldsymbol{\theta}}(\boldsymbol{H}_{\leq K-1} | \boldsymbol{Z}_{\leq K-1}) \nabla_{\boldsymbol{\theta}} \ln p_{\boldsymbol{\theta}}(\boldsymbol{H}_{\leq K-1}, \boldsymbol{Z}_{\leq K-1}) \\
&= \mathbb{E}_{p_{\boldsymbol{\theta}}(\boldsymbol{H}_{\leq K-1} | \boldsymbol{Z}_{\leq K-1})} \Big\{ \nabla_{\boldsymbol{\theta}} \ln p_{\boldsymbol{\theta}}(\boldsymbol{Z}_{\leq K-1}, \boldsymbol{H}_{\leq K-1}) \Big\} . && \text{(A.10)}
\end{aligned}$$

In (a), we marginalize by summing over all possible $\boldsymbol{H}_{\leq K-1}$. In (b), we apply the log-derivative trick $\nabla_x f(x) = f(x) \nabla_x \ln(f(x))$. Finally, we can write $\nabla_{\boldsymbol{\theta}} \mathcal{L}_{\leq K-1}(\boldsymbol{\theta})$ as the expectation of the logarithm of the joint probability $p_{\boldsymbol{\theta}}(\boldsymbol{Z}_{\leq K-1}, \boldsymbol{H}_{\leq K-1})$ w.r.t. the posterior distribution $p_{\boldsymbol{\theta}}(\boldsymbol{H}_{\leq K-1} | \boldsymbol{Z}_{\leq K-1})$ [Sim18, (6.22)]. For all spike patterns $\boldsymbol{H}_{\leq K-1}$ of the hidden neurons, the posterior quantifies their probability when observing $\boldsymbol{Z}_{\leq K-1}$.

We can compute the posterior by

$$\begin{aligned}
p_{\boldsymbol{\theta}}(\boldsymbol{H}_{\leq K-1} | \boldsymbol{Z}_{\leq K-1}) &= \frac{p_{\boldsymbol{\theta}}(\boldsymbol{H}_{\leq K-1}, \boldsymbol{Z}_{\leq K-1})}{p_{\boldsymbol{\theta}}(\boldsymbol{Z}_{\leq K-1})} && \text{(A.11)} \\
&= \frac{p_{\boldsymbol{\theta}}(\boldsymbol{H}_{\leq K-1}, \boldsymbol{Z}_{\leq K-1})}{\sum_{\boldsymbol{H}'_{\leq K-1}} p_{\boldsymbol{\theta}}(\boldsymbol{H}'_{\leq K-1}, \boldsymbol{Z}_{\leq K-1})} ,
\end{aligned}$$

where the marginal distribution $p_{\boldsymbol{\theta}}(\boldsymbol{Z}_{\leq K-1})$ is obtained via marginalization over all $\boldsymbol{H}'_{\leq K-1} \in \{0, 1\}^{N_{\text{hid}} \times \kappa}$. With an increasing number of hidden neurons $N_{\text{hid}} = |\mathcal{H}|$ and time steps $K$ of the SNN, computing the marginal distribution becomes infeasible.

To overcome this problem, we can use *variational inference* [Sim18, Ch. 8]. Instead of maximizing $\ln p_{\boldsymbol{\theta}}(\boldsymbol{Z}_{\leq K-1})$, we can maximize the evidence lower bound (ELBO), which is a lower bound for the log-likelihood of the observations $\ln p_{\boldsymbol{\theta}}(\boldsymbol{Z}_{\leq K-1})$ (also called the *evidence*). We introduce parameterized distribution $q_{\boldsymbol{\phi}}(\boldsymbol{H}_{\leq K-1})$ defined over the latent variable $\boldsymbol{H}_{\leq K-1}$ and parameterized by $\boldsymbol{\phi} \in \mathbb{R}^d$, which approximates the true posterior distribution $p_{\boldsymbol{\theta}}(\boldsymbol{H}_{\leq K-1} | \boldsymbol{Z}_{\leq K-1})$ and hence is called *variational posterior*. For any choice of $q_{\boldsymbol{\phi}}(\boldsymbol{H}_{\leq K-1})$, the following decomposition holds [BB24,

Sec. 15.4]

$$\begin{aligned}\mathcal{L}_{\leq K-1} &= \ln p_{\boldsymbol{\theta}}(\boldsymbol{Z}_{\leq K-1}) \\ &= \mathcal{L}(\boldsymbol{\theta}, \boldsymbol{\phi}) + D(q||p)\,,\end{aligned} \tag{A.12}$$

where

$$\begin{aligned}\mathcal{L}(\boldsymbol{\theta}, \boldsymbol{\phi}) &= \sum_{\boldsymbol{H}_{\leq K-1}} q_{\boldsymbol{\phi}}(\boldsymbol{H}_{\leq K-1}) \ln\left(\frac{p_{\boldsymbol{\theta}}(\boldsymbol{Z}_{\leq K-1}, \boldsymbol{H}_{\leq K-1})}{q_{\boldsymbol{\phi}}(\boldsymbol{H}_{\leq K-1})}\right) \\ &= \mathbb{E}_{q_{\boldsymbol{\phi}}(\boldsymbol{H}_{\leq K-1})}\Big\{ \ln p_{\boldsymbol{\theta}}(\boldsymbol{H}_{\leq K-1}, \boldsymbol{Z}_{\leq K-1}) - \ln q_{\boldsymbol{\phi}}(\boldsymbol{H}_{\leq K-1})\Big\}\end{aligned} \tag{A.13}$$

denotes the ELBO and

$$D(q||p) = \sum_{\boldsymbol{H}_{\leq K-1}} q_{\boldsymbol{\phi}}(\boldsymbol{H}_{\leq K-1}) \ln\left(\frac{q_{\boldsymbol{\phi}}(\boldsymbol{H}_{\leq K-1})}{p_{\boldsymbol{\theta}}(\boldsymbol{H}_{\leq K-1}|\boldsymbol{Z}_{\leq K-1})}\right) \tag{A.14}$$

the Kullback-Leibler (KL) divergence. We can show

$$\begin{aligned}\mathcal{L}(\boldsymbol{\theta}, \boldsymbol{\phi}) &= \mathbb{E}_{q_{\boldsymbol{\phi}}(\boldsymbol{H}_{\leq K-1})}\Big\{ \ln p_{\boldsymbol{\theta}}(\boldsymbol{H}_{\leq K-1}, \boldsymbol{Z}_{\leq K-1}) - \ln q_{\boldsymbol{\phi}}(\boldsymbol{H}_{\leq K-1})\Big\} \\ &= \mathbb{E}_{q_{\boldsymbol{\phi}}(\boldsymbol{H}_{\leq K-1})}\Big\{ \ln p_{\boldsymbol{\theta}}(\boldsymbol{H}_{\leq K-1}|\boldsymbol{Z}_{\leq K-1}) + \ln p_{\boldsymbol{\theta}}(\boldsymbol{Z}_{\leq K-1}) \\ &\quad - \ln q_{\boldsymbol{\phi}}(\boldsymbol{H}_{\leq K-1})\Big\} \\ &= \mathbb{E}_{q_{\boldsymbol{\phi}}(\boldsymbol{H}_{\leq K-1})}\Big\{ \ln p_{\boldsymbol{\theta}}(\boldsymbol{Z}_{\leq K-1})\Big\} \\ &\quad - \mathbb{E}_{q_{\boldsymbol{\phi}}(\boldsymbol{H}_{\leq K-1})}\left\{ \ln\left(\frac{q_{\boldsymbol{\phi}}(\boldsymbol{H}_{\leq K-1})}{p_{\boldsymbol{\theta}}(\boldsymbol{H}_{\leq K-1}|\boldsymbol{Z}_{\leq K-1})}\right)\right\} \\ &= \ln p_{\boldsymbol{\theta}}(\boldsymbol{Z}_{\leq K-1}) - \sum_{\boldsymbol{H}_{\leq K-1}} q_{\boldsymbol{\phi}}(\boldsymbol{H}_{\leq K-1}) \ln\left(\frac{q_{\boldsymbol{\phi}}(\boldsymbol{H}_{\leq K-1})}{p_{\boldsymbol{\theta}}(\boldsymbol{H}_{\leq K-1}|\boldsymbol{Z}_{\leq K-1})}\right) \\ &= \mathcal{L}_{\leq K-1} - D(q||p)\,.\end{aligned}$$

Note that the KL divergence is always non-negative, i.e., $D(q||p) \geq 0$, thus it directly follows that $\mathcal{L}(\boldsymbol{\theta}, \boldsymbol{\phi}) \leq \mathcal{L}_{\leq K-1}$. Since $\mathcal{L}_{\leq K-1}$ is given by the observations and is therefore fixed, by maximizing the ELBO, we minimize the KL divergence and thus the difference between the true posterior and the variational posterior.

An iterative procedure to maximize the ELBO is the expectation maximization (EM) algorithm, which conducts $\max_{\boldsymbol{\phi}} \mathcal{L}(\boldsymbol{\theta}, \boldsymbol{\phi})$ and $\max_{\boldsymbol{\theta}} \mathcal{L}(\boldsymbol{\theta}, \boldsymbol{\phi})$ alternately. Using the gradient of the ELBO w.r.t. to $\boldsymbol{\phi}$ and $\boldsymbol{\theta}$, gradient ascent updates are applied to both parameters alternately. First, we update $\boldsymbol{\phi}$ by

$$\boldsymbol{\phi} \leftarrow \boldsymbol{\phi} + \nu_{\boldsymbol{\phi}} \nabla_{\boldsymbol{\phi}} \mathcal{L}(\boldsymbol{\theta}, \boldsymbol{\phi}) \,, \tag{A.15}$$

where $\nu_{\boldsymbol{\phi}}$ denotes the learning rate of $\boldsymbol{\phi}$. Afterwards, we update $\boldsymbol{\theta}$ by

$$\boldsymbol{\theta} \leftarrow \boldsymbol{\theta} + \nu_{\boldsymbol{\theta}} \nabla_{\boldsymbol{\theta}} \mathcal{L}(\boldsymbol{\theta}, \boldsymbol{\phi}) \,, \tag{A.16}$$

where $\nu_{\boldsymbol{\theta}}$ denotes the learning rate of $\boldsymbol{\theta}$. We repeat both update steps until a predefined number of updates is reached, or a convergence criterion is satisfied.

The gradient of the ELBO w.r.t. $\boldsymbol{\theta}$ can now be directly obtained from (A.13) by

$$\nabla_{\boldsymbol{\theta}} \mathcal{L}(\boldsymbol{\theta}, \boldsymbol{\phi}) = \mathbb{E}_{q_{\boldsymbol{\phi}}(\boldsymbol{H}_{\leq K-1})} \Big\{ \nabla_{\boldsymbol{\theta}} \ln p_{\boldsymbol{\theta}}(\boldsymbol{H}_{\leq K-1}, \boldsymbol{Z}_{\leq K-1}) \Big\} \,. \tag{A.17}$$

Next, we obtain the gradient w.r.t. $\boldsymbol{\phi}$ following [MG14]

$$\begin{aligned} \nabla_{\boldsymbol{\phi}} \mathcal{L}(\boldsymbol{\theta}, \boldsymbol{\phi}) =& \nabla_{\boldsymbol{\phi}} \mathbb{E}_{q_{\boldsymbol{\phi}}(\boldsymbol{H}_{\leq K-1})} \Big\{ \ln p_{\boldsymbol{\theta}}(\boldsymbol{H}_{\leq K-1}, \boldsymbol{Z}_{\leq K-1}) - \ln q_{\boldsymbol{\phi}}(\boldsymbol{H}_{\leq K-1}) \Big\} \\ =& \nabla_{\boldsymbol{\phi}} \sum_{\boldsymbol{H}_{\leq K-1}} q_{\boldsymbol{\phi}}(\boldsymbol{H}_{\leq K-1}) \Big( \ln p_{\boldsymbol{\theta}}(\boldsymbol{H}_{\leq K-1}, \boldsymbol{Z}_{\leq K-1}) - \ln q_{\boldsymbol{\phi}}(\boldsymbol{H}_{\leq K-1}) \Big) \\ =& \nabla_{\boldsymbol{\phi}} \sum_{\boldsymbol{H}_{\leq K-1}} q_{\boldsymbol{\phi}}(\boldsymbol{H}_{\leq K-1}) \ln p_{\boldsymbol{\theta}}(\boldsymbol{H}_{\leq K-1}, \boldsymbol{Z}_{\leq K-1}) \\ & - \nabla_{\boldsymbol{\phi}} \sum_{\boldsymbol{H}_{\leq K-1}} q_{\boldsymbol{\phi}}(\boldsymbol{H}_{\leq K-1}) \ln q_{\boldsymbol{\phi}}(\boldsymbol{H}_{\leq K-1}) \\ =:& \alpha + \beta \,, \end{aligned} \tag{A.18}$$

with

$$\begin{aligned} \alpha &= \nabla_{\boldsymbol{\phi}} \sum_{\boldsymbol{H}_{\leq K-1}} q_{\boldsymbol{\phi}}(\boldsymbol{H}_{\leq K-1}) \ln p_{\boldsymbol{\theta}}(\boldsymbol{H}_{\leq K-1}, \boldsymbol{Z}_{\leq K-1}) \\ &= \sum_{\boldsymbol{H}_{\leq K-1}} \ln p_{\boldsymbol{\theta}}(\boldsymbol{H}_{\leq K-1}, \boldsymbol{Z}_{\leq K-1}) \nabla_{\boldsymbol{\phi}} q_{\boldsymbol{\phi}}(\boldsymbol{H}_{\leq K-1}) \end{aligned} \tag{A.19}$$

$$\begin{aligned} \beta &= -\nabla_{\boldsymbol{\phi}} \sum_{\boldsymbol{H}_{\leq K-1}} q_{\boldsymbol{\phi}}(\boldsymbol{H}_{\leq K-1}) \ln q_{\boldsymbol{\phi}}(\boldsymbol{H}_{\leq K-1}) \\ &= - \sum_{\boldsymbol{H}_{\leq K-1}} \Big( \ln q_{\boldsymbol{\phi}}(\boldsymbol{H}_{\leq K-1}) \nabla_{\boldsymbol{\phi}} q_{\boldsymbol{\phi}}(\boldsymbol{H}_{\leq K-1}) + q_{\boldsymbol{\phi}}(\boldsymbol{H}_{\leq K-1}) \nabla_{\boldsymbol{\phi}} \ln q_{\boldsymbol{\phi}}(\boldsymbol{H}_{\leq K-1}) \Big) \\ &= - \sum_{\boldsymbol{H}_{\leq K-1}} \ln q_{\boldsymbol{\phi}}(\boldsymbol{H}_{\leq K-1}) \nabla_{\boldsymbol{\phi}} q_{\boldsymbol{\phi}}(\boldsymbol{H}_{\leq K-1}) - \sum_{\boldsymbol{H}_{\leq K-1}} \nabla_{\boldsymbol{\phi}} q_{\boldsymbol{\phi}}(\boldsymbol{H}_{\leq K-1}) \end{aligned} \tag{A.20}$$

$$= - \sum_{\boldsymbol{H}_{\leq K-1}} \ln q_{\boldsymbol{\phi}}(\boldsymbol{H}_{\leq K-1}) \nabla_{\boldsymbol{\phi}} q_{\boldsymbol{\phi}}(\boldsymbol{H}_{\leq K-1})\,.$$

For the derivation of $\beta$, we apply

$$\sum_{\boldsymbol{H}_{\leq K-1}} \nabla_{\boldsymbol{\phi}} q_{\boldsymbol{\phi}}(\boldsymbol{H}_{\leq K-1}) = \nabla_{\boldsymbol{\phi}} \sum_{\boldsymbol{H}_{\leq K-1}} q_{\boldsymbol{\phi}}(\boldsymbol{H}_{\leq K-1}) = \nabla_{\boldsymbol{\phi}} 1 = 0\,.$$

Returning back to (A.18):

$$\begin{aligned}
\nabla_{\boldsymbol{\phi}} \mathcal{L}(\boldsymbol{\theta}, \boldsymbol{\phi}) &= \alpha + \beta \qquad \text{(A.21)}\\
&= \sum_{\boldsymbol{H}_{\leq K-1}} \ln p_{\boldsymbol{\theta}}(\boldsymbol{H}_{\leq K-1}, \boldsymbol{Z}_{\leq K-1}) \nabla_{\boldsymbol{\phi}} q_{\boldsymbol{\phi}}(\boldsymbol{H}_{\leq K-1})\\
&\quad + \sum_{\boldsymbol{H}_{\leq K-1}} \ln q_{\boldsymbol{\phi}}(\boldsymbol{H}_{\leq K-1}) \nabla_{\boldsymbol{\phi}} q_{\boldsymbol{\phi}}(\boldsymbol{H}_{\leq K-1})\\
&= \sum_{\boldsymbol{H}_{\leq K-1}} \Big(\ln p_{\boldsymbol{\theta}}(\boldsymbol{H}_{\leq K-1}, \boldsymbol{Z}_{\leq K-1}) - \ln q_{\boldsymbol{\phi}}(\boldsymbol{H}_{\leq K-1})\Big) \cdot \nabla_{\boldsymbol{\phi}} q_{\boldsymbol{\phi}}(\boldsymbol{H}_{\leq K-1})\\
&= \sum_{\boldsymbol{H}_{\leq K-1}} \ell_{\boldsymbol{\theta},\boldsymbol{\phi}}(\boldsymbol{Z}_{\leq K-1}, \boldsymbol{H}_{\leq K-1}) \nabla_{\boldsymbol{\phi}} q_{\boldsymbol{\phi}}(\boldsymbol{H}_{\leq K-1})\\
&= \sum_{\boldsymbol{H}_{\leq K-1}} \ell_{\boldsymbol{\theta},\boldsymbol{\phi}}(\boldsymbol{Z}_{\leq K-1}, \boldsymbol{H}_{\leq K-1}) q_{\boldsymbol{\phi}}(\boldsymbol{H}_{\leq K-1}) \cdot \nabla_{\boldsymbol{\phi}} \ln q_{\boldsymbol{\phi}}(\boldsymbol{H}_{\leq K-1})\\
&= \mathbb{E}_{q_{\boldsymbol{\phi}}(\boldsymbol{H}_{\leq K-1})} \Big\{ \ell_{\boldsymbol{\theta},\boldsymbol{\phi}}(\boldsymbol{Z}_{\leq K-1}, \boldsymbol{H}_{\leq K-1}) \cdot \nabla_{\boldsymbol{\phi}} \ln q_{\boldsymbol{\phi}}(\boldsymbol{H}_{\leq K-1}) \Big\}, \qquad \text{(A.22)}
\end{aligned}$$

where we defined the *learning signal*

$$\ell_{\boldsymbol{\theta},\boldsymbol{\phi}}(\boldsymbol{Z}_{\leq K-1}, \boldsymbol{H}_{\leq K-1}) \coloneqq \ln p_{\boldsymbol{\theta}}(\boldsymbol{H}_{\leq K-1}, \boldsymbol{Z}_{\leq K-1}) - \ln q_{\boldsymbol{\phi}}(\boldsymbol{H}_{\leq K-1})\,. \qquad \text{(A.23)}$$

With both gradients, we can rewrite (A.15) and (A.16) as follows:

$$\boldsymbol{\phi} \leftarrow \boldsymbol{\phi} + \nu_{\boldsymbol{\phi}} \mathbb{E}_{q_{\boldsymbol{\phi}}(\boldsymbol{H}_{\leq K-1})} \Big\{ \ell_{\boldsymbol{\theta},\boldsymbol{\phi}}(\boldsymbol{Z}_{\leq K-1}, \boldsymbol{H}_{\leq K-1}) \nabla_{\boldsymbol{\phi}} \ln q_{\boldsymbol{\phi}}(\boldsymbol{H}_{\leq K-1}) \Big\} \qquad \text{(A.24)}$$

$$\boldsymbol{\theta} \leftarrow \boldsymbol{\theta} + \nu_{\boldsymbol{\theta}} \mathbb{E}_{q_{\boldsymbol{\phi}}(\boldsymbol{H}_{\leq K-1})} \Big\{ \nabla_{\boldsymbol{\theta}} \ln p_{\boldsymbol{\theta}}(\boldsymbol{H}_{\leq K-1}, \boldsymbol{Z}_{\leq K-1}) \Big\} \qquad \text{(A.25)}$$

To obtain a variational posterior that is easy to sample from and assures the predictability of the gradient, we define [JSGG19]

$$q_{\boldsymbol{\phi}}(\boldsymbol{H}_{\leq K-1}) \coloneqq p_{\boldsymbol{\theta}_{\mathcal{H}}}(\boldsymbol{H}_{\leq K-1} | \boldsymbol{Z}_{\leq K-1}) \qquad \text{(A.26)}$$

$$
\begin{aligned}
&= \prod_{\kappa=0}^{K-1} p_{\boldsymbol{\theta}_{\mathcal{H}}}(\boldsymbol{h}_\kappa | \boldsymbol{Z}_{\leq \kappa-1}, \boldsymbol{H}_{\leq \kappa-1,}) \\
&= \prod_{\kappa=0}^{K-1} \prod_{j \in \mathcal{H}} p_{\boldsymbol{\theta}_j}(z_{j,\kappa} | v_{j,\kappa}) \\
&= \prod_{\kappa=0}^{K-1} \prod_{j \in \mathcal{H}} \sigma(v_{j,t}) \,,
\end{aligned}
$$

where $\boldsymbol{\phi} = \boldsymbol{\theta}_{\mathcal{H}}$ with $\boldsymbol{\theta}_{\mathcal{H}}$ being the subset of model parameters linked to hidden neurons. Thus, by optimizing $\boldsymbol{\phi}$, we directly obtain $\boldsymbol{\theta}_{\mathcal{H}}$. It is noteworthy that this choice does not introduce any additional parameters.

Using the definition of $q_{\boldsymbol{\phi}}(\boldsymbol{H}_{\leq K-1})$, we can simplify the learning signal to

$$
\begin{aligned}
\ell_{\boldsymbol{\theta},\boldsymbol{\phi}}(\boldsymbol{Z}_{\leq K-1}, \boldsymbol{H}_{\leq K-1}) &= \ln \frac{p_{\boldsymbol{\theta}}(\boldsymbol{H}_{\leq K-1}, \boldsymbol{Z}_{\leq K-1})}{q_{\boldsymbol{\phi}}(\boldsymbol{H}_{\leq K-1})} \\
&= \ln \frac{p_{\boldsymbol{\theta}}(\boldsymbol{H}_{\leq K-1}, \boldsymbol{Z}_{\leq K-1})}{p_{\boldsymbol{\theta}_{\mathcal{H}}}(\boldsymbol{H}_{\leq K-1} | \boldsymbol{Z}_{\leq K-1})} \\
&= \ln \frac{\prod_{\kappa=0}^{K-1} \prod_{j \in \mathcal{N}} p_{\boldsymbol{\theta}_j}(z_{j,\kappa} | v_{j,\kappa})}{\prod_{\kappa=0}^{K-1} \prod_{j \in \mathcal{H}} p_{\boldsymbol{\theta}_j}(z_{j,\kappa} | v_{j,\kappa})} \\
&= \ln \prod_{\kappa=0}^{K-1} \prod_{j \in \mathcal{Z}} p_{\boldsymbol{\theta}_j}(z_{j,\kappa} | v_{j,\kappa}) \\
&= \sum_{\kappa=0}^{K-1} \sum_{j \in \mathcal{Z}} \ln p_{\boldsymbol{\theta}_j}(z_{j,\kappa} | v_{j,\kappa}) \\
&=: \ell_{\boldsymbol{\theta}_{\mathcal{Z}}}(\boldsymbol{Z}_{\leq K-1}) \,,
\end{aligned}
\tag{A.27}
$$

where we define $\ell_{\boldsymbol{\theta}_{\mathcal{Z}}}(\boldsymbol{Z}_{\leq K-1})$ as the learning signal, which we need to update $\boldsymbol{\phi}$ and $\boldsymbol{\theta}_{\mathcal{H}}$, respectively. Since $\boldsymbol{H}_{\leq K-1}$ and $\boldsymbol{Z}_{\leq K-1}$ denote the spike history of all neurons, we can write $p_{\boldsymbol{\theta}}(\boldsymbol{H}_{\leq K-1}, \boldsymbol{Z}_{\leq K-1}) = \prod_{\kappa=0}^{K-1} \prod_{j \in \mathcal{N}} p_{\boldsymbol{\theta}_j}(z_{j,\kappa} | v_{j,\kappa})$, where the joint probability is expressed by the product of the spike probability of each neuron. Note that with (A.26) as the choice of the variational posterior, the learning signal solely depends on the observable neurons. Furthermore, $\ell_{\boldsymbol{\theta}_{\mathcal{Z}}}(\boldsymbol{Z}_{\leq K-1}) \in \mathbb{R}$.

Analogous to $\boldsymbol{\theta}_{\mathcal{H}}$, we define $\boldsymbol{\theta}_{\mathcal{Z}}$ as the subset of model parameters linked to observable neurons. We can update $\boldsymbol{\theta}_{\mathcal{Z}}$ following the update rule of (A.25). Since we set $\boldsymbol{\theta}_{\mathcal{H}} = \boldsymbol{\phi}$, we can update $\boldsymbol{\theta}_{\mathcal{H}}$ following (A.24). Thus, for batch optimization using $|\mathcal{B}_\mathrm{t}|$ samples from the training set $\mathcal{B}_\mathrm{t}$, we can write

$$\boldsymbol{\theta}_\mathcal{Z} \leftarrow \boldsymbol{\theta}_\mathcal{Z} + \nu_{\boldsymbol{\theta}_\mathcal{Z}} \frac{1}{M} \sum_{m \in \mathcal{B}_\mathrm{t}} \sum_{\kappa=0}^{K-1} \sum_{j \in \mathcal{Z}} \nabla_{\boldsymbol{\theta}_j} \ln p_{\boldsymbol{\theta}_j} \left( Z_j^{(m)}[\kappa] | v_j^{(m)}[\kappa] \right) \tag{A.28}$$

$$\boldsymbol{\theta}_\mathcal{H} \leftarrow \boldsymbol{\theta}_\mathcal{H} + \nu_{\boldsymbol{\theta}_\mathcal{H}} \frac{1}{M} \sum_{m \in \mathcal{B}_\mathrm{t}} \ell_{\boldsymbol{\theta}_\mathcal{Z}}^{(m)}(\boldsymbol{Z}_{\leq K-1}) \sum_{\kappa=0}^{K-1} \sum_{j \in \mathcal{H}} \nabla_{\boldsymbol{\theta}_j} \ln p_{\boldsymbol{\theta}_j} \left( Z_j^{(m)}[\kappa] | v_j^{(m)}[\kappa] \right) . \tag{A.29}$$

Again, the superscript $m$ assigns the respective variable to the $m$-th sample of the training set. To obtain (A.29), the definition of the variational posterior (A.26) is used.

As discussed for fully observable SNNs, using $\nabla_{\boldsymbol{\theta}_j} \ln p_{\boldsymbol{\theta}_j}(Z_j[\kappa]|v_j[\kappa])$, the influence of each parameter on the membrane potential $v_j^{(m)}[\kappa]$ is tracked. By comparing the resulting spike probability $\sigma(v_j[\kappa])$ with the desired spike behavior $Z_j^{(m)}[\kappa]$, the corresponding updates, whether to increase or decrease a parameter, can be obtained. While for the output layer the desired spike behavior $\boldsymbol{Z}_{\leq K-1}$ is known, and therefore the update is similar to the update for the fully observable SNN, the desired hidden layer spike behavior $\boldsymbol{H}_{\leq K-1}$, which most likely results in the desired output spike behavior, is unknown.

We restate $\ell_{\boldsymbol{\theta}_\mathcal{Z}}(\boldsymbol{Z}_{\leq K-1})$ to provide deeper insight into its role and explanatory strength.

$$\begin{aligned} \ell_{\boldsymbol{\theta}_\mathcal{Z}}(\boldsymbol{Z}_{\leq K-1}) &= \ln \frac{p_{\boldsymbol{\theta}}(\boldsymbol{H}_{\leq K-1}, \boldsymbol{Z}_{\leq K-1})}{p_{\boldsymbol{\theta}_\mathcal{H}}(\boldsymbol{H}_{\leq K-1}|\boldsymbol{Z}_{\leq K-1})} \\ &= \ln p_{\boldsymbol{\theta}}(\boldsymbol{Z}_{\leq K-1}|\boldsymbol{H}_{\leq K-1}) + \ln p_{\boldsymbol{\theta}}(\boldsymbol{H}_{\leq K-1}) - \ln p_{\boldsymbol{\theta}_\mathcal{H}}(\boldsymbol{H}_{\leq K-1}|\boldsymbol{Z}_{\leq K-1}) \\ &= \ln p_{\boldsymbol{\theta}}(\boldsymbol{Z}_{\leq K-1}|\boldsymbol{H}_{\leq K-1}) - \ln \frac{p_{\boldsymbol{\theta}_\mathcal{H}}(\boldsymbol{H}_{\leq K-1}|\boldsymbol{Z}_{\leq K-1})}{p_{\boldsymbol{\theta}}(\boldsymbol{H}_{\leq K-1})} \end{aligned} \tag{A.30}$$

The first term indicates how likely it is to observe $\boldsymbol{Z}_{\leq K-1}$ given $\boldsymbol{H}_{\leq K-1}$. Consequently, hidden spike patterns $\boldsymbol{H}_{\leq K-1}$ supporting the desired output spike pattern $\boldsymbol{Z}_{\leq K-1}$ are rewarded. The second term penalizes hidden spike patterns $\boldsymbol{H}_{\leq K-1}$ that occur with a high posterior while having a small prior probability. Thus, we can interpret $\ell_{\boldsymbol{\theta}_\mathcal{Z}}(\boldsymbol{Z}_{\leq K-1})$ as a global reward, indicating the plausibility of hidden layer spike patterns $\boldsymbol{H}_{\leq K-1}$.

# List of Figures

# List of Tables

## Bibliography

[AAB+15] M. Abadi, A. Agarwal, P. Barham, E. Brevdo, Z. Chen, C. Citro, G. S. Corrado, A. Davis, J. Dean, M. Devin *et al.*, “TensorFlow: Large-scale machine learning on heterogeneous systems,” 2015, accessed: 2025-07-02. [Online]. Available: https://www.tensorflow.org/

[ABM+22a] E. Arnold, G. Böcherer, E. Müller, P. Spilger, J. Schemmel, S. Calabrò, and M. Kuschnerow, “Spiking neural network equalization for IM/DD optical communication,” in *Proc. Advanced Photonics Congress (APC): Signal Processing in Photonic Communications (SPPCom)*, Maastricht, NL, 2022.

[ABM+22b] E. Arnold, G. Böcherer, E. Müller, P. Spilger, J. Schemmel, S. Calabrò, and M. Kuschnerov, “Spiking neural network equalization on neuromorphic hardware for IM/DD optical communication,” in *Proc. European Conference on Optical Communication (ECOC)*, Basel, Switzerland, 2022.

[ABS+23] E. Arnold, G. Böcherer, F. Strasser, E. Müller, P. Spilger, S. Billaudelle, J. Weis, J. Schemmel, S. Calabrò, and M. Kuschnerov, “Spiking neural network nonlinear demapping on neuromorphic hardware for IM/DD optical communication,” *Journal of Lightwave Technology*, vol. 41, no. 11, pp. 3424–3431, 2023.

[ACR20] F. Al-Tam, N. Correia, and J. Rodriguez, “Learn to schedule (LEASCH): A deep reinforcement learning approach for radio resource scheduling in the 5G MAC layer,” *IEEE Access*, vol. 8, pp. 108 088–108 101, 2020.

[AEvB+25] E. Arnold, E.-M. Edelmann, A. von Bank, E. Müller, L. Schmalen, and J. Schemmel, “Short-reach optical communications: A real-world task for neuromorphic hardware,” in *Proc. Neuro Inspired Computational Elements Conference (NICE)*, Heidelberg, Germany, Mar. 2025.

[AH18] F. Ait Aoudia and J. Hoydis, “End-to-end learning of communications systems without a channel model,” in *Proc. Asilomar Conference on Signals, Systems, and Computers*, Pacific Grove, CA, USA, 2018, pp. 298–303.

[AHMK21] D. Auge, J. Hille, E. Müller, and A. Knoll, “A survey of encoding techniques for signal processing in spiking neural networks,” *Neural Processing Letters*, vol. 53, pp. 4693–4710, 2021.

[ALL19] Z. Allen-Zhu, Y. Li, and Y. Liang, “Learning and generalization in overparameterized neural networks, going beyond two layers,” in *Proc. Advances in Neural Information Processing Systems (NeurIPS)*, Vancouver, Canada, 2019, pp. 6155–6166.

[ANKC25] M. A. Amirabadi, S. A. Nezamalhosseini, M. H. Kahaei, and L. R. Chen, "A comprehensive survey on machine and deep learning for optical communications," *IEEE Access*, vol. 13, pp. 88 794–88 846, 2025.

[ASC+15] F. Akopyan, J. Sawada, A. Cassidy, R. Alvarez-Icaza, J. Arthur, P. Merolla, N. Imam, Y. Nakamura, P. Datta, G.-J. Nam *et al.*, "Truenorth: Design and tool flow of a 65 mw 1 million neuron programmable neurosynaptic chip," *IEEE Transactions on Computer-Aided Design of Integrated Circuits and Systems*, vol. 34, no. 10, pp. 1537–1557, 2015.

[BB24] C. M. Bishop and H. Bishop, *Deep Learning: Foundations and Concepts.* Cham, Switzerland: Springer International Publishing, 2024.

[BBH+14] T. Bekolay, J. Bergstra, E. Hunsberger, T. DeWolf, T. Stewart, D. Rasmussen, X. Choo, A. Voelker, and C. Eliasmith, "Nengo: A Python tool for building large-scale functional brain models," *Frontiers in Neuroinformatics*, vol. 7, no. 48, 2014.

[BES21] E.-M. Bansbach, V. Eliachevitch, and L. Schmalen, "Deep reinforcement learning for wireless resource allocation using buffer state information," in *Proc. IEEE Global Communications Conference (GLOBECOM)*, Madrid, Spain, 2021.

[Bie20] L. Biewald, "Experiment tracking with weights and biases," 2020, accessed: 2025-04-03. [Online]. Available: https://www.wandb.com/

[BKS22] E.-M. Bansbach, Y. Kiyak, and L. Schmalen, "Deep reinforcement learning for uplink multi-carrier non-orthogonal multiple access resource allocation using buffer state information," in *Proc. European Wireless*, Dresden, Germany, 2022.

[BLK02] S. Bohte, H. La Poutre, and J. Kok, "Unsupervised clustering with spiking neurons by sparse temporal coding and multilayer rbf networks," *IEEE Transactions on Neural Networks*, vol. 13, no. 2, pp. 426–435, 2002.

[BP98] G.-Q. Bi and M.-M. Poo, "Synaptic modifications in cultured hippocampal neurons: dependence on spike timing, synaptic strength, and postsynaptic cell type," *Journal of Neuroscience*, vol. 18, no. 24, pp. 10 464–10 472, 1998.

[Bre15] R. Brette, "Philosophy of the spike: Rate-based vs. spike-based theories of the brain," *Frontiers in Systems Neuroscience*, vol. 9, 2015.

[BSA+23] G. Böcherer, F. Strasser, E. Arnold, Y. Lin, J. Schemmel, S. Calabrò, and M. Kuschnerov, "Spiking neural network linear equalization: Experimental demonstration of 2km 100Gb/s IM/DD PAM4 optical

transmission," in *Proc. Optical Fiber Communications Conference (OFC)*, San Diego, CA, USA, 2023.

[BSS+18] G. Bellec, D. Salaj, A. Subramoney, R. Legenstein, and W. Maass, "Long short-term memory and learning-to-learn in networks of spiking neurons," in *Proc. Advances in Neural Information Processing Systems (NIPS)*, 2018.

[BvBS23] E.-M. Bansbach, A. von Bank, and L. Schmalen, "Spiking neural network decision feedback equalization," in *Proc. International ITG Workshop on Smart Antennas and Conference on Systems, Communications, and Coding (WSA-SCC)*, Braunschweig, Germany, 2023.

[BVM+19] M. Bouvier, A. Valentian, T. Mesquida, F. Rummens, M. Reyboz, E. Vianello, and E. Beigne, "Spiking neural networks hardware implementations and challenges: A survey," *Journal of Emerging Technologies in Computing Systems*, vol. 15, no. 2, Apr. 2019.

[BvR99] N. Brunel and M. C. van Rossum, "Lapicque's introduction of the integrate-and-fire model neuron (1907)," *Brain Research Bulletin*, vol. 50, no. 5-6, pp. 303–304, 1999.

[CB13] S. Chandra and A. K. Bharti, "Speed distribution curves for pedestrians during walking and crossing," *Procedia - Social and Behavioral Sciences*, vol. 104, pp. 660–667, 2013.

[Cha19] M. Chagnon, "Optical communications for short reach," *Journal of Lightwave Technology*, vol. 37, no. 8, pp. 1779–1797, 2019.

[CSSZ22] B. Cramer, Y. Stradmann, J. Schemmel, and F. Zenke, "Surrogate gradients for analog neuromorphic computing," *Proc. National Academy of Sciences of the United States of America*, vol. 119, no. 4, 2022.

[CV93] R. S. Cheng and S. Verdú, "Gaussian multiaccess channels with ISI: Capacity region and multiuser water-filling," *IEEE Transactions on Information Theory*, vol. 39, no. 3, pp. 773–785, May 1993.

[CVYA21] D. F. Carrera, C. Vargas-Rosales, N. M. Yungaicela-Naula, and L. Azpilicueta, "Comparative study of artificial neural network based channel equalization methods for mmWave communications," *IEEE Access*, vol. 9, pp. 41 678–41 687, 2021.

[DA05] P. Dayan and L. F. Abbott, *Theoretical Neuroscience: Computational and Mathematical Modeling of Neural Systems.* Cambridge, MA, USA: MIT Press, 2005.

[DC15] P. U. Diehl and M. Cook, "Unsupervised learning of digit recognition using spike-timing-dependent plasticity," *Frontiers in Computational Neuroscience*, vol. 9, 2015.

[DDZ+20] H. Dong, Z. Ding, S. Zhang, H. Yuan, H. Zhang, J. Zhang, Y. Huang, T. Yu, H. Zhang, and R. Huang, *Deep Reinforcement Learning: Fundamentals, Research, and Applications.* Singapore: Springer Nature, 2020.

[DL07] T. Delbruck and P. Lichtsteiner, “Fast sensory motor control based on event-based hybrid neuromorphic-procedural system,” in *Proc. IEEE International Symposium on Circuits and Systems*, New Orleans, LA, USA, May 2007, pp. 845–848.

[dLST+17] T. F. de Lima, B. J. Shastri, A. N. Tait, M. A. Nahmias, and P. R. Prucnal, “Progress in neuromorphic photonics,” *Nanophotonics*, vol. 6, no. 3, pp. 577–599, 2017.

[DNB+15] P. U. Diehl, D. Neil, J. Binas, M. Cook, S.-C. Liu, and M. Pfeiffer, “Fast-classifying, high-accuracy spiking deep networks through weight and threshold balancing,” in *Proc. 2015 International Joint Conference on Neural Networks (IJCNN)*, Killarney, Ireland, 2015.

[DSL+18] M. Davies, N. Srinivasa, T.-H. Lin, G. Chinya, Y. Cao, S. H. Choday, G. Dimou, P. Joshi, N. Imam, S. Jain *et al.*, “Loihi: A neuromorphic manycore processor with on-chip learning,” *IEEE Micro*, vol. 38, no. 1, pp. 82–99, 2018.

[DWL+20] L. Deng, G. Wang, G. Li, S. Li, L. Liang, M. Zhu, Y. Wu, Z. Yang, Z. Zou, J. Pei *et al.*, “Tianjic: A unified and scalable chip bridging spike-based and continuous neural computation,” *IEEE Journal of Solid-State Circuits*, vol. 55, no. 8, pp. 2228–2246, 2020.

[EvBS25] E.-M. Edelmann, A. von Bank, and L. Schmalen, “Encoding optimization for low-complexity spiking neural network equalizers in IM/DD systems,” in *Proc. European Conference on Optical Communication (ECOC)*, Copenhagen, Denmark, Sep. 2025.

[EWN+23] J. K. Eshraghian, M. Ward, E. Neftci, X. Wang, G. Lenz, G. Dwivedi, M. Bennamoun, D. S. Jeong, and W. D. Lu, “Training spiking neural networks using lessons from deep learning,” *Proc. IEEE*, vol. 111, no. 9, pp. 1016–1054, 2023.

[FCD+23] W. Fang, Y. Chen, J. Ding, Z. Yu, T. Masquelier, D. Chen, L. Huang, H. Zhou, G. Li, and Y. Tian, “SpikingJelly: An open-source machine learning infrastructure platform for spike-based intelligence,” *Science Advances*, vol. 9, no. 40, 2023.

[FG16] N. Frémaux and W. Gerstner, “Neuromodulated spike-timing-dependent plasticity, and theory of three-factor learning rules,” *Frontiers in Neural Circuits*, vol. 9, 2016.

[FLP+13] S. B. Furber, D. R. Lester, L. A. Plana, J. D. Garside, E. Painkras, S. Temple, and A. D. Brown, “Overview of the SpiNNaker system

architecture," *IEEE Transactions on Computers*, vol. 62, no. 12, pp. 2454–2467, 2013.

[FOS+21] P. J. Freire, Y. Osadchuk, B. Spinnler, A. Napoli, W. Schairer, N. Costa, J. E. Prilepsky, and S. K. Turitsyn, "Performance versus complexity study of neural network equalizers in coherent optical systems," *Journal of Lightwave Technology*, vol. 39, no. 19, pp. 6085–6096, 2021.

[FTD+16] C. Finn, X. Y. Tan, Y. Duan, T. Darrell, S. Levine, and P. Abbeel, "Deep spatial autoencoders for visuomotor learning," in *Proc. IEEE International Conference on Robotics and Automation (ICRA)*, Stockholm, Sweden, 2016, pp. 512–519.

[FTR+06] R. C. Froemke, I. A. Tsay, M. Raad, J. D. Long, and Y. Dan, "Contribution of individual spikes in burst-induced long-term synaptic modification," *Journal of Neurophysiology*, vol. 95, no. 3, pp. 1620–1629, March 2006.

[GBC16] I. Goodfellow, Y. Bengio, and A. Courville, *Deep Learning.* Cambridge, MA, USA: MIT Press, 2016.

[GBLB12] I. Grondman, L. Bușoniu, G. A. D. Lopes, and R. Babuška, "A survey of actor-critic reinforcement learning: Standard and natural policy gradients," *IEEE Transactions on Systems, Man, and Cybernetics, Part C: Applications and Reviews*, vol. 42, no. 6, pp. 1291–1307, 2012.

[GD07] M.-O. Gewaltig and M. Diesmann, "NEST (NEural Simulation Tool)," *Scholarpedia*, vol. 2, no. 4, p. 1430, 2007.

[GK02] W. Gerstner and W. M. Kistler, *Spiking Neuron Models: Single Neurons, Populations, Plasticity.* Cambridge, UK: Cambridge University Press, 2002.

[GKNP14] W. Gerstner, W. M. Kistler, R. Naud, and L. Paninski, *Neuronal Dynamics: From Single Neurons to Networks and Models of Cognition.* Cambridge, UK: Cambridge University Press, 2014.

[HDK+99] M. Hough, H. De Garis, M. Korkin, F. Gers, and N. E. Nawa, "Spiker: Analog waveform to digital spiketrain conversion in atr's artificial brain (cam-brain) project," in *Proc. International Conference on Robotics and Artificial Life*, 1999.

[Heb49] D. O. Hebb, *The Organization of Behavior.* New York, NY, USA: Wiley, 1949.

[Hee00] D. J. Heeger, "Poisson model of spike generation," 2000, accessed: 2024-09-04. [Online]. Available: http://www.cns.nyu.edu/~david/handouts/poisson.pdf

[HFND15] M.-R. Hojeij, J. Farah, C. A. Nour, and C. Douillard, "Resource allocation in downlink non-orthogonal multiple access (NOMA) for future radio access," in *Proc. IEEE Vehicular Technology Conference - Spring (VTC Spring)*, Glasgow, Sctoland, 2015.

[HH52] A. L. Hodgkin and A. F. Huxley, "A quantitative description of membrane current and its application to conduction and excitation in nerve," *The Journal of Physiology*, vol. 117, no. 4, pp. 500–544, 1952.

[HSK+18] H. Hazan, D. J. Saunders, H. Khan, D. Patel, D. T. Sanghavi, H. T. Siegelmann, and R. Kozma, "BindsNET: A machine learning-oriented spiking neural networks library in python," *Frontiers in Neuroinformatics*, vol. 12, 2018.

[HZS06] G.-B. Huang, Q.-Y. Zhu, and C.-K. Siew, "Extreme learning machine: Theory and applications," *Neurocomputing*, vol. 70, no. 1-3, pp. 489–501, 2006.

[ICK+22] D. Ivanov, A. Chezhegov, M. Kiselev, A. Grunin, and D. Larionov, "Neuromorphic artificial intelligence systems," *Frontiers in Neuroscience*, vol. 16, 2022.

[Int24] Intel Labs, "Lava software framework," 2024, accessed: 2024-09-27. [Online]. Available: https://lava-nc.org

[Izh03] E. M. Izhikevich, "Simple model of spiking neurons," *IEEE Transactions on Neural Networks*, vol. 14, no. 6, pp. 1569–1572, 2003.

[Izh04] E. M. Izhikevich, "Which model to use for cortical spiking neurons?" *IEEE Transactions on Neural Networks*, vol. 15, no. 5, pp. 1063–1070, Sep. 2004.

[JHHS21] W. Jiang, B. Han, M. A. Habibi, and H. D. Schotten, "The road towards 6G: A comprehensive survey," *IEEE Open Journal of the Communications Society*, vol. 2, pp. 334–366, 2021.

[JSGG19] H. Jang, O. Simeone, B. Gardner, and A. Gruning, "An introduction to probabilistic spiking neural networks: Probabilistic models, learning rules, and applications," *IEEE Signal Processing Magazine*, vol. 36, no. 6, pp. 64–77, 2019.

[JSS21] H. Jang, N. Skatchkovsky, and O. Simeone, "Spiking Neural Networks – Part I: Detecting spatial patterns," *IEEE Communications Letters*, vol. 25, no. 6, pp. 1736–1740, 2021.

[JTG04] R. Jolivet, J. W. Timothy, and W. Gerstner, "The spike response model: A framework to predict neuronal spike trains," *Neurocomputing*, vol. 58–60, pp. 271–276, 2004.

[Kas21] N. Kasabov, *Time-Space, Spiking Neural Networks and Brain-Inspired Artificial Intelligence.* Cham, Switzerland: Springer, 2021.

[Kel97] F. Kelly, "Charging and rate control for elastic traffic," *European Transactions on Telecommunications*, vol. 8, pp. 33–37, 1997.

[KT99] V. Konda and J. Tsitsiklis, "Actor-critic algorithms," in *Proc. Advances in Neural Information Processing Systems (NeurIPS)*, vol. 12. Denver, Colorado, USA: MIT Press, 1999.

[LBCS24] S. Li, G. Böcherer, S. Calabrò, and M. Schädler, "Spiking neural network equalizer with fast and low power decoding for IM/DD optical communication," *IEEE Photonics Technology Letters*, vol. 36, no. 17, pp. 1061–1064, 2024.

[LDP16] J. H. Lee, T. Delbruck, and M. Pfeiffer, "Training deep spiking neural networks using backpropagation," *Frontiers in Neuroscience*, vol. 10, 2016.

[LHG+19] N. C. Luong, D. T. Hoang, S. Gong, D. Niyato, P. Wang, Y.-C. Liang, and D. I. Kim, "Applications of deep reinforcement learning in communications and networking: A survey," *IEEE Commun. Surveys Tuts.*, vol. 21, no. 4, pp. 3133–3174, 2019.

[LHP+16] T. P. Lillicrap, J. J. Hunt, A. Pritzel, N. Heess, T. Erez, Y. Tassa, D. Silver, and D. Wierstra, "Continuous control with deep reinforcement learning," in *Proc. International Conference on Learning Representations (ICLR)*, San Juan, Puerto Rico, USA, May 2016.

[LHS+23] G. Lenz, K. Heckel, S. B. Shrestha, C. Barker, and J. E. Pedersen. (2023) Spiking neural network framework benchmarking. Accessed: 2025-07-07. [Online]. Available: https://open-neuromorphic.org/blog/spiking-neural-network-framework-benchmarking/

[Llo82] S. Lloyd, "Least squares quantization in PCM," *IEEE Transactions on Information Theory*, vol. 28, no. 2, pp. 129–137, 1982.

[LPM+22] R.-Z. Liu, Z.-J. Pang, Z.-Y. Meng, W. Wang, Y. Yu, and T. Lu, "On efficient reinforcement learning for full-length game of StarCraft II," *Journal of Artificial Intelligence Research*, vol. 75, pp. 213–260, 2022.

[Mal25] H. A. Mallot, *Computational Neuroscience: An Essential Guide to Membrane Potentials, Receptive Fields, and Neural Networks*, 2nd ed. Cham, Switzerland: Springer, 2025.

[Mas97] W. Mass, "Networks of spiking neurons: The third generation of neural network models," *Neural Networks*, vol. 10, no. 9, pp. 1659–1671, 1997.

[MG14] A. Mnih and K. Gregor, "Neural variational inference and learning in belief networks," in *Proc. 31st International Conference on Machine Learning*, Bejing, China, Jun 2014, pp. 1791–1799.

[MHF19] C. Mayr, S. Höppner, and S. Furber, "SpiNNaker 2: A 10 million core processor system for brain simulation and machine learning," *arXiv preprint*, 2019. [Online]. Available: https://arxiv.org/abs/1911.02385

[MKS+15] V. Mnih, K. Kavukcuoglu, D. Silver, A. A. Rusu, J. Veness, M. G. Bellemare, A. Graves, M. Riedmiller, A. K. Fidjeland, G. Ostrovski *et al.*, "Human-level control through deep reinforcement learning," *Nature*, vol. 518, no. 7540, pp. 529–533, 2015.

[MM91] M. A. Mahowald and C. Mead, "The silicon retina," *Scientific American*, vol. 264, no. 5, pp. 76–83, May 1991.

[MNHW24] M. Moursi, J. Ney, B. Hammoud, and N. Wehn, "Efficient FPGA implementation of an optimized SNN-based DFE for optical communications," in *Proc. IEEE Middle East Conference on Communications and Networking (MECOM)*, Abu Dhabi, UAE, 2024, pp. 47–52.

[MPC+24] M. W. Mathis, A. Perez Rotondo, E. F. Chang, A. S. Tolias, and A. Mathis, "Decoding the brain: From neural representations to mechanistic models," *Cell*, vol. 187, no. 21, pp. 5814–5832, 2024.

[MTT+20] B. Mukherjee, I. Tomkos, M. Tornatore, P. J. Winzer, and Y. L. Zhao, *Springer Handbook of Optical Networks.* Cham, Switzerland: Springer, 2020.

[Mur12] K. P. Murphy, *Machine Learning: A Probabilistic Perspective.* Cambridge, MA, USA: MIT Press, 2012.

[MVSK+23] D. L. Manna, A. Vicente-Sola, P. Kirkland, T. J. Bihl, and G. D. Caterina, "Frameworks for SNNs: A review of data science-oriented software and an expansion of SpykeTorch," in *Proc. Engineering Applications of Neural Networks (EANN).* Cham, Switzerland: Springer, 2023, pp. 227–238.

[MYC+20] O. Moreira, A. Yousefzadeh, F. Chersi, G. Cinserin, R.-J. Zwartenkot, A. Kapoor, P. Qiao, P. Kievits, M. Khoei, L. Rouillard *et al.*, "Neuronflow: A neuromorphic processor architecture for live AI applications," in *Proc. Design, Automation & Test in Europe Conference & Exhibition (DATE)*, Virtual, 2020, pp. 840–845.

[NMZ19] E. O. Neftci, H. Mostafa, and F. Zenke, "Surrogate gradient learning in spiking neural networks," *IEEE Signal Processing Magazine*, vol. 36, no. 6, 2019.

[Nok20] Nokia, "Wireless-suite," 2020, (accessed on: 17.07.2025). [Online]. Available: https://github.com/nokia/wireless-suite

[OFR+21] G. Orchard, E. P. Frady, D. B. D. Rubin, S. Sanborn, S. Shrestha, F. T. Sommer, and M. Davies, "Efficient neuromorphic signal pro-

cessing with Loihi 2," in *Proc. IEEE Workshop on Signal Processing Systems (SiPS)*, Virtual, 2021, pp. 254–259.

[PBC+22] C. Pehle, S. Billaudelle, B. Cramer, J. Kaiser, K. Schreiber, Y. Stradmann, J. Weis, A. Leibfried, E. Müller, and J. Schemmel, "The BrainScaleS-2 accelerated neuromorphic system with hybrid plasticity," *Frontiers in Neuroscience*, vol. 16, 2022.

[PdFC+25] D. G. S. Pivoto, F. A. P. de Figueiredo, C. Cavdar, G. R. d. Tejerina, and L. L. Mendes, "A comprehensive survey of machine learning applied to resource allocation in wireless communications," *IEEE Communications Surveys & Tutorials*, 2025.

[PGM+19] A. Paszke, S. Gross, F. Massa, A. Lerer, J. Bradbury, G. Chanan, T. Killeen, Z. Lin, N. Gimelshein, L. Antiga *et al.*, "PyTorch: An imperative style, high-performance deep learning library," in *Proc. Advances in Neural Information Processing Systems (NeurIPS)*, Red Hook, NY, USA, 2019, pp. 8024–8035.

[PKK20] B. Petro, N. Kasabov, and R. M. Kiss, "Selection and optimization of temporal spike encoding methods for spiking neural networks," *IEEE Transactions on Neural Networks and Learning Systems*, vol. 31, no. 2, pp. 358–370, 2020.

[PP18] M. Pfeiffer and T. Pfeil, "Deep learning with spiking neurons: Opportunities and challenges," *Frontiers in Neuroscience*, vol. 12, p. 774, 2018.

[PP21] C. Pehle and J. E. Pedersen, "Norse - A deep learning library for spiking neural networks," 2021, accessed: 2024-12-10. [Online]. Available: https://doi.org/10.5281/zenodo.4422025

[PPU+05] J. W. Pillow, L. Paninski, V. J. Uzzell, E. P. Simoncelli, and E. J. Chichilnisky, "Prediction and decoding of retinal ganglion cell responses with a probabilistic spiking model," *Journal of Neuroscience*, vol. 25, no. 47, 2005.

[PS08] J. G. Proakis and M. Salehi, *Digital Communications*, 5th ed. New York, NY, USA: McGraw-Hill, 2008.

[PSHP23] P. Pietrzak, S. Szczęsny, D. Huderek, and L. Przyborowski, "Overview of spiking neural network learning approaches and their computational complexities," *Sensors*, vol. 23, no. 6, p. 3037, 2023.

[Put14] M. L. Puterman, *Markov Decision Processes: Discrete Stochastic Dynamic Programming.* Wiley-Interscience, 2014.

[Ras18] D. Rasmussen, "NengoDL: Combining deep learning and neuromorphic modelling methods," *arXiv*, 2018. [Online]. Available: http://arxiv.org/abs/1805.11144

[RC00] W. Rhee and J. M. Cioffi, "Increase in capacity of multiuser OFDM system using dynamic subchannel allocation," in *Proc. IEEE Vehicular Technology Conference - Spring (VTC Spring)*, Tokyo, Japan, May 2000, pp. 1085–1089.

[RHW86] D. E. Rumelhart, G. E. Hinton, and R. J. Williams, "Learning representations by back-propagating errors," *Nature*, vol. 323, no. 6088, pp. 533–536, 1986.

[Ros60] H. H. Rosenbrock, "An automatic method for finding the greatest or least value of a function," *The Computer Journal*, vol. 3, no. 3, pp. 175–184, 1960.

[RT11] E. T. Rolls and A. Treves, "The neuronal encoding of information in the brain," *Progress in Neurobiology*, vol. 95, no. 3, pp. 448–490, Nov. 2011.

[RVS97] R. C. Reid, J. D. Victor, and R. M. Shapley, "The use of m-sequences in the analysis of visual neurons: Linear receptive field properties," *Visual Neuroscience*, vol. 14, no. 6, pp. 1015–1027, 1997.

[RW02] J. N. Reynolds and J. R. Wickens, "Dopamine-dependent plasticity of corticostriatal synapses," *Neural Networks*, vol. 15, no. 4–6, pp. 507–521, 2002.

[SB18] R. S. Sutton and A. G. Barto, *Reinforcement Learning: An Introduction*, 2nd ed. Cambridge, MA, USA: The MIT Press, 2018.

[SBG19] M. Stimberg, R. Brette, and D. F. M. Goodman, "Brian 2, an intuitive and efficient neural simulator," *eLife*, vol. 8, Aug 2019.

[SBL14] F. Shams, G. Bacci, and M. Luise, "A survey on resource allocation techniques in OFDM(A) networks," *Computer Networks*, vol. 65, pp. 129–150, June 2014.

[SFM$^+$22] A. Shrestha, H. Fang, Z. Mei, D. P. Rider, Q. Wu, and Q. Qiu, "A survey on neuromorphic computing: Models and hardware," *IEEE Circuits and Systems Magazine*, vol. 22, no. 2, pp. 6–35, 2022.

[SHM$^+$16] D. Silver, A. Huang, C. J. Maddison, A. Guez, L. Sifre, G. Van Den Driessche, J. Schrittwieser, I. Antonoglou, V. Panneershelvam, M. Lanctot *et al.*, "Mastering the game of Go with deep neural networks and tree search," *Nature*, vol. 529, no. 7587, pp. 484–489, 2016.

[Sim18] O. Simeone, "A brief introduction to machine learning for engineers," *Found. Trends Signal Process.*, vol. 12, no. 3–4, p. 200–431, Aug. 2018.

[SLH$^+$14] D. Silver, G. Lever, N. Heess, T. Degris, W. Daan, and M. Reidmiller, "Deterministic policy gradient algorithms," in *Proc. International Conference on Machine Learning (ICML)*, Bejing, China, June 2014.

[SLN+25] L. Schmalen, V. Lauinger, J. Ney, N. Wehn, P. Matalla, S. Randel, A. von Bank, and E. M. Edelmann, "Recent advances on machine learning-aided DSP for short-reach and long-haul optical communications," in *Proc. Optical Fiber Communications Conference (OFC)*, San Francisco, CA, USA, Mar. 2025.

[SMSM99] R. S. Sutton, D. McAllester, S. Singh, and Y. Mansour, "Policy gradient methods for reinforcement learning with function approximation," in *Proc. Advances in Neural Information Processing Systems (NeurIPS)*, vol. 12. Denver, CO, USA: MIT Press, 1999, pp. 1057–1063.

[SPP+17] C. D. Schuman, T. E. Potok, R. M. Patton, J. D. Birdwell, M. E. Dean, G. S. Rose, and J. S. Plank, "A survey of neuromorphic computing and neural networks in hardware," *arXiv preprint*, 2017. [Online]. Available: https://arxiv.org/abs/1705.06963

[SV03] B. Schrauwen and J. Van Campenhout, "BSA, a fast and accurate spike train encoding scheme," in *Proc. International Joint Conference on Neural Networks (IJCNN)*, vol. 4. Portland, OR, USA: IEEE, 2003, pp. 2825–2830.

[SvHHS18] S. Schmitt, H. van Hasselt, M. Hessel, and D. Silver, "Kickstarting deep reinforcement learning," in *Proc. Advances in Neural Information Processing Systems (NeurIPS)*, Montréal, Canada, Dec. 2018.

[Ten17] TensorFlow Team, "Introducing TensorFlow Feature Columns," Google Developers Blog, 2017, (accessed on: 10.04.2021). [Online]. Available: https://developers.googleblog.com/2017/11/introducing-tensorflow-feature-columns.html

[TG98] S. Thorpe and J. Gautrais, "Rank order coding," in *Computational Neuroscience*. Boston, MA, USA: Springer US, 1998, pp. 113–118.

[TGK+19] E. Tavanaei, M. Ghodrati, S. R. Kheradpisheh, T. Masquelier, and A. Maida, "Deep learning in spiking neural networks," *Neural Networks*, vol. 111, pp. 47–63, 2019.

[Val20] A. Valcarce, "The TimeFreqResourceAllocation-v0 environment," 2020, (accessed on: 17.07.2025). [Online]. Available: https://github.com/nokia/wireless-suite/blob/master/wireless/doc/TimeFreqResourceAllocation-v0.pdf

[vBEM+25] A. von Bank, E.-M. Edelmann, S. Miao, J. Mandelbaum, and L. Schmalen, "Spiking neural belief propagation decoder for short block length LDPC codes," *IEEE Commun. Lett.*, vol. 29, no. 1, pp. 45–49, Jan. 2025.

[vBEMS25] A. von Bank, E.-M. Edelmann, J. Mandelbaum, and L. Schmalen, "Spiking neural belief propagation decoder for LDPC codes with small variable node degrees," in *Proc. Intl. ITG Conf. on Systems,*

*Communications, and Coding (SCC)*, Karlsruhe, Germany, Mar. 2025.

[vBES23] A. von Bank, E.-M. Edelmann, and L. Schmalen, "Spiking neural network decision feedback equalization for IM/DD systems," in *Proc. Advanced Photonic Congress: Signal Processing in Photonic Communications (SPPCom)*, Busan, South Korea, Jul. 2023.

[vBES24] A. von Bank, E.-M. Edelmann, and L. Schmalen, "Energy-efficient spiking neural network equalization for IM/DD systems with optimized neural encoding," in *Proc. Optical Fiber Communications Conference (OFC)*, San Diego, CA, USA, Mar. 2024.

[WLN+21] S. Wang, T. Lv, W. Ni, N. C. Beaulieu, and Y. J. Guo, "Joint resource management for MC-NOMA: A deep reinforcement learning approach," *IEEE Transactions on Wireless Communications*, vol. 20, no. 9, pp. 5672–5688, 2021.

[WXH+19] J. Wang, C. Xu, Y. Huangfu, R. Li, Y. Ge, and J. Wang, "Deep reinforcement learning for scheduling in cellular networks," in *Proc. IEEE Workshop on Constraint Satisfaction Problems (WCSP)*, Xi'an, China, Oct. 2019.

[XWY+20] C. Xu, J. Wang, T. Yu, C. Kong, Y. Huangfu, R. Li, Y. Ge, and J. Wang, "Buffer-aware wireless scheduling based on deep reinforcement learning," in *Proc. 2020 IEEE Wireless Communications and Networking Conference (WCNC)*, Virtual, 2020.

[XYHZ20] Y.-H. Xu, C.-C. Yang, M. Hua, and W. Zhou, "Deep deterministic policy gradient (DDPG)-based resource allocation scheme for NOMA vehicular communications," *IEEE Access*, vol. 8, pp. 18 797–18 807, 2020.

[YBW24] Z. Yan, Z. Bai, and W.-F. Wong, "Reconsidering the energy efficiency of spiking neural networks," *arXiv preprint*, 2024, [Online]. Available: https://arxiv.org/abs/2409.08290.

[YLZ+21] Y. Yuan, Y. Liu, Z. Zhang, W. Li, Y. Zhang, L. Dai, Z. Chen, and L. Hanzo, "Noma for next-generation massive IoT: Performance potential and technology directions," *IEEE Communications Magazine*, vol. 59, no. 7, pp. 115–121, 2021.

[ZBC+21] F. Zenke, S. M. Bohté, C. Clopath, I. M. Comşa, J. Göltz, W. Maass, T. Masquelier, R. Naud, E. O. Neftci, M. A. Petrovici *et al.*, "Visualizing a joint future of neuroscience and neuromorphic engineering," *Neuron*, vol. 109, no. 4, pp. 571–575, 2021.

[ZG18] F. Zenke and S. Ganguli, "Superspike: Supervised learning in multilayer spiking neural networks," *Neural Computation*, vol. 30, no. 6, pp. 1514–1541, Jun 2018.

[ZLYZ19] R. Zeng, T. Liu, X. Yu, and Z. Zhang, "Novel channel quality indicator prediction scheme for adaptive modulation and coding in high mobility environments," *IEEE Access*, vol. 7, pp. 11 543–11 553, 2019.

[ZN21] F. Zenke and E. O. Neftci, "Brain-inspired learning on neuromorphic substrates," *Proc. IEEE*, vol. 109, no. 5, pp. 935–950, 2021.

[ZVS+18] Y. Zhang, Q. H. Vuong, K. Song, X.-Y. Gong, and K. W. Ross, "Efficient entropy for policy gradient with multi-dimensional action space," in *Proc. International Conference on Learning Representations (ICLR)*, Vancouver, Canada, May 2018.